\documentclass[a4paper,11pt]{article}
\usepackage{jheppub} 

\makeatletter
\def\@fpheader{}
\makeatother

\usepackage{units}
\usepackage{xcolor}
\usepackage{array}
\usepackage{graphicx}
\usepackage{subcaption}
\usepackage{longtable}
\usepackage{multicol}
\usepackage{setspace}
\usepackage{comment}
\usepackage{pdfpages}
\usepackage{slashed}
\usepackage{graphicx}

\usepackage{breqn}
\usepackage{amsmath}

\usepackage{diagbox}

\usepackage{pdflscape}
\usepackage{makecell}
\usepackage{array}

\usepackage{mathrsfs}

\usepackage{tabularx}

\def \L {\mathcal{L}} 

\def \vec#1{{\boldsymbol{#1}}}
\newcommand{\ii}{{\rm i}}
\newcommand{\hc}{\ensuremath{\text{h.c.}}}
\newcommand{\isep}{\mathrel{{.}\,{.}}\nobreak}

\newcolumntype{L}{>{$}l<{$}} 
\newcolumntype{C}{>{$}c<{$}}
\newcolumntype{P}{>{$}p<{$}}

\newcommand{\website}{\href{https://uva.theopenscholar.com/heeck/BNV}{uva.theopenscholar.com/heeck/BNV}}

\allowdisplaybreaks

\title{Opening up derivative baryon-number-violating operators}

\author[1]{Julian Heeck,%
\note{ORCID: \href{https://orcid.org/0000-0003-2653-5962}{0000-0003-2653-5962}}}
\emailAdd{heeck@virginia.edu}

\author[2]{Diana Sokhashvili,%
\note{ORCID: \href{https://orcid.org/0009-0002-0428-6264}{0009-0002-0428-6264}}}
\emailAdd{bkq5jz@virginia.edu}

\affiliation{Department of Physics, University of Virginia,
Charlottesville, Virginia 22904-4714, USA}

\abstract{
Baryon number violation is our most sensitive probe of physics beyond the Standard Model, typically encoded in higher-dimensional effective operators. Operators involving derivatives are often ignored in phenomenological analyses since they are generically suppressed compared to non-derivative operators. We systematically construct tree-level UV completions of derivative baryon-number-violating operators up to mass dimension 12 in the Standard Model effective field theory (extended by right-handed neutrinos) and show that derivative operators need not be suppressed. Our public code can be used to UV-complete other derivative operators.
}

\begin{document}

\include{Tables}
\include{ShortTables}

\maketitle
\flushbottom
\clearpage
\addcontentsline{toc}{section}{\listtablename}
\listoftables
\flushbottom
\clearpage
\addcontentsline{toc}{section}{\listfigurename}
\listoffigures
\flushbottom
\clearpage

\pagebreak

\section{Introduction}
\label{sec:intro}

Baryon number $B$ and lepton number $L$ are classically conserved quantities in the Standard Model of particle physics (SM), but are generally not conserved in extensions such as Grand Unified Theories or supersymmetry~\cite{FileviezPerez:2022ypk}. Whether, and how, $B$ and $L$ are violated is ultimately an experimental question, with many detectors currently searching for such violations. Baryon number violation (BNV) in the form of nuclear decays is a particularly powerful probe due to the relative ease of collecting a large amount of SM-stable material and observing it for years, see for example the massive water-Cherenkov detector Super-Kamiokande~\cite{Super-Kamiokande:2002weg}. In clean decay channels such as $p\to e^+\pi^0$, the lifetime limits exceed $\unit[10^{34}]{yr}$~\cite{Super-Kamiokande:2020wjk}, and even invisible channels such as $n\to \nu\nu\nu$ can be constrained to $\unit[10^{30}]{yr}$~\cite{Heeck:2019kgr,SNO:2022trz}. 

Any positive signal of BNV would be proof for physics beyond the SM, i.e.~new particles that mediate these novel processes. The stringent lifetime limits and correspondingly suppressed amplitudes imply either very heavy new particles or minuscule coupling constants. Calculations are usually performed under the assumption of \textit{heavy} particles, allowing usage of the SM Effective Field Theory (SMEFT, see Refs.~\cite{Isidori:2023pyp,Aebischer:2025qhh} for recent reviews) to map new-physics models onto non-renormalizable operators built from SM fields, simplifying both categorization and calculations~\cite{Weinberg:1979sa,Weinberg:1980bf}. Tools such as \texttt{Sym2Int}~\cite{Fonseca:2017lem,Fonseca:2019yya} or Hilbert-series techniques~\cite{Lehman:2015via,Lehman:2015coa,Henning:2015daa,Henning:2015alf} have made it straightforward to collect BNV SMEFT operators up to high mass dimension~\cite{Heeck:2025btc} and even construct an operator basis~\cite{Harlander:2023ozs,Heeck:2026ked}. 
Unlike in other areas of particle physics, operators with mass dimension $d\gg 6$ are testable because BNV is so sensitive \textit{and} because $B$ and $L$ are violated by different amounts for different $d$~\cite{Weinberg:1980bf,Kobach:2016ami,Heeck:2025btc}, allowing for simple selection rules and symmetries to forbid otherwise-dominant lower-dimensional operators~\cite{Weinberg:1980bf,Heeck:2019kgr}.

As convenient as the SMEFT is in organizing BNV operators and signatures, it is by construction blind to the underlying physics. A positive signal would likely first be mapped onto possible SMEFT operators, which would then be mapped onto renormalizable new-physics models to determine the properties of the new elementary particles. The process of opening up SMEFT operators, or finding UV completions for them, is relatively straightforward as long as we restrict ourselves to tree-level UV completions, the main complication being the sheer number of BNV operators at high mass dimension.\footnote{The same research program has been performed for the (smaller) set of $\Delta L =2$ operators -- motivated by neutrino masses -- starting from an operator basis~\cite{Babu:2001ex,deGouvea:2007qla} followed by tree-level UV completions~\cite{Bonnet:2009ej,Angel:2012ug,Helo:2016vsi,Anamiati:2018cuq,Gargalionis:2020xvt}.} Nevertheless, in Ref.~\cite{Heeck:2026dmh} we provided tree-level UV completions for all \textit{non-derivative} BNV SMEFT operators up to $d=14$, plus the $d=15$ operators with unique $(\Delta B,\Delta L)$. 
The reason for our restriction to non-derivative operators was threefold: 1) derivative operators are technically more complicated to UV-complete; 2) at large $d$, derivative operators vastly outnumber non-derivative operators; 3) derivative operators are often claimed to be suppressed compared to non-derivative operators because the derivatives will suppress the nucleon-decay amplitude by factors of order of the nucleon mass, which is smaller than the Higgs vacuum expectation value in non-derivative operators~\cite{Weinberg:1980bf}.
In the present article, we will provide UV completions for all \textit{derivative} BNV SMEFT operators up to $d=12$ and emphasize that they need not be suppressed compared to non-derivative operators.

\begin{table}[tb]
\centering
{\renewcommand{\arraystretch}{1.1}
\begin{tabular}{c c c c} 
 \hline
 field & chirality & generations & $SU(3)_C\times SU(2)_L\times U(1)_Y$ representation  \\ [0.5ex] 
 \hline\hline
 $Q$ & left & 3 & $\left(\vec{3},\vec{2},\tfrac16\right)$\\
 $u$ & right & 3 & $\left(\vec{3},\vec{1},\tfrac23\right)$\\
 $d$ & right & 3 & $\left(\vec{3},\vec{1},-\tfrac13\right)$\\
 $L$ & left & 3 & $\left(\vec{1},\vec{2},-\tfrac12\right)$\\
 $e$ & right & 3 & $\left(\vec{1},\vec{1},-1\right)$\\
 $\nu$ & right & 3 & $\left(\vec{1},\vec{1},0\right)$\\
 $H$ & scalar & 1 & $\left(\vec{1},\vec{2},\tfrac12\right)$\\
 \hline
\end{tabular}
}
\caption{SM fields (plus right-handed neutrinos $\nu$) and quantum numbers; hypercharge $Y$ is related to electric charge $Q_e$ via $Q_e = Y + T_3$, with diagonal $SU(2)_L$ generator $T_3$.}
\label{tab:fields}
\end{table}

We follow, for the most part, the notation and methods of Ref.~\cite{Heeck:2026dmh}. In particular, we will also generalize from SMEFT to $\nu$SMEFT, i.e.~include three SM-singlet right-handed neutrinos $\nu$ in our effective operators, listed in Tab.~\ref{tab:fields}. Unlike Ref.~\cite{Heeck:2026dmh}, however, we will not include vector particles in our UV completions, because consistent heavy-vector completions depend on their gauge embedding and introduce additional model-building assumptions~\cite{Beltran:2023ymm}. Once we have restricted possible mediators to scalars and spin-$1/2$ fermions, the only place where derivatives can be generated in tree-level diagrams are the propagators of the heavy new particles. In the large-mass expansion of virtual particles, the propagators can be expanded as
\begin{align}
    \frac{1}{m_S^2-p^2} &= \frac{1}{m_S^2}+\frac{p^2}{m_S^4}+\frac{p^4}{m_S^6}+\frac{p^6}{m_S^8}+\dots\,,\label{eq:scalar_propagator}\\
    \frac{\slashed{p}+m_F}{m_F^2-p^2} &= \frac{1}{m_F}+\frac{\slashed{p}}{m_F^2}+\frac{p^2}{m_F^3}+\frac{\slashed{p} p^2}{m_F^4}+\dots\,,   \label{eq:fermion_propagator}
\end{align}
for scalars and fermions, respectively, 
leading to an amplitude expansion in powers of $p$ that can be mapped onto operators with corresponding number of covariant derivatives $D$. Equivalently, one can forego the momentum-space calculation and work with the position-space equations of motion to directly obtain the derivative operators. For example, the $d=6$ BNV operator $uud e$ can be tree-level UV-completed by the scalar leptoquark $\mathcal{S}_1\sim (\bar{\vec{3}},\vec{1}, 1/3)$~\cite{Heeck:2026dmh}, with relevant Lagrangian
\begin{align}
    \L_{\mathcal{S}_1} &\supset |D_\mu \mathcal{S}_1|^2 - m_{\mathcal{S}_1}^2 |\mathcal{S}_1|^2 + \left(
      y\, \bar{u} ^C \mathcal{S}_1 e +
  z\, \bar{u} ^C \mathcal{S}_1^{*} d 
    + \hc\right)  ,
\end{align}
where $y$ and $z$ are dimensionless coupling matrices and we have suppressed all index contractions.
Solving the equation of motion for $\mathcal{S}_1$ in the large-mass limit gives the effective Lagrangian
\begin{align}
    \L_{\mathcal{S}_1} \to \frac{\epsilon_{\alpha\beta\gamma} (z\,\bar{u}^{C}_\alpha d_\beta)(y\,\bar{u}^C_\gamma e)}{m_{\mathcal{S}_1}^2}  - \frac{\epsilon_{\alpha\beta\gamma} (z\,\bar{u}^{C}_\alpha d_\beta) D^2 (y\,\bar{u}^C_\gamma e)}{m_{\mathcal{S}_1}^4} + \mathcal{O}\left(\frac{D^4}{m_{\mathcal{S}_1}^{6}}\right) ,
    \label{eq:Leff_S1}
\end{align}
where we reintroduced the $SU(3)_C$ contractions of the quarks. Evidently, $\mathcal{S}_1$ is not only a UV completion for the operator $uude$ but also for the operators $D^{2n} uude$ with any integer $n$. However, already the $D^2$ term will be suppressed by a factor of order $m_p^2/m_{\mathcal{S}_1}^2$ relative to the $D^0$ term in any nucleon-decay amplitude, which is entirely negligible given the SMEFT assumption that $m_{\mathcal{S}_1}$ should exceed the electroweak scale.
Consequently, even though each UV completion of a non-derivative BNV operator $\mathcal{O}$ from Ref.~\cite{Heeck:2026dmh} \textit{automatically} forms a UV completion for the derivative operators $D^{2n}\mathcal{O}$ by going to higher order in the heavy-mass expansion, the so-generated operators are too suppressed to be relevant. These are precisely the \textbf{trivial UV completions} of derivative operators that are \textit{not interesting}.

How then could derivative operators become relevant? In principle, one could have destructive interference of two amplitudes at the non-derivative level. For example, consider the UV completion of the same $uude$ operator by the leptoquark $\tilde{\mathcal{S}}_1\sim (\bar{\vec{3}},\vec{1}, 4/3)$~\cite{Heeck:2026dmh}:
\begin{align}
    \L_{\tilde{\mathcal{S}}_1} &\supset |D_\mu \tilde{\mathcal{S}}_1|^2 - m_{\tilde{\mathcal{S}}_1}^2 |\tilde{\mathcal{S}}_1|^2 + \left(
      \tilde{y}\, \bar{d} ^C \tilde{\mathcal{S}}_1 e +
  \tilde{z}\, \bar{u} ^C \tilde{\mathcal{S}}_1^{*} u 
    + \hc\right)  \\
    &\to \frac{\epsilon_{\alpha\beta\gamma} (\tilde{y}\,\bar{d}^C_\gamma e) (\tilde{z}\,\bar{u}^{C}_\alpha u_\beta)}{m_{\tilde{\mathcal{S}}_1}^2}  - \frac{\epsilon_{\alpha\beta\gamma} (\tilde{y}\,\bar{d}^C_\gamma e) D^2 (\tilde{z}\,\bar{u}^{C}_\alpha u_\beta) }{m_{\tilde{\mathcal{S}}_1}^4} + \mathcal{O}\left(\frac{D^4}{m_{\tilde{\mathcal{S}}_1}^{6}}\right)\\
    &= \frac{\epsilon_{\alpha\beta\gamma} (2\tilde{z}\,\bar{u}^{C}_\alpha d_\beta)(\tilde{y}\,\bar{u}^C_\gamma e)}{m_{\tilde{\mathcal{S}}_1}^2}  - \frac{\epsilon_{\alpha\beta\gamma} (\tilde{y}\,\bar{d}^C_\gamma e) D^2 (\tilde{z}\,\bar{u}^{C}_\alpha u_\beta) }{m_{\tilde{\mathcal{S}}_1}^4} + \mathcal{O}\left(\frac{D^4}{m_{\tilde{\mathcal{S}}_1}^{6}}\right) ,
\end{align}
using a Fierz identity in the first term of the third line. In the presence of both $\tilde{\mathcal{S}}_1$ and $\mathcal{S}_1$, the $d=6$ operator $uude$ can therefore be arranged to \textit{vanish} if 
\begin{align}
    \frac{z\,y}{m_{\mathcal{S}_1}^2} = -  \frac{2\tilde{z}\,\tilde{y}}{m_{\tilde{\mathcal{S}}_1}^2} \,,
    \label{eq:cancellation}
\end{align}
but the same cancellation does not continue at the $d=8$ level, essentially because the two virtual particles carry different momenta. The derivative operator $D^2 uude$ hence survives and is no longer a subleading correction to some non-derivative operator. For this to work, the cancellation condition in Eq.~\eqref{eq:cancellation} needs to be fine-tuned to a high degree, making this a rather unappealing motivation for derivative operators. (This particular example furthermore suffers from the fact that $\mathcal{S}_1$ will generically induce other $d=6$ BNV operators~\cite{Heeck:2026dmh}, and the cancellation can only work for $uude$ operators with different up-quark generations since $\tilde{z}$ is antisymmetric in the up-quark indices.)

Interestingly, there exist UV completions that lead to potentially-dominant derivative operators $D^k \mathcal{O}$ without the need for any fine-tuned cancellations because the non-derivative version $\mathcal{O}$ is either not a Lorentz singlet or because it \textit{is} a singlet but happens to vanish. We can distinguish two cases:
\begin{enumerate}
    \item Operators with an \textit{odd} number of derivatives, which require at least one heavy fermion in the UV completion, can never have a non-derivative counterpart with the same fields due to $F$ parity~\cite{Weinberg:1980bf}, i.e.~gauge and Lorentz symmetry, essentially because $D_\mu$ is not a singlet. Equivalently, the first two terms in the fermion propagator~\eqref{eq:fermion_propagator} have different gamma-matrix structures and only one survives when they are wedged between chirality operators, which always arise in SMEFT operators since the SM is a chiral theory. Starting from a tree-level UV completion of a non-derivative operator, the $\mathcal{O}(\slashed{p})$ terms in the $1/m_F$ expansion will simply drop out! Vice versa, any tree-level UV completion of an operator with an odd number of derivatives (see Fig.~\ref{fig:Der_Examples} (left)) will not generate the same operator without derivatives, breaking the naive relationship between derivative and non-derivative operators. 
    One-derivative SMEFT operators are therefore automatically interesting and could be the dominant source of BNV; operators with three or five derivatives require a more careful UV-completion-dependent analysis because they \textit{might} just be subleading corrections to one-derivative operators.

\begin{figure}[tb]
    \centering
    \includegraphics[scale=0.2]{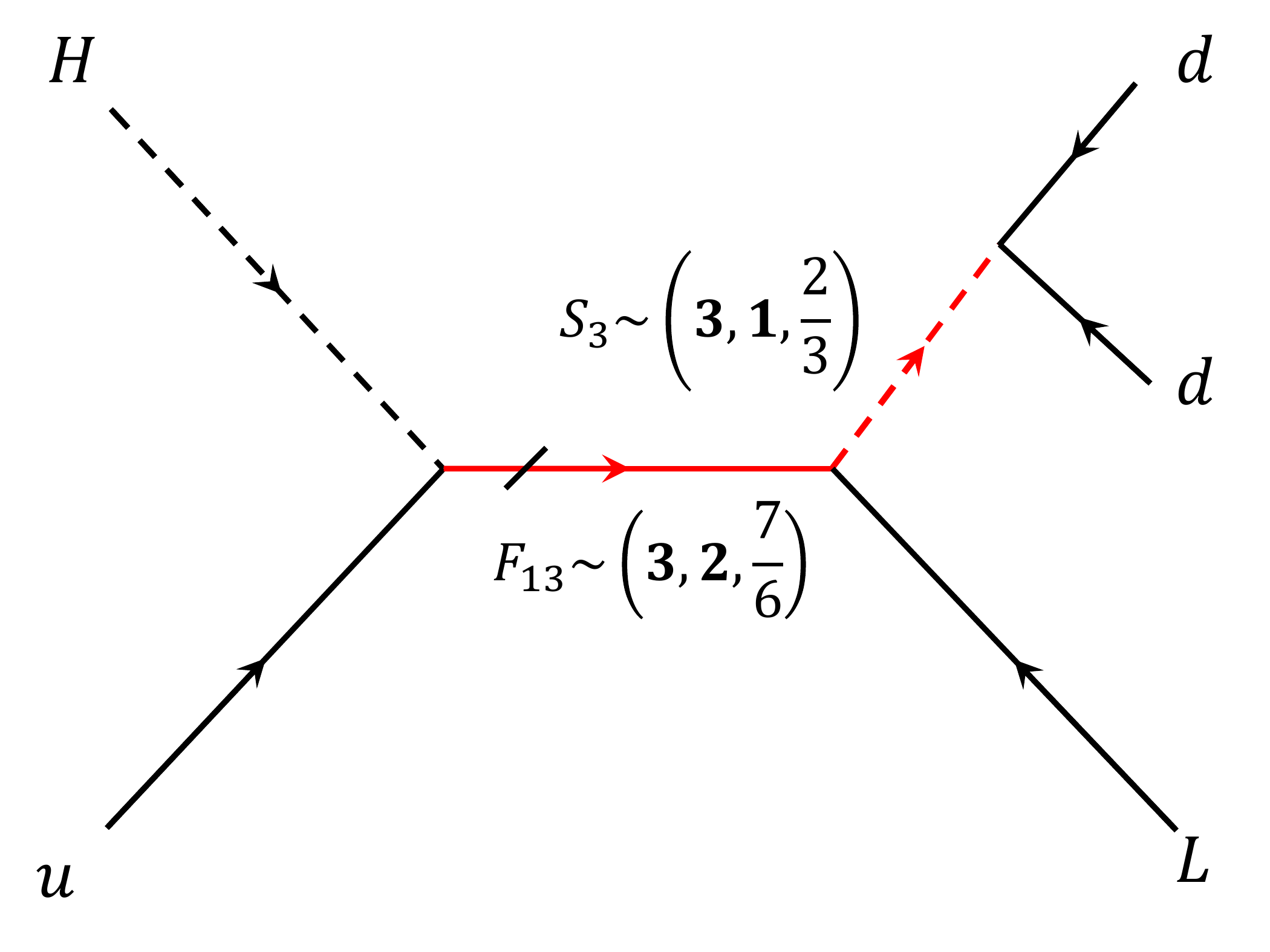}
    \includegraphics[scale=0.2]{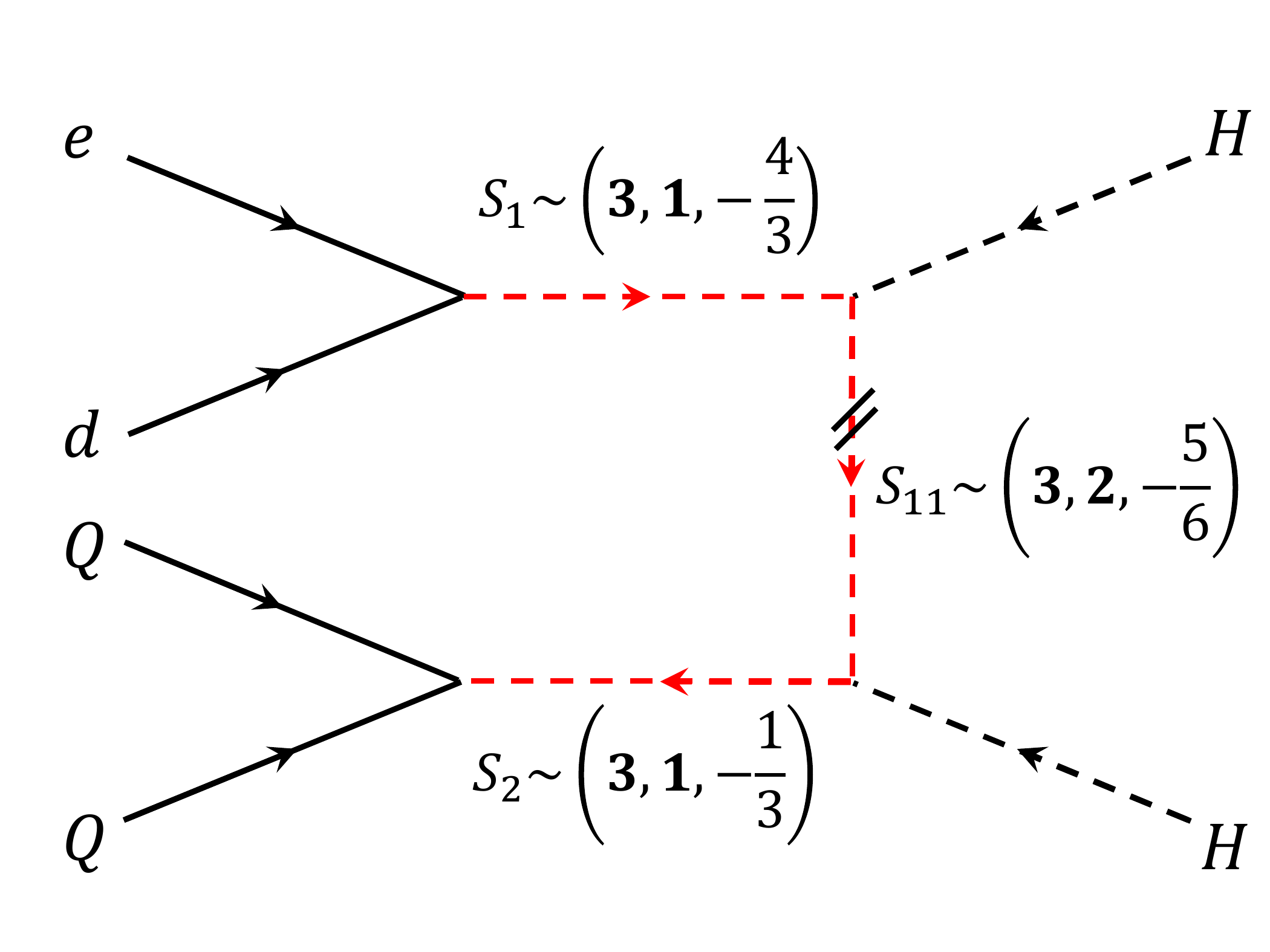}
  \caption{Examples for UV completions of derivative BNV operators that are not subleading corrections to non-derivative operators. Left: UV completion of the $d=8$ operator $DHLduu$, with derivative position indicated by the diagonal line on the fermion propagator. The non-derivative version $HLduu$ is not Lorentz invariant, so is trivially not induced.
   Right: UV completion of the $d = 10$ derivative operator $D^{2} dQQeHH$, both derivatives coming from the $S_{11}$ propagator. The non-derivative version $dQQeHH$ vanishes because the two Higgs are contracted antisymmetrically to an $SU(2)_L$ singlet.}
    \label{fig:Der_Examples}
\end{figure}

    \item Derivative operators with two or more derivatives can still be dominant if they contain at least two Higgs or anti-Higgs fields that are contracted antisymmetrically~\cite{Gargalionis:2020xvt,Heeck:2026dmh}. As an example, take the $d=8$ BNV operator $HHQQde$ with UV completion shown in Fig.~\ref{fig:Der_Examples} (right). Here, $QQ$ are contracted to an $SU(2)_L$ singlet, therefore $HH$ is also forced to contract into an $SU(2)_L$ singlet, which is anti-symmetric under $H$ exchange and hence vanishes for the single Higgs doublet in the SM. Fig.~\ref{fig:Der_Examples} (right) is therefore \textit{not} a UV completion of $HHQQde$, as this leading term simply vanishes. However, expanding the $S_{11}$ propagator to the next order, $p^2/m_{S_{11}}^4$, yields a non-vanishing contribution, since $H D^2 H$ is a non-vanishing $SU(2)_L$ singlet, which corresponds to the $d=10$ derivative operator $D^2 HHQQde$. Starting at $d=10$, we can find such UV completions for two-derivative BNV operators for which the non-derivative analog vanishes due to Higgs asymmetry. In extreme cases, this can occur not just for particular UV completions but at the operator level. For example, the $d=10$ operator $DDHHddue$ exists but the non-derivative version, $HHddue$, simply vanishes because $HH$ is necessarily coupled to an $SU(2)_L$ singlet, seeing as all the fermions in the operator are $SU(2)_L$ singlets. $DDHHddue$ is therefore a potentially-dominant derivative operator, not just a subleading correction.

    Similarly, some UV completions of \textit{one-derivative} operators involving at least two Higgs vanish due to Higgs asymmetry but become non-zero once two more derivatives are inserted between the Higgs fields, leading to dominant three-derivative operators, starting at $d=12$. 
    Operators with even more Higgs fields can require even more derivatives to avoid Higgs asymmetry. Again, in some cases this can occur at the operator level; for example, the $d=11$ operator $D d u u \bar{e} \bar{H} \bar{H} \bar{H} \bar{H}$ is a vanishing gauge \& Lorentz singlet because not only is $(\bar{H})^4 $ zero when coupled to the necessary $SU(2)_L$ singlet, but even  $(D_\mu \bar{H}^4)_{\vec{1}} =0$, no matter where the derivative sits. Consequently, the three-derivative $d=13$ operator $DDD d u u \bar{e} \bar{H} \bar{H} \bar{H} \bar{H}$ is the lowest-dimensional realization, and hence a potentially dominant derivative BNV operator.

\end{enumerate}
More details about our procedure will be given below, but the above arguments and examples provide the conceptual basis for our work to illustrate the importance of derivative SMEFT operators. Hereafter, we are only studying \textbf{non-trivial UV completions} of derivative operators, meaning cases in which the induced operator is the leading term in the heavy-mass expansion and not just a subleading correction, all without fine-tuned cancellations. This does not mean that the operator will be the leading source of BNV, seeing as the mediators could also induce BNV operators with different particle content. Our non-trivial UV completions are nonetheless \textit{potentially dominant}, a necessary pre-selection that will be used in a follow-up article~\cite{Heeck:accidental} to determine actual dominance.
The rest of the article is organized as follows: we explain the general procedure of deriving tree-level UV completions for derivative BNV operators in Sec.~\ref{sec:general_procedure}. BNV operators involving field strength tensors are discussed in detail in Sec.~\ref{sec:field_strength_tensors}. Sec.~\ref{sec:d=7} lists the lowest-dimensional derivative BNV operators, which arise at mass dimension $d=7$ but do not have tree-level UV completions. Sec.~\ref{sec:d=8} lists the operators and UV completions for $d=8$ and provides a dedicated example of the effects of a derivative operator. Sec.~\ref{sec:d=9} presents results for $d=9$. We provide a few examples for novel effects at mass dimension $d>10$ in Sec.~\ref{sec:higher_dims} before concluding in Sec.~\ref{sec:conclusions}. App.~\ref{app:topologies} collects the Feynman-diagram topologies used in our UV completions. App.~\ref{sec:d=10} lists operators and some UV completions for $d=10$; App.~\ref{sec:d=11} does the same for $d=11$; and App.~\ref{sec:d=12} for $d=12$. 
We only provide a relatively small subset of UV completions in this article to keep the number of pages in the triple digits; the complete data tables of topologies and UV completions in multiple formats as well as the \texttt{Mathematica} notebooks used to create and manipulate them can be found online at \website. Our code is in no way specific to BNV operators and can be used to generate tree-level UV completions for any operator at practically any mass dimension.

\section{General procedure}
\label{sec:general_procedure}

Our construction of UV completions for derivative operators is overall similar to the non-derivative case of Ref.~\cite{Heeck:2026dmh}:
\begin{enumerate}
    \item We generate simplified derivative BNV operators up to high mass dimension $d$ using \texttt{Sym2Int}~\cite{Fonseca:2017lem}, but discard operators involving field strength tensors, as justified in Sec.~\ref{sec:field_strength_tensors}. 
    Simplified here refers to operators of the form $D D D d d Q \bar{L}$, suppressing any and all indices as well as the actual position of the derivatives, since these details are not necessary for our UV-completion method. Notice that BNV operators are necessarily complex and we pick our operators to have positive $B$; the conjugate of an operator is generated by the same UV completion.
    We sort operators according to their mass dimension $d$, the baryon number $B$ they carry, the lepton number $L$ they carry, and the number of derivatives $n$ they contain, with an index $j$ that runs over all operators of the same type: $\mathcal{O}_{d, \,(B , L, n)}^{j}$.
    The number of derivative operators grows exponentially with mass dimension, eventually faster than non-derivative operators, see Fig.~\ref{fig:Number_of_derivative_operators}.

\begin{figure}
\begin{center}
    \includegraphics[width=0.95\textwidth]{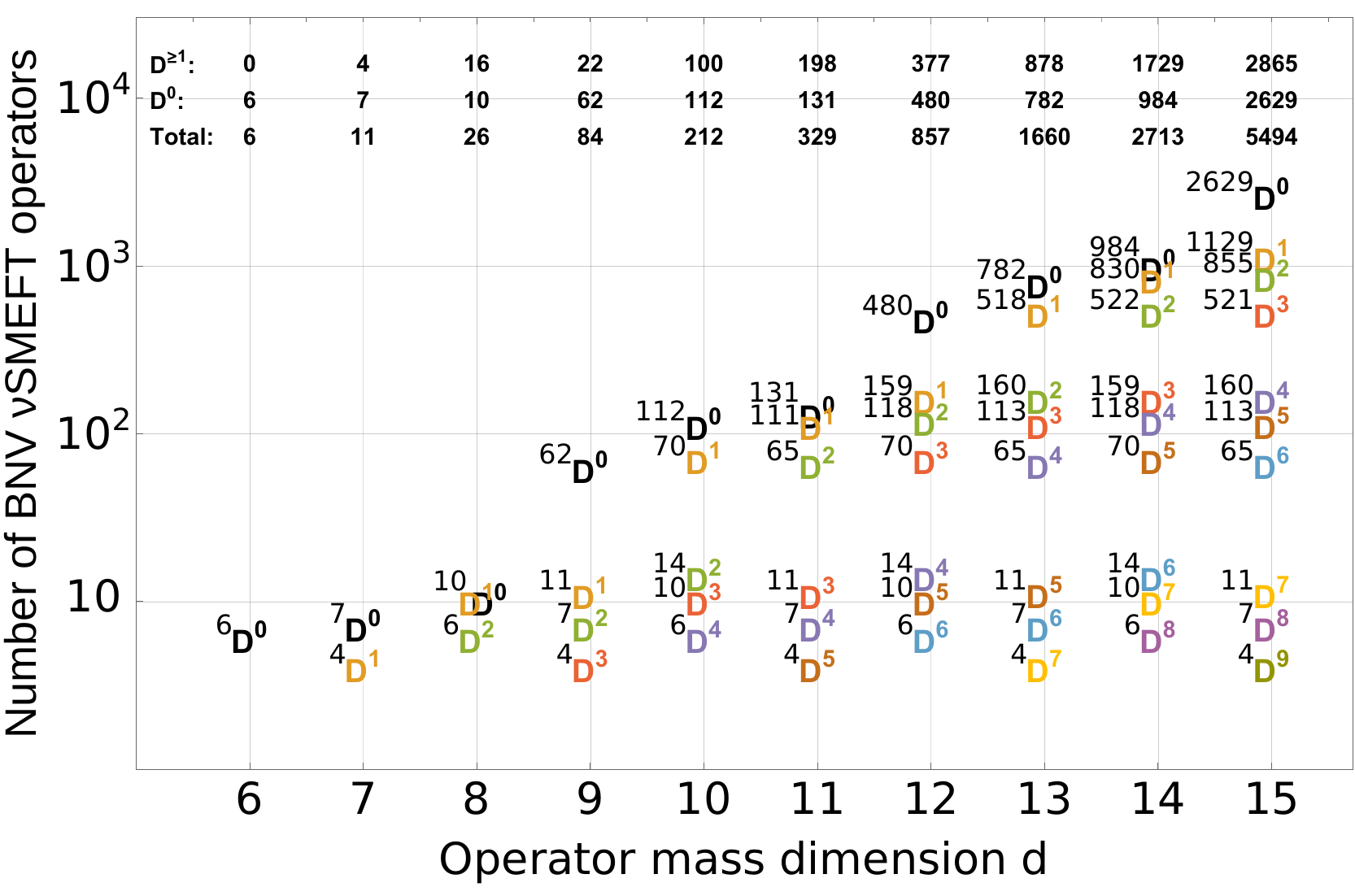}\\[3ex]
    \includegraphics[width=0.95\textwidth]{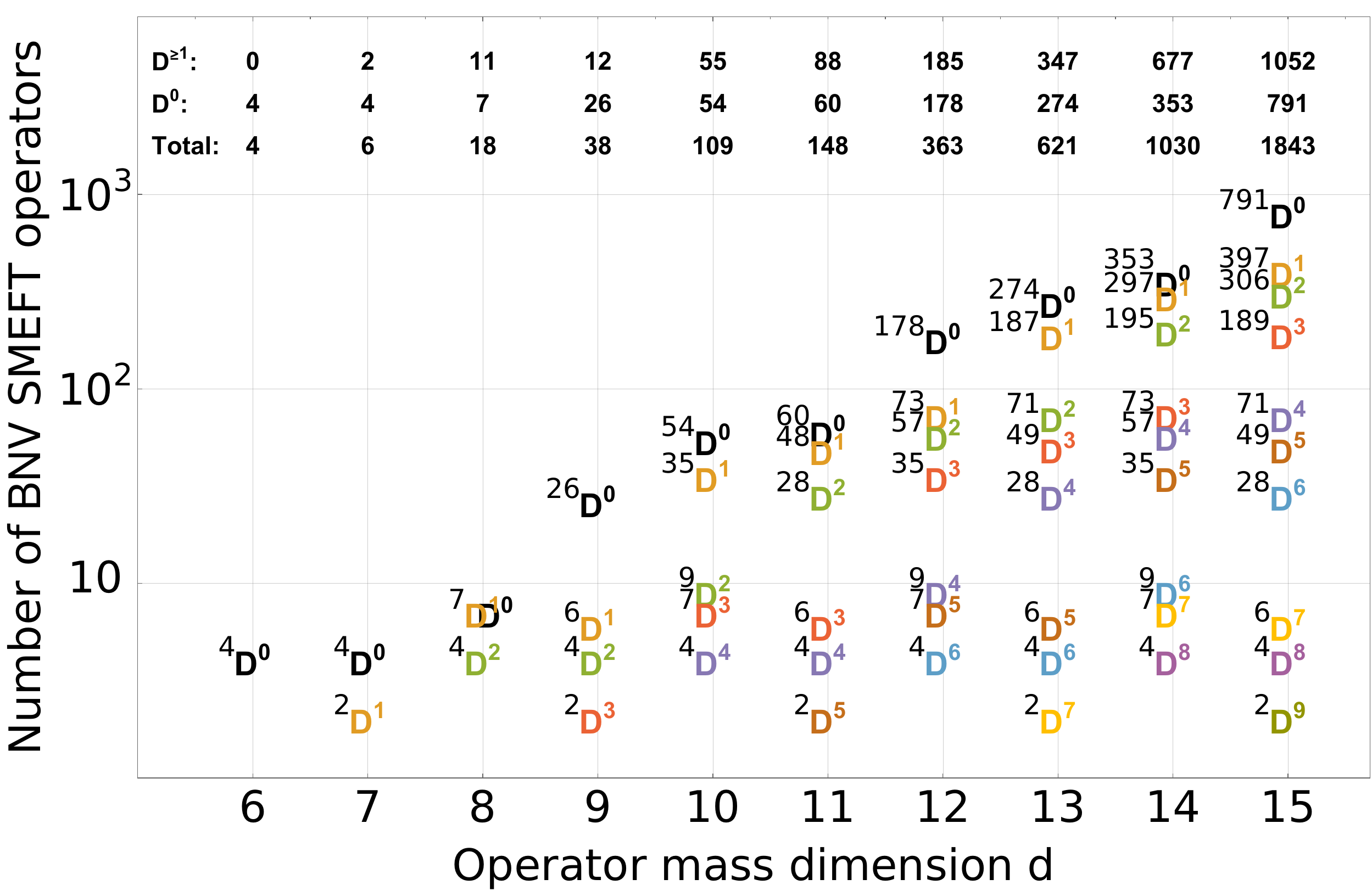}
\end{center}
       \caption{Number of simplified BNV operators in $\nu$SMEFT (top) and SMEFT (bottom) in each mass dimension $d$ grouped by number of derivatives, $D^0$ in black denoting non-derivative operators, $D^1$ in orange one-derivative ones, etc. The total numbers of operators are displayed at the top of the plot for each mass dimension $d$: the total number of operators summed over all derivative orders, including non-derivative operators, and the corresponding total restricted to operators containing at least one derivative. Notice that operators involving field-strength tensors are not included in this counting.
       }
    \label{fig:Number_of_derivative_operators}
\end{figure}

    \item We employ the same topologies $\mathcal{T}_d^j$ and associated files as in Ref.~\cite{Heeck:2026dmh}, where $d$ is the mass dimension of a non-derivative operator and the index $j$ runs over the different distinct topologies. For an operator with $n$ \textit{derivatives} and mass dimension $d$, the appropriate topologies are then $\mathcal{T}_{d-n}^j$. For example, the $d=8$ operator $D H L d u u$ is UV-completed using the same topologies as non-derivative $d=7$ operators, see Fig.~\ref{fig:Der_Examples} (left). Furthermore, operators with an \textit{odd} number of derivatives require at least one internal \textit{fermion}, which restricts the number of possible topologies. Despite these changes, we keep the topology naming of Ref.~\cite{Heeck:2026dmh} and reproduce them in App.~\ref{app:topologies} for the convenience of the reader.

    \item For a given topology, we then distribute the fields of an appropriate operator over the \textit{external} lines and the derivatives over the \textit{internal} lines and determine the Lorentz \& gauge-group $SU(2)_{\text{left}}\times SU(2)_{\text{right}}\times SU(3)_C \times SU(2)_L\times U(1)_Y$ representations of the internal particles via sequential representation products at each vertex~\cite{Heeck:2026dmh} using \texttt{GroupMath}~\cite{Fonseca:2020vke}. 
    Derivatives are distributed so that scalar lines carry an even number, but fermion lines can carry an arbitrary number of derivatives. In our Feynman diagrams, we indicate each derivative by a diagonal slash on the propagator line.
    To find the representations of the internal particles, most of these derivatives are actually irrelevant, the only important rule is that an odd number of derivatives on a fermion line leads to an additional chirality flip, which changes the Lorentz quantum numbers at the following vertex.
    This minor chirality-flip modification of the code developed in Ref.~\cite{Heeck:2026dmh} is sufficient to obtain the possible internal representations that lead to viable gauge- and Lorentz-singlet diagrams exemplified in Fig.~\ref{fig:Der_Examples}.
    This process is repeated for all permutations of external fields and derivative positions for a given topology--operator combination. Due to diagrammatic symmetries, some of the resulting UV completions can be equivalent; we remove these as in Ref.~\cite{Heeck:2026dmh} by turning the Feynman diagrams into directed colored graphs, although now keeping track which internal particles carry how many derivatives, and use \texttt{IGraph/M}~\cite{Horv_t_2023} to identify equivalences.
    Since we did not encounter any representation not already named in Ref.~\cite{Heeck:2026dmh}, we use the same notation for scalars and fermions, collected in Tab.~\ref{tab:newParticles}.

\begingroup
\renewcommand{\arraystretch}{1.2}
\newcommand{\br}{,\linebreak[1]}
\newcolumntype{C}{>{\centering\arraybackslash\setstretch{1.25}}X}

\begin{table}[tb]
\small
\centering
\begin{tabularx}{\textwidth}{|c|C|}
\hline
      $d$ & {\rm Scalars} \\
\hline \hline
    8& $S_{1}=(\bold{3},\bold{1},-\frac{4}{3})$, $S_{2}=(\bold{3},\bold{1},-\frac{1}{3})$, $S_{3}=(\bold{3},\bold{1},\frac{2}{3})$, $S_{4}=(\bold{3},\bold{3},-\frac{1}{3})$.\\
    \hline
    9& $S_{5}=(\bold{3},\bold{2},\frac{1}{6})$, $S_{6}=(\bold{3},\bold{2},\frac{7}{6})$, $S_{7}=(\bold{1},\bold{1},0)$, $S_{8}=(\bold{1},\bold{3},0)$, $S_{9}=(\bold{1},\bold{3},1)$, $S_{11}=(\bold{3},\bold{2},-\frac{5}{6})$, $S_{13}=(\bold{3},\bold{4},\frac{1}{6})$, $S_{19}=(\bold{3},\bold{3},\frac{2}{3})$.\\
    \hline
    10& $S_{10}=(\bold{3},\bold{2},-\frac{11}{6})$, $S_{12}=(\bold{3},\bold{4},-\frac{5}{6})$, $S_{14}=(\bold{1},\bold{1},1)$, $S_{15}=(\bold{1},\bold{1},2)$, $S_{16}=(\bold{1},\bold{2},\frac{1}{2})$, $S_{17}=(\bold{1},\bold{4},\frac{1}{2})$, $S_{21}=(\bold{6},\bold{1},-\frac{4}{3})$, $S_{22}=(\bold{6},\bold{1},-\frac{1}{3})$, $S_{23}=(\bold{6},\bold{1},\frac{2}{3})$, $S_{24}=(\bold{6},\bold{3},-\frac{1}{3})$, $S_{25}=(\bold{8},\bold{2},\frac{1}{2})$, $S_{31}=(\bold{3},\bold{3},-\frac{4}{3})$.\\
    \hline
    11& $S_{18}=(\bold{1},\bold{4},\frac{3}{2})$, $S_{20}=(\bold{3},\bold{3},\frac{5}{3})$, $S_{26}=(\bold{1},\bold{2},\frac{3}{2})$, $S_{29}=(\bold{3},\bold{1},\frac{5}{3})$, $S_{33}=(\bold{3},\bold{5},-\frac{1}{3})$, $S_{34}=(\bold{3},\bold{5},\frac{2}{3})$, $S_{36}=(\bold{6},\bold{2},-\frac{5}{6})$, $S_{37}=(\bold{6},\bold{2},\frac{1}{6})$, $S_{38}=(\bold{6},\bold{2},\frac{7}{6})$, $S_{40}=(\bold{6},\bold{4},\frac{1}{6})$, $S_{41}=(\bold{8},\bold{1},0)$, $S_{42}=(\bold{8},\bold{1},1)$, $S_{43}=(\bold{8},\bold{3},0)$, $S_{44}=(\bold{8},\bold{3},1)$, $S_{45}=(\bold{1},\bold{5},1)$, $S_{47}=(\bold{3},\bold{4},\frac{7}{6})$, $S_{49}=(\bold{6},\bold{3},\frac{2}{3})$, $S_{52}=(\bold{8},\bold{2},\frac{3}{2})$, $S_{53}=(\bold{8},\bold{4},\frac{1}{2})$.\\
    \hline
    12& $S_{27}=(\bold{1},\bold{2},\frac{5}{2})$, $S_{28}=(\bold{1},\bold{5},0)$, $S_{30}=(\bold{3},\bold{3},-\frac{7}{3})$, $S_{32}=(\bold{3},\bold{5},-\frac{4}{3})$, $S_{35}=(\bold{6},\bold{2},-\frac{11}{6})$, $S_{39}=(\bold{6},\bold{4},-\frac{5}{6})$, $S_{55}=(\bold{1},\bold{3},2)$, $S_{59}=(\bold{3},\bold{1},-\frac{7}{3})$, $S_{60}=(\bold{3},\bold{1},\frac{8}{3})$, $S_{61}=(\bold{3},\bold{2},\frac{13}{6})$, $S_{63}=(\bold{3},\bold{4},-\frac{11}{6})$, $S_{71}=(\bold{6},\bold{3},-\frac{4}{3})$, $S_{73}=(\bold{6},\bold{5},-\frac{1}{3})$, $S_{74}=(\bold{8},\bold{1},2)$, $S_{76}=(\bold{8},\bold{5},0)$.\\
    \hline
\end{tabularx}\\[3ex]
\begin{tabularx}{\textwidth}{|c|C|}
\hline
      $d$ & {\rm Fermions} \\
\hline \hline
    8& $F_{1}=(\bold{1},\bold{1},0)$, $F_{2}=(\bold{1},\bold{1},1)$, $F_{3}=(\bold{1},\bold{2},\frac{1}{2})$, $F_{4}=(\bold{1},\bold{3},0)$, $F_{5}=(\bold{3},\bold{1},-\frac{1}{3})$, $F_{6}=(\bold{3},\bold{1},\frac{2}{3})$, $F_{7}=(\bold{3},\bold{2},-\frac{5}{6})$, $F_{8}=(\bold{3},\bold{2},\frac{1}{6})$, $F_{9}=(\bold{3},\bold{3},-\frac{1}{3})$, $F_{10}=(\bold{1},\bold{2},\frac{3}{2})$, $F_{11}=(\bold{1},\bold{3},1)$, $F_{13}=(\bold{3},\bold{2},\frac{7}{6})$, $F_{14}=(\bold{3},\bold{3},\frac{2}{3})$.\\
    \hline
    9& $F_{12}=(\bold{1},\bold{4},\frac{1}{2})$, $F_{17}=(\bold{3},\bold{1},-\frac{4}{3})$, $F_{19}=(\bold{3},\bold{3},-\frac{4}{3})$, $F_{20}=(\bold{3},\bold{4},-\frac{5}{6})$.\\
    \hline
    10& $F_{15}=(\bold{3},\bold{4},\frac{1}{6})$, $F_{16}=(\bold{1},\bold{1},2)$, $F_{18}=(\bold{3},\bold{1},\frac{5}{3})$, $F_{21}=(\bold{6},\bold{1},-\frac{4}{3})$, $F_{22}=(\bold{6},\bold{1},-\frac{1}{3})$, $F_{23}=(\bold{6},\bold{1},\frac{2}{3})$, $F_{24}=(\bold{6},\bold{2},-\frac{5}{6})$, $F_{25}=(\bold{6},\bold{2},\frac{1}{6})$, $F_{26}=(\bold{6},\bold{3},-\frac{1}{3})$, $F_{27}=(\bold{8},\bold{1},0)$, $F_{28}=(\bold{8},\bold{1},1)$, $F_{29}=(\bold{8},\bold{2},\frac{1}{2})$, $F_{30}=(\bold{8},\bold{3},0)$, $F_{31}=(\bold{1},\bold{3},2)$, $F_{32}=(\bold{1},\bold{4},\frac{3}{2})$, $F_{35}=(\bold{3},\bold{1},-\frac{7}{3})$, $F_{36}=(\bold{3},\bold{2},-\frac{11}{6})$, $F_{37}=(\bold{3},\bold{3},\frac{5}{3})$, $F_{38}=(\bold{3},\bold{4},\frac{7}{6})$, $F_{41}=(\bold{6},\bold{1},\frac{5}{3})$, $F_{42}=(\bold{6},\bold{2},\frac{7}{6})$, $F_{43}=(\bold{6},\bold{3},\frac{2}{3})$, $F_{44}=(\bold{6},\bold{4},\frac{1}{6})$, $F_{45}=(\bold{8},\bold{1},2)$, $F_{46}=(\bold{8},\bold{2},\frac{3}{2})$, $F_{47}=(\bold{8},\bold{3},1)$, $F_{48}=(\bold{8},\bold{4},\frac{1}{2})$.\\
    \hline
    11& $F_{33}=(\bold{1},\bold{5},0)$, $F_{34}=(\bold{1},\bold{5},1)$, $F_{39}=(\bold{3},\bold{5},-\frac{1}{3})$, $F_{49}=(\bold{1},\bold{2},\frac{5}{2})$, $F_{50}=(\bold{3},\bold{2},\frac{13}{6})$, $F_{51}=(\bold{3},\bold{4},-\frac{11}{6})$, $F_{52}=(\bold{3},\bold{5},-\frac{4}{3})$, $F_{53}=(\bold{6},\bold{2},-\frac{11}{6})$, $F_{54}=(\bold{6},\bold{3},-\frac{4}{3})$, $F_{55}=(\bold{6},\bold{4},-\frac{5}{6})$.\\
    \hline
    12& $F_{40}=(\bold{3},\bold{5},\frac{2}{3})$, $F_{56}=(\bold{6},\bold{5},-\frac{1}{3})$, $F_{57}=(\bold{8},\bold{5},0)$, $F_{59}=(\bold{1},\bold{4},\frac{5}{2})$, $F_{60}=(\bold{1},\bold{5},2)$, $F_{64}=(\bold{3},\bold{2},-\frac{17}{6})$, $F_{65}=(\bold{3},\bold{3},-\frac{7}{3})$, $F_{66}=(\bold{3},\bold{4},\frac{13}{6})$, $F_{67}=(\bold{3},\bold{5},\frac{5}{3})$, $F_{72}=(\bold{6},\bold{2},\frac{13}{6})$, $F_{73}=(\bold{6},\bold{3},\frac{5}{3})$, $F_{74}=(\bold{6},\bold{4},\frac{7}{6})$, $F_{75}=(\bold{6},\bold{5},\frac{2}{3})$, $F_{76}=(\bold{8},\bold{2},\frac{5}{2})$, $F_{77}=(\bold{8},\bold{3},2)$, $F_{78}=(\bold{8},\bold{4},\frac{3}{2})$, $F_{79}=(\bold{8},\bold{5},1)$.\\
    \hline
\end{tabularx}
\caption{List of scalars (top) and fermions (bottom) appearing in the derivative UV completions of BNV $\nu$SMEFT operators ordered by the lowest operator mass dimension $d$ at which they occur, following the naming of Ref.~\cite{Heeck:2026dmh}.}
\label{tab:newParticles}
\end{table}
\endgroup
    
    Repeating this process for all operators and topologies would be the end of the story if our universe had infinitely many generations of each SM particle. The next step takes into account that fermions only come in three generations and, much more importantly, we only have one Higgs, which will eliminate many UV completions that appear viable at the diagrammatic level.

    \item For a given diagram, we can write down a schematic amplitude or operator expansion by multiplying the singlets of every vertex and integrating out the internal lines by connecting the outgoing mediator-field components to the matching incoming ones, just like in Ref.~\cite{Heeck:2026dmh}. In the non-derivative case, this amounts to the explicit expansion of the operator into products of field components at the classical level, treating fermions as commuting objects for simplicity and without consequence. In the presence of \textit{derivatives}, this procedure is more akin to an amplitude calculation in momentum space, replacing covariant derivatives by momenta.\footnote{At this stage, we are essentially setting the gauge couplings to zero, but will make sure later on that we are not missing anything important coming from the gauge-group matrices in $D_\mu - \partial_\mu$, see Sec.~\ref{sec:field_strength_tensors}.}
    In particular, an odd number of derivatives, $D^{2k+1}$ on a fermion line is replaced by a chirality flip through the insertion of a factor  $(p_F^2)^k \slashed{p}_F$ where the propagator would be.
    An \textit{even} number of derivatives, $D^{2k}$,  on an internal line simply leads to an overall factor of $(p^2)^k$ in the amplitude, with $p$ the four momentum of the internal line. 
    Momentum conservation at each vertex allows us to replace all internal momenta by linear combinations of external momenta, so the final amplitude for a given diagram depends on the components of the external particles -- essentially polarization vectors or spinors -- as well as external momenta, all contracted to a gauge and Lorentz singlet. In our convention, all external momenta are incoming and furthermore square to zero because we are working in the unbroken-phase approximation.

    This amplitude construction is key for the next step: the symmetrization of identical boson fields and antisymmetrization of identical fermion fields. An amplitude $\mathcal{A}$ involving two Higgs fields $H$, with $SU(2)_L$ polarizations $H_j$ and momenta $p_j$ needs to be symmetrized,
    \begin{align}
        \mathcal{A}[H_1,H_2,p_1,p_2] \to \mathcal{A}[H_1,H_2,p_1,p_2] + \mathcal{A}[H_2,H_1,p_2,p_1]  \,,
    \end{align}
    to account for the fact that we cannot tell which $H$ belongs to which vertex. For more Higgs fields, the total symmetrization -- including the Higgs momenta -- is lengthier but straightforward, and the same applies to anti-Higgs. An amplitude that is antisymmetric under Higgs exchange would therefore lead to a vanishing amplitude and hence operator, even though the diagram might be viable and innocent-looking. As mentioned already in the introduction, the underlying reason for such a vanishing amplitude after symmetrization can be traced back to the contraction of $H^2$ or $H^4$ to an $SU(2)_L$ singlet, only saved by wedging derivatives in between (in the notation from above, $\mathcal{A}[H_1,H_2,p_1,p_2]$ might be antisymmetric under $H_1 \leftrightarrow H_2$ exchange, but this can be compensated for by the $p_1\leftrightarrow p_2$ exchange that comes from derivatives). As an example, consider the diagram of Fig.~\ref{fig:Der_Examples} (right). Without derivatives, the two Higgs have to contract into a singlet, $(HH)_{\vec{1}} = \epsilon_{ab}H_1^a H_2^b$, which is antisymmetric under Higgs exchange; the two derivatives on the $S_{11}$ line produce an additional factor $p_{S_{11}}^2 = (p_e + p_d + p_{H_1})^2 = 2(p_e + p_d)\cdot p_{H_1} + 2 p_e\cdot p_d$. Symmetrization of the full amplitude then gives an amplitude proportional to
    \begin{align}
        \epsilon_{ab}H_1^a H_2^b (p_e + p_d + p_{H_1})^2 &+ \epsilon_{ab}H_2^a H_1^b (p_e + p_d + p_{H_2})^2 \nonumber \\
        &\qquad= 2\epsilon_{ab}H_1^a H_2^b (p_e + p_d)_\mu (p_{H_1} - p_{H_2})^\mu \,, \label{eq:momentum_symmetrization}
    \end{align}
    which is not zero thanks to the momentum factor. This is equivalent to the position-space argument that while $(HH)_{\vec{1}} = 0$, the derivative version $(H D^2 H)_{\vec{1}} $ need not vanish. In fact, the above makes it clear that the relevant object is not $(H D^2 H)_{\vec{1}} $ but rather $(H D^\mu H)_{\vec{1}} $, i.e.~the generated operator is $(QQ)_{\vec{1}} (H D^\mu H)_{\vec{1}} D_\mu (d e)$. 
    
    A non-zero symmetrized amplitude is proof that the diagram leads to the operator one obtains by collecting all external fields and internal derivatives.\footnote{Our code performs the symmetrization over identical Higgs and anti-Higgs, but the resulting amplitude is often too complicated to check analytically. Instead, we substitute random rational numbers for the field components, as well as for the momenta, subject to the momentum rules laid out above. If the result is non-zero, so is the amplitude; only if the result is zero for several different random numbers do we check that the analytic expression is indeed zero, implying a diagram with Higgs asymmetry in need of additional derivatives.} On the other hand, a vanishing amplitude really only shows that the partial-derivative piece inside the covariant derivative drops out, whereas the non-trivial gauge matrices inside $D_\mu$ could still lead to a non-zero operator. Luckily, at least for the cases relevant to this article, we can perform the following check for diagrams involving derivatives that lead to vanishing amplitudes: cut the diagram on all lines that carry derivatives and check if any of these non-derivative sub-diagrams vanish due to Higgs asymmetry. If yes, the entire diagram or resulting operator is necessarily zero, including the gauge pieces of the covariant derivatives we have neglected so far. If none of the subdiagrams vanished we would need to investigate further, but in our analysis this only happens for operators involving four Higgs and can be traced back to the equality $(D_\mu H^4)_{\vec{1}} =0$ explained in the introduction, further elaborated on in Sec.~\ref{sec:higher_dims}.

    Notice that in principle the amplitude should also be antisymmetrized over identical fermions, but the three available fermion generations are sufficient to avoid a complete cancellation for the operators under consideration in this article, allowing us to skip this rather tedious step.
\end{enumerate}
The practical workflow is as follows: find the internal quantum numbers for some combination of topology and non- or one-derivative operator $\mathcal{O}$, i.e.~construct a valid Feynman diagram; if the symmetrized amplitude is non-zero, this is a valid UV completion for $\mathcal{O}$; if the amplitude vanishes, we check if we can obtain a non-vanishing amplitude by inserting $D^2$ on any of the internal lines; if yes, we have found a non-trivial UV completion for $D^2\mathcal{O}$, i.e.~one that does not induce the leading $\mathcal{O}$. If the amplitude is still zero no matter where we insert the additional $D^2$, we try putting a $D^2$ on two different lines, corresponding to $D^4\mathcal{O}$, etc. This is an efficient way of finding many non-trivial UV completions for derivative operators, up to two caveats:
\begin{itemize}
    \item Operators that do not have lower-derivative analogs. We found the two-derivative UV completion in Fig.~\ref{fig:Der_Examples} (right) by starting from the non-derivative operator $dQQeHH$ and splitting UV completions into symmetric and antisymmetric pieces. Some operators that cannot be reached in this way because the removal of two or more derivatives would make the entire operator vanish, independent of the UV completion. An example was already provided in the introduction: the $d=10$ operator $DDHHddue$ cannot be UV completed by looking for UV completions of $HHddue$ that suffer from Higgs asymmetry, because $HHddue = 0$ always and is thus not listed by \texttt{Sym2Int}. These somewhat unique operators can easily be obtained from \texttt{Sym2Int} though by looking for operators $D^k \mathcal{O}$ without $D^{k-2} \mathcal{O}$ counterparts. (Removing an \textit{odd} number of derivatives cannot give a viable operator because of Lorentz symmetry.)
    We call operators that do not exist once any derivatives are removed \textbf{minimal-derivative operators}, which trivially include all one-derivative operators and only a few with more derivatives, see Fig.~\ref{fig:minimal_derivative_operator_numbers}. These are the subset of interesting derivative BNV operators that can already be identified at the SMEFT level, i.e.~without going through UV completions. 
    Once identified, these operators can be UV completed like all others, simply by pretending that $HHddue$ is a viable non-derivative operator, feeding it into the code of Ref.~\cite{Heeck:2026dmh} to obtain the quantum numbers of the mediators, and then placing $D^2$ on internal lines until we find a non-zero amplitude.
    \item Diagrams asymmetric under Higgs exchange can potentially be saved by inserting $D^2$ into any internal line, but there is another possibility: inserting $\slashed{D}$ into two different internal fermion lines. The chirality flips typically change the quantum numbers in the diagram, resulting in a new UV completion. In addition to studying the Higgs asymmetry cases, it is hence necessary to separately construct UV completions with single derivatives placed on fermion lines. Our modified code allows for exactly this, going through all permutations of chirality flips.
\end{itemize}

\begin{figure}
    \includegraphics[width=0.95\textwidth]{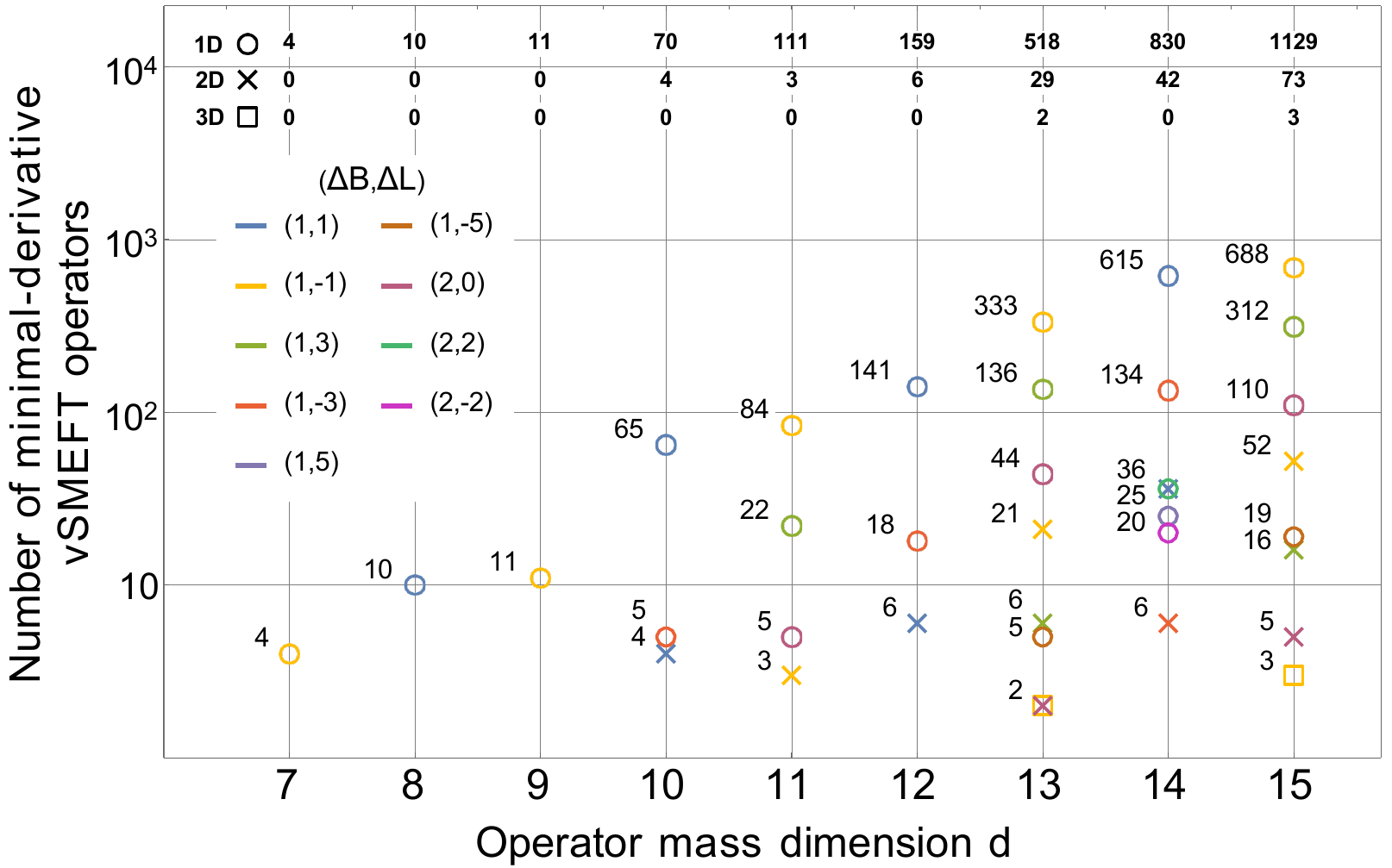}\\[3ex]
    \includegraphics[width=0.95\textwidth]{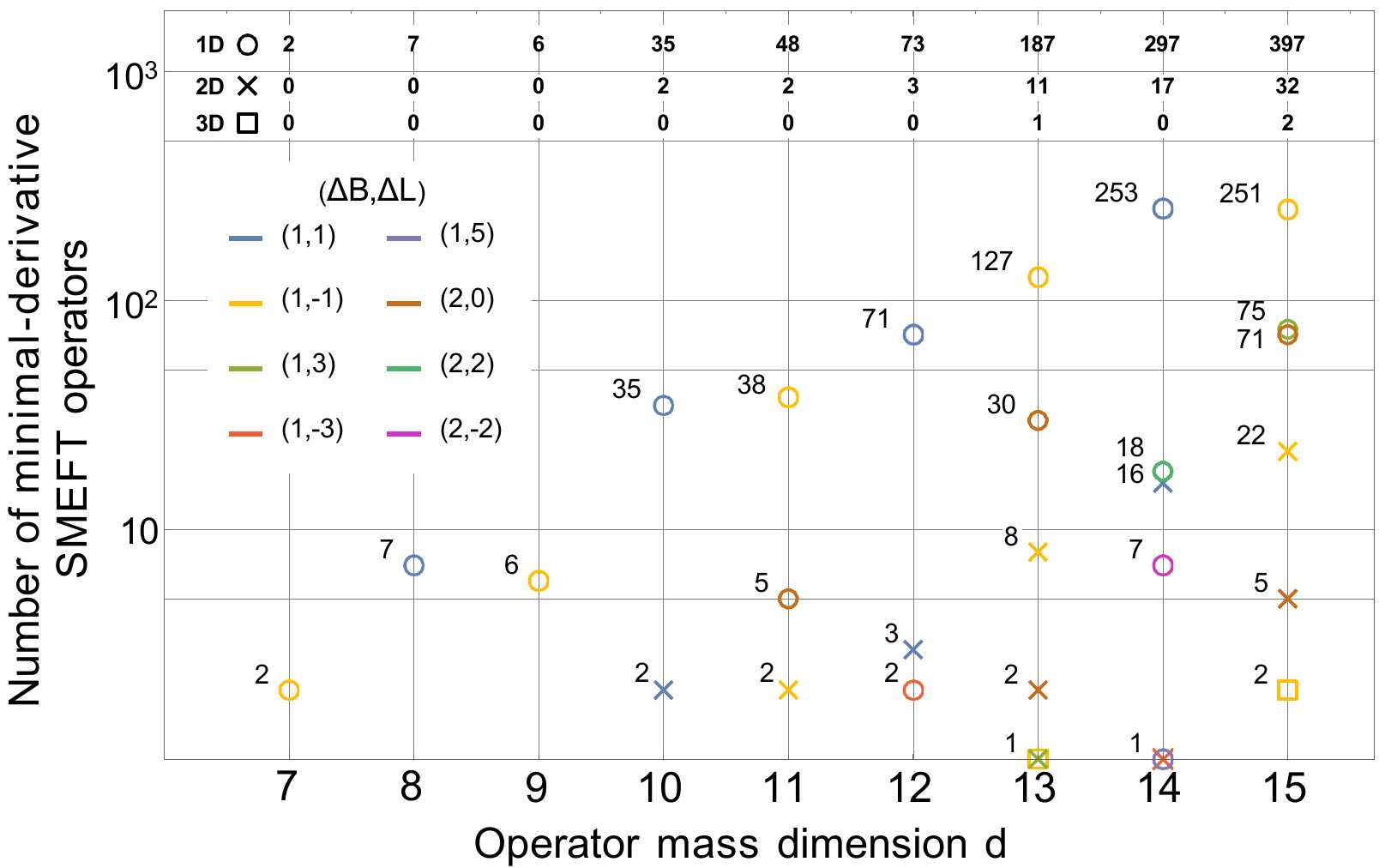}
       \caption{Number of \textit{minimal-derivative} BNV operators in the $\nu$SMEFT (top) and SMEFT (bottom) as a function of mass dimension $d$, grouped according to $(\Delta B,\Delta L)$ (colors) and the number of derivatives (shapes). 
       The total number of one-, two-, and three-derivative operators at each $d$, summed over all $(\Delta B,\Delta L)$ sectors, are given at the top of each plot. 
       }
\label{fig:minimal_derivative_operator_numbers}
\end{figure}

\begin{figure}[tb]
    \includegraphics[width=0.95\textwidth]{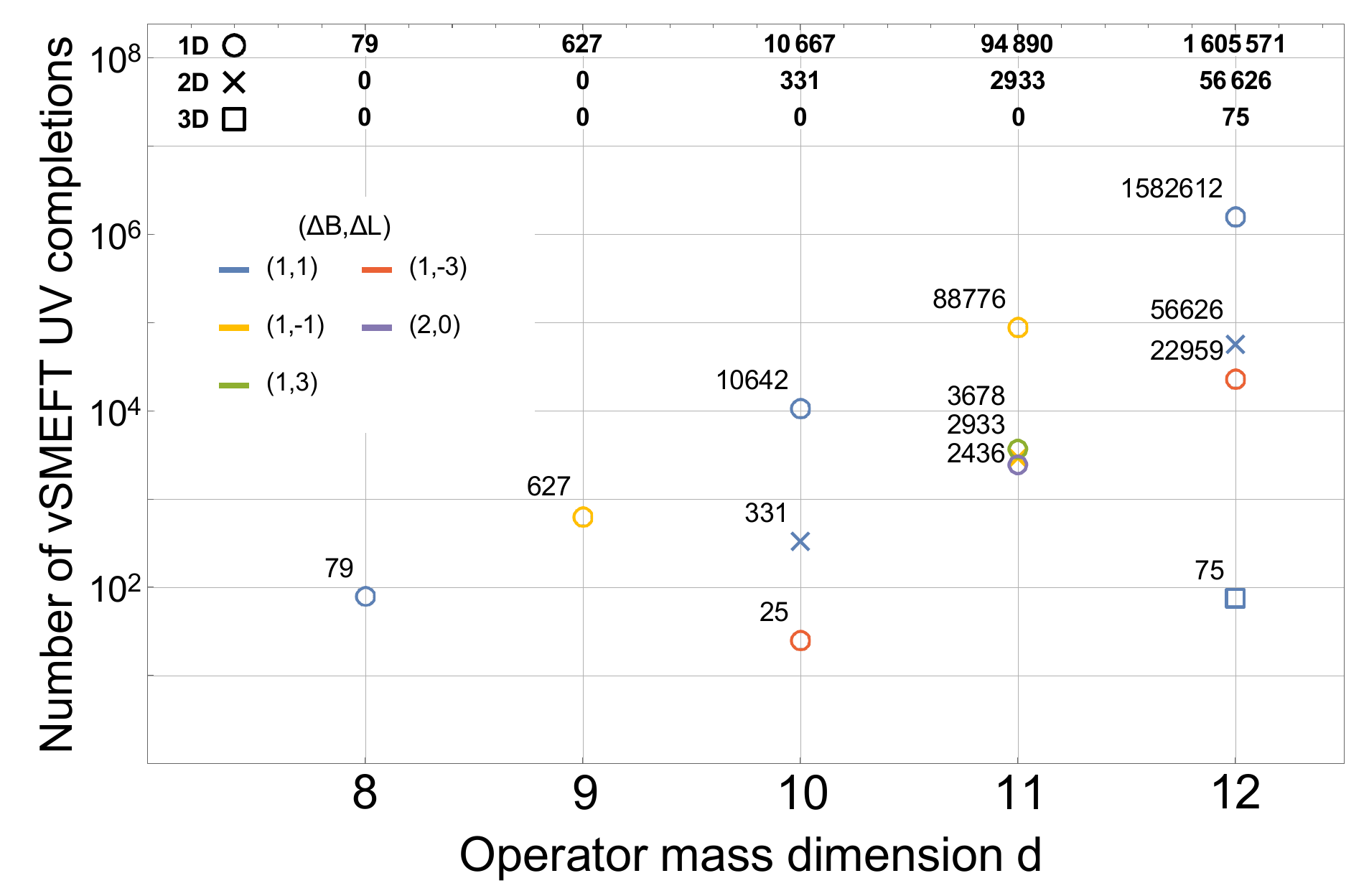}\\[3ex]
    \includegraphics[width=0.95\textwidth]{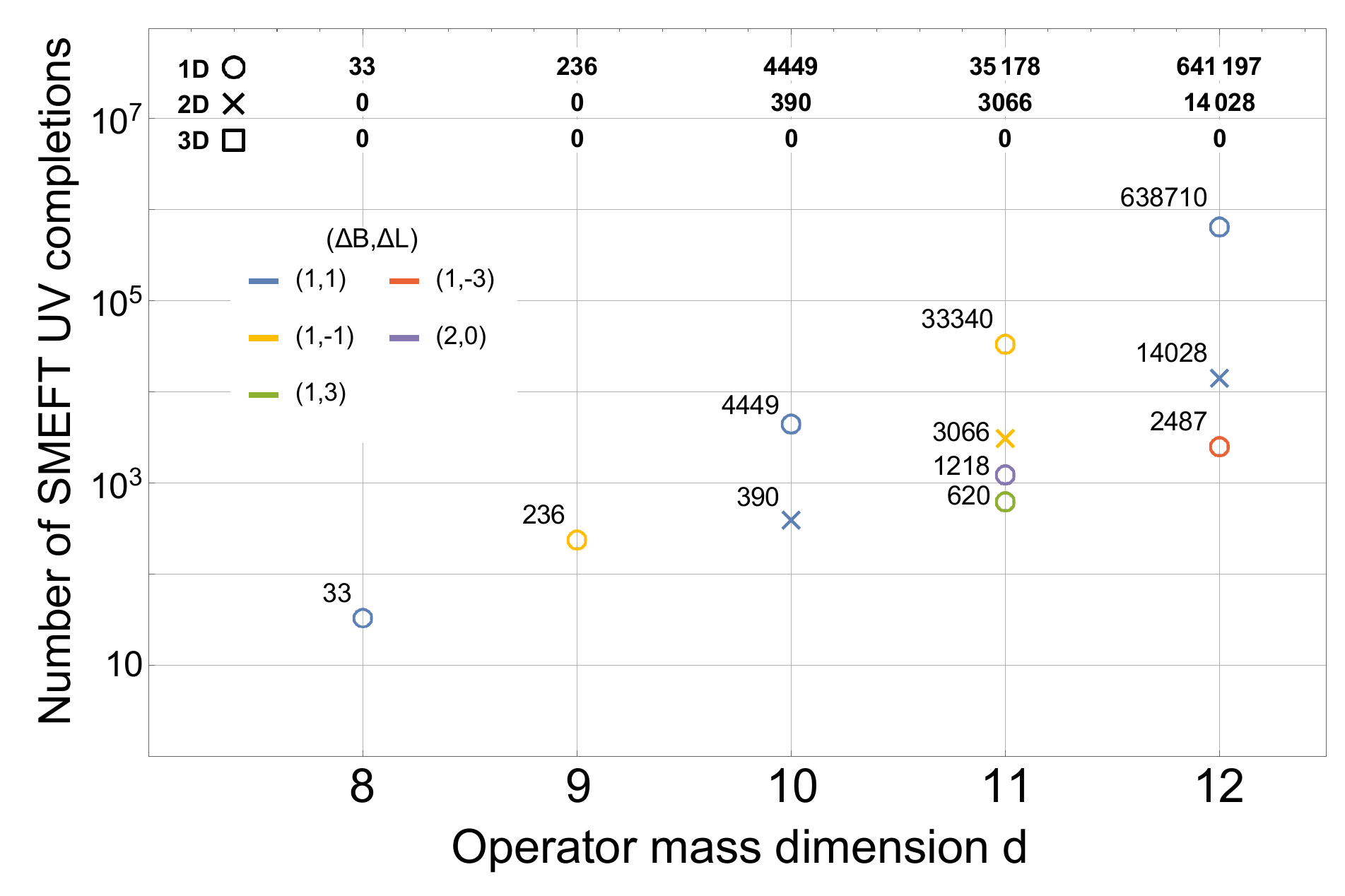}
       \caption{The number of derivative BNV UV completions in the $\nu$SMEFT (top) and SMEFT (bottom) constructed in this article, categorized by mass dimension $d$ (horizontal axis) and ($\Delta B$,$\Delta L$) (colors). The numbers at the very top give the \emph{total} number of UV completions for each mass dimension split by number of derivatives.
       }
    \label{fig:Number_of_UV_completions_Der}
\end{figure}

We followed the above procedure to find all \emph{non-trivial} tree-level UV completions for BNV $\nu$SMEFT operators up to and including mass dimension $d=12$. These supplement the trivial and phenomenologically uninteresting UV completions one obtains by expanding the non-derivative UV completions of Ref.~\cite{Heeck:2026dmh} to higher order in the heavy-mass expansion. We present all the UV completions up to mass dimension $d=9$ in the main text, and only a small selection of UV completions for mass dimensions $d=10$--$12$ in the appendices to keep the length of the paper manageable. For $d\geq10$, we present UV completions for one-derivative symmetry-protected operators, i.e.~operators with $(\Delta B,\Delta L)$ combinations not realized by lower-dimensional derivative BNV operators; two-derivative minimal-derivative operators, i.e.~operators without lower-derivative counterparts; and all three-derivative operators. Our code and the complete set of UV completions can be found on \website. Overall, we have constructed more than a million non-trivial UV completions, see Fig.~\ref{fig:Number_of_UV_completions_Der}.

\clearpage

\section{Field strength tensors}
\label{sec:field_strength_tensors}

Before delving into our lists of UV completions, let us discuss the special subset of derivative operators that contain $SU(3)_C\times SU(2)_L\times U(1)_Y$ field strength tensors $F^{\mu\nu}_{C,L,Y}$. We will argue that UV completions of such BNV operators are automatically included in our list of derivative UV completions and need not be constructed separately. BNV operators involving  $F_{C,L,Y}$ arise for $d\geq 8$, e.g.
\begin{align}
    QQQL F_C\,, \quad QQQL F_L\,,\quad QQQL F_Y\,.
\end{align}
Field strength tensors can be expressed as the commutator of covariant derivatives via
\begin{align}
    [ D^\mu, D^\nu] = -\ii \sum_{j=C,L,Y} g_j T_j^A F^{\mu\nu,A}_j \,,
    \label{eq:commutator}
\end{align}
where $g_j$ are the gauge couplings and $T^A_j$ the appropriate generator matrices for group $j$, with index $A$ running over the number of generators. BNV operators involving field strengths can then be interpreted as derivative operators with a particular contraction. This relation between field-strength and two-derivative operator classes, including its basis dependence, has been discussed more generally in the tree/loop classification of $d=8$ operators in Ref.~\cite{Craig:2019wmo}. Since we work with simplified operators that do not specify contractions, our operator $\mathcal{O}_{8, \, (1,1,2)}^{3} = DD QQQL$ from Sec.~\ref{sec:d=8} also contains $F QQQL$. At the simplified-operator level, we used \texttt{Sym2Int} to verify that every BNV $F\mathcal{O}$ class has a $D^2\mathcal{O}$ counterpart up to at least $d=17$.\footnote{The converse is false. For example, $DDHHHHddde$ exists, but $F_LHHHHddde$ does not.}

Readers familiar with the SMEFT  might be surprised by this, since there are well-known $d=6$ operators such as the dipole $\bar{L} H e F$ that are not accompanied by a two-derivative operator $D^2 \bar{L} H e$~\cite{Grzadkowski:2010es}; this operator $D^2 \bar{L} H e$ was eliminated from the $d=6$ basis using the equations of motion and integration by parts, possible because there are fewer than four particles in the operator~\cite{Craig:2019wmo}. (In momentum space, $D^2 \bar{L} H e$ simply vanishes since all scalar products of momenta $p_i \cdot p_j=0$ due to three-particle kinematics.) In our BNV setup, every operator contains at least four fermions, which in turn saves operators such as $D^2 QQQL$ from being redundant.
In short, simplified BNV field strength operators need not be treated separately but are included in derivative operators. Below, we will illustrate that UV completions for $D^2\mathcal{O}$ generically also generate $F \mathcal{O}$, taking the $D^2 \to F$ correspondence even further.

Let us return to the UV completion of $uude$ from Eq.~\eqref{eq:Leff_S1}, which included a $D^2$  piece that we can further expand as follows using the product rule:
\begin{align}
     \frac{\epsilon_{\alpha\beta\gamma} (z\,\bar{u}^{C}_\alpha d_\beta) D^2 (y\,\bar{u}^C_\gamma e)}{m_{\mathcal{S}_1}^4} &= \frac{\epsilon_{\alpha\beta\gamma} (z\,\bar{u}^{C}_\alpha d_\beta) y}{m_{\mathcal{S}_1}^4} \left[(D^2\bar{u}^C_\gamma) e +\bar{u}^C_\gamma (D^2 e) +2 (D_\mu\bar{u}^C_\gamma) (D^\mu e) \right] .
     \label{eq:Dsqexample}
\end{align}
Now, $D^2 e$ and $D^2 u$ can be rewritten using the equations of motion, for example
\begin{align}
D^2 e &= \left(  \frac{\ii}{2} \sigma_{\mu\nu} [D^\mu,D^\nu] + \slashed{D}^2\right) e \label{eq:DDe}\\
&= \frac12 \sigma_{\mu\nu}  g_Y Y F^{\mu\nu}_Y \, e-\ii \slashed{D}(\ii\slashed{D} e)\\
&=  \frac12\sigma_{\mu\nu}  g_Y (-1) F^{\mu\nu}_Y \, e -\ii \slashed{D}(y_e H^\dagger L)\\
&= -\frac12\sigma_{\mu\nu}  g_Y F^{\mu\nu}_Y \, e-y_e H^\dagger \ii \slashed{D}(L) -\ii y_e D_\mu(H^\dagger) \gamma^\mu L  \\
&= -\frac12\sigma_{\mu\nu}  g_Y F^{\mu\nu}_Y \, e - y_e^2 H^\dagger H \, e  -\ii y_e (D_\mu H^\dagger) \gamma^\mu L
\end{align}
where $\sigma_{\mu\nu}  \equiv \ii\, [\gamma_\mu,\gamma_\nu]/2$. Inserting this expression, and the analogous one for $D^2 u$, into  Eq.~\eqref{eq:Dsqexample} then indeed yields $uude F_Y$ and $uude F_C$. (For $uude F_Y$, the $e$ and $u$ contributions add up and lead to a prefactor $Y(e)+Y(u) = -1+2/3 = -1/3$.) This illustrates how $D^2$ pieces in a position-space operator can lead to field strength tensors. 
Notice that the field strengths did not \textit{replace} the $D^2$ but merely \textit{accompany} them, since the last term in Eq.~\eqref{eq:Dsqexample} remains as a genuine $D^2$ piece.

\begin{figure}
    \includegraphics[width=0.95\textwidth]{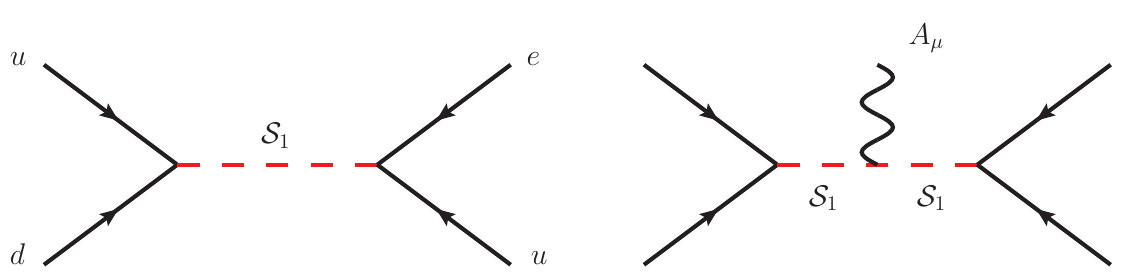}
       \caption{Left: UV completion of the $d=6$ operator $uude$ by the leptoquark $\mathcal{S}_1$, leading to Eq.~\eqref{eq:Leff_S1}, i.e.~a $d=6$ operator $uude/m_{\mathcal{S}_1}^2$.
       Right: emitting a gauge boson $A_\mu$ (hypercharge or color) from the $\mathcal{S}_1$ line generates the $d=8$ operators $g_{Y,C}F_{Y,C}uude/m_{\mathcal{S}_1}^4$.
       }
\label{fig:uude_UV_completion_S1}
\end{figure}

A complementary picture emerges in momentum space through Feynman diagrams, starting again from the $\mathcal{S}_1$ UV-completion of $uude$, illustrated in Fig.~\ref{fig:uude_UV_completion_S1} (left). We had already shown in Eq.~\eqref{eq:Leff_S1} how the propagator expansion in small $1/m_{\mathcal{S}_1}$ generates the chain of operators $D^{2n} uude$. To get field strength tensors, we just need to attach a gluon or hypercharge gauge boson line to the internal propagator, Fig.~\ref{fig:uude_UV_completion_S1} (right),  the Feynman rules for which exist since $\mathcal{S}_1$ is charged under both groups. The Feynman rule for the vertex provides a gauge-coupling factor as well as a momentum, and the two $\mathcal{S}_1$ propagators produce an overall $1/m_{\mathcal{S}_1}^4$, exactly matching the $d=8$ operators $uude F$ we derived above in position space.\footnote{In both pictures,  an $SU(2)_L$ field strength tensor cannot be generated since all particles involved are $SU(2)_L$ singlets, which matches the fact that $uude F_L$ is not an allowed operator, again because the $SU(2)_L$ indices of $F_L$ cannot be contracted with anything else. \textit{Two} field strength tensors, on the other hand, are allowed as a SMEFT operator, e.g.~$uude F_L F_L$, but cannot be generated at tree level because none of the particles involved can emit an $SU(2)_L$ gauge boson, let alone two.} (Here, the prefactor $1/3$ for $uude F_Y$ comes directly from $Y(\mathcal{S}_1)=1/3$.)

An even cleaner connection between $D^2$ and $F$ arises when we consider internal \textit{fermions}. Performing the large-mass expansion of Eq.~\eqref{eq:fermion_propagator} in position space gives~\cite{Gargalionis:2020xvt}
\begin{align}
    \frac{1}{\ii \slashed{D} - m_F} &= -\frac{1}{m_F} - \frac{\ii \slashed{D} }{m_F^2}- \frac{(\ii \slashed{D} )^2}{m_F^3} + \ldots\nonumber\\
    &= -\frac{1}{m_F} - \frac{\ii \slashed{D} }{m_F^2}+\frac{D^2-\frac{\ii}{2} \sigma^{\mu\nu} [D_\mu,D_\nu]}{m_F^3} + \ldots\,,\label{eq:fermion_propagator_positionspace}
\end{align}
which again contains field strength tensors through Eq.~\eqref{eq:commutator} together with $D^2$. In terms of Feynman diagrams, we can again emit gauge bosons from any internal fermion line that is charged under that gauge symmetry, turning a tree-level operator $\mathcal{O}$ into $g_jF_j\mathcal{O}/m_F^2$.
Both in the scalar and fermion mediator case one can hence turn a $D^2$ operator into a $D^2 +F$ combination, all being subleading corrections to a lower-derivative operator in general.

The Feynman diagram picture is particularly helpful, since it not only illustrates how one can obtain tree-level UV completions of field strength BNV operators by simply emitting gauge bosons from internal lines, but immediately shows that all such operators are suppressed compared to the operator without field strength tensor, and that the UV completions are identical. Just like we argued in the introduction that $D^{2n}\mathcal{O}$ operators are generically UV-completed by the same particles that generate $\mathcal{O}$ and therefore doomed to be subleading corrections, the same holds for operators with field strength tensors, should they even have tree-level UV completions, as already argued in Ref.~\cite{Gargalionis:2020xvt}.
And just like for the $D^{2n}$ cases, there are in principle cases for which this generic argument fails, potentially leading to dominant field strength operators. Fine-tuned cancellations aside, interesting BNV operators $F\mathcal{O}$ require a UV completion for which $\mathcal{O} = 0$ due to Higgs asymmetry; the leading operators are then $D^2\mathcal{O}$, which can sometimes be accompanied by $F\mathcal{O}$ using the steps from above. This indeed works when the $D^2$ that eliminates the Higgs asymmetry come from a \textit{fermion} propagator, using Eq.~\eqref{eq:fermion_propagator_positionspace}. For cases with $D^2$ on \textit{scalar} lines, the Higgs asymmetry that cancels the leading term also cancels the field-strength term, leaving a genuine $D^2$ operator. As an example, consider again Fig.~\ref{fig:Der_Examples} (right), a UV completion for $DD QQ ed HH$ via scalar mediators, ultimately producing the leading term $(QQ)_{\vec{1}} H_i \epsilon^{ij} D^2( H_j e d)$; using the product rule, we find terms with factors $D^2 e$ and $D^2 d$ that would produce field strength tensors like in Eq.~\eqref{eq:DDe}, but they are all multiplied by the vanishing $(HH)_{\vec{1}} = H_i \epsilon^{ij} H_j = 0$. The surviving derivative pieces instead take the form $\left[ (QQ){\overset{\leftrightarrow}{D}}{}^\mu (ed)\right] (H_i \epsilon^{ij} D_\mu H_j)$, matching exactly the momentum-space expression of Eq.~\eqref{eq:momentum_symmetrization}. While this is indeed a dominant $D^2$ operator, it cannot be rewritten as a field strength operator; the $(H_i \epsilon^{ij} D_\mu H_j)$ piece still contains a gauge boson term $\sum_{I=1,2,3}H \epsilon T^I H A^I_\mu$, with $HH$ coupled to an $SU(2)_L$ triplet, that describes the process of emitting a gauge boson from the $S_{11}$ mediator, just not in the form of a field strength tensor. In general, $D^2\mathcal{O}$ operators with $\mathcal{O}=0$ and $D^2$ from an internal scalar line do not lead to $F\mathcal{O}$ operators because those vanish by the same Higgs asymmetry that led to $\mathcal{O}=0$; putting $D^2$ on internal fermion lines, on the other hand, does induce field strength operators by means of Eq.~\eqref{eq:fermion_propagator_positionspace}, but only ever \textit{accompanies} $D^2$ structures, never replaces them, and ultimately depend on the chosen basis~\cite{Craig:2019wmo}.

The UV completions for $D^2$ and $D^3$ operators presented in this article therefore contain those that generate BNV operators with one field strength. Similarly, if even $D^2\mathcal{O}$ vanishes due to Higgs asymmetry, the leading operator would be $D^4\mathcal{O}$, which also includes pieces with two field strength operators, although this first occurs at $d=14$. 
The discussion so far shows that, as far as UV completions are concerned, field strength tensors are parasitic to $D^2$. We have not found any cases in our data set that could lead to dominant field strength operators, i.e.~$F\mathcal{O}$ with $\mathcal{O}=0$ and $D^2\mathcal{O}=0$, and conjecture that this remains true for $d>12$.

 \section{Dimension 7: \texorpdfstring{$\Delta B=1, \Delta L=-1$}{Delta B=1,Delta L=-1}}
 \label{sec:d=7}

 Having discussed our procedure for UV-completing derivative operators, we present our results for the lowest-dimensional  BNV operators, $\mathcal{O}_{d, \, (\Delta B,\Delta L,n)}^{j}$. These first arise at mass dimension $d=7$, and carry $\Delta B=1= - \Delta L$, just like in the non-derivative case:
 \begin{multicols}{3}
 \begin{spacing}{1.5}
 \noindent
\begin{math}
\mathcal{O}_{7, \, (1,-1,1)}^{1}\equiv Dddd\bar{e}\,,\\
\mathcal{O}_{7, \, (1,-1,1)}^{2}\equiv DddQ\bar{L}\,,\\
\mathcal{O}_{7, \, (1,-1,1)}^{3}\equiv DdQQ\bar{\nu}\,,\\
\mathcal{O}_{7, \, (1,-1,1)}^{4}\equiv Dddu\bar{\nu}\,.
\label{eq:1Ddequal7basisB1L-1}
 \end{math}
 \end{spacing}
 \end{multicols}
 \noindent
The two SMEFT operators for dimension 7 have already been found and discussed by Weinberg~\cite{Weinberg:1980bf}, and a minimal basis has been constructed in Ref.~\cite{Liao:2016hru}.
Their nucleon-decay phenomenology has recently been revisited in Refs.~\cite{Liao:2026ugl,Arbelaez:2026edo}.
The underlying four-fermion topology for all four operators can only arise via boson exchange in a renormalizable theory, which in turn cannot lead to the desired single derivative. There are hence no tree-level UV completions for these $d=7$ BNV operators~\cite{Beltran:2023ymm}, although they can of course be generated at loop level from other operators with the same $(\Delta B,\Delta L)$.

 \section{Dimension 8: \texorpdfstring{$\Delta B=1, \Delta L=1$}{Delta B=1,Delta L=1}}
 \label{sec:d=8}

At mass dimension $d=8$, there are one-derivative,
 \begin{multicols}{3}
 \begin{spacing}{1.5}
 \noindent
 \begin{math}
\mathcal{O}_{8, \, (1,1,1)}^{1}\equiv D\bar{H}uQQL\,,\\
\mathcal{O}_{8, \, (1,1,1)}^{2}\equiv D\bar{H}duuL\,,\\
\mathcal{O}_{8, \, (1,1,1)}^{3}\equiv D\bar{H}uuQe\,,\\
\mathcal{O}_{8, \, (1,1,1)}^{4}\equiv DHdQQL\,,\\ 
\mathcal{O}_{8, \, (1,1,1)}^{5}\equiv DHQQQe\,,\\
\mathcal{O}_{8, \, (1,1,1)}^{6}\equiv DHdduL\,,\\ 
\mathcal{O}_{8, \, (1,1,1)}^{7}\equiv DHduQe\,,\\
\mathcal{O}_{8, \, (1,1,1)}^{8}\equiv D\bar{H}QQQ\nu\,,\\
\mathcal{O}_{8, \, (1,1,1)}^{9}\equiv D\bar{H}duQ\nu\,,\\
\mathcal{O}_{8, \, (1,1,1)}^{10}\equiv DHddQ\nu\,,
\label{eq:1Ddequal8basisB1L1}
 \end{math}
 \end{spacing}
 \end{multicols}
\noindent
and two-derivative $\nu$SMEFT operators,
 \begin{multicols}{3}
 \begin{spacing}{1.5}
 \noindent
 \begin{math}
\mathcal{O}_{8, \, (1,1,2)}^{1}\equiv DDduQL\,,\\
\mathcal{O}_{8, \, (1,1,2)}^{2}\equiv DDQQue\,,\\
\mathcal{O}_{8, \, (1,1,2)}^{3}\equiv DDQQQL\,,\\
\mathcal{O}_{8, \, (1,1,2)}^{4}\equiv DDduue\,,\\ 
\mathcal{O}_{8, \, (1,1,2)}^{5}\equiv DDQQd\nu\,,\\
\mathcal{O}_{8, \, (1,1,2)}^{6}\equiv DDdud\nu\,,
\label{eq:2Ddequal8basisB1L1}
 \end{math}
 \end{spacing}
 \end{multicols}
\noindent
all carrying $\Delta B = \Delta L = 1$. The two-derivative operators $D^2\mathcal{O}$ are all UV-completed through the leptoquarks that generate the non-derivative $\mathcal{O}$~\cite{Heeck:2026dmh}, and hence always subdominant and not shown here.
The one-derivative operators are more interesting because there is a one-scalar--four-fermion topology that contains an internal fermion, $\mathcal{T}_7^2$ (see Fig.~\ref{fig:d=7_topologies}), and can generate the single derivative. We collect the UV completions for all ten one-derivative operators in Tab.~\ref{tab:T8Top2B1L1D1}; they agree with Ref.~\cite{Li:2023pfw}.

\begin{center}
\begin{longtable}{| c | m{13cm} |}
   \hline
   $\mathcal{O}^j(\mathcal{T}_{7}^{2})$ & New Particles \\
   \hline
   \hline
   \endhead
   \multicolumn{2}{|c|}{\hspace{9cm} Continued on next page} \\
   \endfoot
  $\mathcal{O}_{8,(1,1,1)}^{10}$ & {$S_{2}F_{6}$, $S_{2}F_{8}$, $S_{3}F_{3}$, $S_{3}F_{6}$} \\
   \hline
   \caption{Dimension 8 UV completions for topology $\mathcal{T}_{7}^{2}$ (see Fig.~\ref{fig:d=7_topologies}).}
   \label{tab:T8Top2B1L1D1}\\
   \endlastfoot
  $\mathcal{O}_{8,(1,1,1)}^{1}$ & {$S_{4}F_{9}$, $S_{2}F_{5}$, $S_{4}F_{11}$, $S_{2}F_{2}$, $S_{4}F_{8}$, $S_{2}F_{8}$} \\
   \hline
  $\mathcal{O}_{8,(1,1,1)}^{2}$ & {$S_{2}F_{2}$, $S_{1}F_{2}$, $S_{2}F_{8}$, $S_{1}F_{7}$} \\
   \hline
  $\mathcal{O}_{8,(1,1,1)}^{3}$ & {$S_{2}F_{5}$, $S_{2}F_{8}$, $S_{1}F_{10}$, $S_{1}F_{5}$} \\
   \hline
  $\mathcal{O}_{8,(1,1,1)}^{4}$ & {$S_{4}F_{14}$, $S_{2}F_{6}$, $S_{4}F_{4}$, $S_{2}F_{1}$, $S_{4}F_{8}$, $S_{2}F_{8}$} \\
   \hline
  $\mathcal{O}_{8,(1,1,1)}^{5}$ & {$S_{4}F_{3}$, $S_{2}F_{3}$, $S_{4}F_{14}$, $S_{2}F_{6}$} \\
   \hline
  $\mathcal{O}_{8,(1,1,1)}^{6}$ & {$S_{3}F_{1}$, $S_{2}F_{1}$, $S_{3}F_{13}$, $S_{2}F_{8}$} \\
   \hline
  $\mathcal{O}_{8,(1,1,1)}^{7}$ & {$S_{1}F_{6}$, $S_{2}F_{6}$, $S_{1}F_{13}$, $S_{2}F_{8}$, $S_{2}F_{3}$} \\
   \hline
  $\mathcal{O}_{8,(1,1,1)}^{8}$ & {$S_{4}F_{3}$, $S_{2}F_{3}$, $S_{4}F_{9}$, $S_{2}F_{5}$} \\
   \hline
  $\mathcal{O}_{8,(1,1,1)}^{9}$ & {$S_{2}F_{5}$, $S_{3}F_{5}$, $S_{2}F_{8}$, $S_{3}F_{7}$, $S_{2}F_{3}$} \\
   \hline
\end{longtable}
\end{center}

The UV completions in Tab.~\ref{tab:T8Top2B1L1D1}  generate the one-derivative operators $\mathcal{O}_{8, \, (1,1,1)}^{j}$, which are then the lowest-dimensional tree-level derivative BNV operators that do not have a non-derivative counterpart. This makes them interesting starting points for BNV in principle. However, all UV completions contain a scalar $S_{1,2,3,4}$, which by itself generically induces non-derivative $d=6$ operators~\cite{Heeck:2026dmh}, that are expected to dominate the phenomenology. This is not guaranteed though, since the $d=6$ operators depend on two different $S_j$ couplings to SM fermions, whereas the $d=8$ operators require one $S_j$ coupling to SM fermions and one coupling to a heavy $F_j$, so in principle one can suppress the $d=6$ operators by making the unneeded $S_j$ coupling small. Since there is no symmetry that could eliminate this coupling, this amounts to significant fine-tuning, at least within $\nu$SMEFT.

Restricting ourselves to \textit{SMEFT}, however, we note that $S_3$ ($\bar{\mathcal S}_1$ in the notation of Refs.~\cite{Buchmuller:1986zs,Dorsner:2016wpm}) by itself does not induce any BNV operators because it is a pure diquark, so the UV completions of SMEFT operators in Tab.~\ref{tab:T8Top2B1L1D1} involving $S_3$ could indeed generate dominantly derivative BNV operators without fine-tuning. Only two candidates emerge, both UV completions for $\mathcal{O}_{8, \, (1,1,1)}^{6}$: $S_3 F_1$ and $S_3 F_{13}$; the former generically induces the non-derivative $d=7$ operator $\mathcal{O}_{7, \, (1,-1)}^{3} = \bar{H}ddu\bar{L}$, which should be dominant over $\mathcal{O}_{8, \, (1,1,1)}^{6}$, but the latter is indeed a simple UV completion that dominantly induces the one-derivative $d=8$ operator $\mathcal{O}_{8, \, (1,1,1)}^{6}$ at tree level (Fig.~\ref{fig:Der_Examples} (left)), without any fine-tuning or imposition of symmetries! 
The relevant Lagrangian terms for $S\equiv S_3$ and $F\equiv F_{13}$ are
\begin{align}
    \L \ \supset \  \frac{1}{2} y_{ab} \epsilon_{\alpha\beta\gamma} \overline{(d^\alpha_a)}^C P_R d^\beta_b S^\gamma + \lambda_l S^\dagger_\alpha \overline{(L_l^i)}^C \epsilon_{ij} P_L F^{\alpha\,j} + \eta_r \bar{F}_\alpha H P_R u^\alpha_r + \hc \,,
\end{align}
with color indices $\alpha$, $\beta$, $\gamma$ and $SU(2)_L$ indices $i$ and $j$. The down-quark coupling is antisymmetric in flavor space, $y_{ab} = - y_{ba}$. Assuming $S$ and $F$ to be degenerate for simplicity, $m_F = m_S \equiv M$, the diagram from Fig.~\ref{fig:Der_Examples} (left) leads to the following effective operator with suppressed group contractions
\begin{align}
   \L \ &\to \ \frac{y_{ab} \lambda_l \eta_r}{M^4} (d_a d_b) L_l \ii\slashed{D}(H u_r) + \hc\\
   & = \ \ii\frac{y_{ab} \lambda_l \eta_r}{M^4} (d_a d_b) (L_l \gamma^\mu u_r) D_\mu H +\frac{y_{ab} \lambda_l \eta_r}{M^4} (d_a d_b) L_l H ( \ii\slashed{D}u_r) + \hc\\
   & = \ \ii\frac{y_{ab} \lambda_l \eta_r}{M^4} (d_a d_b) (L_l \gamma^\mu u_r) D_\mu H +\frac{y_{ab} \lambda_l \eta_r}{M^4} (d_a d_b) L_l H ( Y_r \tilde{H}^\dagger Q) + \hc \,,
\end{align}
using the equation of motion for $u$ in the last line with SM up-quark Yukawa $Y_r = m_r/\langle H \rangle$, $r\in \{u,c,t\}$, and $\langle H\rangle \simeq \unit[174]{GeV}$. The operator can also be mapped onto Murphy's basis~\cite{Murphy:2020rsh} operators, but the form above is more convenient. The $D_\mu H$ term leads to BNV operators involving heavy SM bosons (Higgs, $Z$, and $W$) after electroweak symmetry breaking, which is of limited phenomenological interest compared to the second term, which gives the non-derivative four-fermion operator 
\begin{align}
\frac{y_{ab} \lambda_l \eta_r}{M^4} (d_a d_b) L_l H ( Y_r \tilde{H}^\dagger Q)\ \to\  \frac{y_{ab} \lambda_l \eta_r Y_r\langle H\rangle^2}{M^4}    (d_a P_R d_b) (u_r P_L \nu_l)
\end{align}
in the low-energy effective field theory (LEFT). This operator is usually denoted by $\mathcal{O}^{S, RL}_{ddu}$ and is noteworthy because it is not induced by $d=6$ SMEFT operators at tree level~\cite{Jenkins:2017jig,Beneito:2023xbk}. Our model is hence a simple UV completion for the $d=8$ operator $dd LQ HH$ that induces this missing LEFT operator. The most promising flavor combination for nucleon decays is $(d_Rs_R)(u_L\nu_L)$, which gives rise to the approximate decay rate~\cite{Beneito:2023xbk}
\begin{align}
 \Gamma (p\to K^+\nu_l)\simeq    \frac{(m_p^2-m_K^2)^2\alpha^2}{32\pi f_\pi^2 m_p^3}\left(\frac{y_{ds} \lambda_l \eta_u m_u\langle H\rangle}{M^4} \right)^2 \simeq \frac{(y_{ds} \lambda_l \eta_u)^2}{\unit[6\times 10^{33}]{yr}}\left(\frac{\unit[3\times 10^7]{GeV}}{M}\right)^8 ,
\end{align}
ignoring renormalization-group running and using the up-quark mass $m_u = \unit[2]{MeV}$. The current Super-K limit on this decay mode~\cite{Super-Kamiokande:2014otb}, $\Gamma (p\to K^+\nu_l) < (\unit[6\times 10^{33}]{yr})^{-1}$, then probes $S$ and $F$ masses up to $\unit[3\times 10^7]{GeV}$ if the couplings are of order one.
However, in addition to this tree-level nucleon-decay contribution, our model also gives rise to the one-loop $d=6$ SMEFT operator $duQL$ upon attaching $H$ to a down quark in Fig.~\ref{fig:Der_Examples} (left), as also pointed out in Ref.~\cite{Gargalionis:2024nij}. Sticking to our flavor choice $(d_Rs_R)(u_L\nu_L)$, we attach $H$ to the strange quark to benefit from the larger SM Yukawa coupling $Y_s\simeq 5\times 10^{-4}$ and eventually find the one-loop $d=6$ LEFT operator
\begin{align}
\frac{y_{ds} \lambda_l \eta_u Y_s}{32\pi^2 M^2}    (d P_R u) (s P_L \nu_l)\,,
\end{align}
which gives rise to the same nucleon decay but with a faster rate, at least if $M> \unit[450]{GeV}$:
\begin{align}
    \frac{ \Gamma (p\to K^+\nu_l)_{\text{loop}}}{\Gamma (p\to K^+\nu_l)_{\text{tree}}} \simeq \left(\frac{m_s  M^2}{32\pi^2m_u\langle H\rangle^2} \right)^2
    \simeq \left(\frac{M}{\unit[450]{GeV}} \right)^4 .
\end{align}
In this particular case, the loop amplitude is not suppressed by any quark mixing angles, seeing as we did not violate strangeness in the loop. Since our colored particles $S$ and $F$ almost certainly need to be heavier than $\unit[450]{GeV}$ to evade LHC constraints -- even if we made the couplings $y_{ds} \lambda_l \eta_u$ small enough to suppress proton decay -- we must conclude that the one-loop contribution gives the largest phenomenological effect and this model should be classified as a genuine loop-realization of the $d=6$ BNV operator $duQL$ (see Refs.~\cite{Helo:2019yqp,Dorsner:2022twk} for other loop realizations of BNV operators).
We denote these situations in which a UV completion generates tree-level BNV operators starting at some $d > 6$ without fine-tuning or symmetries as \textbf{accidentally protected} scenarios, to be investigated thoroughly in Ref.~\cite{Heeck:accidental} now that we have derived all BNV UV completions up to $d=12$.

 \section{Dimension 9: \texorpdfstring{$\Delta B=1, \Delta L=-1$}{Delta B=1,Delta L=-1}}
 \label{sec:d=9}

At $d=9$, all derivative BNV operators carry $\Delta B = - \Delta L =1$, whereas the non-derivative $d=9$ operators could also have $\Delta B = 2$, $\Delta L =0$ and $\Delta B = 1$, $\Delta L =3$~\cite{Heeck:2026dmh}. We have one-derivative operators,
 \begin{multicols}{3}
 \begin{spacing}{1.5}
 \noindent
 \begin{math}
\mathcal{O}_{9, \, (1,-1,1)}^{1}\equiv DH\bar{H}ddd\bar{e}\,,\\
\mathcal{O}_{9, \, (1,-1,1)}^{2}\equiv D\bar{H}\bar{H}dQQ\bar{e}\,,\\
\mathcal{O}_{9, \, (1,-1,1)}^{3}\equiv D\bar{H}\bar{H}ddu\bar{e}\,,\\
\mathcal{O}_{9, \, (1,-1,1)}^{4}\equiv D\bar{H}\bar{H}QQQ\bar{L}\,,\\
\mathcal{O}_{9, \, (1,-1,1)}^{5}\equiv D\bar{H}\bar{H}duQ\bar{L}\,,\\
\mathcal{O}_{9, \, (1,-1,1)}^{6}\equiv DH\bar{H}ddQ\bar{L}\,,\\
\mathcal{O}_{9, \, (1,-1,1)}^{7}\equiv D\bar{H}\bar{H}uQQ\bar{\nu}\,,\\
\mathcal{O}_{9, \, (1,-1,1)}^{8}\equiv D\bar{H}\bar{H}duu\bar{\nu}\,,\\
\mathcal{O}_{9, \, (1,-1,1)}^{9}\equiv DH\bar{H}dQQ\bar{\nu}\,,\\
\mathcal{O}_{9, \, (1,-1,1)}^{10}\equiv DH\bar{H}ddu\bar{\nu}\,,\\
\mathcal{O}_{9, \, (1,-1,1)}^{11}\equiv DHHddd\bar{\nu}\,,
\label{eq:1Ddequal9basisB1L-1}
 \end{math}
 \end{spacing}
 \end{multicols}
 \noindent
two-derivative operators,
 \begin{multicols}{3}
 \begin{spacing}{1.4}
 \noindent
 \begin{math}
\mathcal{O}_{9, \,(1 , -1,2)}^{1} \equiv DD\bar{H}ddQ\bar{e} \,, \\
\mathcal{O}_{9, \,(1 , -1,2)}^{2} \equiv DD\bar{H}dQQ\bar{L} \,, \\
\mathcal{O}_{9, \,(1 , -1,2)}^{3} \equiv DD\bar{H}ddu\bar{L} \,, \\
\mathcal{O}_{9, \,(1 , -1,2)}^{4} \equiv DDHddd\bar{L} \,, \\
\mathcal{O}_{9, \,(1 , -1,2)}^{5} \equiv DDHddQ\bar{\nu} \,, \\
\mathcal{O}_{9, \,(1 , -1,2)}^{6} \equiv DD\bar{H}QQQ\bar{\nu} \,, \\
\mathcal{O}_{9, \,(1 , -1,2)}^{7} \equiv DD\bar{H}duQ\bar{\nu} \,,
 \label{eq:2Ddequal9basisB1L-1}
 \end{math}
 \end{spacing}
 \end{multicols} 
 \noindent
and three-derivative operators, which are closely related to the $d=7$ operators from Sec.~\ref{sec:d=7} and we inherit their numbering by defining $\mathcal{O}_{9, \, (1,-1,3)}^{j} \equiv D^2 \mathcal{O}_{7, \, (1,-1,1)}^{j}$ with $j\in \{1 \isep 4\}$. Just like the underlying $d=7$ operators, there are no tree-level UV completions for these three-derivative operators. An actual operator basis with explicit contractions for all $d=9$ operators can be found in Refs.~\cite{Li:2020xlh,Liao:2020jmn,Li:2021tsq}.

The two-derivative operators $D^2\mathcal{O}$ are UV completed trivially by the UV completions for the non-derivative $d=7$ operators $\mathcal{O}$ from Ref.~\cite{Heeck:2026dmh} and are hence always subleading and not shown here. 
We list the UV completions for the one-derivative operators in tables~\ref{tab:T9Top3B1L-1D1}--\ref{tab:T9Top8B1L-1D1}.
In principle, some UV completions for the one-derivative operators with two $H$ or two $\bar{H}$ could end up asymmetric under Higgs exchange and hence furnish UV completions for three-derivative $d=11$ operators instead; however, the relevant four-fermion--two-scalar topologies involving at least one internal fermion (Fig.~\ref{fig:d=8_topologies}) all have the derivative wedged between the Higgses, automatically evading any and all Higgs asymmetries.

\begin{center}
\begin{longtable}{| c | m{13cm} |}
   \hline
   $\mathcal{O}^j(\mathcal{T}_{8}^{3})$ & New Particles \\
   \hline
   \hline
   \endhead
   \multicolumn{2}{|c|}{\hspace{9cm} Continued on next page} \\
   \endfoot
   \caption{Dimension 9 UV completions for topology $\mathcal{T}_{8}^{3}$ (see Fig.~\ref{fig:d=8_topologies}).}
   \label{tab:T9Top3B1L-1D1}\\
   \endlastfoot
  $\mathcal{O}_{9,(1,-1,1)}^{1}$ & {$S_{3}F_{3}F_{8}$, $S_{3}F_{7}F_{10}$} \\
   \hline
  $\mathcal{O}_{9,(1,-1,1)}^{2}$ & {$S_{4}F_{3}F_{7}$, $S_{2}F_{3}F_{7}$, $S_{6}F_{7}F_{9}$, $S_{6}F_{5}F_{7}$} \\
   \hline
  $\mathcal{O}_{9,(1,-1,1)}^{3}$ & {$S_{3}F_{3}F_{8}$, $S_{2}F_{3}F_{7}$} \\
   \hline
  $\mathcal{O}_{9,(1,-1,1)}^{4}$ & {$S_{4}F_{4}F_{9}$, $S_{2}F_{4}F_{9}$, $S_{4}F_{4}F_{5}$, $S_{4}F_{1}F_{9}$, $S_{2}F_{1}F_{5}$} \\
   \hline
  $\mathcal{O}_{9,(1,-1,1)}^{5}$ & {$S_{2}F_{4}F_{9}$, $S_{2}F_{1}F_{5}$, $S_{5}F_{8}F_{9}$, $S_{5}F_{5}F_{8}$, $S_{6}F_{7}F_{9}$, $S_{6}F_{5}F_{7}$} \\
   \hline
  $\mathcal{O}_{9,(1,-1,1)}^{6}$ & {$S_{3}F_{4}F_{14}$, $S_{3}F_{1}F_{6}$, $S_{5}F_{7}F_{14}$, $S_{5}F_{6}F_{7}$, $S_{3}F_{9}F_{11}$, $S_{3}F_{2}F_{5}$, $S_{5}F_{8}F_{9}$, $S_{5}F_{5}F_{8}$} \\
   \hline
  $\mathcal{O}_{9,(1,-1,1)}^{7}$ & {$S_{4}F_{3}F_{8}$, $S_{2}F_{3}F_{8}$, $S_{5}F_{8}F_{9}$, $S_{5}F_{5}F_{8}$} \\
   \hline
  $\mathcal{O}_{9,(1,-1,1)}^{8}$ & {$S_{2}F_{3}F_{8}$, $S_{1}F_{3}F_{7}$} \\
   \hline
  $\mathcal{O}_{9,(1,-1,1)}^{9}$ & {$S_{4}F_{3}F_{8}$, $S_{2}F_{3}F_{8}$, $S_{4}F_{3}F_{7}$, $S_{2}F_{3}F_{7}$, $S_{5}F_{7}F_{14}$, $S_{5}F_{6}F_{7}$, $S_{5}F_{8}F_{9}$, $S_{5}F_{5}F_{8}$} \\
   \hline
  $\mathcal{O}_{9,(1,-1,1)}^{10}$ & {$S_{3}F_{3}F_{13}$, $S_{2}F_{3}F_{8}$, $S_{3}F_{3}F_{8}$, $S_{2}F_{3}F_{7}$} \\
   \hline
  $\mathcal{O}_{9,(1,-1,1)}^{11}$ & {$S_{3}F_{3}F_{8}$} \\
   \hline
   \nopagebreak[0]
\end{longtable}
\end{center}

\begin{center}
\begin{longtable}{| c | m{13cm} |}
   \hline
   $\mathcal{O}^j(\mathcal{T}_{8}^{4})$ & New Particles \\
   \hline
   \hline
   \endhead
   \multicolumn{2}{|c|}{\hspace{9cm} Continued on next page} \\
   \endfoot
  $\mathcal{O}_{9,(1,-1,1)}^{11}$ & {$S_{3}F_{6}F_{8}$, $S_{3}F_{2}F_{3}$} \\
   \hline
   \caption{Dimension 9 UV completions for topology $\mathcal{T}_{8}^{4}$ (see Fig.~\ref{fig:d=8_topologies}).}
   \label{tab:T9Top4B1L-1D1}\\
   \endlastfoot
  $\mathcal{O}_{9,(1,-1,1)}^{1}$ & {$S_{3}F_{5}F_{8}$, $S_{3}F_{2}F_{10}$, $S_{3}F_{5}F_{7}$, $S_{3}F_{2}F_{3}$} \\
   \hline
  $\mathcal{O}_{9,(1,-1,1)}^{2}$ & {$S_{4}F_{7}F_{19}$, $S_{2}F_{7}F_{17}$, $S_{4}F_{3}F_{4}$, $S_{2}F_{1}F_{3}$, $S_{6}F_{7}F_{9}$, $S_{6}F_{5}F_{7}$, $S_{6}F_{7}F_{19}$, $S_{6}F_{7}F_{17}$} \\
   \hline
  $\mathcal{O}_{9,(1,-1,1)}^{3}$ & {$S_{3}F_{5}F_{8}$, $S_{2}F_{7}F_{17}$, $S_{3}F_{1}F_{3}$, $S_{2}F_{1}F_{3}$} \\
   \hline
  $\mathcal{O}_{9,(1,-1,1)}^{4}$ & {$S_{4}F_{9}F_{20}$, $S_{4}F_{7}F_{9}$, $S_{4}F_{5}F_{7}$, $S_{2}F_{7}F_{9}$, $S_{2}F_{5}F_{7}$, $S_{4}F_{4}F_{12}$, $S_{4}F_{3}F_{4}$, $S_{4}F_{1}F_{3}$, $S_{2}F_{3}F_{4}$, $S_{2}F_{1}F_{3}$} \\
   \hline
  $\mathcal{O}_{9,(1,-1,1)}^{5}$ & {$S_{2}F_{7}F_{9}$, $S_{2}F_{5}F_{7}$, $S_{5}F_{7}F_{9}$, $S_{5}F_{5}F_{7}$, $S_{6}F_{7}F_{9}$, $S_{6}F_{5}F_{7}$, $S_{2}F_{3}F_{4}$, $S_{2}F_{1}F_{3}$, $S_{5}F_{8}F_{9}$, $S_{5}F_{5}F_{8}$, $S_{6}F_{7}F_{19}$, $S_{6}F_{7}F_{17}$} \\
   \hline
  $\mathcal{O}_{9,(1,-1,1)}^{6}$ & {$S_{3}F_{8}F_{14}$, $S_{3}F_{6}F_{8}$, $S_{5}F_{8}F_{14}$, $S_{5}F_{6}F_{8}$, $S_{3}F_{3}F_{11}$, $S_{3}F_{2}F_{3}$, $S_{5}F_{8}F_{9}$, $S_{5}F_{5}F_{8}$, $S_{3}F_{8}F_{9}$, $S_{3}F_{5}F_{8}$, $S_{3}F_{3}F_{4}$, $S_{3}F_{1}F_{3}$, $S_{5}F_{7}F_{9}$, $S_{5}F_{5}F_{7}$} \\
   \hline
  $\mathcal{O}_{9,(1,-1,1)}^{7}$ & {$S_{4}F_{8}F_{9}$, $S_{2}F_{5}F_{8}$, $S_{4}F_{3}F_{11}$, $S_{2}F_{2}F_{3}$, $S_{5}F_{7}F_{9}$, $S_{5}F_{5}F_{7}$, $S_{5}F_{8}F_{9}$, $S_{5}F_{5}F_{8}$} \\
   \hline
  $\mathcal{O}_{9,(1,-1,1)}^{8}$ & {$S_{2}F_{5}F_{8}$, $S_{1}F_{7}F_{17}$, $S_{2}F_{2}F_{3}$, $S_{1}F_{2}F_{3}$} \\
   \hline
  $\mathcal{O}_{9,(1,-1,1)}^{9}$ & {$S_{4}F_{8}F_{9}$, $S_{2}F_{5}F_{8}$, $S_{4}F_{3}F_{4}$, $S_{2}F_{1}F_{3}$, $S_{5}F_{8}F_{14}$, $S_{5}F_{6}F_{8}$, $S_{5}F_{8}F_{9}$, $S_{5}F_{5}F_{8}$, $S_{4}F_{7}F_{9}$, $S_{2}F_{5}F_{7}$, $S_{5}F_{7}F_{9}$, $S_{5}F_{5}F_{7}$} \\
   \hline
  $\mathcal{O}_{9,(1,-1,1)}^{10}$ & {$S_{3}F_{6}F_{13}$, $S_{2}F_{5}F_{8}$, $S_{3}F_{1}F_{3}$, $S_{2}F_{1}F_{3}$, $S_{3}F_{6}F_{8}$, $S_{2}F_{5}F_{7}$} \\
   \hline
\end{longtable}
\end{center}

\begin{center}
\begin{longtable}{| c | m{13cm} |}
   \hline
   $\mathcal{O}^j(\mathcal{T}_{8}^{5})$ & New Particles \\
   \hline
   \hline
   \endhead
   \multicolumn{2}{|c|}{\hspace{9cm} Continued on next page} \\
   \endfoot
  $\mathcal{O}_{9,(1,-1,1)}^{11}$ & {$S_{3}S_{5}F_{8}$, $S_{3}S_{5}F_{3}$} \\
   \hline
   \caption{Dimension 9 UV completions for topology $\mathcal{T}_{8}^{5}$ (see Fig.~\ref{fig:d=8_topologies}).}
   \label{tab:T9Top5B1L-1D1}\\
   \endlastfoot
  $\mathcal{O}_{9,(1,-1,1)}^{1}$ & {$S_{3}S_{6}F_{8}$, $S_{3}S_{6}F_{10}$, $S_{3}S_{5}F_{7}$, $S_{3}S_{5}F_{3}$} \\
   \hline
  $\mathcal{O}_{9,(1,-1,1)}^{2}$ & {$S_{4}S_{5}F_{7}$, $S_{2}S_{5}F_{7}$, $S_{6}S_{19}F_{9}$, $S_{3}S_{6}F_{5}$, $S_{4}S_{5}F_{3}$, $S_{2}S_{5}F_{3}$, $S_{6}S_{19}F_{7}$, $S_{3}S_{6}F_{7}$} \\
   \hline
  $\mathcal{O}_{9,(1,-1,1)}^{3}$ & {$S_{2}S_{5}F_{7}$, $S_{3}S_{6}F_{8}$, $S_{3}S_{6}F_{3}$, $S_{2}S_{5}F_{3}$} \\
   \hline
  $\mathcal{O}_{9,(1,-1,1)}^{4}$ & {$S_{4}S_{13}F_{9}$, $S_{4}S_{5}F_{9}$, $S_{2}S_{5}F_{9}$, $S_{4}S_{5}F_{5}$, $S_{2}S_{5}F_{5}$, $S_{4}S_{13}F_{4}$, $S_{4}S_{5}F_{4}$, $S_{2}S_{5}F_{4}$, $S_{4}S_{5}F_{1}$, $S_{2}S_{5}F_{1}$} \\
   \hline
  $\mathcal{O}_{9,(1,-1,1)}^{5}$ & {$S_{2}S_{5}F_{9}$, $S_{2}S_{5}F_{5}$, $S_{4}S_{5}F_{9}$, $S_{6}S_{19}F_{9}$, $S_{3}S_{6}F_{5}$, $S_{6}S_{19}F_{7}$, $S_{3}S_{6}F_{7}$, $S_{4}S_{5}F_{8}$, $S_{2}S_{5}F_{8}$, $S_{2}S_{5}F_{4}$, $S_{2}S_{5}F_{1}$} \\
   \hline
  $\mathcal{O}_{9,(1,-1,1)}^{6}$ & {$S_{3}S_{6}F_{14}$, $S_{3}S_{6}F_{6}$, $S_{4}S_{5}F_{14}$, $S_{2}S_{5}F_{6}$, $S_{4}S_{5}F_{8}$, $S_{2}S_{5}F_{8}$, $S_{3}S_{6}F_{11}$, $S_{3}S_{6}F_{2}$, $S_{3}S_{5}F_{9}$, $S_{3}S_{5}F_{5}$, $S_{5}S_{19}F_{9}$, $S_{5}S_{19}F_{7}$, $S_{3}S_{5}F_{7}$, $S_{3}S_{5}F_{4}$, $S_{3}S_{5}F_{1}$} \\
   \hline
  $\mathcal{O}_{9,(1,-1,1)}^{7}$ & {$S_{4}S_{5}F_{8}$, $S_{2}S_{5}F_{8}$, $S_{4}S_{5}F_{9}$, $S_{2}S_{5}F_{5}$, $S_{4}S_{5}F_{3}$, $S_{2}S_{5}F_{3}$} \\
   \hline
  $\mathcal{O}_{9,(1,-1,1)}^{8}$ & {$S_{1}S_{11}F_{7}$, $S_{2}S_{5}F_{8}$, $S_{2}S_{5}F_{3}$, $S_{1}S_{11}F_{3}$} \\
   \hline
  $\mathcal{O}_{9,(1,-1,1)}^{9}$ & {$S_{4}S_{5}F_{8}$, $S_{2}S_{5}F_{8}$, $S_{4}S_{5}F_{14}$, $S_{2}S_{5}F_{6}$, $S_{4}S_{5}F_{3}$, $S_{2}S_{5}F_{3}$, $S_{4}S_{11}F_{7}$, $S_{2}S_{11}F_{7}$, $S_{5}S_{19}F_{9}$, $S_{3}S_{5}F_{5}$, $S_{4}S_{11}F_{3}$, $S_{2}S_{11}F_{3}$, $S_{5}S_{19}F_{7}$, $S_{3}S_{5}F_{7}$} \\
   \hline
  $\mathcal{O}_{9,(1,-1,1)}^{10}$ & {$S_{2}S_{5}F_{8}$, $S_{3}S_{6}F_{13}$, $S_{3}S_{6}F_{3}$, $S_{2}S_{5}F_{3}$, $S_{2}S_{11}F_{7}$, $S_{3}S_{5}F_{8}$, $S_{3}S_{5}F_{3}$, $S_{2}S_{11}F_{3}$} \\
   \hline
\end{longtable}
\end{center}

\begin{center}
\begin{longtable}{| c | m{13cm} |}
   \hline
   $\mathcal{O}^j(\mathcal{T}_{8}^{7})$ & New Particles \\
   \hline
   \hline
   \endhead
   \multicolumn{2}{|c|}{\hspace{9cm} Continued on next page} \\
   \endfoot
  $\mathcal{O}_{9,(1,-1,1)}^{11}$ & {} \\
   \hline
   \caption{Dimension 9 UV completions for topology $\mathcal{T}_{8}^{7}$ (see Fig.~\ref{fig:d=8_topologies}).}
   \label{tab:T9Top7B1L-1D1}\\
   \endlastfoot
  $\mathcal{O}_{9,(1,-1,1)}^{1}$ & {$S_{3}S_{7}F_{2}$, $S_{3}S_{7}F_{5}$} \\
   \hline
  $\mathcal{O}_{9,(1,-1,1)}^{2}$ & {$S_{4}S_{9}F_{4}$, $S_{4}S_{9}F_{19}$, $S_{6}S_{9}F_{19}$, $S_{6}S_{9}F_{7}$} \\
   \hline
  $\mathcal{O}_{9,(1,-1,1)}^{3}$ & {} \\
   \hline
  $\mathcal{O}_{9,(1,-1,1)}^{4}$ & {$S_{4}S_{9}F_{12}$, $S_{4}S_{9}F_{3}$, $S_{2}S_{9}F_{3}$, $S_{4}S_{9}F_{20}$, $S_{4}S_{9}F_{7}$, $S_{2}S_{9}F_{7}$} \\
   \hline
  $\mathcal{O}_{9,(1,-1,1)}^{5}$ & {$S_{2}S_{9}F_{3}$, $S_{5}S_{9}F_{9}$, $S_{6}S_{9}F_{19}$, $S_{2}S_{9}F_{7}$, $S_{5}S_{9}F_{7}$, $S_{6}S_{9}F_{7}$} \\
   \hline
  $\mathcal{O}_{9,(1,-1,1)}^{6}$ & {$S_{3}S_{8}F_{3}$, $S_{3}S_{7}F_{3}$, $S_{5}S_{8}F_{9}$, $S_{5}S_{7}F_{5}$, $S_{3}S_{8}F_{8}$, $S_{3}S_{7}F_{8}$, $S_{5}S_{8}F_{8}$, $S_{5}S_{7}F_{8}$} \\
   \hline
  $\mathcal{O}_{9,(1,-1,1)}^{7}$ & {$S_{4}S_{9}F_{11}$, $S_{4}S_{9}F_{9}$, $S_{5}S_{9}F_{9}$, $S_{5}S_{9}F_{7}$} \\
   \hline
  $\mathcal{O}_{9,(1,-1,1)}^{8}$ & {} \\
   \hline
  $\mathcal{O}_{9,(1,-1,1)}^{9}$ & {$S_{4}S_{8}F_{4}$, $S_{2}S_{7}F_{1}$, $S_{4}S_{8}F_{9}$, $S_{2}S_{7}F_{5}$, $S_{5}S_{8}F_{9}$, $S_{5}S_{7}F_{5}$, $S_{5}S_{8}F_{8}$, $S_{5}S_{7}F_{8}$} \\
   \hline
  $\mathcal{O}_{9,(1,-1,1)}^{10}$ & {$S_{3}S_{7}F_{1}$, $S_{2}S_{7}F_{1}$, $S_{3}S_{7}F_{6}$, $S_{2}S_{7}F_{5}$} \\
   \hline
\end{longtable}
\end{center}

\begin{center}
\begin{longtable}{| c | m{13cm} |}
   \hline
   $\mathcal{O}^j(\mathcal{T}_{8}^{8})$ & New Particles \\
   \hline
   \hline
   \endhead
   \multicolumn{2}{|c|}{\hspace{9cm} Continued on next page} \\
   \endfoot
  $\mathcal{O}_{9,(1,-1,1)}^{11}$ & {$S_{5}F_{8}F_{8}$, $S_{5}F_{3}F_{8}$} \\
   \hline
   \caption{Dimension 9 UV completions for topology $\mathcal{T}_{8}^{8}$ (see Fig.~\ref{fig:d=8_topologies}).}
   \label{tab:T9Top8B1L-1D1}\\
   \endlastfoot
  $\mathcal{O}_{9,(1,-1,1)}^{1}$ & {$S_{5}F_{7}F_{8}$, $S_{6}F_{7}F_{8}$, $S_{5}F_{3}F_{8}$, $S_{6}F_{7}F_{10}$} \\
   \hline
  $\mathcal{O}_{9,(1,-1,1)}^{2}$ & {$S_{19}F_{9}F_{9}$, $S_{3}F_{5}F_{5}$, $S_{5}F_{7}F_{9}$, $S_{5}F_{5}F_{7}$, $S_{19}F_{7}F_{9}$, $S_{3}F_{5}F_{7}$, $S_{5}F_{3}F_{9}$, $S_{5}F_{3}F_{5}$, $S_{19}F_{3}F_{9}$, $S_{3}F_{3}F_{5}$, $S_{19}F_{3}F_{7}$, $S_{3}F_{3}F_{7}$} \\
   \hline
  $\mathcal{O}_{9,(1,-1,1)}^{3}$ & {$S_{5}F_{7}F_{7}$, $S_{6}F_{7}F_{8}$, $S_{5}F_{7}F_{8}$, $S_{6}F_{3}F_{7}$, $S_{5}F_{3}F_{7}$, $S_{5}F_{3}F_{8}$} \\
   \hline
  $\mathcal{O}_{9,(1,-1,1)}^{4}$ & {$S_{13}F_{9}F_{9}$, $S_{5}F_{9}F_{9}$, $S_{5}F_{5}F_{9}$, $S_{5}F_{5}F_{5}$, $S_{13}F_{4}F_{9}$, $S_{5}F_{4}F_{9}$, $S_{5}F_{1}F_{9}$, $S_{5}F_{4}F_{5}$, $S_{5}F_{1}F_{5}$} \\
   \hline
  $\mathcal{O}_{9,(1,-1,1)}^{5}$ & {$S_{4}F_{7}F_{9}$, $S_{2}F_{5}F_{7}$, $S_{5}F_{7}F_{9}$, $S_{5}F_{5}F_{7}$, $S_{19}F_{8}F_{9}$, $S_{3}F_{5}F_{8}$, $S_{5}F_{8}F_{9}$, $S_{5}F_{5}F_{8}$, $S_{19}F_{7}F_{8}$, $S_{3}F_{7}F_{8}$, $S_{4}F_{7}F_{8}$, $S_{2}F_{7}F_{8}$, $S_{19}F_{4}F_{9}$, $S_{3}F_{1}F_{5}$, $S_{4}F_{4}F_{9}$, $S_{2}F_{1}F_{5}$, $S_{19}F_{4}F_{7}$, $S_{3}F_{1}F_{7}$, $S_{5}F_{4}F_{7}$, $S_{5}F_{1}F_{7}$, $S_{4}F_{4}F_{8}$, $S_{2}F_{1}F_{8}$, $S_{5}F_{4}F_{8}$, $S_{5}F_{1}F_{8}$} \\
   \hline
  $\mathcal{O}_{9,(1,-1,1)}^{6}$ & {$S_{4}F_{7}F_{14}$, $S_{2}F_{6}F_{7}$, $S_{5}F_{8}F_{9}$, $S_{5}F_{5}F_{8}$, $S_{4}F_{7}F_{8}$, $S_{2}F_{7}F_{8}$, $S_{6}F_{7}F_{14}$, $S_{6}F_{6}F_{7}$, $S_{19}F_{8}F_{9}$, $S_{3}F_{5}F_{8}$, $S_{4}F_{4}F_{14}$, $S_{2}F_{1}F_{6}$, $S_{19}F_{9}F_{11}$, $S_{3}F_{2}F_{5}$, $S_{4}F_{4}F_{8}$, $S_{2}F_{1}F_{8}$, $S_{6}F_{7}F_{11}$, $S_{6}F_{2}F_{7}$, $S_{19}F_{7}F_{8}$, $S_{3}F_{7}F_{8}$, $S_{5}F_{4}F_{8}$, $S_{5}F_{1}F_{8}$, $S_{19}F_{7}F_{11}$, $S_{3}F_{2}F_{7}$} \\
   \hline
  $\mathcal{O}_{9,(1,-1,1)}^{7}$ & {$S_{4}F_{9}F_{9}$, $S_{2}F_{5}F_{5}$, $S_{5}F_{8}F_{9}$, $S_{5}F_{5}F_{8}$, $S_{4}F_{8}F_{9}$, $S_{2}F_{5}F_{8}$, $S_{5}F_{3}F_{9}$, $S_{5}F_{3}F_{5}$, $S_{4}F_{3}F_{9}$, $S_{2}F_{3}F_{5}$, $S_{4}F_{3}F_{8}$, $S_{2}F_{3}F_{8}$} \\
   \hline
  $\mathcal{O}_{9,(1,-1,1)}^{8}$ & {$S_{5}F_{7}F_{8}$, $S_{11}F_{7}F_{8}$, $S_{5}F_{8}F_{8}$, $S_{5}F_{3}F_{7}$, $S_{5}F_{3}F_{8}$, $S_{11}F_{3}F_{8}$} \\
   \hline
  $\mathcal{O}_{9,(1,-1,1)}^{9}$ & {$S_{4}F_{9}F_{14}$, $S_{2}F_{5}F_{6}$, $S_{11}F_{7}F_{14}$, $S_{11}F_{6}F_{7}$, $S_{4}F_{8}F_{9}$, $S_{2}F_{5}F_{8}$, $S_{19}F_{9}F_{14}$, $S_{3}F_{5}F_{6}$, $S_{11}F_{3}F_{14}$, $S_{11}F_{3}F_{6}$, $S_{19}F_{3}F_{9}$, $S_{3}F_{3}F_{5}$, $S_{19}F_{7}F_{14}$, $S_{3}F_{6}F_{7}$, $S_{5}F_{8}F_{9}$, $S_{5}F_{5}F_{8}$, $S_{4}F_{3}F_{14}$, $S_{2}F_{3}F_{6}$, $S_{5}F_{3}F_{9}$, $S_{5}F_{3}F_{5}$, $S_{4}F_{3}F_{8}$, $S_{2}F_{3}F_{8}$, $S_{19}F_{3}F_{7}$, $S_{3}F_{3}F_{7}$} \\
   \hline
  $\mathcal{O}_{9,(1,-1,1)}^{10}$ & {$S_{11}F_{7}F_{8}$, $S_{5}F_{8}F_{8}$, $S_{11}F_{7}F_{13}$, $S_{5}F_{7}F_{8}$, $S_{5}F_{3}F_{8}$, $S_{5}F_{3}F_{7}$, $S_{6}F_{7}F_{13}$, $S_{11}F_{3}F_{8}$, $S_{6}F_{3}F_{7}$, $S_{11}F_{3}F_{13}$} \\
   \hline
\end{longtable}
\end{center}

 \section{Higher dimensional operators}
 \label{sec:higher_dims}

Above, we have given all non-trivial UV completions for derivative BNV operators up to $d=9$. The number of operators and UV completions grows dramatically for $d\geq 10$, so we only list a small subset of UV completions for $d=10$, $11$, and $12$ in App.~\ref{sec:d=10}, \ref{sec:d=11}, and \ref{sec:d=12}, respectively. In this section, we briefly mention a few interesting novel scenarios that arise for $d>10$, without providing an exhaustive list of UV completions.\footnote{In this spirit, Ref.~\cite{Fonseca:2018ehk} constructs a UV completion for the $d=13$ $(\Delta B,\Delta L) = (1,3)$ operator $D QQuu\bar{d}LLe$.}

Let us start with three-derivative operators. Non-trivial ones first arise at $d=12$ from one-derivative $d=10$ operators that vanish due to Higgs asymmetry. For example, $\mathcal{O}_{10, \, (1,1,1)}^{60}= DHHHdddL$ has UV completions that give a vanishing amplitude and require going to higher order in the heavy mass expansion to obtain the non-vanishing operator $\mathcal{O}_{12, \, (1,1,3)}^{60}= DDDHHHdddL$, see Tabs.~\ref{tab:T12Top12B1L1D3}--\ref{tab:T12Top19B1L1D3}. The first non-trivial $D^3$ BNV operators thus arise at $d=12$; at $d=13$, we find the first \textit{minimal-derivative} $D^3$ operators:
 \begin{align}
\mathcal{O}_{13, \, (1,-1,3)}^{85} \equiv DDD \bar{H} \bar{H} \bar{H} \bar{H} duu \bar{e} \,, &&
\mathcal{O}_{13, \, (1,-1,3)}^{86} \equiv DDD \bar{H} \bar{H} \bar{H} \bar{H} uuu \bar{\nu} \,.
\label{eq:1Ddequal13basisB1L-1}
 \end{align}
These are the lowest-dimensional $D^3$ BNV operators that inevitably vanish upon removal of any derivatives independent of the UV completion.\footnote{We list these operators as 85 and 86 because there are 84 $(\Delta B, \Delta L)=(1,-1)$ operators at $d=11$ that induce 84 $d=13$ operators through $\mathcal{O}_{13,(1,-1,3)} =D^2 \mathcal{O}_{11,(1,-1,1)}$.}
In both cases it is easy to see why their one-derivative analogues vanish: with all fermions being $SU(2)_L$ singlets, the four $\bar{H}$ are forced to form an $SU(2)_L$ singlet by themselves, which vanishes: $(\bar{H}^4)_{\vec{1}} = 0$. The derivative hence needs to act on an $\bar{H}$ if we want a non-zero operator, but this forces the other three $\bar{H}$ to form an $SU(2)_L$ doublet (to contract with the doublet $D_\mu \bar{H}$), which also vanishes for a single Higgs doublet: $(\bar{H}^3)_{\vec{2}} = 0$. At least \textit{two} derivatives are necessary to form an $SU(2)_L$ singlet out of four identical Higgses -- and an odd number to form a Lorentz singlet -- proving the minimal-derivative nature of the operators in Eq.~\eqref{eq:1Ddequal13basisB1L-1}.

\begin{figure}[tb]
    \centering
    \includegraphics[width = 0.99\textwidth]{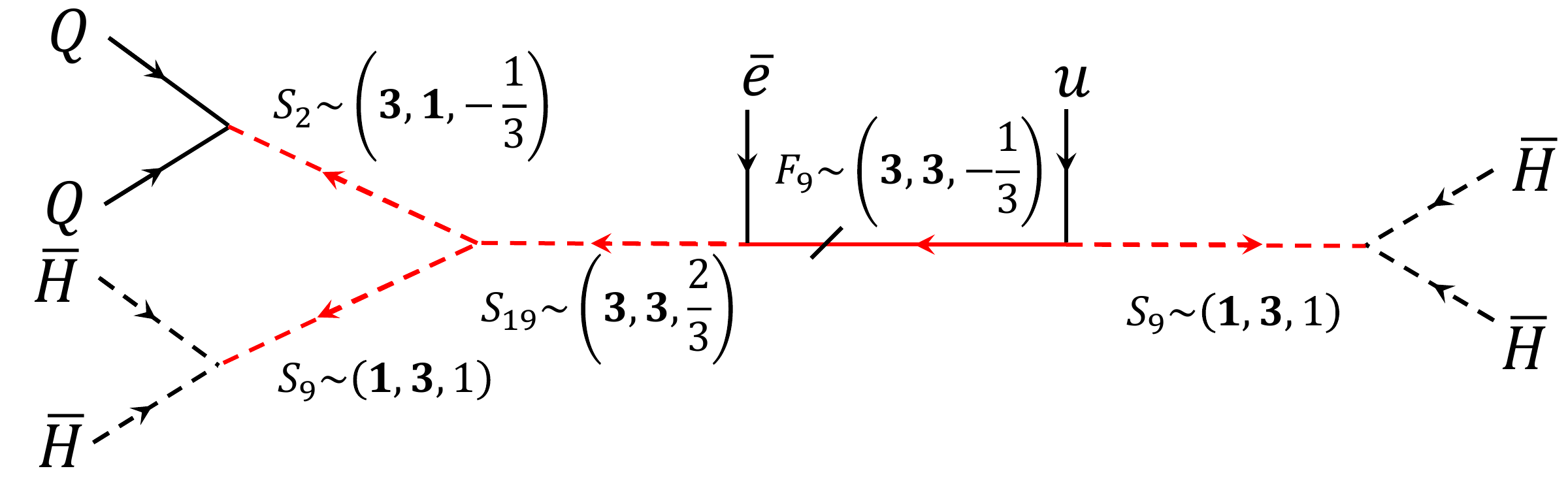}
    \caption{UV completion for the $d=11$ operator $D \bar{H} \bar{H} \bar{H} \bar{H} QQ u\bar{e}$ that vanishes due to the Higgs asymmetry $[  (\bar{H}\bar{H})_{\vec{3}} D_\mu (\bar{H}\bar{H})_{\vec{3}} ]_{\vec{1}}=0$ structure and thus actually UV completes the $d=13$ operator $DDD \bar{H} \bar{H} \bar{H} \bar{H} QQ u\bar{e}$ when two more derivatives are added on any internal line except $S_2$.}
    \label{fig:DHHHH_example}
\end{figure}

Such derivative $H^4$ operators introduce a mild complication to our workflow. As explained in detail above, we check for Higgs asymmetry in a given diagram by going to momentum space, essentially replacing $D_\mu\to \partial_\mu$. Whenever this procedure finds a vanishing amplitude, we perform an additional check to make sure the entire operator vanishes, not just the $\partial_\mu$ term: we cut the diagram at every line with a derivative and study the resulting non-derivative subdiagrams; if any of them vanish due to Higgs asymmetry, the full derivative diagram does as well, as one can integrate out the heavy particles in such an order that we end up with zero before any derivative even enters the operator. This is a powerful check that allows us to infer the full $D_\mu$ result from the $\partial_\mu$ piece. The lowest mass dimension where this procedure is inconclusive is $d=11$, in operators with four Higgses. Take, for example, the UV completion in Fig.~\ref{fig:DHHHH_example}. Our amplitude calculation returns zero in momentum space, but each sub-diagram on either side of the derivative insertion is non-zero. Luckily we can confirm a vanishing amplitude for the covariant-derivative case by appealing once again to the fact that $(D_\mu H^4)_{\vec{1}} = 0$. Tracking only the $SU(2)_L$ charges of the $\bar{H}$ pieces in Fig.~\ref{fig:DHHHH_example}, we see that the resulting structure in position space is $[  (\bar{H}\bar{H})_{\vec{3}} D_\mu (\bar{H}\bar{H})_{\vec{3}} ]_{\vec{1}}$, which indeed vanishes even though $(\bar{H}\bar{H})_{\vec{3}}$ is non-zero. Fig.~\ref{fig:DHHHH_example} is hence zero due to Higgs asymmetry and requires additional derivatives, e.g.~on the fermion line, to generate the leading-order $d=13$ operator  $DDD \bar{H} \bar{H} \bar{H} \bar{H} QQ u\bar{e}$.

Moving on to \textit{four}-derivative operators, we find the first non-trivial ones at $d=14$, all involving four fermions and four scalars. For example, there are UV completions of $HH \bar{H}\bar{H} uude$ where $HH$ and $\bar{H}\bar{H}$ are independently coupled to $SU(2)_L$ singlets, forcing us to expand up to  $(HD^2H)(\bar{H}D^2\bar{H})uude$ to find a non-vanishing term. Finding such non-trivial potentially-dominant $D^4$ operators requires a careful study of UV completions; we did not find any $D^4$ minimal-derivative operators up to $d\leq 17$.

\section{Conclusions}
\label{sec:conclusions}

The exceptional sensitivity of experiments searching for baryon number violation in the form of nuclear decays allows them to probe BNV operators up to high mass dimension $d\gg 6$. Non-derivative BNV $\nu$SMEFT operators up to $d=15$ are potentially detectable and have recently been UV completed exhaustively in Ref.~\cite{Heeck:2026dmh}. Here, we provide the missing tree-level UV completions via heavy scalars and fermions for derivative BNV operators up to $d=12$. Overall, we have constructed more than one million UV completions that are non-trivial, i.e.~generate derivative operators that are not subleading corrections to operators with the same fields but fewer derivatives. The first non-trivial UV completions for operators with one, two, three, and four derivatives appear at $d=8$, $10$, $12$, and $14$, respectively. Antisymmetry under exchange of identical Higgs fields in operators or diagrams is the key mechanism that allows for operators with more than one derivative to become the leading contribution in the heavy-mass expansion, as already pointed out in Refs.~\cite{Gargalionis:2020xvt,Heeck:2026dmh}. Non-trivial UV completions are a necessary ingredient but do not guarantee the phenomenological dominance of the underlying operator, as loop-induced lower-dimensional operators can lead to faster nucleon decay rates, as illustrated by an explicit $d=8$ example in the main text.

In follow-up work, we will 
\begin{itemize}
    \item perform the tree-level matching of each UV completion onto an operator basis, e.g.~the permutation basis of \texttt{AutoEFT}~\cite{Harlander:2023ozs,Harlander:2023psl}, to provide an exhaustive dictionary between UV-complete models and low-energy effective operators that are used for amplitude calculations~\cite{Song:2026gyo};
    \item systematically identify accidentally-protected scenarios, i.e.~$d>6$ BNV operators that are the dominant tree-level term by choice of mediator particles without any imposition of symmetries. As exemplified by the $d=8$ example in the main text, a subset of these accidentally-protected cases will furnish genuine loop realizations of BNV operators.
\end{itemize}

The full list of UV completions can be found at \website; we present only a small selection in the article itself. The website also contains the underlying \texttt{Mathematica} code and various other data files that could be useful for the BNV or SMEFT communities.

\section*{Acknowledgements}
This work was supported in part by the U.S. Department of Energy under Grant No.~DE-SC0007974.
We acknowledge Research Computing at The University of Virginia for providing computational resources that have contributed to the results reported within this publication.

\clearpage

\appendix

\section{Topologies}
\label{app:topologies}

In this appendix, we collect the tree-level topologies relevant to the main text, reproduced from Ref.~\cite{Heeck:2026dmh} and thus following the same non-derivative naming scheme, namely $\mathcal{T}_d^j$ for a non-derivative operator of mass dimension $d$, with $j$ running over the different topologies. For an operator with mass dimension $d$ and $n$ derivatives, the relevant topology is then $\mathcal{T}_{d-n}^j$, so $\mathcal{T}_7^2$ is a possible topology for a non-derivative $d=7$ operator or a one-derivative $d=8$ operator or a two-derivative $d=9$ operator and so on and so forth. \texttt{Mathematica} files for these and other topologies can be found at \website.

\begin{figure}[hb]
    \centering
    \includegraphics[width = 0.7\textwidth]{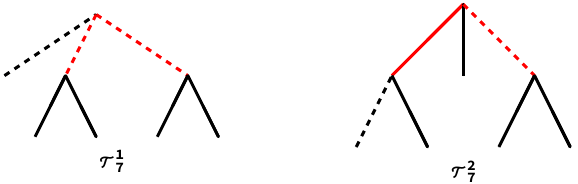}
    \caption{Feynman-diagram topologies for non-derivative  $d=7$ BNV operators.}
    \label{fig:d=7_topologies}
\end{figure}

\begin{figure}[hb]
    \centering
    \includegraphics[width = \textwidth]{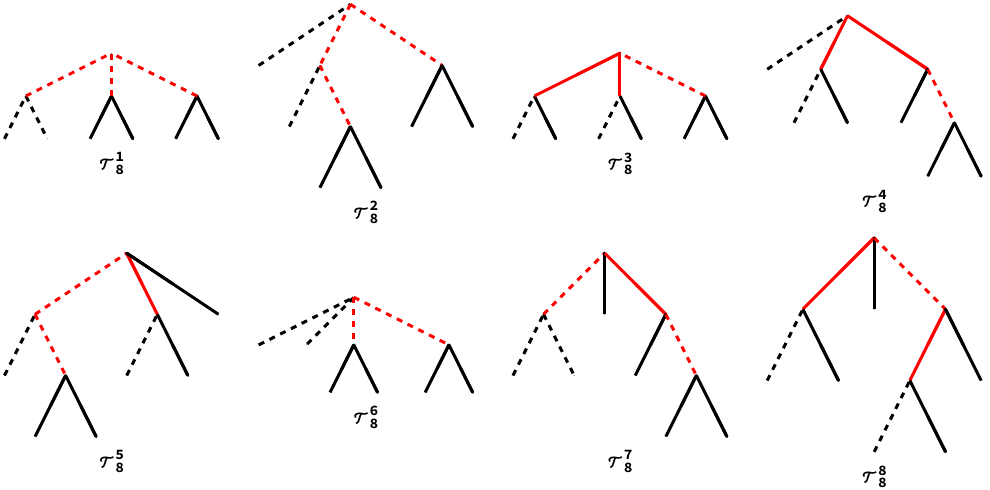}
    \caption{Feynman-diagram topologies for non-derivative  $d=8$ BNV operators.}
    \label{fig:d=8_topologies}
\end{figure}

\begin{figure}[hb]
    \centering
    \includegraphics[width = \textwidth]{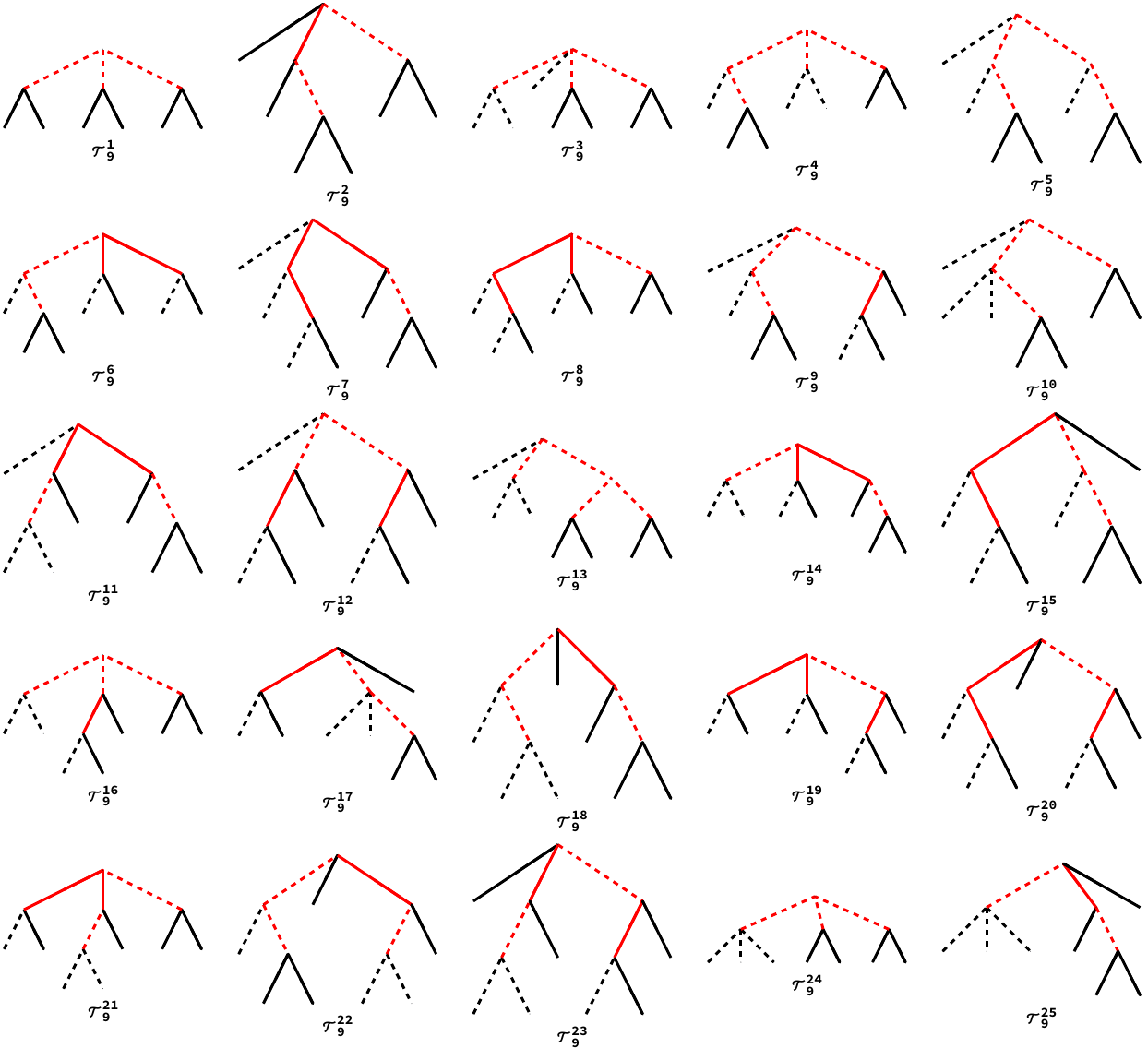}
    \caption{Feynman-diagram topologies for non-derivative  $d=9$ BNV operators.}
    \label{fig:d=9_topologies}
\end{figure}

\begin{figure}[hb]
    \centering
    \includegraphics[width = \textwidth]{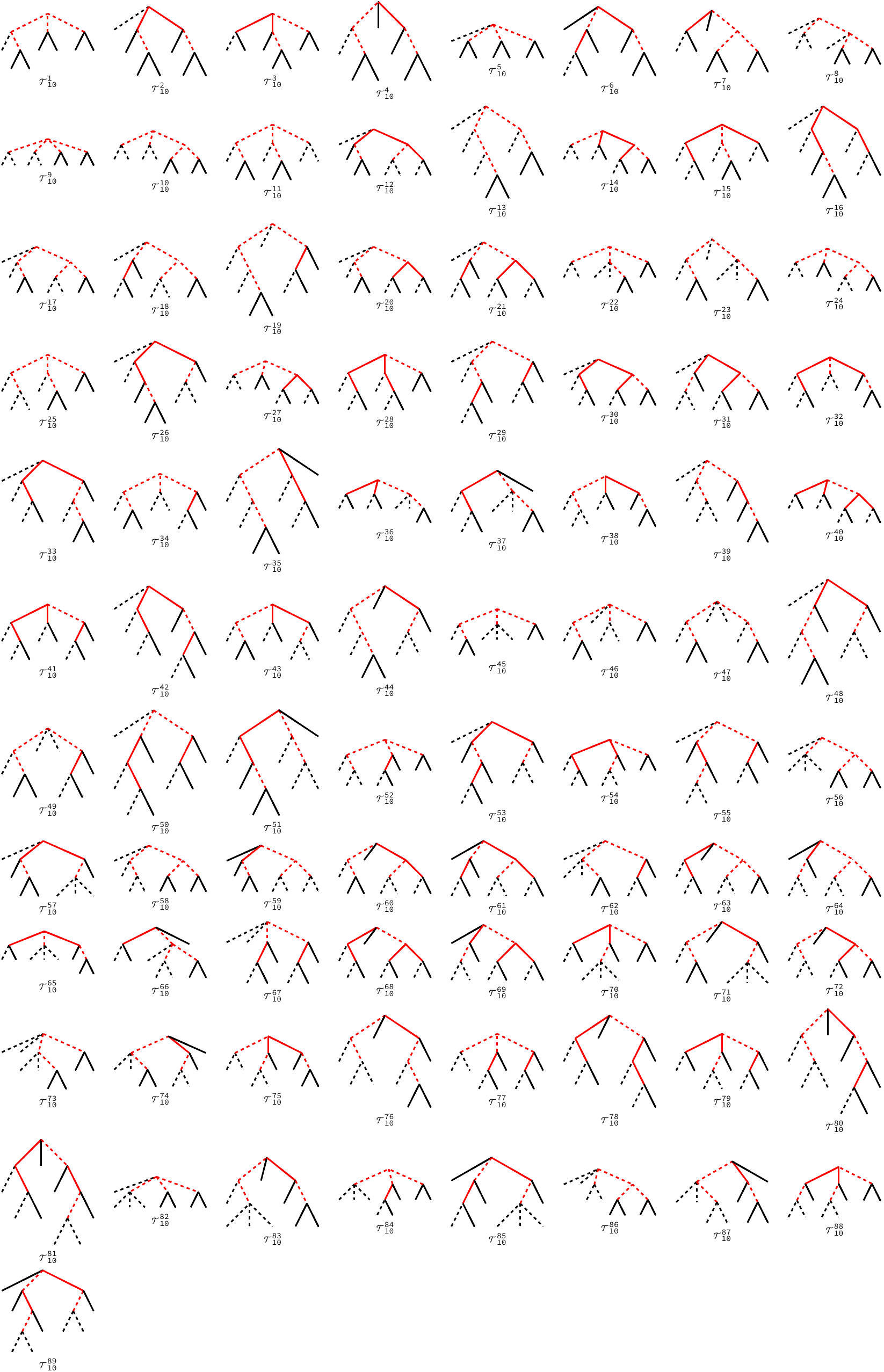}
    \caption{Feynman-diagram topologies for non-derivative  $d=10$ BNV operators.}
    \label{fig:d=10_topologies}
\end{figure}

 \begin{figure}[hb]
    \centering
    \includegraphics[width = \textwidth]{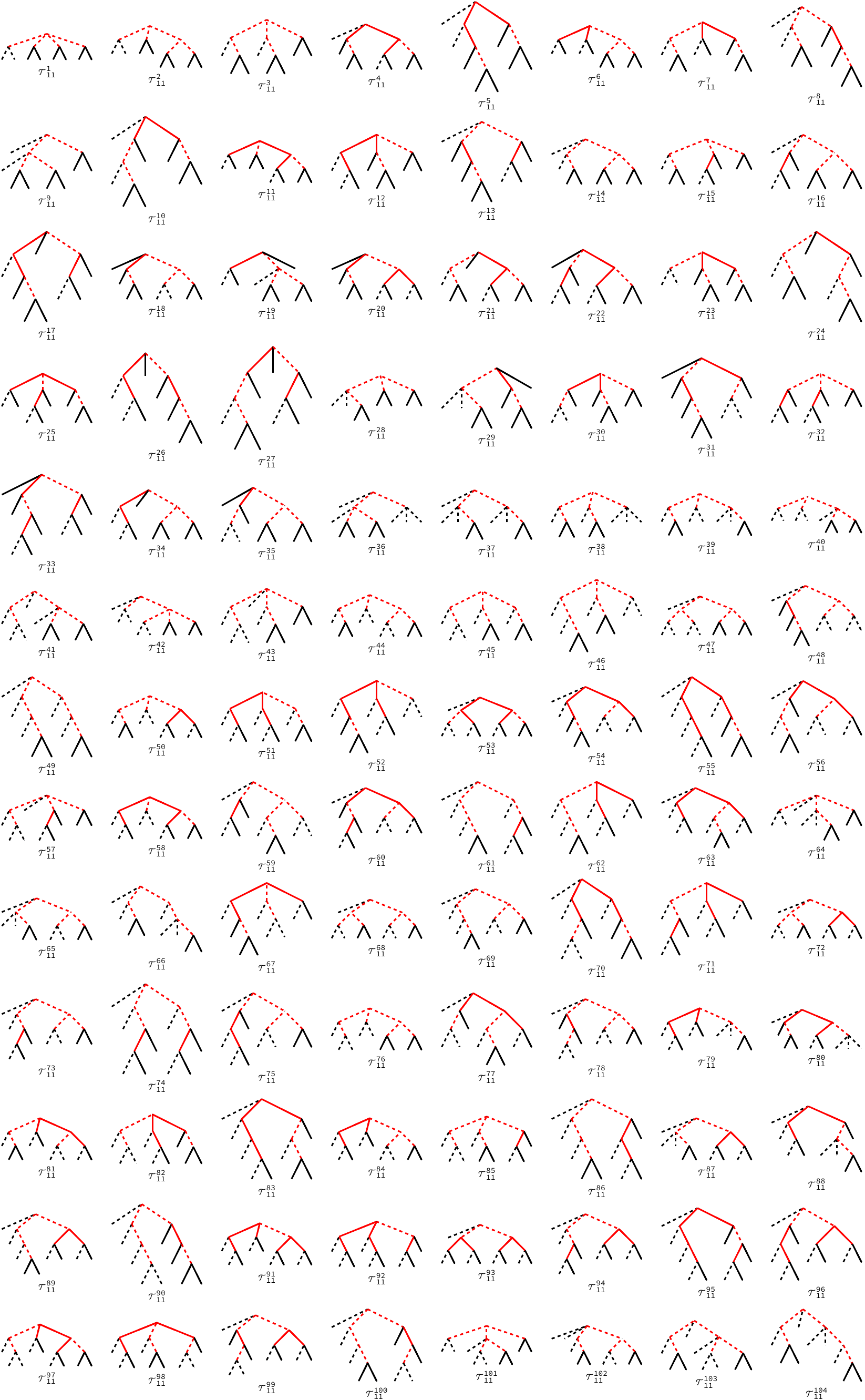}
    \caption{Feynman-diagram topologies for non-derivative  $d=11$ BNV operators.}
    \label{fig:d=11_topologies}
\end{figure}

\begin{figure}[hb]
    \centering
    \includegraphics[width = \textwidth]{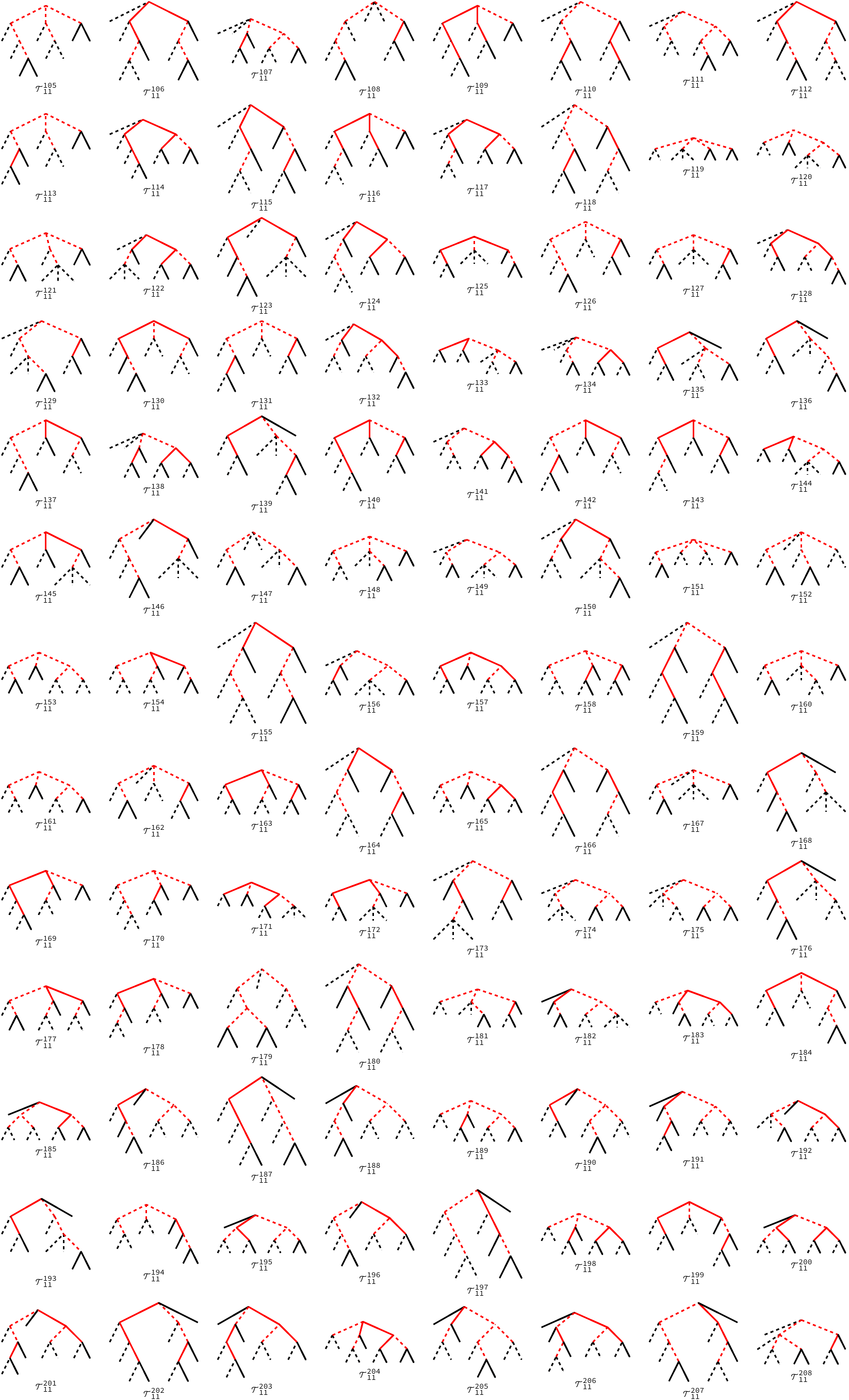}
    \caption*{\hfill Contd.}
\end{figure}

\begin{figure}[hb]
    \centering
    \includegraphics[width = \textwidth]{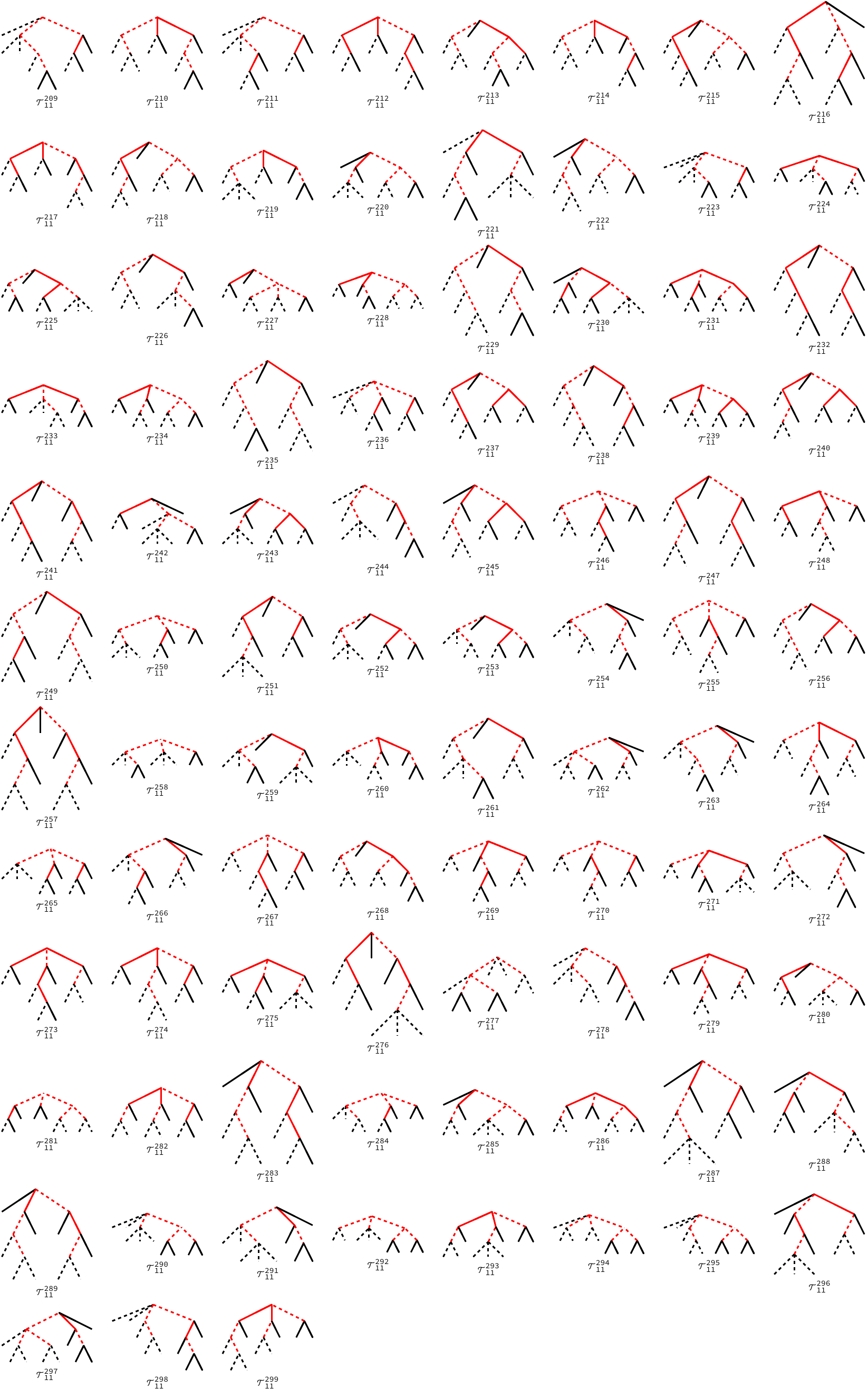}
    \caption*{\hfill Contd.}
\end{figure}

\clearpage

 \section{Dimension 10}
 \label{sec:d=10}

 In this appendix, we collect derivative $d=10$ BNV operators and a subset of UV completions. 
At $d=10$, derivative BNV operators carry either $(\Delta B, \Delta L)=(1,1)$ or $(1,-3)$, which we will split into subsections below.
 
  \subsection{\texorpdfstring{$\Delta B=1, \Delta L=1$}{Delta B=1,Delta L=1}}

For $(\Delta B, \Delta L)=(1,1)$, we have one-derivative operators,
\begin{multicols}{3}
 \begin{spacing}{1.5}
 \noindent
 \begin{math}
\mathcal{O}_{10, \, (1,1,1)}^{1}\equiv Dd\bar{d}QQQL\,,\\
\mathcal{O}_{10, \, (1,1,1)}^{2}\equiv D\bar{d}QQQQe\,,\\
\mathcal{O}_{10, \, (1,1,1)}^{3}\equiv Ddd\bar{d}uQL\,,\\
\mathcal{O}_{10, \, (1,1,1)}^{4}\equiv Dd\bar{d}uQQe\,,\\
\mathcal{O}_{10, \, (1,1,1)}^{5}\equiv Ddd\bar{d}uue\,,\\
\mathcal{O}_{10, \, (1,1,1)}^{6}\equiv DdQQLL\bar{e}\,,\\
\mathcal{O}_{10, \, (1,1,1)}^{7}\equiv DQQQLe\bar{e}\,,\\
\mathcal{O}_{10, \, (1,1,1)}^{8}\equiv DdduLL\bar{e}\,,\\
\mathcal{O}_{10, \, (1,1,1)}^{9}\equiv DduQLe\bar{e}\,,\\
\mathcal{O}_{10, \, (1,1,1)}^{10}\equiv DuQQee\bar{e}\,,\\
\mathcal{O}_{10, \, (1,1,1)}^{11}\equiv Dduuee\bar{e}\,,\\
\mathcal{O}_{10, \, (1,1,1)}^{12}\equiv DQQQLL\bar{L}\,,\\
\mathcal{O}_{10, \, (1,1,1)}^{13}\equiv DduQLL\bar{L}\,,\\
\mathcal{O}_{10, \, (1,1,1)}^{14}\equiv DuQQL\bar{L}e\,,\\
\mathcal{O}_{10, \, (1,1,1)}^{15}\equiv DduuL\bar{L}e\,,\\
\mathcal{O}_{10, \, (1,1,1)}^{16}\equiv DuuQ\bar{L}ee\,,\\
\mathcal{O}_{10, \, (1,1,1)}^{17}\equiv DQQQQ\bar{Q}L\,,\\
\mathcal{O}_{10, \, (1,1,1)}^{18}\equiv DduQQ\bar{Q}L\,,\\
\mathcal{O}_{10, \, (1,1,1)}^{19}\equiv DuQQQ\bar{Q}e\,,\\
\mathcal{O}_{10, \, (1,1,1)}^{20}\equiv Ddduu\bar{Q}L\,,\\
\mathcal{O}_{10, \, (1,1,1)}^{21}\equiv DduuQ\bar{Q}e\,,\\
\mathcal{O}_{10, \, (1,1,1)}^{22}\equiv Du\bar{u}QQQL\,,\\
\mathcal{O}_{10, \, (1,1,1)}^{23}\equiv Dduu\bar{u}QL\,,\\
\mathcal{O}_{10, \, (1,1,1)}^{24}\equiv Duu\bar{u}QQe\,,\\
\mathcal{O}_{10, \, (1,1,1)}^{25}\equiv Dduuu\bar{u}e\,,\\
\mathcal{O}_{10, \, (1,1,1)}^{26}\equiv Ddd\bar{d}QQ\nu\,,\\
\mathcal{O}_{10, \, (1,1,1)}^{27}\equiv Dddd\bar{d}u\nu\,,\\
\mathcal{O}_{10, \, (1,1,1)}^{28}\equiv DddQL\bar{e}\nu\,,\\
\mathcal{O}_{10, \, (1,1,1)}^{29}\equiv DdQQe\bar{e}\nu\,,\\
\mathcal{O}_{10, \, (1,1,1)}^{30}\equiv Dddue\bar{e}\nu\,,\\
\mathcal{O}_{10, \, (1,1,1)}^{31}\equiv DdQQL\bar{L}\nu\,,\\
\mathcal{O}_{10, \, (1,1,1)}^{32}\equiv DQQQ\bar{L}e\nu\,,\\
\mathcal{O}_{10, \, (1,1,1)}^{33}\equiv DdduL\bar{L}\nu\,,\\
\mathcal{O}_{10, \, (1,1,1)}^{34}\equiv DduQ\bar{L}e\nu\,,\\
\mathcal{O}_{10, \, (1,1,1)}^{35}\equiv DuQQLL\bar{\nu}\,,\\
\mathcal{O}_{10, \, (1,1,1)}^{36}\equiv DduuLL\bar{\nu}\,,\\
\mathcal{O}_{10, \, (1,1,1)}^{37}\equiv DuuQLe\bar{\nu}\,,\\
\mathcal{O}_{10, \, (1,1,1)}^{38}\equiv Duuuee\bar{\nu}\,,\\
\mathcal{O}_{10, \, (1,1,1)}^{39}\equiv DdQQQ\bar{Q}\nu\,,\\
\mathcal{O}_{10, \, (1,1,1)}^{40}\equiv DdduQ\bar{Q}\nu\,,\\
\mathcal{O}_{10, \, (1,1,1)}^{41}\equiv D\bar{u}QQQQ\nu\,,\\
\mathcal{O}_{10, \, (1,1,1)}^{42}\equiv Ddu\bar{u}QQ\nu\,,\\
\mathcal{O}_{10, \, (1,1,1)}^{43}\equiv Ddduu\bar{u}\nu\,,\\
\mathcal{O}_{10, \, (1,1,1)}^{44}\equiv Dddd\bar{e}\nu\nu\,,\\
\mathcal{O}_{10, \, (1,1,1)}^{45}\equiv DddQ\bar{L}\nu\nu\,,\\
\mathcal{O}_{10, \, (1,1,1)}^{46}\equiv DQQQL\nu\bar{\nu}\,,\\
\mathcal{O}_{10, \, (1,1,1)}^{47}\equiv DduQL\nu\bar{\nu}\,,\\
\mathcal{O}_{10, \, (1,1,1)}^{48}\equiv DuQQe\nu\bar{\nu}\,,\\
\mathcal{O}_{10, \, (1,1,1)}^{49}\equiv Dduue\nu\bar{\nu}\,,\\
\mathcal{O}_{10, \, (1,1,1)}^{50}\equiv DdQQ\nu\nu\bar{\nu}\,,\\
\mathcal{O}_{10, \, (1,1,1)}^{51}\equiv Dddu\nu\nu\bar{\nu}\,,\\
\mathcal{O}_{10, \, (1,1,1)}^{52}\equiv D\bar{H}\bar{H}\bar{H}uuuL\,,\\
\mathcal{O}_{10, \, (1,1,1)}^{53}\equiv DH\bar{H}\bar{H}uQQL\,,\\
\mathcal{O}_{10, \, (1,1,1)}^{54}\equiv DH\bar{H}\bar{H}duuL\,,\\
\mathcal{O}_{10, \, (1,1,1)}^{55}\equiv DH\bar{H}\bar{H}uuQe\,,\\
\mathcal{O}_{10, \, (1,1,1)}^{56}\equiv DHH\bar{H}dQQL\,,\\
\mathcal{O}_{10, \, (1,1,1)}^{57}\equiv DHH\bar{H}QQQe\,,\\
\mathcal{O}_{10, \, (1,1,1)}^{58}\equiv DHH\bar{H}dduL\,,\\
\mathcal{O}_{10, \, (1,1,1)}^{59}\equiv DHH\bar{H}duQe\,,\\
\mathcal{O}_{10, \, (1,1,1)}^{60}\equiv DHHHdddL\,,\\
\mathcal{O}_{10, \, (1,1,1)}^{61}\equiv DHHHddQe\,,\\
\mathcal{O}_{10, \, (1,1,1)}^{62}\equiv D\bar{H}\bar{H}\bar{H}uuQ\nu\,,\\
\mathcal{O}_{10, \, (1,1,1)}^{63}\equiv DH\bar{H}\bar{H}QQQ\nu\,,\\
\mathcal{O}_{10, \, (1,1,1)}^{64}\equiv DH\bar{H}\bar{H}duQ\nu\,,\\
\mathcal{O}_{10, \, (1,1,1)}^{65}\equiv DHH\bar{H}ddQ\nu\,,
\label{eq:1Ddequal10basisB1L1}
 \end{math}
 \end{spacing}
 \end{multicols}
 \noindent
two-derivative operators,
\begin{multicols}{3}
 \begin{spacing}{1.4}
 \noindent
 \begin{math}
\mathcal{O}_{10, \,(1 , 1,2)}^{1} \equiv DDHHddQL \,, \\
\mathcal{O}_{10, \,(1 , 1,2)}^{2} \equiv DDHHdQQe \,, \\
\mathcal{O}_{10, \,(1 , 1,2)}^{3} \equiv DD\bar{H}\bar{H}uuQL \,, \\
\mathcal{O}_{10, \,(1 , 1,2)}^{4} \equiv DDH\bar{H}QQQL \,, \\
\mathcal{O}_{10, \,(1 , 1,2)}^{5} \equiv DDH\bar{H}duQL \,, \\
\mathcal{O}_{10, \,(1 , 1,2)}^{6} \equiv DDH\bar{H}uQQe \,, \\
\mathcal{O}_{10, \,(1 , 1,2)}^{7} \equiv DDH\bar{H}duue \,, \\
\mathcal{O}_{10, \,(1 , 1,2)}^{8} \equiv DDH\bar{H}dQQ\nu \,, \\
\mathcal{O}_{10, \,(1 , 1,2)}^{9} \equiv DD\bar{H}\bar{H}uQQ\nu \,, \\
\mathcal{O}_{10, \,(1 , 1,2)}^{10} \equiv DDH\bar{H}ddu\nu \,,
 \label{eq:2Ddequal10basisB1L1}
 \end{math}
 \end{spacing}
 \end{multicols}
 \noindent
and minimal-derivative operators,
\begin{multicols}{3}
 \begin{spacing}{1.4}
 \noindent
 \begin{math}
\mathcal{O}_{10, \,(1 , 1,2)}^{11} \equiv DD\bar{H}\bar{H}uuue \,, \\
\mathcal{O}_{10, \,(1 , 1,2)}^{12} \equiv DDHHddue \,, \\
\mathcal{O}_{10, \,(1 , 1,2)}^{13} \equiv DD\bar{H}\bar{H}duu\nu \,, \\
\mathcal{O}_{10, \,(1 , 1,2)}^{14} \equiv DDHHddd\nu \,. 
 \label{eq:2Ddequal10basisB1L1Unique}
 \end{math}
 \end{spacing}
 \end{multicols}
For the two-derivative operators, we depart from our standard convention of listing all SMEFT operators first, followed by the $\nu$SMEFT operators. Instead, the operators $\mathcal{O}_{10,(1,1,2)}^{j}$ are ordered to preserve their correspondence with the dimension-8 operators listed in Sec.~B of Ref.~\cite{Heeck:2026dmh}, i.e.~$\mathcal{O}_{10,(1,1,2)}^{j} \equiv D^2\mathcal{O}_{8,(1,1)}^{j},$ $j\in\{1\isep 10\},$ with the index $j$ following the ordering adopted in Ref.~\cite{Heeck:2026dmh}. As for $j\in\{11\isep 14\}$, the lower-derivative analogues of $\mathcal{O}_{10,(1,1,2)}^{j}$ vanish due to Higgs asymmetry and therefore do not appear in the dimension-8 basis of Ref.~\cite{Heeck:2026dmh}. We retain the established ordering for operators with nonvanishing dimension-8 counterparts and append the minimal-derivative operators.
In the interest of article length, we only present UV completions for these minimal-derivative operators $\mathcal{O}_{10,(1,1,2)}^{11,12,13,14}$, collected in tables~\ref{tab:T10Top2B1L1D2}--\ref{tab:T10Top8B1L1D2}.

All three-derivative operators at mass dimension $d=10$ can be obtained from the dimension-8 one-derivative basis in Eq.~\eqref{eq:1Ddequal8basisB1L1}, according to $\mathcal{O}_{10,(1,1,3)}^{j} \equiv D^2\mathcal{O}_{8,(1,1,1)}^{j},$ $j\in\{1\isep 10\}.$ Similarly, all four-derivative operators at $d=10$ can be obtained from the dimension-8 two-derivative basis in Eq.~\eqref{eq:2Ddequal8basisB1L1}, i.e.~$\mathcal{O}_{10,(1,1,4)}^{j} \equiv D^2\mathcal{O}_{8,(1,1,2)}^{j},$ $j\in\{1\isep 6\}.$

 \begin{center}
 \begin{longtable}{| c  | m{13cm} |} 
   \hline
  $\mathcal{O}^j{(\mathcal{T}_ {8} ^{ 2 })}$ &  New Particles \\
   \hline
   \hline
   \endhead
 \multicolumn{2}{|c|}{\hspace{9cm} Continued on next page}
   \endfoot
  $\mathcal{O}_{10,\;(1,1,2)}^{14}$ & {$S_{2}S_{3}S_{5}$} \\
   \hline
 \caption{Dimension 10 UV completions for topology $\mathcal{T}_ {8} ^{ 2 }$ (see Fig.~\ref{fig:d=8_topologies}).} 
 \label{tab:T10Top2B1L1D2}\\
   \endlastfoot
  $\mathcal{O}_{10,\;(1,1,2)}^{11}$ & {$S_{1}S_{2}S_{11}$} \\
   \hline
  $\mathcal{O}_{10,\;(1,1,2)}^{12}$ & {$S_{2}S_{3}S_{5}$, $S_{1}S_{2}S_{11}$} \\
   \hline
  $\mathcal{O}_{10,\;(1,1,2)}^{13}$ & {$S_{2}S_{3}S_{5}$, $S_{1}S_{2}S_{11}$} \\
   \hline
 \end{longtable}
 \end{center}

 \begin{center}
 \begin{longtable}{| c  | m{13cm} |} 
   \hline
  $\mathcal{O}^j{(\mathcal{T}_ {8} ^{ 3 })}$ &  New Particles \\
   \hline
   \hline
   \endhead
 \multicolumn{2}{|c|}{\hspace{9cm} Continued on next page}
   \endfoot
  $\mathcal{O}_{10,\;(1,1,2)}^{14}$ & {$S_{2}F_{8}F_{8}$, $S_{3}F_{3}F_{8}$} \\
   \hline
 \caption{Dimension 10 UV completions for topology $\mathcal{T}_ {8} ^{ 3 }$ (see Fig.~\ref{fig:d=8_topologies}).} 
 \label{tab:T10Top3B1L1D2}\\
   \endlastfoot
  $\mathcal{O}_{10,\;(1,1,2)}^{11}$ & {$S_{2}F_{8}F_{8}$, $S_{1}F_{8}F_{10}$} \\
   \hline
  $\mathcal{O}_{10,\;(1,1,2)}^{12}$ & {$S_{2}F_{8}F_{8}$, $S_{1}F_{8}F_{13}$, $S_{2}F_{3}F_{8}$, $S_{3}F_{3}F_{13}$} \\
   \hline
  $\mathcal{O}_{10,\;(1,1,2)}^{13}$ & {$S_{3}F_{7}F_{8}$, $S_{1}F_{3}F_{7}$, $S_{2}F_{8}F_{8}$, $S_{2}F_{3}F_{8}$} \\
   \hline
 \end{longtable}
 \end{center}

 \begin{center}
 \begin{longtable}{| c  | m{13cm} |} 
   \hline
  $\mathcal{O}^j{(\mathcal{T}_ {8} ^{ 4 })}$ &  New Particles \\
   \hline
   \hline
   \endhead
 \multicolumn{2}{|c|}{\hspace{9cm} Continued on next page}
   \endfoot
  $\mathcal{O}_{10,\;(1,1,2)}^{14}$ & {$S_{2}F_{6}F_{8}$, $S_{3}F_{2}F_{3}$, $S_{3}F_{6}F_{8}$} \\
   \hline
 \caption{Dimension 10 UV completions for topology $\mathcal{T}_ {8} ^{ 4 }$ (see Fig.~\ref{fig:d=8_topologies}).} 
 \label{tab:T10Top4B1L1D2}\\
   \endlastfoot
  $\mathcal{O}_{10,\;(1,1,2)}^{11}$ & {$S_{2}F_{5}F_{8}$, $S_{1}F_{10}F_{16}$, $S_{1}F_{5}F_{8}$} \\
   \hline
  $\mathcal{O}_{10,\;(1,1,2)}^{12}$ & {$S_{2}F_{6}F_{8}$, $S_{1}F_{13}F_{18}$, $S_{2}F_{1}F_{3}$, $S_{1}F_{6}F_{8}$, $S_{3}F_{1}F_{3}$, $S_{3}F_{13}F_{18}$} \\
   \hline
  $\mathcal{O}_{10,\;(1,1,2)}^{13}$ & {$S_{3}F_{5}F_{8}$, $S_{1}F_{2}F_{3}$, $S_{3}F_{7}F_{17}$, $S_{2}F_{5}F_{8}$, $S_{2}F_{2}F_{3}$, $S_{1}F_{7}F_{17}$} \\
   \hline
 \end{longtable}
 \end{center}

 \begin{center}
 \begin{longtable}{| c  | m{13cm} |} 
   \hline
  $\mathcal{O}^j{(\mathcal{T}_ {8} ^{ 5 })}$ &  New Particles \\
   \hline
   \hline
   \endhead
 \multicolumn{2}{|c|}{\hspace{9cm} Continued on next page}
   \endfoot
  $\mathcal{O}_{10,\;(1,1,2)}^{14}$ & {$S_{3}S_{5}F_{3}$, $S_{2}S_{5}F_{8}$, $S_{3}S_{5}F_{8}$} \\
   \hline
 \caption{Dimension 10 UV completions for topology $\mathcal{T}_ {8} ^{ 5 }$ (see Fig.~\ref{fig:d=8_topologies}).} 
 \label{tab:T10Top5B1L1D2}\\
   \endlastfoot
  $\mathcal{O}_{10,\;(1,1,2)}^{11}$ & {$S_{1}S_{11}F_{10}$, $S_{2}S_{11}F_{8}$, $S_{1}S_{11}F_{8}$} \\
   \hline
  $\mathcal{O}_{10,\;(1,1,2)}^{12}$ & {$S_{2}S_{11}F_{3}$, $S_{1}S_{11}F_{13}$, $S_{2}S_{5}F_{8}$, $S_{3}S_{5}F_{3}$, $S_{1}S_{11}F_{8}$, $S_{3}S_{5}F_{13}$, $S_{2}S_{11}F_{8}$} \\
   \hline
  $\mathcal{O}_{10,\;(1,1,2)}^{13}$ & {$S_{1}S_{11}F_{3}$, $S_{3}S_{5}F_{8}$, $S_{2}S_{5}F_{3}$, $S_{2}S_{11}F_{8}$, $S_{3}S_{5}F_{7}$, $S_{2}S_{5}F_{8}$, $S_{1}S_{11}F_{7}$} \\
   \hline
 \end{longtable}
 \end{center}

 \begin{center}
 \begin{longtable}{| c  | m{13cm} |} 
   \hline
  $\mathcal{O}^j{(\mathcal{T}_ {8} ^{ 8 })}$ &  New Particles \\
   \hline
   \hline
   \endhead
 \multicolumn{2}{|c|}{\hspace{9cm} Continued on next page}
   \endfoot
  $\mathcal{O}_{10,\;(1,1,2)}^{14}$ & {$S_{5}F_{3}F_{8}$, $S_{5}F_{8}F_{8}$} \\
   \hline
 \caption{Dimension 10 UV completions for topology $\mathcal{T}_ {8} ^{ 8 }$ (see Fig.~\ref{fig:d=8_topologies}).} 
 \label{tab:T10Top8B1L1D2}\\
   \endlastfoot
  $\mathcal{O}_{10,\;(1,1,2)}^{11}$ & {$S_{11}F_{8}F_{10}$, $S_{11}F_{8}F_{8}$} \\
   \hline
  $\mathcal{O}_{10,\;(1,1,2)}^{12}$ & {$S_{11}F_{3}F_{13}$, $S_{5}F_{3}F_{8}$, $S_{11}F_{3}F_{8}$, $S_{5}F_{8}F_{13}$, $S_{11}F_{8}F_{13}$, $S_{11}F_{8}F_{8}$} \\
   \hline
  $\mathcal{O}_{10,\;(1,1,2)}^{13}$ & {$S_{5}F_{3}F_{8}$, $S_{11}F_{3}F_{8}$, $S_{5}F_{8}F_{8}$, $S_{5}F_{3}F_{7}$, $S_{5}F_{7}F_{8}$, $S_{11}F_{7}F_{8}$} \\
   \hline
 \end{longtable}
 \end{center}

  \subsection{\texorpdfstring{$\Delta B=1, \Delta L=-3$}{Delta B=1,Delta L=-3}}

For $(\Delta B, \Delta L)=(1,-3)$, we only have one-derivative operators,
 \begin{multicols}{3}
 \begin{spacing}{1.5}
 \noindent
 \begin{math}
\mathcal{O}_{10, \, (1,-3,1)}^{1}\equiv Dddd\bar{L}\bar{L}\bar{\nu}\,,\\
\mathcal{O}_{10, \, (1,-3,1)}^{2}\equiv Dddd\bar{e}\bar{\nu}\bar{\nu}\,,\\
\mathcal{O}_{10, \, (1,-3,1)}^{3}\equiv DddQ\bar{L}\bar{\nu}\bar{\nu}\,,\\
\mathcal{O}_{10, \, (1,-3,1)}^{4}\equiv DdQQ\bar{\nu}\bar{\nu}\bar{\nu}\,,\\
\mathcal{O}_{10, \, (1,-3,1)}^{5}\equiv Dddu\bar{\nu}\bar{\nu}\bar{\nu}\,,
\label{eq:1Ddequal10basisB1L-3}
 \end{math}
 \end{spacing}
 \end{multicols}
 \noindent
and we collect their UV completions in Tab.~\ref{tab:T10Top2B1L-3D1}. Notice that all operators contain right-handed neutrinos, and that there is only one relevant topology.

\begin{center}
\begin{longtable}{| c | m{13cm} |}
   \hline
   $\mathcal{O}^j(\mathcal{T}_{9}^{2})$ & New Particles \\
   \hline
   \hline
   \endhead
   \multicolumn{2}{|c|}{\hspace{9cm} Continued on next page} \\
   \endfoot
  $\mathcal{O}_{10,(1,-3,1)}^{5}$ & {$S_{3}S_{7}F_{1}$, $S_{3}S_{7}F_{6}$, $S_{2}S_{7}F_{1}$, $S_{2}S_{7}F_{5}$} \\
   \hline
   \caption{Dimension 10 UV completions for topology $\mathcal{T}_{9}^{2}$ (see Fig.~\ref{fig:d=9_topologies}).}
   \label{tab:T10Top2B1L-3D1}\\
   \endlastfoot
  $\mathcal{O}_{10,(1,-3,1)}^{1}$ & {$S_{3}S_{5}F_{8}$, $S_{3}S_{5}F_{6}$, $S_{3}S_{14}F_{2}$, $S_{5}S_{5}F_{8}$, $S_{3}S_{14}F_{6}$} \\
   \hline
  $\mathcal{O}_{10,(1,-3,1)}^{2}$ & {$S_{3}S_{14}F_{2}$, $S_{3}S_{14}F_{6}$, $S_{3}S_{7}F_{2}$, $S_{3}S_{7}F_{5}$} \\
   \hline
  $\mathcal{O}_{10,(1,-3,1)}^{3}$ & {$S_{3}S_{5}F_{8}$, $S_{3}S_{5}F_{6}$, $S_{5}S_{5}F_{8}$, $S_{3}S_{7}F_{3}$, $S_{3}S_{7}F_{8}$, $S_{5}S_{7}F_{5}$, $S_{5}S_{7}F_{8}$} \\
   \hline
  $\mathcal{O}_{10,(1,-3,1)}^{4}$ & {$S_{2}S_{7}F_{1}$, $S_{5}S_{5}F_{8}$, $S_{2}S_{7}F_{5}$, $S_{5}S_{7}F_{5}$, $S_{5}S_{7}F_{8}$} \\
   \hline
\end{longtable}
\end{center}

 \section{Dimension 11}
 \label{sec:d=11}
 
 At $d=11$, derivative BNV operators carry either $(\Delta B, \Delta L) = (1, -1)$, $( 1,  3)$ or $(2,  0)$, split into subsections below.
 
   \subsection{\texorpdfstring{$\Delta B=1, \Delta L=-1$}{Delta B=1,Delta L=-1}}

We have one-derivative operators,
\begin{multicols}{2}
 \begin{spacing}{1.5}
 \noindent
 \begin{math}
\mathcal{O}_{11, \, (1,-1,1)}^{1}\equiv DHdddd\bar{Q}\bar{e}\,,\\
\mathcal{O}_{11, \, (1,-1,1)}^{2}\equiv DHddd\bar{u}Q\bar{e}\,,\\
\mathcal{O}_{11, \, (1,-1,1)}^{3}\equiv D\bar{H}ddd\bar{d}Q\bar{e}\,,\\
\mathcal{O}_{11, \, (1,-1,1)}^{4}\equiv D\bar{H}dddL\bar{e}\bar{e}\,,\\
\mathcal{O}_{11, \, (1,-1,1)}^{5}\equiv D\bar{H}ddQe\bar{e}\bar{e}\,,\\
\mathcal{O}_{11, \, (1,-1,1)}^{6}\equiv D\bar{H}ddQQ\bar{Q}\bar{e}\,,\\
\mathcal{O}_{11, \, (1,-1,1)}^{7}\equiv D\bar{H}dddu\bar{Q}\bar{e}\,,\\
\mathcal{O}_{11, \, (1,-1,1)}^{8}\equiv D\bar{H}d\bar{u}QQQ\bar{e}\,,\\
\mathcal{O}_{11, \, (1,-1,1)}^{9}\equiv D\bar{H}ddu\bar{u}Q\bar{e}\,,\\
\mathcal{O}_{11, \, (1,-1,1)}^{10}\equiv D\bar{H}dd\bar{d}QQ\bar{L}\,,\\
\mathcal{O}_{11, \, (1,-1,1)}^{11}\equiv D\bar{H}ddd\bar{d}u\bar{L}\,,\\
\mathcal{O}_{11, \, (1,-1,1)}^{12}\equiv D\bar{H}ddQL\bar{L}\bar{e}\,,\\
\mathcal{O}_{11, \, (1,-1,1)}^{13}\equiv D\bar{H}dQQ\bar{L}e\bar{e}\,,\\
\mathcal{O}_{11, \, (1,-1,1)}^{14}\equiv D\bar{H}ddu\bar{L}e\bar{e}\,,\\
\mathcal{O}_{11, \, (1,-1,1)}^{15}\equiv D\bar{H}dQQL\bar{L}\bar{L}\,,\\
\mathcal{O}_{11, \, (1,-1,1)}^{16}\equiv D\bar{H}QQQ\bar{L}\bar{L}e\,,\\
\mathcal{O}_{11, \, (1,-1,1)}^{17}\equiv D\bar{H}dduL\bar{L}\bar{L}\,,\\
\mathcal{O}_{11, \, (1,-1,1)}^{18}\equiv D\bar{H}duQ\bar{L}\bar{L}e\,,\\
\mathcal{O}_{11, \, (1,-1,1)}^{19}\equiv D\bar{H}dQQQ\bar{Q}\bar{L}\,,\\
\mathcal{O}_{11, \, (1,-1,1)}^{20}\equiv D\bar{H}dduQ\bar{Q}\bar{L}\,,\\
\mathcal{O}_{11, \, (1,-1,1)}^{21}\equiv D\bar{H}\bar{u}QQQQ\bar{L}\,,\\
\mathcal{O}_{11, \, (1,-1,1)}^{22}\equiv D\bar{H}du\bar{u}QQ\bar{L}\,,\\
\mathcal{O}_{11, \, (1,-1,1)}^{23}\equiv D\bar{H}dduu\bar{u}\bar{L}\,,\\
\mathcal{O}_{11, \, (1,-1,1)}^{24}\equiv DHdddd\bar{d}\bar{L}\,,\\
\mathcal{O}_{11, \, (1,-1,1)}^{25}\equiv DHddd\bar{L}e\bar{e}\,,\\
\mathcal{O}_{11, \, (1,-1,1)}^{26}\equiv DHdddL\bar{L}\bar{L}\,,\\
\mathcal{O}_{11, \, (1,-1,1)}^{27}\equiv DHddQ\bar{L}\bar{L}e\,,\\
\mathcal{O}_{11, \, (1,-1,1)}^{28}\equiv DHdddQ\bar{Q}\bar{L}\,,\\
\mathcal{O}_{11, \, (1,-1,1)}^{29}\equiv DHdd\bar{u}QQ\bar{L}\,,\\
\mathcal{O}_{11, \, (1,-1,1)}^{30}\equiv DHdddu\bar{u}\bar{L}\,,\\
\mathcal{O}_{11, \, (1,-1,1)}^{31}\equiv D\bar{H}ddd\bar{L}\bar{e}\nu\,,\\
\mathcal{O}_{11, \, (1,-1,1)}^{32}\equiv D\bar{H}ddQ\bar{L}\bar{L}\nu\,,\\
\mathcal{O}_{11, \, (1,-1,1)}^{33}\equiv DHddd\bar{d}Q\bar{\nu}\,,\\
\mathcal{O}_{11, \, (1,-1,1)}^{34}\equiv DHdddL\bar{e}\bar{\nu}\,,\\
\mathcal{O}_{11, \, (1,-1,1)}^{35}\equiv DHddQe\bar{e}\bar{\nu}\,,\\
\mathcal{O}_{11, \, (1,-1,1)}^{36}\equiv D\bar{H}d\bar{d}QQQ\bar{\nu}\,,\\
\mathcal{O}_{11, \, (1,-1,1)}^{37}\equiv D\bar{H}dd\bar{d}uQ\bar{\nu}\,,\\
\mathcal{O}_{11, \, (1,-1,1)}^{38}\equiv D\bar{H}dQQL\bar{e}\bar{\nu}\,,\\
\mathcal{O}_{11, \, (1,-1,1)}^{39}\equiv D\bar{H}QQQe\bar{e}\bar{\nu}\,,\\
\mathcal{O}_{11, \, (1,-1,1)}^{40}\equiv D\bar{H}dduL\bar{e}\bar{\nu}\,,\\
\mathcal{O}_{11, \, (1,-1,1)}^{41}\equiv D\bar{H}duQe\bar{e}\bar{\nu}\,,\\
\mathcal{O}_{11, \, (1,-1,1)}^{42}\equiv D\bar{H}QQQL\bar{L}\bar{\nu}\,,\\
\mathcal{O}_{11, \, (1,-1,1)}^{43}\equiv D\bar{H}duQL\bar{L}\bar{\nu}\,,\\
\mathcal{O}_{11, \, (1,-1,1)}^{44}\equiv D\bar{H}uQQ\bar{L}e\bar{\nu}\,,\\
\mathcal{O}_{11, \, (1,-1,1)}^{45}\equiv D\bar{H}duu\bar{L}e\bar{\nu}\,,\\
\mathcal{O}_{11, \, (1,-1,1)}^{46}\equiv D\bar{H}QQQQ\bar{Q}\bar{\nu}\,,\\
\mathcal{O}_{11, \, (1,-1,1)}^{47}\equiv D\bar{H}duQQ\bar{Q}\bar{\nu}\,,\\
\mathcal{O}_{11, \, (1,-1,1)}^{48}\equiv D\bar{H}dduu\bar{Q}\bar{\nu}\,,\\
\mathcal{O}_{11, \, (1,-1,1)}^{49}\equiv D\bar{H}u\bar{u}QQQ\bar{\nu}\,,\\
\mathcal{O}_{11, \, (1,-1,1)}^{50}\equiv D\bar{H}duu\bar{u}Q\bar{\nu}\,,\\
\mathcal{O}_{11, \, (1,-1,1)}^{51}\equiv DHddQL\bar{L}\bar{\nu}\,,\\
\mathcal{O}_{11, \, (1,-1,1)}^{52}\equiv DHdQQ\bar{L}e\bar{\nu}\,,\\
\mathcal{O}_{11, \, (1,-1,1)}^{53}\equiv DHddu\bar{L}e\bar{\nu}\,,\\
\mathcal{O}_{11, \, (1,-1,1)}^{54}\equiv DHddQQ\bar{Q}\bar{\nu}\,,\\
\mathcal{O}_{11, \, (1,-1,1)}^{55}\equiv DHdddu\bar{Q}\bar{\nu}\,,\\
\mathcal{O}_{11, \, (1,-1,1)}^{56}\equiv DHd\bar{u}QQQ\bar{\nu}\,,\\
\mathcal{O}_{11, \, (1,-1,1)}^{57}\equiv DHddu\bar{u}Q\bar{\nu}\,,\\
\mathcal{O}_{11, \, (1,-1,1)}^{58}\equiv D\bar{H}ddQ\bar{e}\nu\bar{\nu}\,,\\
\mathcal{O}_{11, \, (1,-1,1)}^{59}\equiv D\bar{H}dQQ\bar{L}\nu\bar{\nu}\,,\\
\mathcal{O}_{11, \, (1,-1,1)}^{60}\equiv D\bar{H}ddu\bar{L}\nu\bar{\nu}\,,\\
\mathcal{O}_{11, \, (1,-1,1)}^{61}\equiv DHddd\bar{L}\nu\bar{\nu}\,,\\
\mathcal{O}_{11, \, (1,-1,1)}^{62}\equiv D\bar{H}uQQL\bar{\nu}\bar{\nu}\,,\\
\mathcal{O}_{11, \, (1,-1,1)}^{63}\equiv D\bar{H}duuL\bar{\nu}\bar{\nu}\,,\\
\mathcal{O}_{11, \, (1,-1,1)}^{64}\equiv D\bar{H}uuQe\bar{\nu}\bar{\nu}\,,\\
\mathcal{O}_{11, \, (1,-1,1)}^{65}\equiv DHdQQL\bar{\nu}\bar{\nu}\,,\\
\mathcal{O}_{11, \, (1,-1,1)}^{66}\equiv DHQQQe\bar{\nu}\bar{\nu}\,,\\
\mathcal{O}_{11, \, (1,-1,1)}^{67}\equiv DHdduL\bar{\nu}\bar{\nu}\,,\\
\mathcal{O}_{11, \, (1,-1,1)}^{68}\equiv DHduQe\bar{\nu}\bar{\nu}\,,\\
\mathcal{O}_{11, \, (1,-1,1)}^{69}\equiv D\bar{H}QQQ\nu\bar{\nu}\bar{\nu}\,,\\
\mathcal{O}_{11, \, (1,-1,1)}^{70}\equiv D\bar{H}duQ\nu\bar{\nu}\bar{\nu}\,,\\
\mathcal{O}_{11, \, (1,-1,1)}^{71}\equiv DHddQ\nu\bar{\nu}\bar{\nu}\,,\\
\mathcal{O}_{11, \, (1,-1,1)}^{72}\equiv DHH\bar{H}\bar{H}ddd\bar{e}\,,\\
\mathcal{O}_{11, \, (1,-1,1)}^{73}\equiv DH\bar{H}\bar{H}\bar{H}dQQ\bar{e}\,,\\
\mathcal{O}_{11, \, (1,-1,1)}^{74}\equiv DH\bar{H}\bar{H}\bar{H}ddu\bar{e}\,,\\
\mathcal{O}_{11, \, (1,-1,1)}^{75}\equiv DH\bar{H}\bar{H}\bar{H}QQQ\bar{L}\,,\\
\mathcal{O}_{11, \, (1,-1,1)}^{76}\equiv DH\bar{H}\bar{H}\bar{H}duQ\bar{L}\,,\\
\mathcal{O}_{11, \, (1,-1,1)}^{77}\equiv DHH\bar{H}\bar{H}ddQ\bar{L}\,,\\
\mathcal{O}_{11, \, (1,-1,1)}^{78}\equiv D\bar{H}\bar{H}\bar{H}\bar{H}uQQ\bar{e}\,,\\
\mathcal{O}_{11, \, (1,-1,1)}^{79}\equiv D\bar{H}\bar{H}\bar{H}\bar{H}uuQ\bar{L}\,,\\
\mathcal{O}_{11, \, (1,-1,1)}^{80}\equiv DH\bar{H}\bar{H}\bar{H}uQQ\bar{\nu}\,,\\
\mathcal{O}_{11, \, (1,-1,1)}^{81}\equiv DH\bar{H}\bar{H}\bar{H}duu\bar{\nu}\,,\\
\mathcal{O}_{11, \, (1,-1,1)}^{82}\equiv DHH\bar{H}\bar{H}dQQ\bar{\nu}\,,\\
\mathcal{O}_{11, \, (1,-1,1)}^{83}\equiv DHH\bar{H}\bar{H}ddu\bar{\nu}\,,\\
\mathcal{O}_{11, \, (1,-1,1)}^{84}\equiv DHHH\bar{H}ddd\bar{\nu}\,,
\label{eq:1Ddequal11basisB1L-1}
 \end{math}
 \end{spacing}
 \end{multicols}
 \noindent
two-derivative operators,
\begin{multicols}{2}
 \begin{spacing}{1.5}
 \noindent
 \begin{math}
\mathcal{O}_{11, \,(1 , -1 ,2)}^{1}  \equiv DDdddd\bar{d}\bar{e} \,, \\
 \mathcal{O}_{11, \,(1 , -1 ,2)}^{2}  \equiv DDddde\bar{e}\bar{e} \,, \\
 \mathcal{O}_{11, \,(1 , -1 ,2)}^{3}  \equiv DDdddQ\bar{Q}\bar{e} \,, \\
 \mathcal{O}_{11, \,(1 , -1 ,2)}^{4}  \equiv DDdd\bar{u}QQ\bar{e} \,, \\
 \mathcal{O}_{11, \,(1 , -1 ,2)}^{5}  \equiv DDdddu\bar{u}\bar{e} \,, \\
 \mathcal{O}_{11, \,(1 , -1 ,2)}^{6}  \equiv DDddd\bar{d}Q\bar{L} \,, \\
 \mathcal{O}_{11, \,(1 , -1 ,2)}^{7}  \equiv DDddd\bar{e}L\bar{L} \,, \\
 \mathcal{O}_{11, \,(1 , -1 ,2)}^{8}  \equiv DDddQe\bar{e}\bar{L} \,, \\
 \mathcal{O}_{11, \,(1 , -1 ,2)}^{9}  \equiv DDddQL\bar{L}\bar{L} \,, \\
 \mathcal{O}_{11, \,(1 , -1 ,2)}^{10}  \equiv DDdQQe\bar{L}\bar{L} \,, \\
 \mathcal{O}_{11, \,(1 , -1 ,2)}^{11}  \equiv DDddue\bar{L}\bar{L} \,, \\
 \mathcal{O}_{11, \,(1 , -1 ,2)}^{12}  \equiv DDddQQ\bar{Q}\bar{L} \,, \\
 \mathcal{O}_{11, \,(1 , -1 ,2)}^{13}  \equiv DDdddu\bar{Q}\bar{L} \,, \\
 \mathcal{O}_{11, \,(1 , -1 ,2)}^{14}  \equiv DDd\bar{u}QQQ\bar{L} \,, \\
 \mathcal{O}_{11, \,(1 , -1 ,2)}^{15}  \equiv DDddu\bar{u}Q\bar{L} \,, \\
 \mathcal{O}_{11, \,(1 , -1 ,2)}^{16}  \equiv DDddd\bar{L}\bar{L}\nu \,, \\
 \mathcal{O}_{11, \,(1 , -1 ,2)}^{17}  \equiv DDdd\bar{d}QQ\bar{\nu} \,, \\
 \mathcal{O}_{11, \,(1 , -1 ,2)}^{18}  \equiv DDddd\bar{d}u\bar{\nu} \,, \\
 \mathcal{O}_{11, \,(1 , -1 ,2)}^{19}  \equiv DDddQ\bar{e}L\bar{\nu} \,, \\
 \mathcal{O}_{11, \,(1 , -1 ,2)}^{20}  \equiv DDdQQe\bar{e}\bar{\nu} \,, \\
 \mathcal{O}_{11, \,(1 , -1 ,2)}^{21}  \equiv DDddue\bar{e}\bar{\nu} \,, \\
 \mathcal{O}_{11, \,(1 , -1 ,2)}^{22}  \equiv DDdQQL\bar{L}\bar{\nu} \,, \\
 \mathcal{O}_{11, \,(1 , -1 ,2)}^{23}  \equiv DDQQQe\bar{L}\bar{\nu} \,, \\
 \mathcal{O}_{11, \,(1 , -1 ,2)}^{24}  \equiv DDdduL\bar{L}\bar{\nu} \,, \\
 \mathcal{O}_{11, \,(1 , -1 ,2)}^{25}  \equiv DDduQe\bar{L}\bar{\nu} \,, \\
 \mathcal{O}_{11, \,(1 , -1 ,2)}^{26}  \equiv DDdQQQ\bar{Q}\bar{\nu} \,, \\
 \mathcal{O}_{11, \,(1 , -1 ,2)}^{27}  \equiv DDdduQ\bar{Q}\bar{\nu} \,, \\
 \mathcal{O}_{11, \,(1 , -1 ,2)}^{28}  \equiv DD\bar{u}QQQQ\bar{\nu} \,, \\
 \mathcal{O}_{11, \,(1 , -1 ,2)}^{29}  \equiv DDdu\bar{u}QQ\bar{\nu} \,, \\
 \mathcal{O}_{11, \,(1 , -1 ,2)}^{30}  \equiv DDdduu\bar{u}\bar{\nu} \,, \\
 \mathcal{O}_{11, \,(1 , -1 ,2)}^{31}  \equiv DDddd\bar{e}\nu\bar{\nu} \,, \\
 \mathcal{O}_{11, \,(1 , -1 ,2)}^{32}  \equiv DDddQ\bar{L}\nu\bar{\nu} \,, \\
 \mathcal{O}_{11, \,(1 , -1 ,2)}^{33}  \equiv DDQQQL\bar{\nu}\bar{\nu} \,, \\
 \mathcal{O}_{11, \,(1 , -1 ,2)}^{34}  \equiv DDduQL\bar{\nu}\bar{\nu} \,, \\
 \mathcal{O}_{11, \,(1 , -1 ,2)}^{35}  \equiv DDuQQe\bar{\nu}\bar{\nu} \,, \\
 \mathcal{O}_{11, \,(1 , -1 ,2)}^{36}  \equiv DDduue\bar{\nu}\bar{\nu} \,, \\
 \mathcal{O}_{11, \,(1 , -1 ,2)}^{37}  \equiv DDdQQ\nu\bar{\nu}\bar{\nu} \,, \\
 \mathcal{O}_{11, \,(1 , -1 ,2)}^{38}  \equiv DDddu\nu\bar{\nu}\bar{\nu} \,, \\
 \mathcal{O}_{11, \,(1 , -1 ,2)}^{39}  \equiv DDH\bar{H}\bar{H}ddQ\bar{e} \,, \\
 \mathcal{O}_{11, \,(1 , -1 ,2)}^{40}  \equiv DDH\bar{H}\bar{H}dQQ\bar{L} \,, \\
 \mathcal{O}_{11, \,(1 , -1 ,2)}^{41}  \equiv DDH\bar{H}\bar{H}ddu\bar{L} \,, \\
 \mathcal{O}_{11, \,(1 , -1 ,2)}^{42}  \equiv DDHH\bar{H}ddd\bar{L} \,, \\
 \mathcal{O}_{11, \,(1 , -1 ,2)}^{43}  \equiv DD\bar{H}\bar{H}\bar{H}QQQ\bar{e} \,, \\
 \mathcal{O}_{11, \,(1 , -1 ,2)}^{44}  \equiv DD\bar{H}\bar{H}\bar{H}uQQ\bar{L} \,, \\
 \mathcal{O}_{11, \,(1 , -1 ,2)}^{45}  \equiv DDHH\bar{H}ddQ\bar{\nu} \,, \\
 \mathcal{O}_{11, \,(1 , -1 ,2)}^{46}  \equiv DDH\bar{H}\bar{H}QQQ\bar{\nu} \,, \\
 \mathcal{O}_{11, \,(1 , -1 ,2)}^{47}  \equiv DDH\bar{H}\bar{H}duQ\bar{\nu} \,,
 \label{eq:2Ddequal11basisB1L-1}
 \end{math}
 \end{spacing}
 \end{multicols}
\noindent
and minimal-derivative operators,
\begin{multicols}{2}
 \begin{spacing}{1.5}
 \noindent
 \begin{math}
\mathcal{O}_{11, \,(1 , -1 ,2)}^{48}  \equiv DD\bar{H}\bar{H}\bar{H}duQ\bar{e} \,, \\
 \mathcal{O}_{11, \,(1 , -1 ,2)}^{49}  \equiv DD\bar{H}\bar{H}\bar{H}duu\bar{L} \,, \\
 \mathcal{O}_{11, \,(1 , -1 ,2)}^{50}  \equiv DD\bar{H}\bar{H}\bar{H}uuQ\bar{\nu} \,.
 \label{eq:2Ddequal11basisB1L-1Unique}
 \end{math}
 \end{spacing}
 \end{multicols}
For the two-derivative operators at $d=11$, we again depart from our standard convention of listing all SMEFT operators first, followed by the $\nu$SMEFT operators. $\mathcal{O}_{11,(1,-1,2)}^{j}$ for $j\in\{1\isep 47\}$ correspond to the dimension-9 operators listed in Sec.~D of Ref.~\cite{Heeck:2026dmh}, according to $\mathcal{O}_{11,(1,-1,2)}^{j}\equiv D^2\mathcal{O}_{9,(1,-1)}^{j}$. 
The minimal-derivative operators $\mathcal{O}_{11,(1,-1,2)}^{48,49,50}$  vanish due to Higgs asymmetry without derivatives and  therefore do not appear in the dimension-9 basis of Ref.~\cite{Heeck:2026dmh}. Below, we only present the UV completions for these minimal-derivative operators in tables~\ref{tab:T11Top10B1L-1D2}--\ref{tab:T11Top23B1L-1D2}.

All three-derivative operators at mass dimension $d=11$ can be obtained from the dimension-9 one-derivative basis in Eq.~\eqref{eq:1Ddequal9basisB1L-1}, according to $\mathcal{O}_{11,(1,-1,3)}^{j} \equiv D^2\mathcal{O}_{9,(1,-1,1)}^{j},$ $j\in\{1\isep 11\}.$ Similarly, all four-derivative operators at $d=11$ can be obtained from the dimension-9 two-derivative basis in Eq.~\eqref{eq:2Ddequal9basisB1L-1}, according to $\mathcal{O}_{11,(1,-1,4)}^{j} \equiv D^2\mathcal{O}_{9,(1,-1,2)}^{j},$ $j\in\{1\isep 7\}.$ Finally, all five-derivative operators at $d=11$ can be obtained from the dimension-7 one-derivative basis in Eq.~\eqref{eq:1Ddequal7basisB1L-1}, according to $\mathcal{O}_{11,(1,-1,5)}^{j} \equiv D^4\mathcal{O}_{7,(1,-1,1)}^{j},$ $j\in\{1\isep 4\}.$

 \begin{center}
\begin{longtable}{| c  | m{13cm} |} 
   \hline
  $\mathcal{O}^j{(\mathcal{T}_ {9} ^{ 3 })}$ &  New Particles \\
   \hline
   \hline
   \endhead
 \multicolumn{2}{|c|}{\hspace{9cm} Continued on next page}
   \endfoot
  $\mathcal{O}_{11,\;(1,-1,2)}^{50}$ & {$S_{1}S_{5}S_{9}$} \\
   \hline
 \caption{Dimension 11 UV completions for topology $\mathcal{T}_ {9} ^{ 3 }$ (see Fig.~\ref{fig:d=9_topologies}).} 
 \label{tab:T11Top3B1L-1D2}\\
   \endlastfoot
  $\mathcal{O}_{11,\;(1,-1,2)}^{48}$ & {$S_{2}S_{6}S_{9}$} \\
   \hline
  $\mathcal{O}_{11,\;(1,-1,2)}^{49}$ & {$S_{2}S_{6}S_{9}$, $S_{1}S_{5}S_{9}$} \\
   \hline
 \end{longtable}
 \end{center}

 \begin{center}
 \begin{longtable}{| c  | m{13cm} |} 
   \hline
  $\mathcal{O}^j{(\mathcal{T}_ {9} ^{ 4 })}$ &  New Particles \\
   \hline
   \hline
   \endhead
 \multicolumn{2}{|c|}{\hspace{9cm} Continued on next page}
   \endfoot
  $\mathcal{O}_{11,\;(1,-1,2)}^{50}$ & {$S_{1}S_{4}S_{5}S_{9}$, $S_{1}S_{5}S_{9}S_{11}$} \\
   \hline
 \caption{Dimension 11 UV completions for topology $\mathcal{T}_ {9} ^{ 4 }$ (see Fig.~\ref{fig:d=9_topologies}).} 
 \label{tab:T11Top4B1L-1D2}\\
   \endlastfoot
  $\mathcal{O}_{11,\;(1,-1,2)}^{48}$ & {$S_{2}S_{6}S_{9}S_{19}$, $S_{2}S_{5}S_{6}S_{9}$} \\
   \hline
  $\mathcal{O}_{11,\;(1,-1,2)}^{49}$ & {$S_{2}S_{6}S_{9}S_{19}$, $S_{1}S_{5}S_{9}S_{11}$, $S_{1}S_{4}S_{5}S_{9}$, $S_{2}S_{5}S_{6}S_{9}$} \\
   \hline
 \end{longtable}
 \end{center}

 \begin{center}
 \begin{longtable}{| c  | m{13cm} |} 
   \hline
  $\mathcal{O}^j{(\mathcal{T}_ {9} ^{ 5 })}$ &  New Particles \\
   \hline
   \hline
   \endhead
 \multicolumn{2}{|c|}{\hspace{9cm} Continued on next page}
   \endfoot
  $\mathcal{O}_{11,\;(1,-1,2)}^{50}$ & {$S_{1}S_{4}S_{5}S_{11}$, $S_{1}S_{2}S_{5}S_{11}$} \\
   \hline
 \caption{Dimension 11 UV completions for topology $\mathcal{T}_ {9} ^{ 5 }$ (see Fig.~\ref{fig:d=9_topologies}).} 
 \label{tab:T11Top5B1L-1D2}\\
   \endlastfoot
  $\mathcal{O}_{11,\;(1,-1,2)}^{48}$ & {$S_{2}S_{5}S_{6}S_{19}$, $S_{2}S_{3}S_{5}S_{6}$} \\
   \hline
  $\mathcal{O}_{11,\;(1,-1,2)}^{49}$ & {$S_{2}S_{5}S_{6}S_{19}$, $S_{2}S_{3}S_{5}S_{6}$, $S_{1}S_{4}S_{5}S_{11}$, $S_{1}S_{2}S_{5}S_{11}$} \\
   \hline
 \end{longtable}
 \end{center}

 \begin{center}
 \begin{longtable}{| c  | m{13cm} |} 
   \hline
  $\mathcal{O}^j{(\mathcal{T}_ {9} ^{ 6 })}$ &  New Particles \\
   \hline
   \hline
   \endhead
 \multicolumn{2}{|c|}{\hspace{9cm} Continued on next page}
   \endfoot
  $\mathcal{O}_{11,\;(1,-1,2)}^{50}$ & {$S_{4}S_{5}F_{8}F_{8}$, $S_{2}S_{5}F_{8}F_{8}$, $S_{1}S_{11}F_{3}F_{9}$, $S_{1}S_{11}F_{3}F_{5}$} \\
   \hline
 \caption{Dimension 11 UV completions for topology $\mathcal{T}_ {9} ^{ 6 }$ (see Fig.~\ref{fig:d=9_topologies}).} 
 \label{tab:T11Top6B1L-1D2}\\
   \endlastfoot
  $\mathcal{O}_{11,\;(1,-1,2)}^{48}$ & {$S_{6}S_{19}F_{7}F_{8}$, $S_{3}S_{6}F_{7}F_{8}$, $S_{2}S_{5}F_{3}F_{9}$, $S_{2}S_{5}F_{3}F_{5}$} \\
   \hline
  $\mathcal{O}_{11,\;(1,-1,2)}^{49}$ & {$S_{6}S_{19}F_{7}F_{8}$, $S_{3}S_{6}F_{7}F_{8}$, $S_{1}S_{11}F_{4}F_{7}$, $S_{1}S_{11}F_{1}F_{7}$, $S_{4}S_{5}F_{8}F_{8}$, $S_{2}S_{5}F_{8}F_{8}$, $S_{2}S_{5}F_{4}F_{8}$, $S_{2}S_{5}F_{1}F_{8}$} \\
   \hline
 \end{longtable}
 \end{center}

 \begin{center}
 \begin{longtable}{| c  | m{13cm} |} 
   \hline
  $\mathcal{O}^j{(\mathcal{T}_ {9} ^{ 7 })}$ &  New Particles \\
   \hline
   \hline
   \endhead
 \multicolumn{2}{|c|}{\hspace{9cm} Continued on next page}
   \endfoot
  $\mathcal{O}_{11,\;(1,-1,2)}^{50}$ & {$S_{5}F_{7}F_{8}F_{9}$, $S_{5}F_{5}F_{7}F_{8}$, $S_{1}F_{3}F_{10}F_{11}$, $S_{1}F_{2}F_{3}F_{10}$, $S_{1}F_{7}F_{9}F_{17}$, $S_{1}F_{5}F_{7}F_{17}$} \\
   \hline
 \caption{Dimension 11 UV completions for topology $\mathcal{T}_ {9} ^{ 7 }$ (see Fig.~\ref{fig:d=9_topologies}).} 
 \label{tab:T11Top7B1L-1D2}\\
   \endlastfoot
  $\mathcal{O}_{11,\;(1,-1,2)}^{48}$ & {$S_{6}F_{7}F_{8}F_{9}$, $S_{6}F_{5}F_{7}F_{8}$, $S_{6}F_{7}F_{19}F_{36}$, $S_{6}F_{7}F_{17}F_{36}$, $S_{2}F_{3}F_{3}F_{4}$, $S_{2}F_{1}F_{3}F_{3}$, $S_{2}F_{7}F_{9}F_{17}$, $S_{2}F_{5}F_{7}F_{17}$} \\
   \hline
  $\mathcal{O}_{11,\;(1,-1,2)}^{49}$ & {$S_{6}F_{7}F_{8}F_{9}$, $S_{6}F_{5}F_{7}F_{8}$, $S_{1}F_{2}F_{3}F_{4}$, $S_{1}F_{1}F_{2}F_{3}$, $S_{6}F_{7}F_{19}F_{36}$, $S_{6}F_{7}F_{17}F_{36}$, $S_{5}F_{7}F_{8}F_{9}$, $S_{5}F_{5}F_{7}F_{8}$, $S_{2}F_{2}F_{3}F_{4}$, $S_{2}F_{1}F_{2}F_{3}$, $S_{1}F_{7}F_{19}F_{36}$, $S_{1}F_{7}F_{17}F_{36}$, $S_{2}F_{7}F_{8}F_{9}$, $S_{2}F_{5}F_{7}F_{8}$} \\
   \hline
 \end{longtable}
 \end{center}

 \begin{center}
 \begin{longtable}{| c  | m{13cm} |} 
   \hline
  $\mathcal{O}^j{(\mathcal{T}_ {9} ^{ 8 })}$ &  New Particles \\
   \hline
   \hline
   \endhead
 \multicolumn{2}{|c|}{\hspace{9cm} Continued on next page}
   \endfoot
  $\mathcal{O}_{11,\;(1,-1,2)}^{50}$ & {$S_{5}F_{8}F_{8}F_{9}$, $S_{5}F_{5}F_{8}F_{8}$, $S_{1}F_{3}F_{9}F_{11}$, $S_{1}F_{2}F_{3}F_{5}$, $S_{1}F_{3}F_{7}F_{9}$, $S_{1}F_{3}F_{5}F_{7}$} \\
   \hline
 \caption{Dimension 11 UV completions for topology $\mathcal{T}_ {9} ^{ 8 }$ (see Fig.~\ref{fig:d=9_topologies}).} 
 \label{tab:T11Top8B1L-1D2}\\
   \endlastfoot
  $\mathcal{O}_{11,\;(1,-1,2)}^{48}$ & {$S_{6}F_{7}F_{8}F_{9}$, $S_{6}F_{5}F_{7}F_{8}$, $S_{6}F_{7}F_{8}F_{19}$, $S_{6}F_{7}F_{8}F_{17}$, $S_{2}F_{3}F_{4}F_{9}$, $S_{2}F_{1}F_{3}F_{5}$, $S_{2}F_{3}F_{7}F_{9}$, $S_{2}F_{3}F_{5}F_{7}$} \\
   \hline
  $\mathcal{O}_{11,\;(1,-1,2)}^{49}$ & {$S_{1}F_{3}F_{4}F_{7}$, $S_{1}F_{1}F_{3}F_{7}$, $S_{6}F_{7}F_{8}F_{9}$, $S_{6}F_{5}F_{7}F_{8}$, $S_{2}F_{3}F_{4}F_{8}$, $S_{2}F_{1}F_{3}F_{8}$, $S_{5}F_{8}F_{8}F_{9}$, $S_{5}F_{5}F_{8}F_{8}$, $S_{6}F_{7}F_{8}F_{19}$, $S_{6}F_{7}F_{8}F_{17}$, $S_{2}F_{4}F_{8}F_{9}$, $S_{2}F_{1}F_{5}F_{8}$, $S_{1}F_{4}F_{7}F_{19}$, $S_{1}F_{1}F_{7}F_{17}$} \\
   \hline
 \end{longtable}
 \end{center}

 \begin{center}
 \begin{longtable}{| c  | m{13cm} |} 
   \hline
  $\mathcal{O}^j{(\mathcal{T}_ {9} ^{ 9 })}$ &  New Particles \\
   \hline
   \hline
   \endhead
 \multicolumn{2}{|c|}{\hspace{9cm} Continued on next page}
   \endfoot
  $\mathcal{O}_{11,\;(1,-1,2)}^{50}$ & {$S_{4}S_{5}S_{11}F_{8}$, $S_{2}S_{5}S_{11}F_{8}$, $S_{1}S_{4}S_{11}F_{3}$, $S_{1}S_{2}S_{11}F_{3}$, $S_{1}S_{4}S_{11}F_{9}$, $S_{1}S_{2}S_{11}F_{5}$} \\
   \hline
 \caption{Dimension 11 UV completions for topology $\mathcal{T}_ {9} ^{ 9 }$ (see Fig.~\ref{fig:d=9_topologies}).} 
 \label{tab:T11Top9B1L-1D2}\\
   \endlastfoot
  $\mathcal{O}_{11,\;(1,-1,2)}^{48}$ & {$S_{5}S_{6}S_{19}F_{8}$, $S_{3}S_{5}S_{6}F_{8}$, $S_{5}S_{6}S_{19}F_{7}$, $S_{3}S_{5}S_{6}F_{7}$, $S_{2}S_{5}S_{19}F_{3}$, $S_{2}S_{3}S_{5}F_{3}$, $S_{2}S_{5}S_{19}F_{9}$, $S_{2}S_{3}S_{5}F_{5}$} \\
   \hline
  $\mathcal{O}_{11,\;(1,-1,2)}^{49}$ & {$S_{1}S_{4}S_{11}F_{4}$, $S_{1}S_{2}S_{11}F_{1}$, $S_{5}S_{6}S_{19}F_{8}$, $S_{3}S_{5}S_{6}F_{8}$, $S_{2}S_{5}S_{19}F_{4}$, $S_{2}S_{3}S_{5}F_{1}$, $S_{4}S_{5}S_{11}F_{8}$, $S_{2}S_{5}S_{11}F_{8}$, $S_{5}S_{6}S_{19}F_{7}$, $S_{3}S_{5}S_{6}F_{7}$, $S_{2}S_{5}S_{19}F_{8}$, $S_{2}S_{3}S_{5}F_{8}$, $S_{1}S_{4}S_{11}F_{7}$, $S_{1}S_{2}S_{11}F_{7}$} \\
   \hline
 \end{longtable}
 \end{center}

 \begin{center}
 \begin{longtable}{| c  | m{13cm} |} 
   \hline
  $\mathcal{O}^j{(\mathcal{T}_ {9} ^{ 10 })}$ &  New Particles \\
   \hline
   \hline
   \endhead
 \multicolumn{2}{|c|}{\hspace{9cm} Continued on next page}
   \endfoot
  $\mathcal{O}_{11,\;(1,-1,2)}^{50}$ & {$S_{1}S_{5}S_{11}$, $S_{1}S_{4}S_{5}$} \\
   \hline
 \caption{Dimension 11 UV completions for topology $\mathcal{T}_ {9} ^{ 10 }$ (see Fig.~\ref{fig:d=9_topologies}).} 
 \label{tab:T11Top10B1L-1D2}\\
   \endlastfoot
  $\mathcal{O}_{11,\;(1,-1,2)}^{48}$ & {$S_{2}S_{5}S_{6}$, $S_{2}S_{6}S_{19}$} \\
   \hline
  $\mathcal{O}_{11,\;(1,-1,2)}^{49}$ & {$S_{2}S_{5}S_{6}$, $S_{1}S_{4}S_{5}$, $S_{1}S_{5}S_{11}$, $S_{2}S_{6}S_{19}$} \\
   \hline
 \end{longtable}
 \end{center}

 \begin{center}
 \begin{longtable}{| c  | m{13cm} |} 
   \hline
  $\mathcal{O}^j{(\mathcal{T}_ {9} ^{ 11 })}$ &  New Particles \\
   \hline
   \hline
   \endhead
 \multicolumn{2}{|c|}{\hspace{9cm} Continued on next page}
   \endfoot
  $\mathcal{O}_{11,\;(1,-1,2)}^{50}$ & {$S_{5}S_{9}F_{7}F_{9}$, $S_{1}S_{9}F_{7}F_{17}$, $S_{1}S_{9}F_{10}F_{11}$} \\
   \hline
 \caption{Dimension 11 UV completions for topology $\mathcal{T}_ {9} ^{ 11 }$ (see Fig.~\ref{fig:d=9_topologies}).} 
 \label{tab:T11Top11B1L-1D2}\\
   \endlastfoot
  $\mathcal{O}_{11,\;(1,-1,2)}^{48}$ & {$S_{6}S_{9}F_{19}F_{36}$, $S_{6}S_{9}F_{7}F_{9}$, $S_{2}S_{9}F_{7}F_{17}$, $S_{2}S_{9}F_{3}F_{4}$} \\
   \hline
  $\mathcal{O}_{11,\;(1,-1,2)}^{49}$ & {$S_{6}S_{9}F_{19}F_{36}$, $S_{1}S_{9}F_{19}F_{36}$, $S_{6}S_{9}F_{7}F_{9}$, $S_{5}S_{9}F_{7}F_{9}$, $S_{2}S_{9}F_{7}F_{9}$, $S_{1}S_{9}F_{2}F_{3}$, $S_{2}S_{9}F_{2}F_{3}$} \\
   \hline
 \end{longtable}
 \end{center}

 \begin{center}
 \begin{longtable}{| c  | m{13cm} |} 
   \hline
  $\mathcal{O}^j{(\mathcal{T}_ {9} ^{ 12 })}$ &  New Particles \\
   \hline
   \hline
   \endhead
 \multicolumn{2}{|c|}{\hspace{9cm} Continued on next page}
   \endfoot
  $\mathcal{O}_{11,\;(1,-1,2)}^{50}$ & {$S_{4}S_{11}F_{3}F_{8}$, $S_{2}S_{11}F_{3}F_{8}$, $S_{4}S_{11}F_{8}F_{9}$, $S_{2}S_{11}F_{5}F_{8}$, $S_{4}S_{5}F_{3}F_{9}$, $S_{2}S_{5}F_{3}F_{5}$, $S_{4}S_{5}F_{3}F_{8}$, $S_{2}S_{5}F_{3}F_{8}$, $S_{4}S_{5}F_{8}F_{9}$, $S_{2}S_{5}F_{5}F_{8}$, $S_{4}S_{5}F_{8}F_{8}$, $S_{2}S_{5}F_{8}F_{8}$} \\
   \hline
 \caption{Dimension 11 UV completions for topology $\mathcal{T}_ {9} ^{ 12 }$ (see Fig.~\ref{fig:d=9_topologies}).} 
 \label{tab:T11Top12B1L-1D2}\\
   \endlastfoot
  $\mathcal{O}_{11,\;(1,-1,2)}^{48}$ & {$S_{5}S_{19}F_{3}F_{8}$, $S_{3}S_{5}F_{3}F_{8}$, $S_{5}S_{19}F_{8}F_{9}$, $S_{3}S_{5}F_{5}F_{8}$, $S_{5}S_{19}F_{3}F_{7}$, $S_{3}S_{5}F_{3}F_{7}$, $S_{5}S_{19}F_{7}F_{9}$, $S_{3}S_{5}F_{5}F_{7}$, $S_{6}S_{19}F_{3}F_{9}$, $S_{3}S_{6}F_{3}F_{5}$, $S_{4}S_{5}F_{3}F_{9}$, $S_{2}S_{5}F_{3}F_{5}$, $S_{4}S_{5}F_{3}F_{8}$, $S_{2}S_{5}F_{3}F_{8}$, $S_{6}S_{19}F_{8}F_{9}$, $S_{3}S_{6}F_{5}F_{8}$, $S_{6}S_{19}F_{3}F_{7}$, $S_{3}S_{6}F_{3}F_{7}$, $S_{4}S_{5}F_{7}F_{9}$, $S_{2}S_{5}F_{5}F_{7}$, $S_{6}S_{19}F_{7}F_{8}$, $S_{3}S_{6}F_{7}F_{8}$, $S_{4}S_{5}F_{7}F_{8}$, $S_{2}S_{5}F_{7}F_{8}$} \\
   \hline
  $\mathcal{O}_{11,\;(1,-1,2)}^{49}$ & {$S_{5}S_{19}F_{4}F_{8}$, $S_{3}S_{5}F_{1}F_{8}$, $S_{4}S_{11}F_{4}F_{8}$, $S_{2}S_{11}F_{1}F_{8}$, $S_{5}S_{19}F_{8}F_{8}$, $S_{3}S_{5}F_{8}F_{8}$, $S_{5}S_{19}F_{4}F_{7}$, $S_{3}S_{5}F_{1}F_{7}$, $S_{5}S_{19}F_{7}F_{8}$, $S_{3}S_{5}F_{7}F_{8}$, $S_{4}S_{11}F_{7}F_{8}$, $S_{2}S_{11}F_{7}F_{8}$} \\
   \hline
 \end{longtable}
 \end{center}

 \begin{center}
 \begin{longtable}{| c  | m{13cm} |} 
   \hline
  $\mathcal{O}^j{(\mathcal{T}_ {9} ^{ 13 })}$ &  New Particles \\
   \hline
   \hline
   \endhead
 \multicolumn{2}{|c|}{\hspace{9cm} Continued on next page}
   \endfoot
  $\mathcal{O}_{11,\;(1,-1,2)}^{50}$ & {$S_{1}S_{5}S_{9}S_{26}$} \\
   \hline
 \caption{Dimension 11 UV completions for topology $\mathcal{T}_ {9} ^{ 13 }$ (see Fig.~\ref{fig:d=9_topologies}).} 
 \label{tab:T11Top13B1L-1D2}\\
   \endlastfoot
  $\mathcal{O}_{11,\;(1,-1,2)}^{48}$ & {$S_{2}S_{6}S_{9}S_{26}$} \\
   \hline
  $\mathcal{O}_{11,\;(1,-1,2)}^{49}$ & {$S_{2}S_{6}S_{9}S_{26}$, $S_{1}S_{5}S_{9}S_{26}$} \\
   \hline
 \end{longtable}
 \end{center}

 \begin{center}
 \begin{longtable}{| c  | m{13cm} |} 
   \hline
  $\mathcal{O}^j{(\mathcal{T}_ {9} ^{ 14 })}$ &  New Particles \\
   \hline
   \hline
   \endhead
 \multicolumn{2}{|c|}{\hspace{9cm} Continued on next page}
   \endfoot
  $\mathcal{O}_{11,\;(1,-1,2)}^{50}$ & {$S_{5}S_{9}F_{7}F_{8}$, $S_{1}S_{9}F_{3}F_{10}$, $S_{1}S_{9}F_{9}F_{17}$} \\
   \hline
 \caption{Dimension 11 UV completions for topology $\mathcal{T}_ {9} ^{ 14 }$ (see Fig.~\ref{fig:d=9_topologies}).} 
 \label{tab:T11Top14B1L-1D2}\\
   \endlastfoot
  $\mathcal{O}_{11,\;(1,-1,2)}^{48}$ & {$S_{6}S_{9}F_{7}F_{8}$, $S_{6}S_{9}F_{7}F_{36}$, $S_{2}S_{9}F_{3}F_{3}$, $S_{2}S_{9}F_{9}F_{17}$} \\
   \hline
  $\mathcal{O}_{11,\;(1,-1,2)}^{49}$ & {$S_{6}S_{9}F_{7}F_{8}$, $S_{1}S_{9}F_{2}F_{4}$, $S_{6}S_{9}F_{7}F_{36}$, $S_{5}S_{9}F_{7}F_{8}$, $S_{2}S_{9}F_{2}F_{4}$, $S_{1}S_{9}F_{7}F_{36}$, $S_{2}S_{9}F_{7}F_{8}$} \\
   \hline
 \end{longtable}
 \end{center}

 \begin{center}
 \begin{longtable}{| c  | m{13cm} |} 
   \hline
  $\mathcal{O}^j{(\mathcal{T}_ {9} ^{ 15 })}$ &  New Particles \\
   \hline
   \hline
   \endhead
 \multicolumn{2}{|c|}{\hspace{9cm} Continued on next page}
   \endfoot
  $\mathcal{O}_{11,\;(1,-1,2)}^{50}$ & {$S_{4}S_{5}F_{8}F_{9}$, $S_{2}S_{5}F_{5}F_{8}$, $S_{1}S_{11}F_{3}F_{11}$, $S_{1}S_{11}F_{2}F_{3}$, $S_{1}S_{11}F_{7}F_{9}$, $S_{1}S_{11}F_{5}F_{7}$} \\
   \hline
 \caption{Dimension 11 UV completions for topology $\mathcal{T}_ {9} ^{ 15 }$ (see Fig.~\ref{fig:d=9_topologies}).} 
 \label{tab:T11Top15B1L-1D2}\\
   \endlastfoot
  $\mathcal{O}_{11,\;(1,-1,2)}^{48}$ & {$S_{6}S_{19}F_{8}F_{9}$, $S_{3}S_{6}F_{5}F_{8}$, $S_{6}S_{19}F_{7}F_{19}$, $S_{3}S_{6}F_{7}F_{17}$, $S_{2}S_{5}F_{3}F_{4}$, $S_{2}S_{5}F_{1}F_{3}$, $S_{2}S_{5}F_{7}F_{9}$, $S_{2}S_{5}F_{5}F_{7}$} \\
   \hline
  $\mathcal{O}_{11,\;(1,-1,2)}^{49}$ & {$S_{6}S_{19}F_{8}F_{9}$, $S_{3}S_{6}F_{5}F_{8}$, $S_{1}S_{11}F_{3}F_{4}$, $S_{1}S_{11}F_{1}F_{3}$, $S_{6}S_{19}F_{7}F_{19}$, $S_{3}S_{6}F_{7}F_{17}$, $S_{4}S_{5}F_{8}F_{9}$, $S_{2}S_{5}F_{5}F_{8}$, $S_{2}S_{5}F_{3}F_{4}$, $S_{2}S_{5}F_{1}F_{3}$, $S_{1}S_{11}F_{7}F_{19}$, $S_{1}S_{11}F_{7}F_{17}$, $S_{2}S_{5}F_{8}F_{9}$} \\
   \hline
 \end{longtable}
 \end{center}

 \begin{center}
 \begin{longtable}{| c  | m{13cm} |} 
   \hline
  $\mathcal{O}^j{(\mathcal{T}_ {9} ^{ 16 })}$ &  New Particles \\
   \hline
   \hline
   \endhead
 \multicolumn{2}{|c|}{\hspace{9cm} Continued on next page}
   \endfoot
  $\mathcal{O}_{11,\;(1,-1,2)}^{50}$ & {$S_{5}S_{9}S_{11}F_{8}$, $S_{1}S_{4}S_{9}F_{3}$, $S_{1}S_{4}S_{9}F_{9}$} \\
   \hline
 \caption{Dimension 11 UV completions for topology $\mathcal{T}_ {9} ^{ 16 }$ (see Fig.~\ref{fig:d=9_topologies}).} 
 \label{tab:T11Top16B1L-1D2}\\
   \endlastfoot
  $\mathcal{O}_{11,\;(1,-1,2)}^{48}$ & {$S_{5}S_{6}S_{9}F_{8}$, $S_{5}S_{6}S_{9}F_{7}$, $S_{2}S_{9}S_{19}F_{3}$, $S_{2}S_{9}S_{19}F_{9}$} \\
   \hline
  $\mathcal{O}_{11,\;(1,-1,2)}^{49}$ & {$S_{1}S_{4}S_{9}F_{4}$, $S_{5}S_{6}S_{9}F_{8}$, $S_{2}S_{9}S_{19}F_{4}$, $S_{5}S_{9}S_{11}F_{8}$, $S_{5}S_{6}S_{9}F_{7}$, $S_{2}S_{9}S_{19}F_{8}$, $S_{1}S_{4}S_{9}F_{7}$} \\
   \hline
 \end{longtable}
 \end{center}

 \begin{center}
 \begin{longtable}{| c  | m{13cm} |} 
   \hline
  $\mathcal{O}^j{(\mathcal{T}_ {9} ^{ 17 })}$ &  New Particles \\
   \hline
   \hline
   \endhead
 \multicolumn{2}{|c|}{\hspace{9cm} Continued on next page}
   \endfoot
  $\mathcal{O}_{11,\;(1,-1,2)}^{50}$ & {$S_{5}S_{11}F_{8}$, $S_{1}S_{4}F_{3}$, $S_{1}S_{4}F_{9}$} \\
   \hline
 \caption{Dimension 11 UV completions for topology $\mathcal{T}_ {9} ^{ 17 }$ (see Fig.~\ref{fig:d=9_topologies}).} 
 \label{tab:T11Top17B1L-1D2}\\
   \endlastfoot
  $\mathcal{O}_{11,\;(1,-1,2)}^{48}$ & {$S_{5}S_{6}F_{8}$, $S_{5}S_{6}F_{7}$, $S_{2}S_{19}F_{3}$, $S_{2}S_{19}F_{9}$} \\
   \hline
  $\mathcal{O}_{11,\;(1,-1,2)}^{49}$ & {$S_{5}S_{6}F_{8}$, $S_{1}S_{4}F_{4}$, $S_{5}S_{6}F_{7}$, $S_{5}S_{11}F_{8}$, $S_{2}S_{19}F_{4}$, $S_{1}S_{4}F_{7}$, $S_{2}S_{19}F_{8}$} \\
   \hline
 \end{longtable}
 \end{center}

 \begin{center}
 \begin{longtable}{| c  | m{13cm} |} 
   \hline
  $\mathcal{O}^j{(\mathcal{T}_ {9} ^{ 18 })}$ &  New Particles \\
   \hline
   \hline
   \endhead
 \multicolumn{2}{|c|}{\hspace{9cm} Continued on next page}
   \endfoot
  $\mathcal{O}_{11,\;(1,-1,2)}^{50}$ & {$S_{5}S_{9}S_{26}F_{7}$, $S_{1}S_{9}S_{26}F_{17}$, $S_{1}S_{9}S_{26}F_{10}$} \\
   \hline
 \caption{Dimension 11 UV completions for topology $\mathcal{T}_ {9} ^{ 18 }$ (see Fig.~\ref{fig:d=9_topologies}).} 
 \label{tab:T11Top18B1L-1D2}\\
   \endlastfoot
  $\mathcal{O}_{11,\;(1,-1,2)}^{48}$ & {$S_{6}S_{9}S_{26}F_{36}$, $S_{6}S_{9}S_{26}F_{7}$, $S_{2}S_{9}S_{26}F_{17}$, $S_{2}S_{9}S_{26}F_{3}$} \\
   \hline
  $\mathcal{O}_{11,\;(1,-1,2)}^{49}$ & {$S_{6}S_{9}S_{26}F_{36}$, $S_{1}S_{9}S_{26}F_{36}$, $S_{6}S_{9}S_{26}F_{7}$, $S_{5}S_{9}S_{26}F_{7}$, $S_{2}S_{9}S_{26}F_{7}$, $S_{1}S_{9}S_{26}F_{2}$, $S_{2}S_{9}S_{26}F_{2}$} \\
   \hline
 \end{longtable}
 \end{center}

 \begin{center}
 \begin{longtable}{| c  | m{13cm} |} 
   \hline
  $\mathcal{O}^j{(\mathcal{T}_ {9} ^{ 19 })}$ &  New Particles \\
   \hline
   \hline
   \endhead
 \multicolumn{2}{|c|}{\hspace{9cm} Continued on next page}
   \endfoot
  $\mathcal{O}_{11,\;(1,-1,2)}^{50}$ & {$S_{11}F_{3}F_{8}F_{9}$, $S_{11}F_{3}F_{5}F_{8}$, $S_{4}F_{3}F_{8}F_{8}$, $S_{2}F_{3}F_{8}F_{8}$, $S_{4}F_{8}F_{8}F_{9}$, $S_{2}F_{5}F_{8}F_{8}$, $S_{5}F_{3}F_{8}F_{9}$, $S_{5}F_{3}F_{5}F_{8}$, $S_{4}F_{3}F_{8}F_{9}$, $S_{2}F_{3}F_{5}F_{8}$, $S_{5}F_{8}F_{8}F_{9}$, $S_{5}F_{5}F_{8}F_{8}$} \\
   \hline
 \caption{Dimension 11 UV completions for topology $\mathcal{T}_ {9} ^{ 19 }$ (see Fig.~\ref{fig:d=9_topologies}).} 
 \label{tab:T11Top19B1L-1D2}\\
   \endlastfoot
  $\mathcal{O}_{11,\;(1,-1,2)}^{48}$ & {$S_{5}F_{3}F_{8}F_{9}$, $S_{5}F_{3}F_{5}F_{8}$, $S_{5}F_{3}F_{7}F_{9}$, $S_{5}F_{3}F_{5}F_{7}$, $S_{19}F_{3}F_{7}F_{8}$, $S_{3}F_{3}F_{7}F_{8}$, $S_{19}F_{7}F_{8}F_{9}$, $S_{3}F_{5}F_{7}F_{8}$, $S_{19}F_{3}F_{8}F_{9}$, $S_{3}F_{3}F_{5}F_{8}$, $S_{6}F_{3}F_{7}F_{9}$, $S_{6}F_{3}F_{5}F_{7}$, $S_{4}F_{3}F_{7}F_{9}$, $S_{2}F_{3}F_{5}F_{7}$, $S_{4}F_{3}F_{7}F_{8}$, $S_{2}F_{3}F_{7}F_{8}$, $S_{6}F_{7}F_{8}F_{9}$, $S_{6}F_{5}F_{7}F_{8}$, $S_{5}F_{7}F_{8}F_{9}$, $S_{5}F_{5}F_{7}F_{8}$} \\
   \hline
  $\mathcal{O}_{11,\;(1,-1,2)}^{49}$ & {$S_{4}F_{4}F_{8}F_{8}$, $S_{2}F_{1}F_{8}F_{8}$, $S_{5}F_{4}F_{8}F_{8}$, $S_{5}F_{1}F_{8}F_{8}$, $S_{19}F_{4}F_{7}F_{8}$, $S_{3}F_{1}F_{7}F_{8}$, $S_{11}F_{4}F_{7}F_{8}$, $S_{11}F_{1}F_{7}F_{8}$, $S_{5}F_{4}F_{7}F_{8}$, $S_{5}F_{1}F_{7}F_{8}$, $S_{19}F_{7}F_{8}F_{8}$, $S_{3}F_{7}F_{8}F_{8}$, $S_{4}F_{7}F_{8}F_{8}$, $S_{2}F_{7}F_{8}F_{8}$} \\
   \hline
 \end{longtable}
 \end{center}

 \begin{center}
 \begin{longtable}{| c  | m{13cm} |} 
   \hline
  $\mathcal{O}^j{(\mathcal{T}_ {9} ^{ 20 })}$ &  New Particles \\
   \hline
   \hline
   \endhead
 \multicolumn{2}{|c|}{\hspace{9cm} Continued on next page}
   \endfoot
  $\mathcal{O}_{11,\;(1,-1,2)}^{50}$ & {$S_{4}F_{3}F_{8}F_{9}$, $S_{2}F_{3}F_{5}F_{8}$, $S_{4}F_{8}F_{9}F_{9}$, $S_{2}F_{5}F_{5}F_{8}$, $S_{11}F_{3}F_{8}F_{11}$, $S_{11}F_{2}F_{3}F_{8}$, $S_{11}F_{7}F_{8}F_{9}$, $S_{11}F_{5}F_{7}F_{8}$, $S_{5}F_{3}F_{7}F_{9}$, $S_{5}F_{3}F_{5}F_{7}$, $S_{4}F_{3}F_{9}F_{11}$, $S_{2}F_{2}F_{3}F_{5}$, $S_{4}F_{3}F_{8}F_{11}$, $S_{2}F_{2}F_{3}F_{8}$, $S_{5}F_{7}F_{8}F_{9}$, $S_{5}F_{5}F_{7}F_{8}$, $S_{5}F_{3}F_{8}F_{9}$, $S_{5}F_{3}F_{5}F_{8}$, $S_{5}F_{8}F_{8}F_{9}$, $S_{5}F_{5}F_{8}F_{8}$, $S_{4}F_{8}F_{8}F_{9}$, $S_{2}F_{5}F_{8}F_{8}$} \\
   \hline
 \caption{Dimension 11 UV completions for topology $\mathcal{T}_ {9} ^{ 20 }$ (see Fig.~\ref{fig:d=9_topologies}).} 
 \label{tab:T11Top20B1L-1D2}\\
   \endlastfoot
  $\mathcal{O}_{11,\;(1,-1,2)}^{48}$ & {$S_{19}F_{3}F_{8}F_{9}$, $S_{3}F_{3}F_{5}F_{8}$, $S_{19}F_{8}F_{9}F_{9}$, $S_{3}F_{5}F_{5}F_{8}$, $S_{19}F_{3}F_{7}F_{19}$, $S_{3}F_{3}F_{7}F_{17}$, $S_{19}F_{7}F_{9}F_{19}$, $S_{3}F_{5}F_{7}F_{17}$, $S_{5}F_{3}F_{4}F_{8}$, $S_{5}F_{1}F_{3}F_{8}$, $S_{5}F_{3}F_{4}F_{7}$, $S_{5}F_{1}F_{3}F_{7}$, $S_{5}F_{7}F_{8}F_{9}$, $S_{5}F_{5}F_{7}F_{8}$, $S_{5}F_{7}F_{7}F_{9}$, $S_{5}F_{5}F_{7}F_{7}$, $S_{6}F_{3}F_{7}F_{9}$, $S_{6}F_{3}F_{5}F_{7}$, $S_{4}F_{3}F_{4}F_{9}$, $S_{2}F_{1}F_{3}F_{5}$, $S_{4}F_{3}F_{4}F_{8}$, $S_{2}F_{1}F_{3}F_{8}$, $S_{6}F_{7}F_{8}F_{9}$, $S_{6}F_{5}F_{7}F_{8}$, $S_{5}F_{3}F_{7}F_{9}$, $S_{5}F_{3}F_{5}F_{7}$, $S_{19}F_{3}F_{4}F_{9}$, $S_{3}F_{1}F_{3}F_{5}$, $S_{19}F_{3}F_{4}F_{7}$, $S_{3}F_{1}F_{3}F_{7}$, $S_{5}F_{3}F_{8}F_{9}$, $S_{5}F_{3}F_{5}F_{8}$, $S_{6}F_{3}F_{7}F_{19}$, $S_{6}F_{3}F_{7}F_{17}$, $S_{6}F_{7}F_{8}F_{19}$, $S_{6}F_{7}F_{8}F_{17}$, $S_{4}F_{7}F_{9}F_{19}$, $S_{2}F_{5}F_{7}F_{17}$, $S_{4}F_{7}F_{8}F_{19}$, $S_{2}F_{7}F_{8}F_{17}$, $S_{19}F_{7}F_{8}F_{9}$, $S_{3}F_{5}F_{7}F_{8}$} \\
   \hline
  $\mathcal{O}_{11,\;(1,-1,2)}^{49}$ & {$S_{19}F_{4}F_{8}F_{9}$, $S_{3}F_{1}F_{5}F_{8}$, $S_{11}F_{3}F_{4}F_{8}$, $S_{11}F_{1}F_{3}F_{8}$, $S_{19}F_{8}F_{8}F_{9}$, $S_{3}F_{5}F_{8}F_{8}$, $S_{4}F_{4}F_{8}F_{9}$, $S_{2}F_{1}F_{5}F_{8}$, $S_{5}F_{3}F_{4}F_{8}$, $S_{5}F_{1}F_{3}F_{8}$, $S_{19}F_{4}F_{7}F_{19}$, $S_{3}F_{1}F_{7}F_{17}$, $S_{5}F_{3}F_{4}F_{7}$, $S_{5}F_{1}F_{3}F_{7}$, $S_{19}F_{7}F_{8}F_{19}$, $S_{3}F_{7}F_{8}F_{17}$, $S_{4}F_{7}F_{8}F_{9}$, $S_{2}F_{5}F_{7}F_{8}$, $S_{5}F_{8}F_{8}F_{9}$, $S_{5}F_{5}F_{8}F_{8}$, $S_{11}F_{7}F_{8}F_{19}$, $S_{11}F_{7}F_{8}F_{17}$, $S_{5}F_{7}F_{8}F_{9}$, $S_{5}F_{5}F_{7}F_{8}$} \\
\hline
 \end{longtable}
 \end{center}

 \begin{center}
 \begin{longtable}{| c  | m{13cm} |} 
   \hline
  $\mathcal{O}^j{(\mathcal{T}_ {9} ^{ 21 })}$ &  New Particles \\
   \hline
   \hline
   \endhead
 \multicolumn{2}{|c|}{\hspace{9cm} Continued on next page}
   \endfoot
  $\mathcal{O}_{11,\;(1,-1,2)}^{50}$ & {$S_{5}S_{9}F_{8}F_{9}$, $S_{1}S_{9}F_{3}F_{7}$, $S_{1}S_{9}F_{9}F_{11}$} \\
   \hline
 \caption{Dimension 11 UV completions for topology $\mathcal{T}_ {9} ^{ 21 }$ (see Fig.~\ref{fig:d=9_topologies}).} 
 \label{tab:T11Top21B1L-1D2}\\
   \endlastfoot
  $\mathcal{O}_{11,\;(1,-1,2)}^{48}$ & {$S_{6}S_{9}F_{8}F_{19}$, $S_{6}S_{9}F_{7}F_{9}$, $S_{2}S_{9}F_{3}F_{7}$, $S_{2}S_{9}F_{4}F_{9}$} \\
   \hline
  $\mathcal{O}_{11,\;(1,-1,2)}^{49}$ & {$S_{6}S_{9}F_{8}F_{19}$, $S_{1}S_{9}F_{4}F_{19}$, $S_{6}S_{9}F_{7}F_{9}$, $S_{5}S_{9}F_{8}F_{9}$, $S_{2}S_{9}F_{4}F_{9}$, $S_{1}S_{9}F_{3}F_{7}$, $S_{2}S_{9}F_{3}F_{8}$} \\
   \hline
 \end{longtable}
 \end{center}

 \begin{center}
 \begin{longtable}{| c  | m{13cm} |} 
   \hline
  $\mathcal{O}^j{(\mathcal{T}_ {9} ^{ 22 })}$ &  New Particles \\
   \hline
   \hline
   \endhead
 \multicolumn{2}{|c|}{\hspace{9cm} Continued on next page}
   \endfoot
  $\mathcal{O}_{11,\;(1,-1,2)}^{50}$ & {$S_{4}S_{5}S_{9}F_{9}$, $S_{1}S_{9}S_{11}F_{11}$, $S_{1}S_{9}S_{11}F_{7}$} \\
   \hline
 \caption{Dimension 11 UV completions for topology $\mathcal{T}_ {9} ^{ 22 }$ (see Fig.~\ref{fig:d=9_topologies}).} 
 \label{tab:T11Top22B1L-1D2}\\
   \endlastfoot
  $\mathcal{O}_{11,\;(1,-1,2)}^{48}$ & {$S_{6}S_{9}S_{19}F_{9}$, $S_{6}S_{9}S_{19}F_{19}$, $S_{2}S_{5}S_{9}F_{4}$, $S_{2}S_{5}S_{9}F_{7}$} \\
   \hline
  $\mathcal{O}_{11,\;(1,-1,2)}^{49}$ & {$S_{6}S_{9}S_{19}F_{9}$, $S_{1}S_{9}S_{11}F_{3}$, $S_{6}S_{9}S_{19}F_{19}$, $S_{4}S_{5}S_{9}F_{9}$, $S_{2}S_{5}S_{9}F_{3}$, $S_{1}S_{9}S_{11}F_{19}$, $S_{2}S_{5}S_{9}F_{9}$} \\
   \hline
 \end{longtable}
 \end{center}

 \begin{center}
 \begin{longtable}{| c  | m{13cm} |} 
   \hline
  $\mathcal{O}^j{(\mathcal{T}_ {9} ^{ 23 })}$ &  New Particles \\
   \hline
   \hline
   \endhead
 \multicolumn{2}{|c|}{\hspace{9cm} Continued on next page}
   \endfoot
  $\mathcal{O}_{11,\;(1,-1,2)}^{50}$ & {$S_{4}S_{9}F_{3}F_{9}$, $S_{4}S_{9}F_{9}F_{9}$, $S_{9}S_{11}F_{8}F_{11}$, $S_{9}S_{11}F_{7}F_{8}$, $S_{5}S_{9}F_{3}F_{7}$, $S_{5}S_{9}F_{7}F_{8}$, $S_{4}S_{9}F_{9}F_{11}$, $S_{4}S_{9}F_{8}F_{11}$, $S_{5}S_{9}F_{3}F_{9}$, $S_{5}S_{9}F_{8}F_{9}$, $S_{4}S_{9}F_{8}F_{9}$} \\
   \hline
 \caption{Dimension 11 UV completions for topology $\mathcal{T}_ {9} ^{ 23 }$ (see Fig.~\ref{fig:d=9_topologies}).} 
 \label{tab:T11Top23B1L-1D2}\\
   \endlastfoot
  $\mathcal{O}_{11,\;(1,-1,2)}^{48}$ & {$S_{9}S_{19}F_{3}F_{9}$, $S_{9}S_{19}F_{9}F_{9}$, $S_{9}S_{19}F_{3}F_{19}$, $S_{9}S_{19}F_{9}F_{19}$, $S_{5}S_{9}F_{4}F_{8}$, $S_{5}S_{9}F_{4}F_{7}$, $S_{5}S_{9}F_{7}F_{8}$, $S_{5}S_{9}F_{7}F_{7}$, $S_{6}S_{9}F_{3}F_{7}$, $S_{6}S_{9}F_{7}F_{8}$, $S_{4}S_{9}F_{4}F_{9}$, $S_{4}S_{9}F_{4}F_{8}$, $S_{5}S_{9}F_{3}F_{7}$, $S_{9}S_{19}F_{4}F_{9}$, $S_{9}S_{19}F_{4}F_{7}$, $S_{6}S_{9}F_{3}F_{19}$, $S_{6}S_{9}F_{8}F_{19}$, $S_{5}S_{9}F_{3}F_{9}$, $S_{5}S_{9}F_{7}F_{9}$, $S_{4}S_{9}F_{9}F_{19}$, $S_{4}S_{9}F_{8}F_{19}$, $S_{9}S_{19}F_{7}F_{9}$} \\
   \hline
  $\mathcal{O}_{11,\;(1,-1,2)}^{49}$ & {$S_{9}S_{19}F_{4}F_{9}$, $S_{9}S_{19}F_{8}F_{9}$, $S_{9}S_{11}F_{3}F_{8}$, $S_{9}S_{19}F_{4}F_{19}$, $S_{9}S_{19}F_{8}F_{19}$, $S_{4}S_{9}F_{4}F_{9}$, $S_{4}S_{9}F_{7}F_{9}$, $S_{5}S_{9}F_{3}F_{8}$, $S_{5}S_{9}F_{3}F_{7}$, $S_{9}S_{11}F_{8}F_{19}$, $S_{5}S_{9}F_{8}F_{9}$, $S_{5}S_{9}F_{7}F_{9}$} \\
   \hline
 \end{longtable}
 \end{center}

\subsection{\texorpdfstring{$\Delta B=1, \Delta L=3$}{Delta B=1,Delta L=3}}

For $(\Delta B,\Delta L)=(1,3)$, we have one-derivative operators,
\begin{multicols}{3}
 \begin{spacing}{1.5}
 \noindent
 \begin{math}
\mathcal{O}_{11, \, (1,3,1)}^{1}\equiv D\bar{H}uuuLLL\,,\\
\mathcal{O}_{11, \, (1,3,1)}^{2}\equiv DHuQQLLL\,,\\
\mathcal{O}_{11, \, (1,3,1)}^{3}\equiv DHduuLLL\,,\\
\mathcal{O}_{11, \, (1,3,1)}^{4}\equiv DHuuQLLe\,,\\
\mathcal{O}_{11, \, (1,3,1)}^{5}\equiv DHuuuLee\,,\\
\mathcal{O}_{11, \, (1,3,1)}^{6}\equiv D\bar{H}uuQLL\nu\,,\\
\mathcal{O}_{11, \, (1,3,1)}^{7}\equiv D\bar{H}uuuLe\nu\,,\\
\mathcal{O}_{11, \, (1,3,1)}^{8}\equiv DHQQQLL\nu\,,\\
\mathcal{O}_{11, \, (1,3,1)}^{9}\equiv DHduQLL\nu\,,\\
\mathcal{O}_{11, \, (1,3,1)}^{10}\equiv DHuQQLe\nu\,,\\
\mathcal{O}_{11, \, (1,3,1)}^{11}\equiv DHduuLe\nu\,,\\
\mathcal{O}_{11, \, (1,3,1)}^{12}\equiv DHuuQee\nu\,,\\
\mathcal{O}_{11, \, (1,3,1)}^{13}\equiv D\bar{H}uQQL\nu\nu\,,\\
\mathcal{O}_{11, \, (1,3,1)}^{14}\equiv D\bar{H}duuL\nu\nu\,,\\
\mathcal{O}_{11, \, (1,3,1)}^{15}\equiv D\bar{H}uuQe\nu\nu\,,\\
\mathcal{O}_{11, \, (1,3,1)}^{16}\equiv DHdQQL\nu\nu\,,\\
\mathcal{O}_{11, \, (1,3,1)}^{17}\equiv DHQQQe\nu\nu\,,\\
\mathcal{O}_{11, \, (1,3,1)}^{18}\equiv DHdduL\nu\nu\,,\\
\mathcal{O}_{11, \, (1,3,1)}^{19}\equiv DHduQe\nu\nu\,,\\
\mathcal{O}_{11, \, (1,3,1)}^{20}\equiv D\bar{H}QQQ\nu\nu\nu\,,\\
\mathcal{O}_{11, \, (1,3,1)}^{21}\equiv D\bar{H}duQ\nu\nu\nu\,,\\
\mathcal{O}_{11, \, (1,3,1)}^{22}\equiv DHddQ\nu\nu\nu\,,
\label{eq:1Ddequal11basisB1L3}
 \end{math}
 \end{spacing}
 \end{multicols}
\noindent
and two-derivative operators,
  \begin{multicols}{3}
 \begin{spacing}{1.5}
 \noindent
 \begin{math}
\mathcal{O}_{11, \,(1 , 3 ,2)}^{1}  \equiv DDuuQLLL \,, \\
 \mathcal{O}_{11, \,(1 , 3 ,2)}^{2}  \equiv DDuuueLL \,, \\
 \mathcal{O}_{11, \,(1 , 3 ,2)}^{3}  \equiv DDuQQLL\nu \,, \\
 \mathcal{O}_{11, \,(1 , 3 ,2)}^{4}  \equiv DDduuLL\nu \,, \\
 \mathcal{O}_{11, \,(1 , 3 ,2)}^{5}  \equiv DDuuQeL\nu \,, \\
 \mathcal{O}_{11, \,(1 , 3 ,2)}^{6}  \equiv DDuuuee\nu \,, \\
 \mathcal{O}_{11, \,(1 , 3 ,2)}^{7}  \equiv DDQQQL\nu\nu \,, \\
 \mathcal{O}_{11, \,(1 , 3 ,2)}^{8}  \equiv DDduQL\nu\nu \,, \\
 \mathcal{O}_{11, \,(1 , 3 ,2)}^{9}  \equiv DDuQQe\nu\nu \,, \\
 \mathcal{O}_{11, \,(1 , 3 ,2)}^{10}  \equiv DDduue\nu\nu \,, \\
 \mathcal{O}_{11, \,(1 , 3 ,2)}^{11}  \equiv DDdQQ\nu\nu\nu \,, \\
 \mathcal{O}_{11, \,(1 , 3 ,2)}^{12}  \equiv DDddu\nu\nu\nu \,. 
 \label{eq:2Ddequal11basisB1L3}
 \end{math}
 \end{spacing}
 \end{multicols}
 \noindent
 We present only the $1D$ UV completions, see tables~\ref{tab:T11Top2B1L3D1}--\ref{tab:T11Top7B1L3D1}. A one-loop realization of $\mathcal{O}_{11, \, (1,3,1)}^{4}$ was given in Ref.~\cite{Fonseca:2018ehk}.

\begin{center}
\begin{longtable}{| c | m{13cm} |}
   \hline
   $\mathcal{O}^j(\mathcal{T}_{10}^{2})$ & New Particles \\
   \hline
   \hline
   \endhead
   \multicolumn{2}{|c|}{\hspace{9cm} Continued on next page} \\
   \endfoot
  $\mathcal{O}_{11,(1,3,1)}^{22}$ & {$S_{2}S_{7}F_{6}F_{8}$, $S_{2}S_{7}F_{5}F_{8}$, $S_{3}S_{7}F_{1}F_{3}$, $S_{2}S_{2}F_{5}F_{8}$, $S_{3}S_{7}F_{6}F_{8}$} \\
   \hline
   \caption{Dimension 11 UV completions for topology $\mathcal{T}_{10}^{2}$ (see Fig.~\ref{fig:d=10_topologies}).}
   \label{tab:T11Top2B1L3D1}\\
   \endlastfoot
  $\mathcal{O}_{11,(1,3,1)}^{1}$ & {$S_{1}S_{9}F_{10}F_{16}$, $S_{1}S_{14}F_{10}F_{16}$, $S_{1}S_{9}F_{7}F_{9}$, $S_{1}S_{14}F_{5}F_{7}$} \\
   \hline
  $\mathcal{O}_{11,(1,3,1)}^{2}$ & {$S_{4}S_{9}F_{9}F_{20}$, $S_{4}S_{9}F_{7}F_{9}$, $S_{4}S_{14}F_{7}F_{9}$, $S_{2}S_{9}F_{5}F_{7}$, $S_{2}S_{14}F_{5}F_{7}$, $S_{4}S_{9}F_{9}F_{15}$, $S_{4}S_{9}F_{8}F_{9}$, $S_{2}S_{9}F_{8}F_{9}$, $S_{4}S_{14}F_{5}F_{8}$, $S_{2}S_{14}F_{5}F_{8}$, $S_{4}S_{9}F_{11}F_{32}$, $S_{4}S_{9}F_{10}F_{11}$, $S_{4}S_{14}F_{10}F_{11}$, $S_{2}S_{9}F_{2}F_{10}$, $S_{2}S_{14}F_{2}F_{10}$, $S_{4}S_{4}F_{9}F_{20}$, $S_{4}S_{4}F_{7}F_{9}$, $S_{2}S_{4}F_{7}F_{9}$, $S_{2}S_{4}F_{5}F_{7}$, $S_{2}S_{2}F_{5}F_{7}$} \\
   \hline
  $\mathcal{O}_{11,(1,3,1)}^{3}$ & {$S_{2}S_{9}F_{2}F_{10}$, $S_{2}S_{14}F_{2}F_{10}$, $S_{2}S_{9}F_{8}F_{9}$, $S_{2}S_{14}F_{5}F_{8}$, $S_{1}S_{9}F_{2}F_{10}$, $S_{1}S_{14}F_{2}F_{10}$, $S_{1}S_{9}F_{7}F_{19}$, $S_{1}S_{14}F_{7}F_{17}$} \\
   \hline
  $\mathcal{O}_{11,(1,3,1)}^{4}$ & {$S_{2}S_{9}F_{5}F_{7}$, $S_{2}S_{14}F_{5}F_{7}$, $S_{2}S_{9}F_{8}F_{9}$, $S_{2}S_{14}F_{5}F_{8}$, $S_{2}S_{4}F_{5}F_{7}$, $S_{2}S_{2}F_{5}F_{7}$, $S_{2}S_{4}F_{7}F_{9}$, $S_{1}S_{9}F_{10}F_{31}$, $S_{1}S_{14}F_{10}F_{16}$, $S_{1}S_{9}F_{5}F_{7}$, $S_{1}S_{14}F_{5}F_{7}$, $S_{1}S_{4}F_{5}F_{7}$, $S_{1}S_{2}F_{5}F_{7}$, $S_{1}S_{4}F_{7}F_{19}$, $S_{1}S_{2}F_{7}F_{17}$} \\
   \hline
  $\mathcal{O}_{11,(1,3,1)}^{5}$ & {$S_{1}S_{15}F_{16}F_{49}$, $S_{2}S_{2}F_{5}F_{7}$, $S_{1}S_{15}F_{7}F_{17}$, $S_{1}S_{2}F_{5}F_{7}$, $S_{1}S_{2}F_{7}F_{17}$} \\
   \hline
  $\mathcal{O}_{11,(1,3,1)}^{6}$ & {$S_{3}S_{9}F_{7}F_{17}$, $S_{3}S_{14}F_{7}F_{17}$, $S_{3}S_{9}F_{7}F_{9}$, $S_{3}S_{14}F_{5}F_{7}$, $S_{3}S_{4}F_{7}F_{17}$, $S_{2}S_{3}F_{7}F_{17}$, $S_{3}S_{4}F_{8}F_{9}$, $S_{2}S_{3}F_{5}F_{8}$, $S_{1}S_{9}F_{10}F_{11}$, $S_{1}S_{14}F_{2}F_{10}$, $S_{1}S_{9}F_{7}F_{17}$, $S_{1}S_{14}F_{7}F_{17}$, $S_{1}S_{4}F_{7}F_{9}$, $S_{1}S_{2}F_{5}F_{7}$, $S_{1}S_{4}F_{7}F_{17}$, $S_{1}S_{2}F_{7}F_{17}$} \\
\hline
\end{longtable}
\end{center}

\begin{center}
\begin{longtable}{| c | m{13cm} |}
   \hline
   $\mathcal{O}^j(\mathcal{T}_{10}^{3})$ & New Particles \\
   \hline
   \hline
   \endhead
   \multicolumn{2}{|c|}{\hspace{9cm} Continued on next page} \\
   \endfoot
  $\mathcal{O}_{11,(1,3,1)}^{22}$ & {$S_{2}S_{7}F_{5}F_{6}$, $S_{2}S_{7}F_{6}F_{6}$, $S_{2}S_{7}F_{8}F_{8}$, $S_{3}S_{7}F_{3}F_{8}$, $S_{2}S_{2}F_{3}F_{8}$, $S_{3}S_{7}F_{3}F_{3}$, $S_{3}S_{7}F_{1}F_{6}$, $S_{2}S_{2}F_{5}F_{6}$, $S_{3}S_{7}F_{6}F_{6}$} \\
   \hline
   \caption{Dimension 11 UV completions for topology $\mathcal{T}_{10}^{3}$ (see Fig.~\ref{fig:d=10_topologies}).}
   \label{tab:T11Top3B1L3D1}\\
   \endlastfoot
  $\mathcal{O}_{11,(1,3,1)}^{1}$ & {$S_{1}S_{9}F_{9}F_{11}$, $S_{1}S_{14}F_{2}F_{5}$, $S_{1}S_{9}F_{11}F_{16}$, $S_{1}S_{14}F_{2}F_{16}$, $S_{1}S_{9}F_{8}F_{10}$, $S_{1}S_{14}F_{8}F_{10}$, $S_{1}S_{9}F_{7}F_{8}$, $S_{1}S_{14}F_{7}F_{8}$} \\
   \hline
  $\mathcal{O}_{11,(1,3,1)}^{2}$ & {$S_{4}S_{9}F_{9}F_{14}$, $S_{2}S_{9}F_{9}F_{14}$, $S_{4}S_{9}F_{6}F_{9}$, $S_{4}S_{14}F_{5}F_{14}$, $S_{2}S_{14}F_{5}F_{6}$, $S_{4}S_{14}F_{9}F_{14}$, $S_{2}S_{9}F_{5}F_{14}$, $S_{4}S_{9}F_{4}F_{9}$, $S_{2}S_{9}F_{4}F_{9}$, $S_{4}S_{9}F_{1}F_{9}$, $S_{4}S_{14}F_{4}F_{5}$, $S_{2}S_{14}F_{1}F_{5}$, $S_{4}S_{4}F_{4}F_{9}$, $S_{2}S_{4}F_{4}F_{9}$, $S_{4}S_{4}F_{1}F_{9}$, $S_{2}S_{4}F_{4}F_{5}$, $S_{2}S_{2}F_{1}F_{5}$, $S_{4}S_{9}F_{4}F_{11}$, $S_{4}S_{14}F_{4}F_{11}$, $S_{4}S_{9}F_{1}F_{11}$, $S_{2}S_{9}F_{2}F_{4}$, $S_{2}S_{14}F_{1}F_{2}$, $S_{4}S_{9}F_{13}F_{20}$, $S_{4}S_{9}F_{7}F_{13}$, $S_{2}S_{9}F_{7}F_{13}$, $S_{4}S_{14}F_{7}F_{13}$, $S_{2}S_{14}F_{7}F_{13}$, $S_{4}S_{9}F_{13}F_{15}$, $S_{4}S_{9}F_{8}F_{13}$, $S_{4}S_{14}F_{8}F_{13}$, $S_{2}S_{9}F_{8}F_{13}$, $S_{2}S_{14}F_{8}F_{13}$, $S_{4}S_{9}F_{13}F_{32}$, $S_{4}S_{9}F_{10}F_{13}$, $S_{2}S_{9}F_{10}F_{13}$, $S_{4}S_{14}F_{10}F_{13}$, $S_{2}S_{14}F_{10}F_{13}$, $S_{4}S_{4}F_{13}F_{20}$, $S_{4}S_{4}F_{7}F_{13}$, $S_{2}S_{4}F_{7}F_{13}$, $S_{2}S_{2}F_{7}F_{13}$} \\
   \hline
  $\mathcal{O}_{11,(1,3,1)}^{3}$ & {$S_{2}S_{9}F_{4}F_{9}$, $S_{2}S_{14}F_{1}F_{5}$, $S_{2}S_{9}F_{2}F_{4}$, $S_{2}S_{14}F_{1}F_{2}$, $S_{1}S_{9}F_{4}F_{19}$, $S_{1}S_{14}F_{1}F_{17}$, $S_{1}S_{9}F_{2}F_{4}$, $S_{1}S_{14}F_{1}F_{2}$, $S_{2}S_{9}F_{10}F_{13}$, $S_{2}S_{14}F_{10}F_{13}$, $S_{2}S_{9}F_{8}F_{13}$, $S_{2}S_{14}F_{8}F_{13}$, $S_{1}S_{9}F_{8}F_{10}$, $S_{1}S_{14}F_{8}F_{10}$, $S_{1}S_{9}F_{7}F_{8}$, $S_{1}S_{14}F_{7}F_{8}$} \\
\hline
\end{longtable}
\end{center}

\begin{center}
\begin{longtable}{| c | m{13cm} |}
   \hline
   $\mathcal{O}^j(\mathcal{T}_{10}^{4})$ & New Particles \\
   \hline
   \hline
   \endhead
   \multicolumn{2}{|c|}{\hspace{9cm} Continued on next page} \\
   \endfoot
  $\mathcal{O}_{11,(1,3,1)}^{22}$ & {$S_{2}S_{5}S_{7}F_{8}$, $S_{2}S_{7}S_{16}F_{8}$, $S_{2}S_{5}S_{7}F_{5}$, $S_{2}S_{7}S_{16}F_{6}$, $S_{3}S_{5}S_{7}F_{1}$, $S_{2}S_{2}S_{5}F_{5}$, $S_{3}S_{7}S_{16}F_{6}$, $S_{3}S_{5}S_{7}F_{8}$, $S_{2}S_{2}S_{5}F_{8}$, $S_{3}S_{7}S_{16}F_{3}$} \\
   \hline
   \caption{Dimension 11 UV completions for topology $\mathcal{T}_{10}^{4}$ (see Fig.~\ref{fig:d=10_topologies}).}
   \label{tab:T11Top4B1L3D1}\\
   \endlastfoot
  $\mathcal{O}_{11,(1,3,1)}^{1}$ & {$S_{1}S_{9}S_{11}F_{10}$, $S_{1}S_{11}S_{14}F_{10}$, $S_{1}S_{9}S_{26}F_{7}$, $S_{1}S_{14}S_{26}F_{7}$, $S_{1}S_{9}S_{11}F_{9}$, $S_{1}S_{11}S_{14}F_{5}$, $S_{1}S_{9}S_{26}F_{16}$, $S_{1}S_{14}S_{26}F_{16}$} \\
   \hline
  $\mathcal{O}_{11,(1,3,1)}^{2}$ & {$S_{4}S_{9}S_{13}F_{20}$, $S_{4}S_{5}S_{9}F_{7}$, $S_{4}S_{5}S_{14}F_{7}$, $S_{2}S_{5}S_{9}F_{7}$, $S_{2}S_{5}S_{14}F_{7}$, $S_{4}S_{9}S_{17}F_{15}$, $S_{4}S_{9}S_{16}F_{8}$, $S_{2}S_{9}S_{16}F_{8}$, $S_{4}S_{14}S_{16}F_{8}$, $S_{2}S_{14}S_{16}F_{8}$, $S_{4}S_{9}S_{12}F_{32}$, $S_{4}S_{9}S_{11}F_{10}$, $S_{4}S_{11}S_{14}F_{10}$, $S_{2}S_{9}S_{11}F_{10}$, $S_{2}S_{11}S_{14}F_{10}$, $S_{4}S_{4}S_{13}F_{20}$, $S_{4}S_{4}S_{5}F_{7}$, $S_{2}S_{4}S_{5}F_{7}$, $S_{2}S_{2}S_{5}F_{7}$, $S_{4}S_{9}S_{13}F_{9}$, $S_{4}S_{5}S_{9}F_{9}$, $S_{4}S_{5}S_{14}F_{5}$, $S_{2}S_{5}S_{9}F_{9}$, $S_{2}S_{5}S_{14}F_{5}$, $S_{4}S_{9}S_{17}F_{9}$, $S_{4}S_{9}S_{16}F_{9}$, $S_{2}S_{9}S_{16}F_{5}$, $S_{4}S_{14}S_{16}F_{9}$, $S_{2}S_{14}S_{16}F_{5}$, $S_{4}S_{9}S_{12}F_{9}$, $S_{4}S_{9}S_{11}F_{9}$, $S_{4}S_{11}S_{14}F_{5}$, $S_{2}S_{9}S_{11}F_{9}$, $S_{2}S_{11}S_{14}F_{5}$, $S_{4}S_{4}S_{13}F_{9}$, $S_{4}S_{4}S_{5}F_{9}$, $S_{2}S_{4}S_{5}F_{5}$, $S_{2}S_{4}S_{5}F_{9}$, $S_{2}S_{2}S_{5}F_{5}$, $S_{4}S_{9}S_{17}F_{11}$, $S_{4}S_{9}S_{16}F_{11}$, $S_{2}S_{9}S_{16}F_{2}$, $S_{4}S_{14}S_{16}F_{11}$, $S_{2}S_{14}S_{16}F_{2}$} \\
   \hline
  $\mathcal{O}_{11,(1,3,1)}^{3}$ & {$S_{2}S_{9}S_{11}F_{10}$, $S_{2}S_{11}S_{14}F_{10}$, $S_{2}S_{9}S_{16}F_{8}$, $S_{2}S_{14}S_{16}F_{8}$, $S_{1}S_{9}S_{10}F_{10}$, $S_{1}S_{10}S_{14}F_{10}$, $S_{1}S_{9}S_{16}F_{7}$, $S_{1}S_{14}S_{16}F_{7}$, $S_{2}S_{9}S_{11}F_{9}$, $S_{2}S_{11}S_{14}F_{5}$, $S_{2}S_{9}S_{16}F_{2}$, $S_{2}S_{14}S_{16}F_{2}$, $S_{1}S_{9}S_{10}F_{19}$, $S_{1}S_{10}S_{14}F_{17}$, $S_{1}S_{9}S_{16}F_{2}$, $S_{1}S_{14}S_{16}F_{2}$} \\
\hline
\end{longtable}
\end{center}

\begin{center}
\begin{longtable}{| c | m{13cm} |}
   \hline
   $\mathcal{O}^j(\mathcal{T}_{10}^{6})$ & New Particles \\
   \hline
   \hline
   \endhead
   \multicolumn{2}{|c|}{\hspace{9cm} Continued on next page} \\
   \endfoot
  $\mathcal{O}_{11,(1,3,1)}^{22}$ & {$S_{5}S_{7}F_{3}F_{8}$, $S_{2}S_{7}F_{1}F_{6}$, $S_{3}S_{7}F_{3}F_{5}$, $S_{2}S_{7}F_{1}F_{8}$, $S_{2}S_{3}F_{3}F_{5}$, $S_{3}S_{7}F_{5}F_{6}$, $S_{5}S_{7}F_{8}F_{8}$, $S_{2}S_{16}F_{3}F_{8}$, $S_{2}S_{3}F_{5}F_{6}$, $S_{5}S_{7}F_{3}F_{5}$, $S_{5}S_{7}F_{1}F_{8}$, $S_{2}S_{5}F_{3}F_{5}$, $S_{5}S_{7}F_{5}F_{8}$, $S_{2}S_{16}F_{3}F_{6}$, $S_{2}S_{5}F_{5}F_{8}$, $S_{3}S_{16}F_{3}F_{6}$, $S_{2}S_{7}F_{5}F_{6}$, $S_{2}S_{5}F_{3}F_{8}$, $S_{2}S_{2}F_{5}F_{6}$, $S_{2}S_{7}F_{5}F_{8}$, $S_{2}S_{3}F_{3}F_{6}$, $S_{2}S_{2}F_{5}F_{8}$, $S_{3}S_{3}F_{3}F_{6}$, $S_{2}S_{3}F_{6}F_{6}$, $S_{2}S_{5}F_{8}F_{8}$, $S_{3}S_{16}F_{3}F_{3}$, $S_{3}S_{3}F_{6}F_{6}$} \\
   \hline
   \caption{Dimension 11 UV completions for topology $\mathcal{T}_{10}^{6}$ (see Fig.~\ref{fig:d=10_topologies}).}
   \label{tab:T11Top6B1L3D1}\\
   \endlastfoot
  $\mathcal{O}_{11,(1,3,1)}^{1}$ & {$S_{4}S_{9}F_{9}F_{11}$, $S_{2}S_{14}F_{2}F_{5}$, $S_{9}S_{11}F_{8}F_{10}$, $S_{11}S_{14}F_{8}F_{10}$, $S_{4}S_{9}F_{8}F_{9}$, $S_{2}S_{14}F_{5}F_{8}$, $S_{1}S_{26}F_{7}F_{11}$, $S_{1}S_{26}F_{2}F_{7}$, $S_{9}S_{11}F_{8}F_{9}$, $S_{11}S_{14}F_{5}F_{8}$, $S_{1}S_{4}F_{7}F_{11}$, $S_{1}S_{2}F_{2}F_{7}$, $S_{1}S_{26}F_{11}F_{16}$, $S_{1}S_{26}F_{2}F_{16}$, $S_{1}S_{4}F_{7}F_{8}$, $S_{1}S_{2}F_{7}F_{8}$} \\
\hline
\end{longtable}
\end{center}

\begin{center}
\begin{longtable}{| c | m{13cm} |}
   \hline
   $\mathcal{O}^j(\mathcal{T}_{10}^{7})$ & New Particles \\
   \hline
   \hline
   \endhead
   \multicolumn{2}{|c|}{\hspace{9cm} Continued on next page} \\
   \endfoot
  $\mathcal{O}_{11,(1,3,1)}^{22}$ & {$S_{2}S_{2}S_{7}F_{6}$, $S_{3}S_{3}S_{7}F_{3}$, $S_{2}S_{2}S_{3}F_{3}$, $S_{2}S_{2}S_{7}F_{8}$, $S_{3}S_{3}S_{7}F_{6}$, $S_{2}S_{2}S_{3}F_{6}$} \\
   \hline
   \caption{Dimension 11 UV completions for topology $\mathcal{T}_{10}^{7}$ (see Fig.~\ref{fig:d=10_topologies}).}
   \label{tab:T11Top7B1L3D1}\\
   \endlastfoot
  $\mathcal{O}_{11,(1,3,1)}^{1}$ & {$S_{1}S_{4}S_{9}F_{11}$, $S_{1}S_{2}S_{14}F_{2}$, $S_{1}S_{4}S_{9}F_{8}$, $S_{1}S_{2}S_{14}F_{8}$} \\
   \hline
  $\mathcal{O}_{11,(1,3,1)}^{2}$ & {$S_{4}S_{9}S_{19}F_{4}$, $S_{4}S_{14}S_{19}F_{4}$, $S_{2}S_{9}S_{19}F_{4}$, $S_{3}S_{4}S_{9}F_{1}$, $S_{2}S_{3}S_{14}F_{1}$, $S_{4}S_{4}S_{19}F_{4}$, $S_{2}S_{4}S_{19}F_{4}$, $S_{3}S_{4}S_{4}F_{1}$, $S_{2}S_{2}S_{3}F_{1}$, $S_{4}S_{9}S_{31}F_{14}$, $S_{4}S_{14}S_{31}F_{14}$, $S_{2}S_{9}S_{31}F_{14}$, $S_{1}S_{4}S_{9}F_{6}$, $S_{1}S_{2}S_{14}F_{6}$, $S_{4}S_{9}S_{31}F_{13}$, $S_{4}S_{14}S_{31}F_{13}$, $S_{2}S_{9}S_{31}F_{13}$, $S_{1}S_{4}S_{9}F_{13}$, $S_{1}S_{2}S_{14}F_{13}$, $S_{4}S_{9}S_{19}F_{13}$, $S_{4}S_{14}S_{19}F_{13}$, $S_{2}S_{9}S_{19}F_{13}$, $S_{3}S_{4}S_{9}F_{13}$, $S_{2}S_{3}S_{14}F_{13}$, $S_{4}S_{4}S_{19}F_{13}$, $S_{2}S_{4}S_{19}F_{13}$, $S_{3}S_{4}S_{4}F_{13}$, $S_{2}S_{2}S_{3}F_{13}$} \\
   \hline
  $\mathcal{O}_{11,(1,3,1)}^{3}$ & {$S_{2}S_{9}S_{19}F_{4}$, $S_{2}S_{3}S_{14}F_{1}$, $S_{1}S_{4}S_{9}F_{4}$, $S_{1}S_{2}S_{14}F_{1}$, $S_{1}S_{4}S_{9}F_{8}$, $S_{1}S_{2}S_{14}F_{8}$, $S_{2}S_{9}S_{19}F_{13}$, $S_{2}S_{3}S_{14}F_{13}$} \\
   \hline
  $\mathcal{O}_{11,(1,3,1)}^{4}$ & {$S_{2}S_{4}S_{19}F_{4}$, $S_{2}S_{2}S_{3}F_{1}$, $S_{2}S_{9}S_{31}F_{14}$, $S_{1}S_{2}S_{14}F_{6}$, $S_{1}S_{4}S_{9}F_{3}$, $S_{1}S_{2}S_{14}F_{3}$, $S_{2}S_{9}S_{31}F_{13}$, $S_{1}S_{2}S_{14}F_{13}$, $S_{1}S_{4}S_{9}F_{4}$, $S_{1}S_{2}S_{14}F_{1}$, $S_{1}S_{4}S_{9}F_{14}$, $S_{2}S_{4}S_{19}F_{13}$, $S_{2}S_{2}S_{3}F_{13}$} \\
   \hline
  $\mathcal{O}_{11,(1,3,1)}^{5}$ & {$S_{1}S_{3}S_{15}F_{1}$, $S_{2}S_{2}S_{3}F_{1}$, $S_{1}S_{2}S_{14}F_{1}$, $S_{1}S_{2}S_{14}F_{3}$, $S_{1}S_{3}S_{15}F_{13}$, $S_{2}S_{2}S_{3}F_{13}$} \\
\hline
\end{longtable}
\end{center}

  \subsection{\texorpdfstring{$\Delta B=2, \Delta L=0$}{Delta B=2,Delta L=0}}

For $(\Delta B,\Delta L)=(2,0)$, we have one-derivative operators,
\begin{multicols}{3}
 \begin{spacing}{1.5}
 \noindent
 \begin{math}
\mathcal{O}_{11, \, (2,0,1)}^{1}\equiv D\bar{H}dQQQQQ\,,\\
\mathcal{O}_{11, \, (2,0,1)}^{2}\equiv D\bar{H}dduQQQ\,,\\
\mathcal{O}_{11, \, (2,0,1)}^{3}\equiv D\bar{H}ddduuQ\,,\\
\mathcal{O}_{11, \, (2,0,1)}^{4}\equiv DHdddQQQ\,,\\
\mathcal{O}_{11, \, (2,0,1)}^{5}\equiv DHdddduQ\,,
\label{eq:1Ddequal11basisB2L0}
 \end{math}
 \end{spacing}
 \end{multicols}
\noindent
and two-derivative operators,
  \begin{multicols}{3}
 \begin{spacing}{1.4}
 \noindent
 \begin{math}
\mathcal{O}_{11, \,(2 , 0 ,2)}^{1}  \equiv DDddQQQQ \,, \\
 \mathcal{O}_{11, \,(2 , 0 ,2)}^{2}  \equiv DDddduQQ \,, \\
 \mathcal{O}_{11, \,(2 , 0 ,2)}^{3}  \equiv DDdddduu \,.
 \label{eq:2Ddequal11basisB2L0}
 \end{math}
 \end{spacing}
 \end{multicols}
 \noindent
 We collect the one-derivative UV completions in tables~\ref{tab:T11Top2B2L0D1}--\ref{tab:T11Top7B2L0D1}.

\begin{center}
\begin{longtable}{| c | m{13cm} |}
   \hline
   $\mathcal{O}^j(\mathcal{T}_{10}^{2})$ & New Particles \\
   \hline
   \hline
   \endhead
   \multicolumn{2}{|c|}{\hspace{9cm} Continued on next page} \\
   \endfoot
  $\mathcal{O}_{11,(2,0,1)}^{5}$ & {$S_{23}S_{23}F_{27}F_{29}$, $S_{3}S_{23}F_{27}F_{29}$, $S_{3}S_{3}F_{27}F_{29}$, $S_{3}S_{3}F_{1}F_{3}$, $S_{22}S_{23}F_{27}F_{29}$, $S_{3}S_{22}F_{27}F_{29}$, $S_{2}S_{23}F_{27}F_{29}$, $S_{2}S_{3}F_{27}F_{29}$, $S_{2}S_{3}F_{1}F_{3}$, $S_{22}S_{23}F_{28}F_{29}$, $S_{2}S_{23}F_{28}F_{29}$, $S_{3}S_{22}F_{28}F_{29}$, $S_{2}S_{3}F_{28}F_{29}$, $S_{2}S_{3}F_{2}F_{3}$} \\
   \hline
   \caption{Dimension 11 UV completions for topology $\mathcal{T}_{10}^{2}$ (see Fig.~\ref{fig:d=10_topologies}).}
   \label{tab:T11Top2B2L0D1}\\
   \endlastfoot
  $\mathcal{O}_{11,(2,0,1)}^{1}$ & {$S_{24}S_{24}F_{30}F_{48}$, $S_{4}S_{24}F_{30}F_{48}$, $S_{24}S_{24}F_{29}F_{30}$, $S_{22}S_{24}F_{29}F_{30}$, $S_{4}S_{24}F_{29}F_{30}$, $S_{2}S_{24}F_{29}F_{30}$, $S_{22}S_{24}F_{27}F_{29}$, $S_{22}S_{22}F_{27}F_{29}$, $S_{4}S_{22}F_{27}F_{29}$, $S_{2}S_{22}F_{27}F_{29}$, $S_{4}S_{4}F_{30}F_{48}$, $S_{4}S_{22}F_{29}F_{30}$, $S_{4}S_{4}F_{29}F_{30}$, $S_{2}S_{4}F_{29}F_{30}$, $S_{4}S_{4}F_{4}F_{12}$, $S_{4}S_{4}F_{3}F_{4}$, $S_{2}S_{4}F_{3}F_{4}$, $S_{2}S_{24}F_{27}F_{29}$, $S_{2}S_{4}F_{27}F_{29}$, $S_{2}S_{2}F_{27}F_{29}$, $S_{2}S_{4}F_{1}F_{3}$, $S_{2}S_{2}F_{1}F_{3}$} \\
   \hline
  $\mathcal{O}_{11,(2,0,1)}^{2}$ & {$S_{23}S_{24}F_{27}F_{29}$, $S_{22}S_{23}F_{27}F_{29}$, $S_{4}S_{23}F_{27}F_{29}$, $S_{2}S_{23}F_{27}F_{29}$, $S_{3}S_{24}F_{27}F_{29}$, $S_{3}S_{22}F_{27}F_{29}$, $S_{3}S_{4}F_{27}F_{29}$, $S_{2}S_{3}F_{27}F_{29}$, $S_{3}S_{4}F_{1}F_{3}$, $S_{2}S_{3}F_{1}F_{3}$, $S_{23}S_{24}F_{29}F_{47}$, $S_{3}S_{24}F_{29}F_{47}$, $S_{22}S_{23}F_{28}F_{29}$, $S_{3}S_{22}F_{28}F_{29}$, $S_{4}S_{23}F_{29}F_{47}$, $S_{3}S_{4}F_{29}F_{47}$, $S_{3}S_{4}F_{3}F_{11}$, $S_{2}S_{23}F_{28}F_{29}$, $S_{2}S_{3}F_{28}F_{29}$, $S_{2}S_{3}F_{2}F_{3}$, $S_{22}S_{24}F_{27}F_{29}$, $S_{22}S_{22}F_{27}F_{29}$, $S_{4}S_{22}F_{27}F_{29}$, $S_{2}S_{22}F_{27}F_{29}$, $S_{2}S_{24}F_{27}F_{29}$, $S_{2}S_{4}F_{27}F_{29}$, $S_{2}S_{2}F_{27}F_{29}$, $S_{2}S_{4}F_{1}F_{3}$, $S_{2}S_{2}F_{1}F_{3}$, $S_{22}S_{24}F_{29}F_{30}$, $S_{2}S_{24}F_{29}F_{30}$, $S_{4}S_{22}F_{29}F_{30}$, $S_{2}S_{4}F_{29}F_{30}$, $S_{2}S_{4}F_{3}F_{4}$} \\
   \hline
  $\mathcal{O}_{11,(2,0,1)}^{3}$ & {$S_{22}S_{23}F_{28}F_{29}$, $S_{3}S_{22}F_{28}F_{29}$, $S_{2}S_{23}F_{28}F_{29}$, $S_{2}S_{3}F_{28}F_{29}$, $S_{2}S_{3}F_{2}F_{3}$, $S_{22}S_{23}F_{27}F_{29}$, $S_{2}S_{23}F_{27}F_{29}$, $S_{3}S_{22}F_{27}F_{29}$, $S_{2}S_{3}F_{27}F_{29}$, $S_{2}S_{3}F_{1}F_{3}$, $S_{21}S_{23}F_{28}F_{29}$, $S_{3}S_{21}F_{28}F_{29}$, $S_{1}S_{23}F_{28}F_{29}$, $S_{1}S_{3}F_{28}F_{29}$, $S_{1}S_{3}F_{2}F_{3}$, $S_{22}S_{22}F_{27}F_{29}$, $S_{2}S_{22}F_{27}F_{29}$, $S_{2}S_{2}F_{27}F_{29}$, $S_{2}S_{2}F_{1}F_{3}$, $S_{21}S_{23}F_{28}F_{46}$, $S_{1}S_{23}F_{28}F_{46}$, $S_{3}S_{21}F_{28}F_{46}$, $S_{1}S_{3}F_{28}F_{46}$, $S_{1}S_{3}F_{2}F_{10}$} \\
   \hline
  $\mathcal{O}_{11,(2,0,1)}^{4}$ & {$S_{23}S_{24}F_{28}F_{29}$, $S_{22}S_{23}F_{28}F_{29}$, $S_{4}S_{23}F_{28}F_{29}$, $S_{2}S_{23}F_{28}F_{29}$, $S_{3}S_{24}F_{28}F_{29}$, $S_{3}S_{22}F_{28}F_{29}$, $S_{3}S_{4}F_{28}F_{29}$, $S_{2}S_{3}F_{28}F_{29}$, $S_{3}S_{4}F_{2}F_{3}$, $S_{2}S_{3}F_{2}F_{3}$, $S_{23}S_{24}F_{29}F_{30}$, $S_{3}S_{24}F_{29}F_{30}$, $S_{22}S_{23}F_{27}F_{29}$, $S_{3}S_{22}F_{27}F_{29}$, $S_{4}S_{23}F_{29}F_{30}$, $S_{3}S_{4}F_{29}F_{30}$, $S_{3}S_{4}F_{3}F_{4}$, $S_{2}S_{23}F_{27}F_{29}$, $S_{2}S_{3}F_{27}F_{29}$, $S_{2}S_{3}F_{1}F_{3}$} \\
   \hline
\end{longtable}
\end{center}

\begin{center}
\begin{longtable}{| c | m{13cm} |}
   \hline
   $\mathcal{O}^j(\mathcal{T}_{10}^{3})$ & New Particles \\
   \hline
   \hline
   \endhead
   \multicolumn{2}{|c|}{\hspace{9cm} Continued on next page} \\
   \endfoot
  $\mathcal{O}_{11,(2,0,1)}^{5}$ & {$S_{23}S_{23}F_{6}F_{27}$, $S_{3}S_{23}F_{6}F_{27}$, $S_{3}S_{3}F_{6}F_{27}$, $S_{3}S_{3}F_{1}F_{6}$, $S_{22}S_{23}F_{6}F_{28}$, $S_{2}S_{23}F_{6}F_{28}$, $S_{3}S_{22}F_{6}F_{28}$, $S_{2}S_{3}F_{6}F_{28}$, $S_{2}S_{3}F_{2}F_{6}$, $S_{22}S_{23}F_{6}F_{27}$, $S_{3}S_{22}F_{6}F_{27}$, $S_{2}S_{23}F_{6}F_{27}$, $S_{2}S_{3}F_{6}F_{27}$, $S_{2}S_{3}F_{1}F_{6}$, $S_{23}S_{23}F_{13}F_{29}$, $S_{3}S_{23}F_{13}F_{29}$, $S_{3}S_{3}F_{13}F_{29}$, $S_{3}S_{3}F_{3}F_{13}$, $S_{22}S_{23}F_{8}F_{29}$, $S_{2}S_{23}F_{8}F_{29}$, $S_{3}S_{22}F_{8}F_{29}$, $S_{2}S_{3}F_{8}F_{29}$, $S_{2}S_{3}F_{3}F_{8}$} \\
   \hline
   \caption{Dimension 11 UV completions for topology $\mathcal{T}_{10}^{3}$ (see Fig.~\ref{fig:d=10_topologies}).}
   \label{tab:T11Top3B2L0D1}\\
   \endlastfoot
  $\mathcal{O}_{11,(2,0,1)}^{1}$ & {$S_{24}S_{24}F_{9}F_{30}$, $S_{22}S_{24}F_{9}F_{30}$, $S_{4}S_{24}F_{9}F_{30}$, $S_{2}S_{24}F_{9}F_{30}$, $S_{24}S_{24}F_{5}F_{30}$, $S_{4}S_{24}F_{5}F_{30}$, $S_{22}S_{24}F_{9}F_{27}$, $S_{4}S_{22}F_{9}F_{27}$, $S_{22}S_{22}F_{5}F_{27}$, $S_{2}S_{22}F_{5}F_{27}$, $S_{4}S_{22}F_{9}F_{30}$, $S_{4}S_{4}F_{9}F_{30}$, $S_{2}S_{4}F_{9}F_{30}$, $S_{4}S_{4}F_{5}F_{30}$, $S_{4}S_{4}F_{4}F_{9}$, $S_{2}S_{4}F_{4}F_{9}$, $S_{4}S_{4}F_{4}F_{5}$, $S_{2}S_{24}F_{9}F_{27}$, $S_{2}S_{4}F_{9}F_{27}$, $S_{2}S_{2}F_{5}F_{27}$, $S_{2}S_{4}F_{1}F_{9}$, $S_{2}S_{2}F_{1}F_{5}$, $S_{24}S_{24}F_{7}F_{48}$, $S_{4}S_{24}F_{7}F_{48}$, $S_{24}S_{24}F_{7}F_{29}$, $S_{22}S_{24}F_{7}F_{29}$, $S_{4}S_{24}F_{7}F_{29}$, $S_{2}S_{24}F_{7}F_{29}$, $S_{22}S_{22}F_{7}F_{29}$, $S_{4}S_{22}F_{7}F_{29}$, $S_{2}S_{22}F_{7}F_{29}$, $S_{4}S_{4}F_{7}F_{48}$, $S_{4}S_{4}F_{7}F_{29}$, $S_{2}S_{4}F_{7}F_{29}$, $S_{4}S_{4}F_{7}F_{12}$, $S_{4}S_{4}F_{3}F_{7}$, $S_{2}S_{4}F_{3}F_{7}$, $S_{2}S_{2}F_{7}F_{29}$, $S_{2}S_{2}F_{3}F_{7}$} \\
   \hline
  $\mathcal{O}_{11,(2,0,1)}^{2}$ & {$S_{23}S_{24}F_{9}F_{47}$, $S_{3}S_{24}F_{9}F_{47}$, $S_{22}S_{23}F_{5}F_{28}$, $S_{3}S_{22}F_{5}F_{28}$, $S_{4}S_{23}F_{9}F_{47}$, $S_{3}S_{4}F_{9}F_{47}$, $S_{3}S_{4}F_{9}F_{11}$, $S_{2}S_{23}F_{5}F_{28}$, $S_{2}S_{3}F_{5}F_{28}$, $S_{2}S_{3}F_{2}F_{5}$, $S_{23}S_{24}F_{9}F_{27}$, $S_{4}S_{23}F_{9}F_{27}$, $S_{22}S_{23}F_{5}F_{27}$, $S_{2}S_{23}F_{5}F_{27}$, $S_{3}S_{24}F_{9}F_{27}$, $S_{3}S_{4}F_{9}F_{27}$, $S_{3}S_{22}F_{5}F_{27}$, $S_{2}S_{3}F_{5}F_{27}$, $S_{3}S_{4}F_{1}F_{9}$, $S_{2}S_{3}F_{1}F_{5}$, $S_{22}S_{24}F_{9}F_{30}$, $S_{2}S_{24}F_{9}F_{30}$, $S_{22}S_{22}F_{5}F_{27}$, $S_{2}S_{22}F_{5}F_{27}$, $S_{4}S_{22}F_{9}F_{30}$, $S_{2}S_{4}F_{9}F_{30}$, $S_{2}S_{4}F_{4}F_{9}$, $S_{2}S_{2}F_{5}F_{27}$, $S_{2}S_{2}F_{1}F_{5}$, $S_{22}S_{24}F_{9}F_{27}$, $S_{4}S_{22}F_{9}F_{27}$, $S_{2}S_{24}F_{9}F_{27}$, $S_{2}S_{4}F_{9}F_{27}$, $S_{2}S_{4}F_{1}F_{9}$, $S_{23}S_{24}F_{8}F_{29}$, $S_{3}S_{24}F_{8}F_{29}$, $S_{22}S_{23}F_{8}F_{29}$, $S_{3}S_{22}F_{8}F_{29}$, $S_{4}S_{23}F_{8}F_{29}$, $S_{3}S_{4}F_{8}F_{29}$, $S_{3}S_{4}F_{3}F_{8}$, $S_{2}S_{23}F_{8}F_{29}$, $S_{2}S_{3}F_{8}F_{29}$, $S_{2}S_{3}F_{3}F_{8}$, $S_{22}S_{24}F_{7}F_{29}$, $S_{2}S_{24}F_{7}F_{29}$, $S_{22}S_{22}F_{7}F_{29}$, $S_{2}S_{22}F_{7}F_{29}$, $S_{4}S_{22}F_{7}F_{29}$, $S_{2}S_{4}F_{7}F_{29}$, $S_{2}S_{4}F_{3}F_{7}$, $S_{2}S_{2}F_{7}F_{29}$, $S_{2}S_{2}F_{3}F_{7}$} \\
   \hline
  $\mathcal{O}_{11,(2,0,1)}^{3}$ & {$S_{22}S_{23}F_{5}F_{27}$, $S_{2}S_{23}F_{5}F_{27}$, $S_{3}S_{22}F_{5}F_{27}$, $S_{2}S_{3}F_{5}F_{27}$, $S_{2}S_{3}F_{1}F_{5}$, $S_{22}S_{23}F_{5}F_{28}$, $S_{3}S_{22}F_{5}F_{28}$, $S_{2}S_{23}F_{5}F_{28}$, $S_{2}S_{3}F_{5}F_{28}$, $S_{2}S_{3}F_{2}F_{5}$, $S_{21}S_{23}F_{5}F_{28}$, $S_{1}S_{23}F_{5}F_{28}$, $S_{3}S_{21}F_{5}F_{28}$, $S_{1}S_{3}F_{5}F_{28}$, $S_{1}S_{3}F_{2}F_{5}$, $S_{22}S_{22}F_{5}F_{27}$, $S_{2}S_{22}F_{5}F_{27}$, $S_{2}S_{2}F_{5}F_{27}$, $S_{2}S_{2}F_{1}F_{5}$, $S_{22}S_{23}F_{8}F_{29}$, $S_{2}S_{23}F_{8}F_{29}$, $S_{3}S_{22}F_{8}F_{29}$, $S_{2}S_{3}F_{8}F_{29}$, $S_{2}S_{3}F_{3}F_{8}$, $S_{21}S_{23}F_{7}F_{29}$, $S_{1}S_{23}F_{7}F_{29}$, $S_{3}S_{21}F_{7}F_{29}$, $S_{1}S_{3}F_{7}F_{29}$, $S_{1}S_{3}F_{3}F_{7}$, $S_{22}S_{22}F_{7}F_{29}$, $S_{2}S_{22}F_{7}F_{29}$, $S_{2}S_{2}F_{7}F_{29}$, $S_{2}S_{2}F_{3}F_{7}$, $S_{21}S_{23}F_{7}F_{46}$, $S_{3}S_{21}F_{7}F_{46}$, $S_{1}S_{23}F_{7}F_{46}$, $S_{1}S_{3}F_{7}F_{46}$, $S_{1}S_{3}F_{7}F_{10}$} \\
   \hline
  $\mathcal{O}_{11,(2,0,1)}^{4}$ & {$S_{23}S_{24}F_{14}F_{30}$, $S_{3}S_{24}F_{14}F_{30}$, $S_{22}S_{23}F_{6}F_{27}$, $S_{3}S_{22}F_{6}F_{27}$, $S_{4}S_{23}F_{14}F_{30}$, $S_{3}S_{4}F_{14}F_{30}$, $S_{3}S_{4}F_{4}F_{14}$, $S_{2}S_{23}F_{6}F_{27}$, $S_{2}S_{3}F_{6}F_{27}$, $S_{2}S_{3}F_{1}F_{6}$, $S_{23}S_{24}F_{14}F_{28}$, $S_{4}S_{23}F_{14}F_{28}$, $S_{22}S_{23}F_{6}F_{28}$, $S_{2}S_{23}F_{6}F_{28}$, $S_{3}S_{24}F_{14}F_{28}$, $S_{3}S_{4}F_{14}F_{28}$, $S_{3}S_{22}F_{6}F_{28}$, $S_{2}S_{3}F_{6}F_{28}$, $S_{3}S_{4}F_{2}F_{14}$, $S_{2}S_{3}F_{2}F_{6}$, $S_{23}S_{24}F_{8}F_{29}$, $S_{3}S_{24}F_{8}F_{29}$, $S_{22}S_{23}F_{8}F_{29}$, $S_{3}S_{22}F_{8}F_{29}$, $S_{4}S_{23}F_{8}F_{29}$, $S_{3}S_{4}F_{8}F_{29}$, $S_{3}S_{4}F_{3}F_{8}$, $S_{2}S_{23}F_{8}F_{29}$, $S_{2}S_{3}F_{8}F_{29}$, $S_{2}S_{3}F_{3}F_{8}$} \\
   \hline
\end{longtable}
\end{center}

\begin{center}
\begin{longtable}{| c | m{13cm} |}
   \hline
   $\mathcal{O}^j(\mathcal{T}_{10}^{4})$ & New Particles \\
   \hline
   \hline
   \endhead
   \multicolumn{2}{|c|}{\hspace{9cm} Continued on next page} \\
   \endfoot
  $\mathcal{O}_{11,(2,0,1)}^{5}$ & {$S_{23}S_{23}S_{37}F_{29}$, $S_{3}S_{23}S_{37}F_{29}$, $S_{3}S_{5}S_{23}F_{29}$, $S_{3}S_{3}S_{5}F_{29}$, $S_{3}S_{3}S_{5}F_{3}$, $S_{22}S_{23}S_{36}F_{29}$, $S_{3}S_{22}S_{36}F_{29}$, $S_{2}S_{11}S_{23}F_{29}$, $S_{2}S_{3}S_{11}F_{29}$, $S_{2}S_{3}S_{11}F_{3}$, $S_{22}S_{23}S_{37}F_{29}$, $S_{2}S_{23}S_{37}F_{29}$, $S_{3}S_{5}S_{22}F_{29}$, $S_{2}S_{3}S_{5}F_{29}$, $S_{2}S_{3}S_{5}F_{3}$, $S_{23}S_{23}S_{37}F_{27}$, $S_{3}S_{23}S_{37}F_{27}$, $S_{3}S_{5}S_{23}F_{27}$, $S_{3}S_{3}S_{5}F_{27}$, $S_{3}S_{3}S_{5}F_{1}$, $S_{22}S_{23}S_{36}F_{28}$, $S_{3}S_{22}S_{36}F_{28}$, $S_{2}S_{11}S_{23}F_{28}$, $S_{2}S_{3}S_{11}F_{28}$, $S_{2}S_{3}S_{11}F_{2}$, $S_{22}S_{23}S_{37}F_{27}$, $S_{2}S_{23}S_{37}F_{27}$, $S_{3}S_{5}S_{22}F_{27}$, $S_{2}S_{3}S_{5}F_{27}$, $S_{2}S_{3}S_{5}F_{1}$} \\
   \hline
   \caption{Dimension 11 UV completions for topology $\mathcal{T}_{10}^{4}$ (see Fig.~\ref{fig:d=10_topologies}).}
   \label{tab:T11Top4B2L0D1}\\
   \endlastfoot
  $\mathcal{O}_{11,(2,0,1)}^{1}$ & {$S_{24}S_{24}S_{40}F_{48}$, $S_{4}S_{24}S_{40}F_{48}$, $S_{24}S_{24}S_{37}F_{29}$, $S_{22}S_{24}S_{37}F_{29}$, $S_{4}S_{24}S_{37}F_{29}$, $S_{2}S_{24}S_{37}F_{29}$, $S_{22}S_{22}S_{37}F_{29}$, $S_{4}S_{22}S_{37}F_{29}$, $S_{2}S_{22}S_{37}F_{29}$, $S_{4}S_{13}S_{24}F_{48}$, $S_{4}S_{4}S_{13}F_{48}$, $S_{4}S_{4}S_{13}F_{12}$, $S_{4}S_{5}S_{24}F_{29}$, $S_{4}S_{5}S_{22}F_{29}$, $S_{4}S_{4}S_{5}F_{29}$, $S_{2}S_{4}S_{5}F_{29}$, $S_{4}S_{4}S_{5}F_{3}$, $S_{2}S_{4}S_{5}F_{3}$, $S_{2}S_{5}S_{24}F_{29}$, $S_{2}S_{5}S_{22}F_{29}$, $S_{2}S_{2}S_{5}F_{29}$, $S_{2}S_{2}S_{5}F_{3}$, $S_{24}S_{24}S_{40}F_{30}$, $S_{4}S_{24}S_{40}F_{30}$, $S_{24}S_{24}S_{37}F_{30}$, $S_{4}S_{24}S_{37}F_{30}$, $S_{22}S_{24}S_{37}F_{27}$, $S_{2}S_{24}S_{37}F_{27}$, $S_{22}S_{24}S_{37}F_{30}$, $S_{4}S_{22}S_{37}F_{30}$, $S_{22}S_{22}S_{37}F_{27}$, $S_{2}S_{22}S_{37}F_{27}$, $S_{4}S_{13}S_{24}F_{30}$, $S_{4}S_{4}S_{13}F_{30}$, $S_{4}S_{4}S_{13}F_{4}$, $S_{4}S_{5}S_{24}F_{30}$, $S_{4}S_{4}S_{5}F_{30}$, $S_{4}S_{5}S_{22}F_{27}$, $S_{2}S_{4}S_{5}F_{27}$, $S_{4}S_{4}S_{5}F_{4}$, $S_{2}S_{4}S_{5}F_{1}$, $S_{2}S_{5}S_{24}F_{30}$, $S_{2}S_{4}S_{5}F_{30}$, $S_{2}S_{5}S_{22}F_{27}$, $S_{2}S_{2}S_{5}F_{27}$, $S_{2}S_{4}S_{5}F_{4}$, $S_{2}S_{2}S_{5}F_{1}$} \\
   \hline
  $\mathcal{O}_{11,(2,0,1)}^{2}$ & {$S_{23}S_{24}S_{38}F_{29}$, $S_{22}S_{23}S_{38}F_{29}$, $S_{4}S_{23}S_{38}F_{29}$, $S_{2}S_{23}S_{38}F_{29}$, $S_{3}S_{6}S_{24}F_{29}$, $S_{3}S_{6}S_{22}F_{29}$, $S_{3}S_{4}S_{6}F_{29}$, $S_{2}S_{3}S_{6}F_{29}$, $S_{3}S_{4}S_{6}F_{3}$, $S_{2}S_{3}S_{6}F_{3}$, $S_{23}S_{24}S_{37}F_{29}$, $S_{3}S_{24}S_{37}F_{29}$, $S_{22}S_{23}S_{37}F_{29}$, $S_{3}S_{22}S_{37}F_{29}$, $S_{4}S_{5}S_{23}F_{29}$, $S_{3}S_{4}S_{5}F_{29}$, $S_{3}S_{4}S_{5}F_{3}$, $S_{2}S_{5}S_{23}F_{29}$, $S_{2}S_{3}S_{5}F_{29}$, $S_{2}S_{3}S_{5}F_{3}$, $S_{22}S_{24}S_{37}F_{29}$, $S_{22}S_{22}S_{37}F_{29}$, $S_{4}S_{22}S_{37}F_{29}$, $S_{2}S_{22}S_{37}F_{29}$, $S_{2}S_{5}S_{24}F_{29}$, $S_{2}S_{5}S_{22}F_{29}$, $S_{2}S_{4}S_{5}F_{29}$, $S_{2}S_{2}S_{5}F_{29}$, $S_{2}S_{4}S_{5}F_{3}$, $S_{2}S_{2}S_{5}F_{3}$, $S_{2}S_{24}S_{37}F_{29}$, $S_{4}S_{5}S_{22}F_{29}$, $S_{23}S_{24}S_{38}F_{47}$, $S_{4}S_{23}S_{38}F_{47}$, $S_{22}S_{23}S_{38}F_{28}$, $S_{2}S_{23}S_{38}F_{28}$, $S_{3}S_{6}S_{24}F_{47}$, $S_{3}S_{4}S_{6}F_{47}$, $S_{3}S_{6}S_{22}F_{28}$, $S_{2}S_{3}S_{6}F_{28}$, $S_{3}S_{4}S_{6}F_{11}$, $S_{2}S_{3}S_{6}F_{2}$, $S_{23}S_{24}S_{37}F_{27}$, $S_{3}S_{24}S_{37}F_{27}$, $S_{22}S_{23}S_{37}F_{27}$, $S_{3}S_{22}S_{37}F_{27}$, $S_{4}S_{5}S_{23}F_{27}$, $S_{3}S_{4}S_{5}F_{27}$, $S_{3}S_{4}S_{5}F_{1}$, $S_{2}S_{5}S_{23}F_{27}$, $S_{2}S_{3}S_{5}F_{27}$, $S_{2}S_{3}S_{5}F_{1}$, $S_{22}S_{24}S_{37}F_{30}$, $S_{4}S_{22}S_{37}F_{30}$, $S_{22}S_{22}S_{37}F_{27}$, $S_{2}S_{22}S_{37}F_{27}$, $S_{2}S_{5}S_{24}F_{30}$, $S_{2}S_{4}S_{5}F_{30}$, $S_{2}S_{5}S_{22}F_{27}$, $S_{2}S_{2}S_{5}F_{27}$, $S_{2}S_{4}S_{5}F_{4}$, $S_{2}S_{2}S_{5}F_{1}$, $S_{22}S_{24}S_{37}F_{27}$, $S_{2}S_{24}S_{37}F_{27}$, $S_{4}S_{5}S_{22}F_{27}$, $S_{2}S_{4}S_{5}F_{27}$, $S_{2}S_{4}S_{5}F_{1}$} \\
   \hline
  $\mathcal{O}_{11,(2,0,1)}^{3}$ & {$S_{22}S_{23}S_{37}F_{29}$, $S_{3}S_{22}S_{37}F_{29}$, $S_{2}S_{5}S_{23}F_{29}$, $S_{2}S_{3}S_{5}F_{29}$, $S_{2}S_{3}S_{5}F_{3}$, $S_{22}S_{23}S_{38}F_{29}$, $S_{2}S_{23}S_{38}F_{29}$, $S_{3}S_{6}S_{22}F_{29}$, $S_{2}S_{3}S_{6}F_{29}$, $S_{2}S_{3}S_{6}F_{3}$, $S_{21}S_{23}S_{36}F_{29}$, $S_{3}S_{21}S_{36}F_{29}$, $S_{1}S_{11}S_{23}F_{29}$, $S_{1}S_{3}S_{11}F_{29}$, $S_{1}S_{3}S_{11}F_{3}$, $S_{22}S_{22}S_{37}F_{29}$, $S_{2}S_{22}S_{37}F_{29}$, $S_{2}S_{5}S_{22}F_{29}$, $S_{2}S_{2}S_{5}F_{29}$, $S_{2}S_{2}S_{5}F_{3}$, $S_{21}S_{23}S_{38}F_{46}$, $S_{1}S_{23}S_{38}F_{46}$, $S_{3}S_{6}S_{21}F_{46}$, $S_{1}S_{3}S_{6}F_{46}$, $S_{1}S_{3}S_{6}F_{10}$, $S_{22}S_{23}S_{37}F_{27}$, $S_{3}S_{22}S_{37}F_{27}$, $S_{2}S_{5}S_{23}F_{27}$, $S_{2}S_{3}S_{5}F_{27}$, $S_{2}S_{3}S_{5}F_{1}$, $S_{22}S_{23}S_{38}F_{28}$, $S_{2}S_{23}S_{38}F_{28}$, $S_{3}S_{6}S_{22}F_{28}$, $S_{2}S_{3}S_{6}F_{28}$, $S_{2}S_{3}S_{6}F_{2}$, $S_{21}S_{23}S_{36}F_{28}$, $S_{3}S_{21}S_{36}F_{28}$, $S_{1}S_{11}S_{23}F_{28}$, $S_{1}S_{3}S_{11}F_{28}$, $S_{1}S_{3}S_{11}F_{2}$, $S_{22}S_{22}S_{37}F_{27}$, $S_{2}S_{22}S_{37}F_{27}$, $S_{2}S_{5}S_{22}F_{27}$, $S_{2}S_{2}S_{5}F_{27}$, $S_{2}S_{2}S_{5}F_{1}$, $S_{21}S_{23}S_{38}F_{28}$, $S_{1}S_{23}S_{38}F_{28}$, $S_{3}S_{6}S_{21}F_{28}$, $S_{1}S_{3}S_{6}F_{28}$, $S_{1}S_{3}S_{6}F_{2}$} \\
   \hline
  $\mathcal{O}_{11,(2,0,1)}^{4}$ & {$S_{23}S_{24}S_{37}F_{29}$, $S_{22}S_{23}S_{37}F_{29}$, $S_{4}S_{23}S_{37}F_{29}$, $S_{2}S_{23}S_{37}F_{29}$, $S_{3}S_{5}S_{24}F_{29}$, $S_{3}S_{5}S_{22}F_{29}$, $S_{3}S_{4}S_{5}F_{29}$, $S_{2}S_{3}S_{5}F_{29}$, $S_{3}S_{4}S_{5}F_{3}$, $S_{2}S_{3}S_{5}F_{3}$, $S_{23}S_{24}S_{36}F_{29}$, $S_{3}S_{24}S_{36}F_{29}$, $S_{22}S_{23}S_{36}F_{29}$, $S_{3}S_{22}S_{36}F_{29}$, $S_{4}S_{11}S_{23}F_{29}$, $S_{3}S_{4}S_{11}F_{29}$, $S_{3}S_{4}S_{11}F_{3}$, $S_{2}S_{11}S_{23}F_{29}$, $S_{2}S_{3}S_{11}F_{29}$, $S_{2}S_{3}S_{11}F_{3}$, $S_{23}S_{24}S_{37}F_{30}$, $S_{4}S_{23}S_{37}F_{30}$, $S_{22}S_{23}S_{37}F_{27}$, $S_{2}S_{23}S_{37}F_{27}$, $S_{3}S_{5}S_{24}F_{30}$, $S_{3}S_{4}S_{5}F_{30}$, $S_{3}S_{5}S_{22}F_{27}$, $S_{2}S_{3}S_{5}F_{27}$, $S_{3}S_{4}S_{5}F_{4}$, $S_{2}S_{3}S_{5}F_{1}$, $S_{23}S_{24}S_{36}F_{28}$, $S_{3}S_{24}S_{36}F_{28}$, $S_{22}S_{23}S_{36}F_{28}$, $S_{3}S_{22}S_{36}F_{28}$, $S_{4}S_{11}S_{23}F_{28}$, $S_{3}S_{4}S_{11}F_{28}$, $S_{3}S_{4}S_{11}F_{2}$, $S_{2}S_{11}S_{23}F_{28}$, $S_{2}S_{3}S_{11}F_{28}$, $S_{2}S_{3}S_{11}F_{2}$} \\
   \hline
\end{longtable}
\end{center}

\begin{center}
\begin{longtable}{| c | m{13cm} |}
   \hline
   $\mathcal{O}^j(\mathcal{T}_{10}^{6})$ & New Particles \\
   \hline
   \hline
   \endhead
   \multicolumn{2}{|c|}{\hspace{9cm} Continued on next page} \\
   \endfoot
  $\mathcal{O}_{11,(2,0,1)}^{5}$ & {$S_{22}S_{23}F_{6}F_{28}$, $S_{3}S_{22}F_{6}F_{28}$, $S_{2}S_{23}F_{6}F_{28}$, $S_{2}S_{3}F_{6}F_{28}$, $S_{2}S_{3}F_{2}F_{6}$, $S_{23}S_{37}F_{8}F_{29}$, $S_{3}S_{37}F_{8}F_{29}$, $S_{5}S_{23}F_{8}F_{29}$, $S_{3}S_{5}F_{8}F_{29}$, $S_{3}S_{5}F_{3}F_{8}$, $S_{22}S_{23}F_{8}F_{28}$, $S_{3}S_{22}F_{8}F_{28}$, $S_{2}S_{23}F_{8}F_{28}$, $S_{2}S_{3}F_{8}F_{28}$, $S_{2}S_{3}F_{2}F_{8}$, $S_{21}S_{23}F_{6}F_{28}$, $S_{3}S_{21}F_{6}F_{28}$, $S_{1}S_{23}F_{6}F_{28}$, $S_{1}S_{3}F_{6}F_{28}$, $S_{1}S_{3}F_{2}F_{6}$, $S_{23}S_{36}F_{8}F_{29}$, $S_{3}S_{36}F_{8}F_{29}$, $S_{11}S_{23}F_{8}F_{29}$, $S_{3}S_{11}F_{8}F_{29}$, $S_{3}S_{11}F_{3}F_{8}$, $S_{22}S_{23}F_{6}F_{27}$, $S_{3}S_{22}F_{6}F_{27}$, $S_{2}S_{23}F_{6}F_{27}$, $S_{2}S_{3}F_{6}F_{27}$, $S_{2}S_{3}F_{1}F_{6}$, $S_{22}S_{22}F_{6}F_{27}$, $S_{2}S_{22}F_{6}F_{27}$, $S_{2}S_{2}F_{6}F_{27}$, $S_{2}S_{2}F_{1}F_{6}$, $S_{23}S_{36}F_{13}F_{29}$, $S_{3}S_{36}F_{13}F_{29}$, $S_{11}S_{23}F_{13}F_{29}$, $S_{3}S_{11}F_{13}F_{29}$, $S_{3}S_{11}F_{3}F_{13}$, $S_{22}S_{37}F_{8}F_{29}$, $S_{2}S_{37}F_{8}F_{29}$, $S_{5}S_{22}F_{8}F_{29}$, $S_{2}S_{5}F_{8}F_{29}$, $S_{2}S_{5}F_{3}F_{8}$, $S_{22}S_{23}F_{8}F_{27}$, $S_{3}S_{22}F_{8}F_{27}$, $S_{2}S_{23}F_{8}F_{27}$, $S_{2}S_{3}F_{8}F_{27}$, $S_{2}S_{3}F_{1}F_{8}$, $S_{21}S_{23}F_{13}F_{28}$, $S_{3}S_{21}F_{13}F_{28}$, $S_{1}S_{23}F_{13}F_{28}$, $S_{1}S_{3}F_{13}F_{28}$, $S_{1}S_{3}F_{2}F_{13}$, $S_{22}S_{22}F_{8}F_{27}$, $S_{2}S_{22}F_{8}F_{27}$, $S_{2}S_{2}F_{8}F_{27}$, $S_{2}S_{2}F_{1}F_{8}$, $S_{23}S_{36}F_{8}F_{28}$, $S_{3}S_{36}F_{8}F_{28}$, $S_{11}S_{23}F_{8}F_{28}$, $S_{3}S_{11}F_{8}F_{28}$, $S_{3}S_{11}F_{2}F_{8}$, $S_{23}S_{37}F_{8}F_{27}$, $S_{3}S_{37}F_{8}F_{27}$, $S_{5}S_{23}F_{8}F_{27}$, $S_{3}S_{5}F_{8}F_{27}$, $S_{3}S_{5}F_{1}F_{8}$, $S_{23}S_{36}F_{13}F_{28}$, $S_{3}S_{36}F_{13}F_{28}$, $S_{11}S_{23}F_{13}F_{28}$, $S_{3}S_{11}F_{13}F_{28}$, $S_{3}S_{11}F_{2}F_{13}$, $S_{22}S_{37}F_{8}F_{27}$, $S_{2}S_{37}F_{8}F_{27}$, $S_{5}S_{22}F_{8}F_{27}$, $S_{2}S_{5}F_{8}F_{27}$, $S_{2}S_{5}F_{1}F_{8}$} \\
   \hline
   \caption{Dimension 11 UV completions for topology $\mathcal{T}_{10}^{6}$ (see Fig.~\ref{fig:d=10_topologies}).}
   \label{tab:T11Top6B2L0D1}\\
   \endlastfoot
  $\mathcal{O}_{11,(2,0,1)}^{1}$ & {$S_{24}S_{40}F_{9}F_{48}$, $S_{4}S_{40}F_{9}F_{48}$, $S_{24}S_{37}F_{9}F_{29}$, $S_{22}S_{37}F_{9}F_{29}$, $S_{4}S_{37}F_{9}F_{29}$, $S_{2}S_{37}F_{9}F_{29}$, $S_{13}S_{24}F_{9}F_{48}$, $S_{4}S_{13}F_{9}F_{48}$, $S_{4}S_{13}F_{9}F_{12}$, $S_{5}S_{24}F_{9}F_{29}$, $S_{5}S_{22}F_{9}F_{29}$, $S_{4}S_{5}F_{9}F_{29}$, $S_{2}S_{5}F_{9}F_{29}$, $S_{4}S_{5}F_{3}F_{9}$, $S_{2}S_{5}F_{3}F_{9}$, $S_{24}S_{37}F_{5}F_{29}$, $S_{22}S_{37}F_{5}F_{29}$, $S_{4}S_{37}F_{5}F_{29}$, $S_{2}S_{37}F_{5}F_{29}$, $S_{5}S_{24}F_{5}F_{29}$, $S_{5}S_{22}F_{5}F_{29}$, $S_{4}S_{5}F_{5}F_{29}$, $S_{2}S_{5}F_{5}F_{29}$, $S_{4}S_{5}F_{3}F_{5}$, $S_{2}S_{5}F_{3}F_{5}$, $S_{24}S_{49}F_{9}F_{48}$, $S_{4}S_{49}F_{9}F_{48}$, $S_{24}S_{49}F_{9}F_{29}$, $S_{22}S_{49}F_{9}F_{29}$, $S_{4}S_{49}F_{9}F_{29}$, $S_{2}S_{49}F_{9}F_{29}$, $S_{19}S_{24}F_{9}F_{48}$, $S_{4}S_{19}F_{9}F_{48}$, $S_{19}S_{24}F_{9}F_{29}$, $S_{19}S_{22}F_{9}F_{29}$, $S_{4}S_{19}F_{9}F_{29}$, $S_{2}S_{19}F_{9}F_{29}$, $S_{4}S_{19}F_{9}F_{12}$, $S_{4}S_{19}F_{3}F_{9}$, $S_{2}S_{19}F_{3}F_{9}$, $S_{23}S_{24}F_{5}F_{29}$, $S_{22}S_{23}F_{5}F_{29}$, $S_{4}S_{23}F_{5}F_{29}$, $S_{2}S_{23}F_{5}F_{29}$, $S_{3}S_{24}F_{5}F_{29}$, $S_{3}S_{22}F_{5}F_{29}$, $S_{3}S_{4}F_{5}F_{29}$, $S_{2}S_{3}F_{5}F_{29}$, $S_{3}S_{4}F_{3}F_{5}$, $S_{2}S_{3}F_{3}F_{5}$, $S_{24}S_{40}F_{9}F_{30}$, $S_{4}S_{40}F_{9}F_{30}$, $S_{24}S_{37}F_{9}F_{30}$, $S_{4}S_{37}F_{9}F_{30}$, $S_{22}S_{37}F_{9}F_{27}$, $S_{2}S_{37}F_{9}F_{27}$, $S_{13}S_{24}F_{9}F_{30}$, $S_{4}S_{13}F_{9}F_{30}$, $S_{4}S_{13}F_{4}F_{9}$, $S_{5}S_{24}F_{9}F_{30}$, $S_{4}S_{5}F_{9}F_{30}$, $S_{5}S_{22}F_{9}F_{27}$, $S_{2}S_{5}F_{9}F_{27}$, $S_{4}S_{5}F_{4}F_{9}$, $S_{2}S_{5}F_{1}F_{9}$, $S_{24}S_{37}F_{5}F_{30}$, $S_{4}S_{37}F_{5}F_{30}$, $S_{22}S_{37}F_{5}F_{27}$, $S_{2}S_{37}F_{5}F_{27}$, $S_{5}S_{24}F_{5}F_{30}$, $S_{4}S_{5}F_{5}F_{30}$, $S_{5}S_{22}F_{5}F_{27}$, $S_{2}S_{5}F_{5}F_{27}$, $S_{4}S_{5}F_{4}F_{5}$, $S_{2}S_{5}F_{1}F_{5}$, $S_{24}S_{49}F_{7}F_{48}$, $S_{4}S_{49}F_{7}F_{48}$, $S_{24}S_{49}F_{7}F_{29}$, $S_{22}S_{49}F_{7}F_{29}$, $S_{4}S_{49}F_{7}F_{29}$, $S_{2}S_{49}F_{7}F_{29}$, $S_{23}S_{24}F_{7}F_{29}$, $S_{22}S_{23}F_{7}F_{29}$, $S_{4}S_{23}F_{7}F_{29}$, $S_{2}S_{23}F_{7}F_{29}$, $S_{19}S_{24}F_{7}F_{48}$, $S_{4}S_{19}F_{7}F_{48}$, $S_{19}S_{24}F_{7}F_{29}$, $S_{19}S_{22}F_{7}F_{29}$, $S_{4}S_{19}F_{7}F_{29}$, $S_{2}S_{19}F_{7}F_{29}$, $S_{4}S_{19}F_{7}F_{12}$, $S_{4}S_{19}F_{3}F_{7}$, $S_{2}S_{19}F_{3}F_{7}$, $S_{3}S_{24}F_{7}F_{29}$, $S_{3}S_{22}F_{7}F_{29}$, $S_{3}S_{4}F_{7}F_{29}$, $S_{2}S_{3}F_{7}F_{29}$, $S_{3}S_{4}F_{3}F_{7}$, $S_{2}S_{3}F_{3}F_{7}$} \\
   \hline
  $\mathcal{O}_{11,(2,0,1)}^{2}$ & {$S_{24}S_{49}F_{9}F_{30}$, $S_{4}S_{49}F_{9}F_{30}$, $S_{19}S_{24}F_{9}F_{30}$, $S_{4}S_{19}F_{9}F_{30}$, $S_{4}S_{19}F_{4}F_{9}$, $S_{22}S_{23}F_{5}F_{27}$, $S_{2}S_{23}F_{5}F_{27}$, $S_{3}S_{22}F_{5}F_{27}$, $S_{2}S_{3}F_{5}F_{27}$, $S_{2}S_{3}F_{1}F_{5}$, $S_{24}S_{38}F_{7}F_{29}$, $S_{22}S_{38}F_{7}F_{29}$, $S_{4}S_{38}F_{7}F_{29}$, $S_{2}S_{38}F_{7}F_{29}$, $S_{6}S_{24}F_{7}F_{29}$, $S_{6}S_{22}F_{7}F_{29}$, $S_{4}S_{6}F_{7}F_{29}$, $S_{2}S_{6}F_{7}F_{29}$, $S_{4}S_{6}F_{3}F_{7}$, $S_{2}S_{6}F_{3}F_{7}$, $S_{24}S_{49}F_{7}F_{30}$, $S_{4}S_{49}F_{7}F_{30}$, $S_{22}S_{23}F_{7}F_{27}$, $S_{2}S_{23}F_{7}F_{27}$, $S_{19}S_{24}F_{7}F_{30}$, $S_{4}S_{19}F_{7}F_{30}$, $S_{4}S_{19}F_{4}F_{7}$, $S_{3}S_{22}F_{7}F_{27}$, $S_{2}S_{3}F_{7}F_{27}$, $S_{2}S_{3}F_{1}F_{7}$, $S_{23}S_{37}F_{9}F_{29}$, $S_{3}S_{37}F_{9}F_{29}$, $S_{5}S_{23}F_{9}F_{29}$, $S_{3}S_{5}F_{9}F_{29}$, $S_{3}S_{5}F_{3}F_{9}$, $S_{23}S_{37}F_{5}F_{29}$, $S_{3}S_{37}F_{5}F_{29}$, $S_{5}S_{23}F_{5}F_{29}$, $S_{3}S_{5}F_{5}F_{29}$, $S_{3}S_{5}F_{3}F_{5}$, $S_{24}S_{24}F_{9}F_{30}$, $S_{4}S_{24}F_{9}F_{30}$, $S_{4}S_{4}F_{9}F_{30}$, $S_{4}S_{4}F_{4}F_{9}$, $S_{22}S_{22}F_{5}F_{27}$, $S_{2}S_{22}F_{5}F_{27}$, $S_{2}S_{2}F_{5}F_{27}$, $S_{2}S_{2}F_{1}F_{5}$, $S_{24}S_{37}F_{7}F_{29}$, $S_{22}S_{37}F_{7}F_{29}$, $S_{4}S_{37}F_{7}F_{29}$, $S_{2}S_{37}F_{7}F_{29}$, $S_{5}S_{24}F_{7}F_{29}$, $S_{5}S_{22}F_{7}F_{29}$, $S_{4}S_{5}F_{7}F_{29}$, $S_{2}S_{5}F_{7}F_{29}$, $S_{4}S_{5}F_{3}F_{7}$, $S_{2}S_{5}F_{3}F_{7}$, $S_{24}S_{49}F_{9}F_{47}$, $S_{4}S_{49}F_{9}F_{47}$, $S_{19}S_{24}F_{9}F_{47}$, $S_{4}S_{19}F_{9}F_{47}$, $S_{4}S_{19}F_{9}F_{11}$, $S_{22}S_{23}F_{5}F_{28}$, $S_{2}S_{23}F_{5}F_{28}$, $S_{3}S_{22}F_{5}F_{28}$, $S_{2}S_{3}F_{5}F_{28}$, $S_{2}S_{3}F_{2}F_{5}$, $S_{24}S_{37}F_{8}F_{29}$, $S_{22}S_{37}F_{8}F_{29}$, $S_{4}S_{37}F_{8}F_{29}$, $S_{2}S_{37}F_{8}F_{29}$, $S_{5}S_{24}F_{8}F_{29}$, $S_{5}S_{22}F_{8}F_{29}$, $S_{4}S_{5}F_{8}F_{29}$, $S_{2}S_{5}F_{8}F_{29}$, $S_{4}S_{5}F_{3}F_{8}$, $S_{2}S_{5}F_{3}F_{8}$, $S_{24}S_{49}F_{7}F_{47}$, $S_{4}S_{49}F_{7}F_{47}$, $S_{22}S_{23}F_{7}F_{28}$, $S_{2}S_{23}F_{7}F_{28}$, $S_{19}S_{24}F_{7}F_{47}$, $S_{4}S_{19}F_{7}F_{47}$, $S_{4}S_{19}F_{7}F_{11}$, $S_{3}S_{22}F_{7}F_{28}$, $S_{2}S_{3}F_{7}F_{28}$, $S_{2}S_{3}F_{2}F_{7}$, $S_{24}S_{24}F_{8}F_{30}$, $S_{4}S_{24}F_{8}F_{30}$, $S_{22}S_{22}F_{8}F_{27}$, $S_{2}S_{22}F_{8}F_{27}$, $S_{4}S_{4}F_{8}F_{30}$, $S_{4}S_{4}F_{4}F_{8}$, $S_{2}S_{2}F_{8}F_{27}$, $S_{2}S_{2}F_{1}F_{8}$, $S_{22}S_{37}F_{9}F_{29}$, $S_{2}S_{37}F_{9}F_{29}$, $S_{5}S_{22}F_{9}F_{29}$, $S_{2}S_{5}F_{9}F_{29}$, $S_{2}S_{5}F_{3}F_{9}$, $S_{22}S_{37}F_{5}F_{29}$, $S_{2}S_{37}F_{5}F_{29}$, $S_{5}S_{22}F_{5}F_{29}$, $S_{2}S_{5}F_{5}F_{29}$, $S_{2}S_{5}F_{3}F_{5}$, $S_{24}S_{37}F_{7}F_{30}$, $S_{4}S_{37}F_{7}F_{30}$, $S_{22}S_{37}F_{7}F_{27}$, $S_{2}S_{37}F_{7}F_{27}$, $S_{5}S_{24}F_{7}F_{30}$, $S_{4}S_{5}F_{7}F_{30}$, $S_{5}S_{22}F_{7}F_{27}$, $S_{2}S_{5}F_{7}F_{27}$, $S_{4}S_{5}F_{4}F_{7}$, $S_{2}S_{5}F_{1}F_{7}$, $S_{23}S_{24}F_{9}F_{29}$, $S_{3}S_{24}F_{9}F_{29}$, $S_{4}S_{23}F_{9}F_{29}$, $S_{3}S_{4}F_{9}F_{29}$, $S_{3}S_{4}F_{3}F_{9}$, $S_{22}S_{23}F_{5}F_{29}$, $S_{3}S_{22}F_{5}F_{29}$, $S_{2}S_{23}F_{5}F_{29}$, $S_{2}S_{3}F_{5}F_{29}$, $S_{2}S_{3}F_{3}F_{5}$, $S_{24}S_{38}F_{7}F_{47}$, $S_{4}S_{38}F_{7}F_{47}$, $S_{22}S_{38}F_{7}F_{28}$, $S_{2}S_{38}F_{7}F_{28}$, $S_{6}S_{24}F_{7}F_{47}$, $S_{4}S_{6}F_{7}F_{47}$, $S_{6}S_{22}F_{7}F_{28}$, $S_{2}S_{6}F_{7}F_{28}$, $S_{4}S_{6}F_{7}F_{11}$, $S_{2}S_{6}F_{2}F_{7}$, $S_{24}S_{37}F_{8}F_{30}$, $S_{4}S_{37}F_{8}F_{30}$, $S_{22}S_{37}F_{8}F_{27}$, $S_{2}S_{37}F_{8}F_{27}$, $S_{5}S_{24}F_{8}F_{30}$, $S_{4}S_{5}F_{8}F_{30}$, $S_{5}S_{22}F_{8}F_{27}$, $S_{2}S_{5}F_{8}F_{27}$, $S_{4}S_{5}F_{4}F_{8}$, $S_{2}S_{5}F_{1}F_{8}$, $S_{22}S_{49}F_{9}F_{29}$, $S_{2}S_{49}F_{9}F_{29}$, $S_{19}S_{22}F_{9}F_{29}$, $S_{2}S_{19}F_{9}F_{29}$, $S_{2}S_{19}F_{3}F_{9}$, $S_{23}S_{37}F_{9}F_{27}$, $S_{3}S_{37}F_{9}F_{27}$, $S_{5}S_{23}F_{9}F_{27}$, $S_{3}S_{5}F_{9}F_{27}$, $S_{3}S_{5}F_{1}F_{9}$, $S_{23}S_{37}F_{5}F_{27}$, $S_{3}S_{37}F_{5}F_{27}$, $S_{5}S_{23}F_{5}F_{27}$, $S_{3}S_{5}F_{5}F_{27}$, $S_{3}S_{5}F_{1}F_{5}$, $S_{22}S_{37}F_{9}F_{27}$, $S_{2}S_{37}F_{9}F_{27}$, $S_{5}S_{22}F_{9}F_{27}$, $S_{2}S_{5}F_{9}F_{27}$, $S_{2}S_{5}F_{1}F_{9}$, $S_{22}S_{37}F_{5}F_{27}$, $S_{2}S_{37}F_{5}F_{27}$, $S_{5}S_{22}F_{5}F_{27}$, $S_{2}S_{5}F_{5}F_{27}$, $S_{2}S_{5}F_{1}F_{5}$, $S_{23}S_{24}F_{8}F_{29}$, $S_{3}S_{24}F_{8}F_{29}$, $S_{22}S_{23}F_{8}F_{29}$, $S_{3}S_{22}F_{8}F_{29}$, $S_{4}S_{23}F_{8}F_{29}$, $S_{3}S_{4}F_{8}F_{29}$, $S_{3}S_{4}F_{3}F_{8}$, $S_{2}S_{23}F_{8}F_{29}$, $S_{2}S_{3}F_{8}F_{29}$, $S_{2}S_{3}F_{3}F_{8}$, $S_{22}S_{49}F_{7}F_{29}$, $S_{2}S_{49}F_{7}F_{29}$, $S_{22}S_{23}F_{7}F_{29}$, $S_{2}S_{23}F_{7}F_{29}$, $S_{19}S_{22}F_{7}F_{29}$, $S_{2}S_{19}F_{7}F_{29}$, $S_{2}S_{19}F_{3}F_{7}$, $S_{3}S_{22}F_{7}F_{29}$, $S_{2}S_{3}F_{7}F_{29}$, $S_{2}S_{3}F_{3}F_{7}$} \\
   \hline
  $\mathcal{O}_{11,(2,0,1)}^{3}$ & {$S_{22}S_{23}F_{5}F_{28}$, $S_{3}S_{22}F_{5}F_{28}$, $S_{2}S_{23}F_{5}F_{28}$, $S_{2}S_{3}F_{5}F_{28}$, $S_{2}S_{3}F_{2}F_{5}$, $S_{23}S_{37}F_{7}F_{29}$, $S_{3}S_{37}F_{7}F_{29}$, $S_{5}S_{23}F_{7}F_{29}$, $S_{3}S_{5}F_{7}F_{29}$, $S_{3}S_{5}F_{3}F_{7}$, $S_{23}S_{23}F_{5}F_{27}$, $S_{3}S_{23}F_{5}F_{27}$, $S_{3}S_{3}F_{5}F_{27}$, $S_{3}S_{3}F_{1}F_{5}$, $S_{22}S_{23}F_{5}F_{27}$, $S_{2}S_{23}F_{5}F_{27}$, $S_{3}S_{22}F_{5}F_{27}$, $S_{2}S_{3}F_{5}F_{27}$, $S_{2}S_{3}F_{1}F_{5}$, $S_{23}S_{37}F_{8}F_{29}$, $S_{3}S_{37}F_{8}F_{29}$, $S_{5}S_{23}F_{8}F_{29}$, $S_{3}S_{5}F_{8}F_{29}$, $S_{3}S_{5}F_{3}F_{8}$, $S_{22}S_{38}F_{7}F_{29}$, $S_{2}S_{38}F_{7}F_{29}$, $S_{6}S_{22}F_{7}F_{29}$, $S_{2}S_{6}F_{7}F_{29}$, $S_{2}S_{6}F_{3}F_{7}$, $S_{23}S_{23}F_{7}F_{27}$, $S_{3}S_{23}F_{7}F_{27}$, $S_{3}S_{3}F_{7}F_{27}$, $S_{3}S_{3}F_{1}F_{7}$, $S_{22}S_{23}F_{8}F_{28}$, $S_{3}S_{22}F_{8}F_{28}$, $S_{2}S_{23}F_{8}F_{28}$, $S_{2}S_{3}F_{8}F_{28}$, $S_{2}S_{3}F_{2}F_{8}$, $S_{22}S_{23}F_{7}F_{27}$, $S_{2}S_{23}F_{7}F_{27}$, $S_{3}S_{22}F_{7}F_{27}$, $S_{2}S_{3}F_{7}F_{27}$, $S_{2}S_{3}F_{1}F_{7}$, $S_{22}S_{22}F_{5}F_{27}$, $S_{2}S_{22}F_{5}F_{27}$, $S_{2}S_{2}F_{5}F_{27}$, $S_{2}S_{2}F_{1}F_{5}$, $S_{23}S_{36}F_{8}F_{29}$, $S_{3}S_{36}F_{8}F_{29}$, $S_{11}S_{23}F_{8}F_{29}$, $S_{3}S_{11}F_{8}F_{29}$, $S_{3}S_{11}F_{3}F_{8}$, $S_{22}S_{37}F_{7}F_{29}$, $S_{2}S_{37}F_{7}F_{29}$, $S_{5}S_{22}F_{7}F_{29}$, $S_{2}S_{5}F_{7}F_{29}$, $S_{2}S_{5}F_{3}F_{7}$, $S_{21}S_{23}F_{5}F_{28}$, $S_{1}S_{23}F_{5}F_{28}$, $S_{3}S_{21}F_{5}F_{28}$, $S_{1}S_{3}F_{5}F_{28}$, $S_{1}S_{3}F_{2}F_{5}$, $S_{22}S_{37}F_{8}F_{29}$, $S_{2}S_{37}F_{8}F_{29}$, $S_{5}S_{22}F_{8}F_{29}$, $S_{2}S_{5}F_{8}F_{29}$, $S_{2}S_{5}F_{3}F_{8}$, $S_{21}S_{38}F_{7}F_{46}$, $S_{1}S_{38}F_{7}F_{46}$, $S_{6}S_{21}F_{7}F_{46}$, $S_{1}S_{6}F_{7}F_{46}$, $S_{1}S_{6}F_{7}F_{10}$, $S_{22}S_{23}F_{8}F_{27}$, $S_{3}S_{22}F_{8}F_{27}$, $S_{2}S_{23}F_{8}F_{27}$, $S_{2}S_{3}F_{8}F_{27}$, $S_{2}S_{3}F_{1}F_{8}$, $S_{22}S_{23}F_{7}F_{28}$, $S_{2}S_{23}F_{7}F_{28}$, $S_{3}S_{22}F_{7}F_{28}$, $S_{2}S_{3}F_{7}F_{28}$, $S_{2}S_{3}F_{2}F_{7}$, $S_{22}S_{22}F_{8}F_{27}$, $S_{2}S_{22}F_{8}F_{27}$, $S_{2}S_{2}F_{8}F_{27}$, $S_{2}S_{2}F_{1}F_{8}$, $S_{21}S_{23}F_{7}F_{28}$, $S_{1}S_{23}F_{7}F_{28}$, $S_{3}S_{21}F_{7}F_{28}$, $S_{1}S_{3}F_{7}F_{28}$, $S_{1}S_{3}F_{2}F_{7}$, $S_{23}S_{37}F_{7}F_{27}$, $S_{3}S_{37}F_{7}F_{27}$, $S_{5}S_{23}F_{7}F_{27}$, $S_{3}S_{5}F_{7}F_{27}$, $S_{3}S_{5}F_{1}F_{7}$, $S_{23}S_{36}F_{8}F_{28}$, $S_{3}S_{36}F_{8}F_{28}$, $S_{11}S_{23}F_{8}F_{28}$, $S_{3}S_{11}F_{8}F_{28}$, $S_{3}S_{11}F_{2}F_{8}$, $S_{22}S_{37}F_{7}F_{27}$, $S_{2}S_{37}F_{7}F_{27}$, $S_{5}S_{22}F_{7}F_{27}$, $S_{2}S_{5}F_{7}F_{27}$, $S_{2}S_{5}F_{1}F_{7}$, $S_{23}S_{37}F_{8}F_{27}$, $S_{3}S_{37}F_{8}F_{27}$, $S_{5}S_{23}F_{8}F_{27}$, $S_{3}S_{5}F_{8}F_{27}$, $S_{3}S_{5}F_{1}F_{8}$, $S_{22}S_{38}F_{7}F_{28}$, $S_{2}S_{38}F_{7}F_{28}$, $S_{6}S_{22}F_{7}F_{28}$, $S_{2}S_{6}F_{7}F_{28}$, $S_{2}S_{6}F_{2}F_{7}$, $S_{22}S_{37}F_{8}F_{27}$, $S_{2}S_{37}F_{8}F_{27}$, $S_{5}S_{22}F_{8}F_{27}$, $S_{2}S_{5}F_{8}F_{27}$, $S_{2}S_{5}F_{1}F_{8}$, $S_{21}S_{38}F_{7}F_{28}$, $S_{1}S_{38}F_{7}F_{28}$, $S_{6}S_{21}F_{7}F_{28}$, $S_{1}S_{6}F_{7}F_{28}$, $S_{1}S_{6}F_{2}F_{7}$} \\
   \hline
  $\mathcal{O}_{11,(2,0,1)}^{4}$ & {$S_{24}S_{24}F_{14}F_{30}$, $S_{4}S_{24}F_{14}F_{30}$, $S_{4}S_{4}F_{14}F_{30}$, $S_{4}S_{4}F_{4}F_{14}$, $S_{22}S_{22}F_{6}F_{27}$, $S_{2}S_{22}F_{6}F_{27}$, $S_{2}S_{2}F_{6}F_{27}$, $S_{2}S_{2}F_{1}F_{6}$, $S_{24}S_{37}F_{8}F_{29}$, $S_{22}S_{37}F_{8}F_{29}$, $S_{4}S_{37}F_{8}F_{29}$, $S_{2}S_{37}F_{8}F_{29}$, $S_{5}S_{24}F_{8}F_{29}$, $S_{5}S_{22}F_{8}F_{29}$, $S_{4}S_{5}F_{8}F_{29}$, $S_{2}S_{5}F_{8}F_{29}$, $S_{4}S_{5}F_{3}F_{8}$, $S_{2}S_{5}F_{3}F_{8}$, $S_{24}S_{24}F_{8}F_{30}$, $S_{4}S_{24}F_{8}F_{30}$, $S_{22}S_{22}F_{8}F_{27}$, $S_{2}S_{22}F_{8}F_{27}$, $S_{4}S_{4}F_{8}F_{30}$, $S_{4}S_{4}F_{4}F_{8}$, $S_{2}S_{2}F_{8}F_{27}$, $S_{2}S_{2}F_{1}F_{8}$, $S_{23}S_{36}F_{14}F_{29}$, $S_{3}S_{36}F_{14}F_{29}$, $S_{11}S_{23}F_{14}F_{29}$, $S_{3}S_{11}F_{14}F_{29}$, $S_{3}S_{11}F_{3}F_{14}$, $S_{23}S_{36}F_{6}F_{29}$, $S_{3}S_{36}F_{6}F_{29}$, $S_{11}S_{23}F_{6}F_{29}$, $S_{3}S_{11}F_{6}F_{29}$, $S_{3}S_{11}F_{3}F_{6}$, $S_{24}S_{37}F_{8}F_{30}$, $S_{4}S_{37}F_{8}F_{30}$, $S_{22}S_{37}F_{8}F_{27}$, $S_{2}S_{37}F_{8}F_{27}$, $S_{5}S_{24}F_{8}F_{30}$, $S_{4}S_{5}F_{8}F_{30}$, $S_{5}S_{22}F_{8}F_{27}$, $S_{2}S_{5}F_{8}F_{27}$, $S_{4}S_{5}F_{4}F_{8}$, $S_{2}S_{5}F_{1}F_{8}$, $S_{23}S_{24}F_{14}F_{29}$, $S_{3}S_{24}F_{14}F_{29}$, $S_{4}S_{23}F_{14}F_{29}$, $S_{3}S_{4}F_{14}F_{29}$, $S_{3}S_{4}F_{3}F_{14}$, $S_{22}S_{23}F_{6}F_{29}$, $S_{3}S_{22}F_{6}F_{29}$, $S_{2}S_{23}F_{6}F_{29}$, $S_{2}S_{3}F_{6}F_{29}$, $S_{2}S_{3}F_{3}F_{6}$, $S_{23}S_{36}F_{14}F_{28}$, $S_{3}S_{36}F_{14}F_{28}$, $S_{11}S_{23}F_{14}F_{28}$, $S_{3}S_{11}F_{14}F_{28}$, $S_{3}S_{11}F_{2}F_{14}$, $S_{23}S_{36}F_{6}F_{28}$, $S_{3}S_{36}F_{6}F_{28}$, $S_{11}S_{23}F_{6}F_{28}$, $S_{3}S_{11}F_{6}F_{28}$, $S_{3}S_{11}F_{2}F_{6}$, $S_{23}S_{24}F_{8}F_{29}$, $S_{3}S_{24}F_{8}F_{29}$, $S_{22}S_{23}F_{8}F_{29}$, $S_{3}S_{22}F_{8}F_{29}$, $S_{4}S_{23}F_{8}F_{29}$, $S_{3}S_{4}F_{8}F_{29}$, $S_{3}S_{4}F_{3}F_{8}$, $S_{2}S_{23}F_{8}F_{29}$, $S_{2}S_{3}F_{8}F_{29}$, $S_{2}S_{3}F_{3}F_{8}$} \\
   \hline
\end{longtable}
\end{center}

\begin{center}
\begin{longtable}{| c | m{13cm} |}
   \hline
   $\mathcal{O}^j(\mathcal{T}_{10}^{7})$ & New Particles \\
   \hline
   \hline
   \endhead
   \multicolumn{2}{|c|}{\hspace{9cm} Continued on next page} \\
   \endfoot
  $\mathcal{O}_{11,(2,0,1)}^{5}$ & {$S_{21}S_{23}S_{23}F_{6}$, $S_{3}S_{3}S_{21}F_{6}$, $S_{1}S_{3}S_{23}F_{6}$, $S_{1}S_{3}S_{3}F_{6}$, $S_{22}S_{22}S_{23}F_{6}$, $S_{2}S_{3}S_{22}F_{6}$, $S_{2}S_{2}S_{23}F_{6}$, $S_{2}S_{2}S_{3}F_{6}$, $S_{22}S_{22}S_{23}F_{8}$, $S_{2}S_{3}S_{22}F_{8}$, $S_{2}S_{2}S_{23}F_{8}$, $S_{2}S_{2}S_{3}F_{8}$, $S_{21}S_{23}S_{23}F_{13}$, $S_{3}S_{3}S_{21}F_{13}$, $S_{1}S_{3}S_{23}F_{13}$, $S_{1}S_{3}S_{3}F_{13}$} \\
   \hline
   \caption{Dimension 11 UV completions for topology $\mathcal{T}_{10}^{7}$ (see Fig.~\ref{fig:d=10_topologies}).}
   \label{tab:T11Top7B2L0D1}\\
   \endlastfoot
  $\mathcal{O}_{11,(2,0,1)}^{1}$ & {$S_{24}S_{24}S_{49}F_{9}$, $S_{22}S_{24}S_{49}F_{9}$, $S_{4}S_{4}S_{49}F_{9}$, $S_{2}S_{4}S_{49}F_{9}$, $S_{4}S_{19}S_{24}F_{9}$, $S_{2}S_{19}S_{24}F_{9}$, $S_{4}S_{19}S_{22}F_{9}$, $S_{4}S_{4}S_{19}F_{9}$, $S_{2}S_{4}S_{19}F_{9}$, $S_{23}S_{24}S_{24}F_{5}$, $S_{22}S_{22}S_{23}F_{5}$, $S_{4}S_{4}S_{23}F_{5}$, $S_{2}S_{2}S_{23}F_{5}$, $S_{3}S_{4}S_{24}F_{5}$, $S_{2}S_{3}S_{22}F_{5}$, $S_{3}S_{4}S_{4}F_{5}$, $S_{2}S_{2}S_{3}F_{5}$, $S_{24}S_{24}S_{49}F_{7}$, $S_{22}S_{24}S_{49}F_{7}$, $S_{4}S_{4}S_{49}F_{7}$, $S_{2}S_{4}S_{49}F_{7}$, $S_{23}S_{24}S_{24}F_{7}$, $S_{22}S_{22}S_{23}F_{7}$, $S_{4}S_{4}S_{23}F_{7}$, $S_{2}S_{2}S_{23}F_{7}$, $S_{4}S_{19}S_{24}F_{7}$, $S_{2}S_{19}S_{24}F_{7}$, $S_{4}S_{19}S_{22}F_{7}$, $S_{4}S_{4}S_{19}F_{7}$, $S_{2}S_{4}S_{19}F_{7}$, $S_{3}S_{4}S_{24}F_{7}$, $S_{2}S_{3}S_{22}F_{7}$, $S_{3}S_{4}S_{4}F_{7}$, $S_{2}S_{2}S_{3}F_{7}$} \\
   \hline
  $\mathcal{O}_{11,(2,0,1)}^{2}$ & {$S_{23}S_{24}S_{24}F_{9}$, $S_{3}S_{4}S_{24}F_{9}$, $S_{4}S_{4}S_{23}F_{9}$, $S_{3}S_{4}S_{4}F_{9}$, $S_{22}S_{22}S_{23}F_{5}$, $S_{2}S_{3}S_{22}F_{5}$, $S_{2}S_{2}S_{23}F_{5}$, $S_{2}S_{2}S_{3}F_{5}$, $S_{22}S_{24}S_{49}F_{9}$, $S_{2}S_{4}S_{49}F_{9}$, $S_{4}S_{19}S_{22}F_{9}$, $S_{2}S_{19}S_{24}F_{9}$, $S_{2}S_{4}S_{19}F_{9}$, $S_{22}S_{24}S_{49}F_{7}$, $S_{2}S_{4}S_{49}F_{7}$, $S_{22}S_{22}S_{23}F_{7}$, $S_{2}S_{2}S_{23}F_{7}$, $S_{4}S_{19}S_{22}F_{7}$, $S_{2}S_{19}S_{24}F_{7}$, $S_{2}S_{4}S_{19}F_{7}$, $S_{2}S_{3}S_{22}F_{7}$, $S_{2}S_{2}S_{3}F_{7}$, $S_{23}S_{24}S_{24}F_{8}$, $S_{3}S_{4}S_{24}F_{8}$, $S_{22}S_{22}S_{23}F_{8}$, $S_{2}S_{3}S_{22}F_{8}$, $S_{4}S_{4}S_{23}F_{8}$, $S_{3}S_{4}S_{4}F_{8}$, $S_{2}S_{2}S_{23}F_{8}$, $S_{2}S_{2}S_{3}F_{8}$} \\
   \hline
  $\mathcal{O}_{11,(2,0,1)}^{3}$ & {$S_{22}S_{22}S_{23}F_{5}$, $S_{2}S_{3}S_{22}F_{5}$, $S_{2}S_{2}S_{23}F_{5}$, $S_{2}S_{2}S_{3}F_{5}$, $S_{21}S_{23}S_{23}F_{5}$, $S_{1}S_{3}S_{23}F_{5}$, $S_{3}S_{3}S_{21}F_{5}$, $S_{1}S_{3}S_{3}F_{5}$, $S_{21}S_{23}S_{23}F_{7}$, $S_{1}S_{3}S_{23}F_{7}$, $S_{3}S_{3}S_{21}F_{7}$, $S_{1}S_{3}S_{3}F_{7}$, $S_{22}S_{22}S_{23}F_{7}$, $S_{2}S_{2}S_{23}F_{7}$, $S_{2}S_{3}S_{22}F_{7}$, $S_{2}S_{2}S_{3}F_{7}$, $S_{22}S_{22}S_{23}F_{8}$, $S_{2}S_{3}S_{22}F_{8}$, $S_{2}S_{2}S_{23}F_{8}$, $S_{2}S_{2}S_{3}F_{8}$} \\
   \hline
  $\mathcal{O}_{11,(2,0,1)}^{4}$ & {$S_{23}S_{24}S_{24}F_{14}$, $S_{3}S_{4}S_{24}F_{14}$, $S_{4}S_{4}S_{23}F_{14}$, $S_{3}S_{4}S_{4}F_{14}$, $S_{22}S_{22}S_{23}F_{6}$, $S_{2}S_{3}S_{22}F_{6}$, $S_{2}S_{2}S_{23}F_{6}$, $S_{2}S_{2}S_{3}F_{6}$, $S_{23}S_{24}S_{24}F_{8}$, $S_{3}S_{4}S_{24}F_{8}$, $S_{22}S_{22}S_{23}F_{8}$, $S_{2}S_{3}S_{22}F_{8}$, $S_{4}S_{4}S_{23}F_{8}$, $S_{3}S_{4}S_{4}F_{8}$, $S_{2}S_{2}S_{23}F_{8}$, $S_{2}S_{2}S_{3}F_{8}$} \\
   \hline
\end{longtable}
\end{center}

 \section{Dimension 12}
 \label{sec:d=12}

At $d=12$, derivative BNV operators carry either $(\Delta B, \Delta L ) = (1,1)$ or $( 1, -3)$, split into subsections below.
 
  \subsection{\texorpdfstring{$\Delta B=1, \Delta L=1$}{Delta B=1,Delta L=1}}

For $(\Delta B, \Delta L ) = (1,1)$, we have one-derivative operators,
   \begin{multicols}{2}
 \begin{spacing}{1.5}
 \noindent
 \begin{math}
\mathcal{O}_{12, \, (1,1,1)}^{1}\equiv DHHddd\bar{d}QL\,,\\
\mathcal{O}_{12, \, (1,1,1)}^{2}\equiv DHHdd\bar{d}QQe\,,\\
\mathcal{O}_{12, \, (1,1,1)}^{3}\equiv DHHddd\bar{d}ue\,,\\
\mathcal{O}_{12, \, (1,1,1)}^{4}\equiv DHHdddLL\bar{e}\,,\\
\mathcal{O}_{12, \, (1,1,1)}^{5}\equiv DHHddQLe\bar{e}\,,\\
\mathcal{O}_{12, \, (1,1,1)}^{6}\equiv DHHdQQee\bar{e}\,,\\
\mathcal{O}_{12, \, (1,1,1)}^{7}\equiv DHHdduee\bar{e}\,,\\
\mathcal{O}_{12, \, (1,1,1)}^{8}\equiv DH\bar{H}d\bar{d}QQQL\,,\\
\mathcal{O}_{12, \, (1,1,1)}^{9}\equiv DH\bar{H}\bar{d}QQQQe\,,\\
\mathcal{O}_{12, \, (1,1,1)}^{10}\equiv DH\bar{H}dd\bar{d}uQL\,,\\
\mathcal{O}_{12, \, (1,1,1)}^{11}\equiv DH\bar{H}d\bar{d}uQQe\,,\\
\mathcal{O}_{12, \, (1,1,1)}^{12}\equiv DH\bar{H}dd\bar{d}uue\,,\\
\mathcal{O}_{12, \, (1,1,1)}^{13}\equiv DH\bar{H}dQQLL\bar{e}\,,\\
\mathcal{O}_{12, \, (1,1,1)}^{14}\equiv DH\bar{H}QQQLe\bar{e}\,,\\
\mathcal{O}_{12, \, (1,1,1)}^{15}\equiv DH\bar{H}dduLL\bar{e}\,,\\
\mathcal{O}_{12, \, (1,1,1)}^{16}\equiv DH\bar{H}duQLe\bar{e}\,,\\
\mathcal{O}_{12, \, (1,1,1)}^{17}\equiv DH\bar{H}uQQee\bar{e}\,,\\
\mathcal{O}_{12, \, (1,1,1)}^{18}\equiv DH\bar{H}duuee\bar{e}\,,\\
\mathcal{O}_{12, \, (1,1,1)}^{19}\equiv D\bar{H}\bar{H}\bar{d}uQQQL\,,\\
\mathcal{O}_{12, \, (1,1,1)}^{20}\equiv D\bar{H}\bar{H}d\bar{d}uuQL\,,\\
\mathcal{O}_{12, \, (1,1,1)}^{21}\equiv D\bar{H}\bar{H}\bar{d}uuQQe\,,\\
\mathcal{O}_{12, \, (1,1,1)}^{22}\equiv D\bar{H}\bar{H}d\bar{d}uuue\,,\\
\mathcal{O}_{12, \, (1,1,1)}^{23}\equiv D\bar{H}\bar{H}uQQLL\bar{e}\,,\\
\mathcal{O}_{12, \, (1,1,1)}^{24}\equiv D\bar{H}\bar{H}duuLL\bar{e}\,,\\
\mathcal{O}_{12, \, (1,1,1)}^{25}\equiv D\bar{H}\bar{H}uuQLe\bar{e}\,,\\
\mathcal{O}_{12, \, (1,1,1)}^{26}\equiv D\bar{H}\bar{H}uuuee\bar{e}\,,\\
\mathcal{O}_{12, \, (1,1,1)}^{27}\equiv D\bar{H}\bar{H}uuQLL\bar{L}\,,\\
\mathcal{O}_{12, \, (1,1,1)}^{28}\equiv D\bar{H}\bar{H}uuuL\bar{L}e\,,\\
\mathcal{O}_{12, \, (1,1,1)}^{29}\equiv D\bar{H}\bar{H}uuQQ\bar{Q}L\,,\\
\mathcal{O}_{12, \, (1,1,1)}^{30}\equiv D\bar{H}\bar{H}duuu\bar{Q}L\,,\\
\mathcal{O}_{12, \, (1,1,1)}^{31}\equiv D\bar{H}\bar{H}uuuQ\bar{Q}e\,,\\
\mathcal{O}_{12, \, (1,1,1)}^{32}\equiv D\bar{H}\bar{H}uuu\bar{u}QL\,,\\
\mathcal{O}_{12, \, (1,1,1)}^{33}\equiv D\bar{H}\bar{H}uuuu\bar{u}e\,,\\
\mathcal{O}_{12, \, (1,1,1)}^{34}\equiv DH\bar{H}QQQLL\bar{L}\,,\\
\mathcal{O}_{12, \, (1,1,1)}^{35}\equiv DH\bar{H}duQLL\bar{L}\,,\\
\mathcal{O}_{12, \, (1,1,1)}^{36}\equiv DH\bar{H}uQQL\bar{L}e\,,\\
\mathcal{O}_{12, \, (1,1,1)}^{37}\equiv DH\bar{H}duuL\bar{L}e\,,\\
\mathcal{O}_{12, \, (1,1,1)}^{38}\equiv DH\bar{H}uuQ\bar{L}ee\,,\\
\mathcal{O}_{12, \, (1,1,1)}^{39}\equiv DH\bar{H}QQQQ\bar{Q}L\,,\\
\mathcal{O}_{12, \, (1,1,1)}^{40}\equiv DH\bar{H}duQQ\bar{Q}L\,,\\
\mathcal{O}_{12, \, (1,1,1)}^{41}\equiv DH\bar{H}uQQQ\bar{Q}e\,,\\
\mathcal{O}_{12, \, (1,1,1)}^{42}\equiv DH\bar{H}dduu\bar{Q}L\,,\\
\mathcal{O}_{12, \, (1,1,1)}^{43}\equiv DH\bar{H}duuQ\bar{Q}e\,,\\
\mathcal{O}_{12, \, (1,1,1)}^{44}\equiv DH\bar{H}u\bar{u}QQQL\,,\\
\mathcal{O}_{12, \, (1,1,1)}^{45}\equiv DH\bar{H}duu\bar{u}QL\,,\\
\mathcal{O}_{12, \, (1,1,1)}^{46}\equiv DH\bar{H}uu\bar{u}QQe\,,\\
\mathcal{O}_{12, \, (1,1,1)}^{47}\equiv DH\bar{H}duuu\bar{u}e\,,\\
\mathcal{O}_{12, \, (1,1,1)}^{48}\equiv DHHddQLL\bar{L}\,,\\
\mathcal{O}_{12, \, (1,1,1)}^{49}\equiv DHHdQQL\bar{L}e\,,\\
\mathcal{O}_{12, \, (1,1,1)}^{50}\equiv DHHQQQ\bar{L}ee\,,\\
\mathcal{O}_{12, \, (1,1,1)}^{51}\equiv DHHdduL\bar{L}e\,,\\
\mathcal{O}_{12, \, (1,1,1)}^{52}\equiv DHHduQ\bar{L}ee\,,\\
\mathcal{O}_{12, \, (1,1,1)}^{53}\equiv DHHddQQ\bar{Q}L\,,\\
\mathcal{O}_{12, \, (1,1,1)}^{54}\equiv DHHdQQQ\bar{Q}e\,,\\
\mathcal{O}_{12, \, (1,1,1)}^{55}\equiv DHHdddu\bar{Q}L\,,\\
\mathcal{O}_{12, \, (1,1,1)}^{56}\equiv DHHdduQ\bar{Q}e\,,\\
\mathcal{O}_{12, \, (1,1,1)}^{57}\equiv DHHd\bar{u}QQQL\,,\\
\mathcal{O}_{12, \, (1,1,1)}^{58}\equiv DHH\bar{u}QQQQe\,,\\
\mathcal{O}_{12, \, (1,1,1)}^{59}\equiv DHHddu\bar{u}QL\,,\\
\mathcal{O}_{12, \, (1,1,1)}^{60}\equiv DHHdu\bar{u}QQe\,,\\
\mathcal{O}_{12, \, (1,1,1)}^{61}\equiv DHHdduu\bar{u}e\,,\\
\mathcal{O}_{12, \, (1,1,1)}^{62}\equiv DHHdddd\bar{d}\nu\,,\\
\mathcal{O}_{12, \, (1,1,1)}^{63}\equiv DHHddde\bar{e}\nu\,,\\
\mathcal{O}_{12, \, (1,1,1)}^{64}\equiv DH\bar{H}dd\bar{d}QQ\nu\,,\\
\mathcal{O}_{12, \, (1,1,1)}^{65}\equiv DH\bar{H}ddd\bar{d}u\nu\,,\\
\mathcal{O}_{12, \, (1,1,1)}^{66}\equiv DH\bar{H}ddQL\bar{e}\nu\,,\\
\mathcal{O}_{12, \, (1,1,1)}^{67}\equiv DH\bar{H}dQQe\bar{e}\nu\,,\\
\mathcal{O}_{12, \, (1,1,1)}^{68}\equiv DH\bar{H}ddue\bar{e}\nu\,,\\
\mathcal{O}_{12, \, (1,1,1)}^{69}\equiv D\bar{H}\bar{H}\bar{d}QQQQ\nu\,,\\
\mathcal{O}_{12, \, (1,1,1)}^{70}\equiv D\bar{H}\bar{H}d\bar{d}uQQ\nu\,,\\
\mathcal{O}_{12, \, (1,1,1)}^{71}\equiv D\bar{H}\bar{H}dd\bar{d}uu\nu\,,\\
\mathcal{O}_{12, \, (1,1,1)}^{72}\equiv D\bar{H}\bar{H}QQQL\bar{e}\nu\,,\\
\mathcal{O}_{12, \, (1,1,1)}^{73}\equiv D\bar{H}\bar{H}duQL\bar{e}\nu\,,\\
\mathcal{O}_{12, \, (1,1,1)}^{74}\equiv D\bar{H}\bar{H}uQQe\bar{e}\nu\,,\\
\mathcal{O}_{12, \, (1,1,1)}^{75}\equiv D\bar{H}\bar{H}duue\bar{e}\nu\,,\\
\mathcal{O}_{12, \, (1,1,1)}^{76}\equiv D\bar{H}\bar{H}uQQL\bar{L}\nu\,,\\
\mathcal{O}_{12, \, (1,1,1)}^{77}\equiv D\bar{H}\bar{H}duuL\bar{L}\nu\,,\\
\mathcal{O}_{12, \, (1,1,1)}^{78}\equiv D\bar{H}\bar{H}uuQ\bar{L}e\nu\,,\\
\mathcal{O}_{12, \, (1,1,1)}^{79}\equiv D\bar{H}\bar{H}uQQQ\bar{Q}\nu\,,\\
\mathcal{O}_{12, \, (1,1,1)}^{80}\equiv D\bar{H}\bar{H}duuQ\bar{Q}\nu\,,\\
\mathcal{O}_{12, \, (1,1,1)}^{81}\equiv D\bar{H}\bar{H}uu\bar{u}QQ\nu\,,\\
\mathcal{O}_{12, \, (1,1,1)}^{82}\equiv D\bar{H}\bar{H}duuu\bar{u}\nu\,,\\
\mathcal{O}_{12, \, (1,1,1)}^{83}\equiv DH\bar{H}dQQL\bar{L}\nu\,,\\
\mathcal{O}_{12, \, (1,1,1)}^{84}\equiv DH\bar{H}QQQ\bar{L}e\nu\,,\\
\mathcal{O}_{12, \, (1,1,1)}^{85}\equiv DH\bar{H}dduL\bar{L}\nu\,,\\
\mathcal{O}_{12, \, (1,1,1)}^{86}\equiv DH\bar{H}duQ\bar{L}e\nu\,,\\
\mathcal{O}_{12, \, (1,1,1)}^{87}\equiv DH\bar{H}dQQQ\bar{Q}\nu\,,\\
\mathcal{O}_{12, \, (1,1,1)}^{88}\equiv DH\bar{H}dduQ\bar{Q}\nu\,,\\
\mathcal{O}_{12, \, (1,1,1)}^{89}\equiv DH\bar{H}\bar{u}QQQQ\nu\,,\\
\mathcal{O}_{12, \, (1,1,1)}^{90}\equiv DH\bar{H}du\bar{u}QQ\nu\,,\\
\mathcal{O}_{12, \, (1,1,1)}^{91}\equiv DH\bar{H}dduu\bar{u}\nu\,,\\
\mathcal{O}_{12, \, (1,1,1)}^{92}\equiv DHHdddL\bar{L}\nu\,,\\
\mathcal{O}_{12, \, (1,1,1)}^{93}\equiv DHHddQ\bar{L}e\nu\,,\\
\mathcal{O}_{12, \, (1,1,1)}^{94}\equiv D\bar{H}\bar{H}uuuLL\bar{\nu}\,,\\
\mathcal{O}_{12, \, (1,1,1)}^{95}\equiv DH\bar{H}uQQLL\bar{\nu}\,,\\
\mathcal{O}_{12, \, (1,1,1)}^{96}\equiv DH\bar{H}duuLL\bar{\nu}\,,\\
\mathcal{O}_{12, \, (1,1,1)}^{97}\equiv DH\bar{H}uuQLe\bar{\nu}\,,\\
\mathcal{O}_{12, \, (1,1,1)}^{98}\equiv DH\bar{H}uuuee\bar{\nu}\,,\\
\mathcal{O}_{12, \, (1,1,1)}^{99}\equiv DHHdQQLL\bar{\nu}\,,\\
\mathcal{O}_{12, \, (1,1,1)}^{100}\equiv DHHQQQLe\bar{\nu}\,,\\
\mathcal{O}_{12, \, (1,1,1)}^{101}\equiv DHHdduLL\bar{\nu}\,,\\
\mathcal{O}_{12, \, (1,1,1)}^{102}\equiv DHHduQLe\bar{\nu}\,,\\
\mathcal{O}_{12, \, (1,1,1)}^{103}\equiv DHHuQQee\bar{\nu}\,,\\
\mathcal{O}_{12, \, (1,1,1)}^{104}\equiv DHHduuee\bar{\nu}\,,\\
\mathcal{O}_{12, \, (1,1,1)}^{105}\equiv DHHdddQ\bar{Q}\nu\,,\\
\mathcal{O}_{12, \, (1,1,1)}^{106}\equiv DHHdd\bar{u}QQ\nu\,,\\
\mathcal{O}_{12, \, (1,1,1)}^{107}\equiv DHHdddu\bar{u}\nu\,,\\
\mathcal{O}_{12, \, (1,1,1)}^{108}\equiv DH\bar{H}ddd\bar{e}\nu\nu\,,\\
\mathcal{O}_{12, \, (1,1,1)}^{109}\equiv D\bar{H}\bar{H}dQQ\bar{e}\nu\nu\,,\\
\mathcal{O}_{12, \, (1,1,1)}^{110}\equiv D\bar{H}\bar{H}ddu\bar{e}\nu\nu\,,\\
\mathcal{O}_{12, \, (1,1,1)}^{111}\equiv D\bar{H}\bar{H}QQQ\bar{L}\nu\nu\,,\\
\mathcal{O}_{12, \, (1,1,1)}^{112}\equiv D\bar{H}\bar{H}duQ\bar{L}\nu\nu\,,\\
\mathcal{O}_{12, \, (1,1,1)}^{113}\equiv DH\bar{H}ddQ\bar{L}\nu\nu\,,\\
\mathcal{O}_{12, \, (1,1,1)}^{114}\equiv D\bar{H}\bar{H}uuQL\nu\bar{\nu}\,,\\
\mathcal{O}_{12, \, (1,1,1)}^{115}\equiv D\bar{H}\bar{H}uuue\nu\bar{\nu}\,,\\
\mathcal{O}_{12, \, (1,1,1)}^{116}\equiv DH\bar{H}QQQL\nu\bar{\nu}\,,\\
\mathcal{O}_{12, \, (1,1,1)}^{117}\equiv DH\bar{H}duQL\nu\bar{\nu}\,,\\
\mathcal{O}_{12, \, (1,1,1)}^{118}\equiv DH\bar{H}uQQe\nu\bar{\nu}\,,\\
\mathcal{O}_{12, \, (1,1,1)}^{119}\equiv DH\bar{H}duue\nu\bar{\nu}\,,\\
\mathcal{O}_{12, \, (1,1,1)}^{120}\equiv DHHddQL\nu\bar{\nu}\,,\\
\mathcal{O}_{12, \, (1,1,1)}^{121}\equiv DHHdQQe\nu\bar{\nu}\,,\\
\mathcal{O}_{12, \, (1,1,1)}^{122}\equiv DHHddue\nu\bar{\nu}\,,\\
\mathcal{O}_{12, \, (1,1,1)}^{123}\equiv D\bar{H}\bar{H}uQQ\nu\nu\bar{\nu}\,,\\
\mathcal{O}_{12, \, (1,1,1)}^{124}\equiv D\bar{H}\bar{H}duu\nu\nu\bar{\nu}\,,\\
\mathcal{O}_{12, \, (1,1,1)}^{125}\equiv DH\bar{H}dQQ\nu\nu\bar{\nu}\,,\\
\mathcal{O}_{12, \, (1,1,1)}^{126}\equiv DH\bar{H}ddu\nu\nu\bar{\nu}\,,\\
\mathcal{O}_{12, \, (1,1,1)}^{127}\equiv DHHddd\nu\nu\bar{\nu}\,,\\
\mathcal{O}_{12, \, (1,1,1)}^{128}\equiv DH\bar{H}\bar{H}\bar{H}\bar{H}uuuL\,,\\
\mathcal{O}_{12, \, (1,1,1)}^{129}\equiv DHH\bar{H}\bar{H}\bar{H}uQQL\,,\\
\mathcal{O}_{12, \, (1,1,1)}^{130}\equiv DHH\bar{H}\bar{H}\bar{H}duuL\,,\\
\mathcal{O}_{12, \, (1,1,1)}^{131}\equiv DHH\bar{H}\bar{H}\bar{H}uuQe\,,\\
\mathcal{O}_{12, \, (1,1,1)}^{132}\equiv DHHH\bar{H}\bar{H}dQQL\,,\\
\mathcal{O}_{12, \, (1,1,1)}^{133}\equiv DHHH\bar{H}\bar{H}QQQe\,,\\
\mathcal{O}_{12, \, (1,1,1)}^{134}\equiv DHHH\bar{H}\bar{H}dduL\,,\\
\mathcal{O}_{12, \, (1,1,1)}^{135}\equiv DHHH\bar{H}\bar{H}duQe\,,\\
\mathcal{O}_{12, \, (1,1,1)}^{136}\equiv DHHHH\bar{H}dddL\,,\\
\mathcal{O}_{12, \, (1,1,1)}^{137}\equiv DHHHH\bar{H}ddQe\,,\\
\mathcal{O}_{12, \, (1,1,1)}^{138}\equiv DH\bar{H}\bar{H}\bar{H}\bar{H}uuQ\nu\,,\\
\mathcal{O}_{12, \, (1,1,1)}^{139}\equiv DHH\bar{H}\bar{H}\bar{H}QQQ\nu\,,\\
\mathcal{O}_{12, \, (1,1,1)}^{140}\equiv DHH\bar{H}\bar{H}\bar{H}duQ\nu\,,\\
\mathcal{O}_{12, \, (1,1,1)}^{141}\equiv DHHH\bar{H}\bar{H}ddQ\nu\,,
\label{eq:1Ddequal12basisB1L1}
 \end{math}
 \end{spacing}
 \end{multicols}
\noindent
two-derivative operators,
\begin{multicols}{2}
 \begin{spacing}{1.5}
 \noindent
 \begin{math}
\mathcal{O}_{12, \,(1 , 1 ,2)}^{1} \equiv DDHdd\bar{d}QQL \,, \\
 \mathcal{O}_{12, \,(1 , 1 ,2)}^{2} \equiv DDHd\bar{d}QQQe \,, \\
 \mathcal{O}_{12, \,(1 , 1 ,2)}^{3} \equiv DDHddd\bar{d}uL \,, \\
 \mathcal{O}_{12, \,(1 , 1 ,2)}^{4} \equiv DDHdd\bar{d}uQe \,, \\
 \mathcal{O}_{12, \,(1 , 1 ,2)}^{5} \equiv DDHddQ\bar{e}LL \,, \\
 \mathcal{O}_{12, \,(1 , 1 ,2)}^{6} \equiv DDHdQQe\bar{e}L \,, \\
 \mathcal{O}_{12, \,(1 , 1 ,2)}^{7} \equiv DDHQQQee\bar{e} \,, \\
 \mathcal{O}_{12, \,(1 , 1 ,2)}^{8} \equiv DDHddue\bar{e}L \,, \\
 \mathcal{O}_{12, \,(1 , 1 ,2)}^{9} \equiv DDHduQee\bar{e} \,, \\
 \mathcal{O}_{12, \,(1 , 1 ,2)}^{10} \equiv DD\bar{H}\bar{d}QQQQL \,, \\
 \mathcal{O}_{12, \,(1 , 1 ,2)}^{11} \equiv DD\bar{H}d\bar{d}uQQL \,, \\
 \mathcal{O}_{12, \,(1 , 1 ,2)}^{12} \equiv DD\bar{H}\bar{d}uQQQe \,, \\
 \mathcal{O}_{12, \,(1 , 1 ,2)}^{13} \equiv DD\bar{H}dd\bar{d}uuL \,, \\
 \mathcal{O}_{12, \,(1 , 1 ,2)}^{14} \equiv DD\bar{H}d\bar{d}uuQe \,, \\
 \mathcal{O}_{12, \,(1 , 1 ,2)}^{15} \equiv DD\bar{H}QQQ\bar{e}LL \,, \\
 \mathcal{O}_{12, \,(1 , 1 ,2)}^{16} \equiv DD\bar{H}duQ\bar{e}LL \,, \\
 \mathcal{O}_{12, \,(1 , 1 ,2)}^{17} \equiv DD\bar{H}uQQe\bar{e}L \,, \\
 \mathcal{O}_{12, \,(1 , 1 ,2)}^{18} \equiv DD\bar{H}duue\bar{e}L \,, \\
 \mathcal{O}_{12, \,(1 , 1 ,2)}^{19} \equiv DD\bar{H}uuQee\bar{e} \,, \\
 \mathcal{O}_{12, \,(1 , 1 ,2)}^{20} \equiv DD\bar{H}uQQLL\bar{L} \,, \\
 \mathcal{O}_{12, \,(1 , 1 ,2)}^{21} \equiv DD\bar{H}duuLL\bar{L} \,, \\
 \mathcal{O}_{12, \,(1 , 1 ,2)}^{22} \equiv DD\bar{H}uuQeL\bar{L} \,, \\
 \mathcal{O}_{12, \,(1 , 1 ,2)}^{23} \equiv DD\bar{H}uuuee\bar{L} \,, \\
 \mathcal{O}_{12, \,(1 , 1 ,2)}^{24} \equiv DD\bar{H}uQQQ\bar{Q}L \,, \\
 \mathcal{O}_{12, \,(1 , 1 ,2)}^{25} \equiv DD\bar{H}duuQ\bar{Q}L \,, \\
 \mathcal{O}_{12, \,(1 , 1 ,2)}^{26} \equiv DD\bar{H}uuQQ\bar{Q}e \,, \\
 \mathcal{O}_{12, \,(1 , 1 ,2)}^{27} \equiv DD\bar{H}duuu\bar{Q}e \,, \\
 \mathcal{O}_{12, \,(1 , 1 ,2)}^{28} \equiv DD\bar{H}uu\bar{u}QQL \,, \\
 \mathcal{O}_{12, \,(1 , 1 ,2)}^{29} \equiv DD\bar{H}duuu\bar{u}L \,, \\
 \mathcal{O}_{12, \,(1 , 1 ,2)}^{30} \equiv DD\bar{H}uuu\bar{u}Qe \,, \\
 \mathcal{O}_{12, \,(1 , 1 ,2)}^{31} \equiv DDHdQQLL\bar{L} \,, \\
 \mathcal{O}_{12, \,(1 , 1 ,2)}^{32} \equiv DDHQQQeL\bar{L} \,, \\
 \mathcal{O}_{12, \,(1 , 1 ,2)}^{33} \equiv DDHdduLL\bar{L} \,, \\
 \mathcal{O}_{12, \,(1 , 1 ,2)}^{34} \equiv DDHduQeL\bar{L} \,, \\
 \mathcal{O}_{12, \,(1 , 1 ,2)}^{35} \equiv DDHuQQee\bar{L} \,, \\
 \mathcal{O}_{12, \,(1 , 1 ,2)}^{36} \equiv DDHduuee\bar{L} \,, \\
 \mathcal{O}_{12, \,(1 , 1 ,2)}^{37} \equiv DDHdQQQ\bar{Q}L \,, \\
 \mathcal{O}_{12, \,(1 , 1 ,2)}^{38} \equiv DDHQQQQ\bar{Q}e \,, \\
 \mathcal{O}_{12, \,(1 , 1 ,2)}^{39} \equiv DDHdduQ\bar{Q}L \,, \\
 \mathcal{O}_{12, \,(1 , 1 ,2)}^{40} \equiv DDHduQQ\bar{Q}e \,, \\
 \mathcal{O}_{12, \,(1 , 1 ,2)}^{41} \equiv DDHdduu\bar{Q}e \,, \\
 \mathcal{O}_{12, \,(1 , 1 ,2)}^{42} \equiv DDH\bar{u}QQQQL \,, \\
 \mathcal{O}_{12, \,(1 , 1 ,2)}^{43} \equiv DDHdu\bar{u}QQL \,, \\
 \mathcal{O}_{12, \,(1 , 1 ,2)}^{44} \equiv DDHu\bar{u}QQQe \,, \\
 \mathcal{O}_{12, \,(1 , 1 ,2)}^{45} \equiv DDHdduu\bar{u}L \,, \\
 \mathcal{O}_{12, \,(1 , 1 ,2)}^{46} \equiv DDHduu\bar{u}Qe \,, \\
 \mathcal{O}_{12, \,(1 , 1 ,2)}^{47} \equiv DDHddd\bar{d}Q\nu \,, \\
 \mathcal{O}_{12, \,(1 , 1 ,2)}^{48} \equiv DDHddd\bar{e}L\nu \,, \\
 \mathcal{O}_{12, \,(1 , 1 ,2)}^{49} \equiv DDHddQe\bar{e}\nu \,, \\
 \mathcal{O}_{12, \,(1 , 1 ,2)}^{50} \equiv DD\bar{H}d\bar{d}QQQ\nu \,, \\
 \mathcal{O}_{12, \,(1 , 1 ,2)}^{51} \equiv DD\bar{H}dd\bar{d}uQ\nu \,, \\
 \mathcal{O}_{12, \,(1 , 1 ,2)}^{52} \equiv DD\bar{H}dQQ\bar{e}L\nu \,, \\
 \mathcal{O}_{12, \,(1 , 1 ,2)}^{53} \equiv DD\bar{H}QQQe\bar{e}\nu \,, \\
 \mathcal{O}_{12, \,(1 , 1 ,2)}^{54} \equiv DD\bar{H}ddu\bar{e}L\nu \,, \\
 \mathcal{O}_{12, \,(1 , 1 ,2)}^{55} \equiv DD\bar{H}duQe\bar{e}\nu \,, \\
 \mathcal{O}_{12, \,(1 , 1 ,2)}^{56} \equiv DD\bar{H}QQQL\bar{L}\nu \,, \\
 \mathcal{O}_{12, \,(1 , 1 ,2)}^{57} \equiv DD\bar{H}duQL\bar{L}\nu \,, \\
 \mathcal{O}_{12, \,(1 , 1 ,2)}^{58} \equiv DD\bar{H}uQQe\bar{L}\nu \,, \\
 \mathcal{O}_{12, \,(1 , 1 ,2)}^{59} \equiv DD\bar{H}duue\bar{L}\nu \,, \\
 \mathcal{O}_{12, \,(1 , 1 ,2)}^{60} \equiv DD\bar{H}QQQQ\bar{Q}\nu \,, \\
 \mathcal{O}_{12, \,(1 , 1 ,2)}^{61} \equiv DD\bar{H}duQQ\bar{Q}\nu \,, \\
 \mathcal{O}_{12, \,(1 , 1 ,2)}^{62} \equiv DD\bar{H}dduu\bar{Q}\nu \,, \\
 \mathcal{O}_{12, \,(1 , 1 ,2)}^{63} \equiv DD\bar{H}u\bar{u}QQQ\nu \,, \\
 \mathcal{O}_{12, \,(1 , 1 ,2)}^{64} \equiv DD\bar{H}duu\bar{u}Q\nu \,, \\
 \mathcal{O}_{12, \,(1 , 1 ,2)}^{65} \equiv DDHddQL\bar{L}\nu \,, \\
 \mathcal{O}_{12, \,(1 , 1 ,2)}^{66} \equiv DDHdQQe\bar{L}\nu \,, \\
 \mathcal{O}_{12, \,(1 , 1 ,2)}^{67} \equiv DDHddue\bar{L}\nu \,, \\
 \mathcal{O}_{12, \,(1 , 1 ,2)}^{68} \equiv DD\bar{H}uuQLL\bar{\nu} \,, \\
 \mathcal{O}_{12, \,(1 , 1 ,2)}^{69} \equiv DD\bar{H}uuueL\bar{\nu} \,, \\
 \mathcal{O}_{12, \,(1 , 1 ,2)}^{70} \equiv DDHQQQLL\bar{\nu} \,, \\
 \mathcal{O}_{12, \,(1 , 1 ,2)}^{71} \equiv DDHduQLL\bar{\nu} \,, \\
 \mathcal{O}_{12, \,(1 , 1 ,2)}^{72} \equiv DDHuQQeL\bar{\nu} \,, \\
 \mathcal{O}_{12, \,(1 , 1 ,2)}^{73} \equiv DDHduueL\bar{\nu} \,, \\
 \mathcal{O}_{12, \,(1 , 1 ,2)}^{74} \equiv DDHuuQee\bar{\nu} \,, \\
 \mathcal{O}_{12, \,(1 , 1 ,2)}^{75} \equiv DDHddQQ\bar{Q}\nu \,, \\
 \mathcal{O}_{12, \,(1 , 1 ,2)}^{76} \equiv DDHdddu\bar{Q}\nu \,, \\
 \mathcal{O}_{12, \,(1 , 1 ,2)}^{77} \equiv DDHd\bar{u}QQQ\nu \,, \\
 \mathcal{O}_{12, \,(1 , 1 ,2)}^{78} \equiv DDHddu\bar{u}Q\nu \,, \\
 \mathcal{O}_{12, \,(1 , 1 ,2)}^{79} \equiv DD\bar{H}ddQ\bar{e}\nu\nu \,, \\
 \mathcal{O}_{12, \,(1 , 1 ,2)}^{80} \equiv DD\bar{H}dQQ\bar{L}\nu\nu \,, \\
 \mathcal{O}_{12, \,(1 , 1 ,2)}^{81} \equiv DD\bar{H}ddu\bar{L}\nu\nu \,, \\
 \mathcal{O}_{12, \,(1 , 1 ,2)}^{82} \equiv DDHddd\bar{L}\nu\nu \,, \\
 \mathcal{O}_{12, \,(1 , 1 ,2)}^{83} \equiv DDHdQQL\nu\bar{\nu} \,, \\
 \mathcal{O}_{12, \,(1 , 1 ,2)}^{84} \equiv DD\bar{H}uQQL\nu\bar{\nu} \,, \\
 \mathcal{O}_{12, \,(1 , 1 ,2)}^{85} \equiv DD\bar{H}duuL\nu\bar{\nu} \,, \\
 \mathcal{O}_{12, \,(1 , 1 ,2)}^{86} \equiv DD\bar{H}uuQe\nu\bar{\nu} \,, \\
 \mathcal{O}_{12, \,(1 , 1 ,2)}^{87} \equiv DDHQQQe\nu\bar{\nu} \,, \\
 \mathcal{O}_{12, \,(1 , 1 ,2)}^{88} \equiv DDHdduL\nu\bar{\nu} \,, \\
 \mathcal{O}_{12, \,(1 , 1 ,2)}^{89} \equiv DDHduQe\nu\bar{\nu} \,, \\
 \mathcal{O}_{12, \,(1 , 1 ,2)}^{90} \equiv DDHddQ\nu\nu\bar{\nu} \,, \\
 \mathcal{O}_{12, \,(1 , 1 ,2)}^{91} \equiv DD\bar{H}QQQ\nu\nu\bar{\nu} \,, \\
 \mathcal{O}_{12, \,(1 , 1 ,2)}^{92} \equiv DD\bar{H}duQ\nu\nu\bar{\nu} \,, \\
 \mathcal{O}_{12, \,(1 , 1 ,2)}^{93} \equiv DDHHH\bar{H}ddQL \,, \\
 \mathcal{O}_{12, \,(1 , 1 ,2)}^{94} \equiv DDHHH\bar{H}dQQe \,, \\
 \mathcal{O}_{12, \,(1 , 1 ,2)}^{95} \equiv DDH\bar{H}\bar{H}\bar{H}uuQL \,, \\
 \mathcal{O}_{12, \,(1 , 1 ,2)}^{96} \equiv DDHH\bar{H}\bar{H}QQQL \,, \\
 \mathcal{O}_{12, \,(1 , 1 ,2)}^{97} \equiv DDHH\bar{H}\bar{H}duQL \,, \\
 \mathcal{O}_{12, \,(1 , 1 ,2)}^{98} \equiv DDHH\bar{H}\bar{H}uQQe \,, \\
 \mathcal{O}_{12, \,(1 , 1 ,2)}^{99} \equiv DDHH\bar{H}\bar{H}duue \,, \\
 \mathcal{O}_{12, \,(1 , 1 ,2)}^{100} \equiv DDHH\bar{H}\bar{H}dQQ\nu \,, \\
 \mathcal{O}_{12, \,(1 , 1 ,2)}^{101} \equiv DDH\bar{H}\bar{H}\bar{H}uQQ\nu \,, \\
 \mathcal{O}_{12, \,(1 , 1 ,2)}^{102} \equiv DDHH\bar{H}\bar{H}ddu\nu \,,
 \label{eq:2Ddequal12basisB1L1}
 \end{math}
 \end{spacing}
 \end{multicols}
\noindent
and minimal-derivative operators
 \begin{multicols}{2}
 \begin{spacing}{1.5}
 \noindent
 \begin{math}
 \mathcal{O}_{12, \,(1 , 1 ,2)}^{103} \equiv DDH\bar{H}\bar{H}\bar{H}uuue \,, \\
 \mathcal{O}_{12, \,(1 , 1 ,2)}^{104} \equiv DDHHH\bar{H}ddue \,, \\
 \mathcal{O}_{12, \,(1 , 1 ,2)}^{105} \equiv DDHHHHddde \,, \\
 \mathcal{O}_{12, \,(1 , 1 ,2)}^{106} \equiv DD\bar{H}\bar{H}\bar{H}\bar{H}uuu\nu \,, \\
 \mathcal{O}_{12, \,(1 , 1 ,2)}^{107} \equiv DDH\bar{H}\bar{H}\bar{H}duu\nu \,, \\
 \mathcal{O}_{12, \,(1 , 1 ,2)}^{108} \equiv DDHHH\bar{H}ddd\nu \,. 
 \label{eq:2Ddequal12basisB1L1Unique}
 \end{math}
 \end{spacing}
 \end{multicols}
\noindent
 For the two-derivative operators, we once again depart from our standard convention of listing all SMEFT operators first, followed by the pure $\nu$SMEFT operators. The operators $\mathcal{O}_{12,(1,1,2)}^{j}$ for $j\in\{1\isep 102\}$ correspond to the dimension-10 operators listed in Sec.~D of Ref.~\cite{Heeck:2026dmh}, according to $\mathcal{O}_{12,(1,1,2)}^{j}\equiv D^2\mathcal{O}_{10,(1,1)}^{j}$. We preserve the established ordering for operators with non-vanishing dimension-10 counterparts and append the minimal-derivative operators. Their UV completions are listed in tables~\ref{tab:T12Top8B1L1D2}--\ref{tab:T12Top89B1L1D2}.

 All $d=12$ three-derivative operators can be obtained by connection to one-derivative operators listed in App.~\ref{sec:d=10} as follows $\mathcal{O}_{12,(1,1,3)}^{j}\equiv D^2\mathcal{O}_{10,(1,1,1)}^{j}$. Since there were $d=10$ UV completions with vanishing $\mathcal{O}_{10,(1,1,1)}^{j}$ due to Higgs asymmetry, we find non-trivial $D^3$ UV completions here at $d=12$, listed in tables~\ref{tab:T12Top12B1L1D3}--\ref{tab:T12Top19B1L1D3}.

As for four-derivative operators at mass dimension $d=12$, they can be obtained from the dimension-10 two-derivative operators listed in Eqs.~\eqref{eq:2Ddequal10basisB1L1} and~\eqref{eq:2Ddequal10basisB1L1Unique}, according to $\mathcal{O}_{12,(1,1,4)}^{j}\equiv D^2\mathcal{O}_{10,(1,1,2)}^{j},$ $j\in\{1\isep 14\}.$ Similarly, all five-derivative operators at $d=12$ can be obtained from the dimension-8 one-derivative basis in Eq.~\eqref{eq:1Ddequal8basisB1L1}, according to $\mathcal{O}_{12,(1,1,5)}^{j}\equiv D^4\mathcal{O}_{8,(1,1,1)}^{j},$ $j\in\{1\isep 4\}.$ Finally, all six-derivative operators at $d=12$ can be obtained from the dimension-8 two-derivative basis in Eq.~\eqref{eq:2Ddequal8basisB1L1}, according to $\mathcal{O}_{12,(1,1,6)}^{j}\equiv D^4\mathcal{O}_{8,(1,1,2)}^{j},$ $j\in\{1\isep 6\}.$

 \begin{center}
 \begin{longtable}{| c  | m{13cm} |} 
   \hline
  $\mathcal{O}^j{(\mathcal{T}_ {10} ^{ 8 })}$ &  New Particles \\
   \hline
   \hline
   \endhead
 \multicolumn{2}{|c|}{\hspace{9cm} Continued on next page}
   \endfoot
  $\mathcal{O}_{12,\;(1,1,2)}^{108}$ & {$S_{2}S_{3}S_{9}S_{26}$, $S_{2}S_{3}S_{8}S_{16}$, $S_{2}S_{3}S_{7}S_{16}$, $S_{2}S_{3}S_{9}S_{16}$} \\
   \hline
 \caption{Dimension 12 UV completions for topology $\mathcal{T}_ {10} ^{ 8 }$ (see Fig.~\ref{fig:d=10_topologies}).} 
 \label{tab:T12Top8B1L1D2}\\
   \endlastfoot
  $\mathcal{O}_{12,\;(1,1,2)}^{103}$ & {$S_{1}S_{2}S_{9}S_{16}$, $S_{1}S_{2}S_{8}S_{16}$, $S_{1}S_{2}S_{7}S_{16}$, $S_{1}S_{2}S_{9}S_{26}$} \\
   \hline
  $\mathcal{O}_{12,\;(1,1,2)}^{104}$ & {$S_{1}S_{2}S_{9}S_{26}$, $S_{2}S_{3}S_{9}S_{26}$, $S_{1}S_{2}S_{8}S_{16}$, $S_{1}S_{2}S_{7}S_{16}$, $S_{2}S_{3}S_{8}S_{16}$, $S_{2}S_{3}S_{7}S_{16}$, $S_{1}S_{2}S_{9}S_{16}$, $S_{2}S_{3}S_{9}S_{16}$} \\
   \hline
  $\mathcal{O}_{12,\;(1,1,2)}^{105}$ & {$S_{1}S_{3}S_{9}S_{26}$} \\
   \hline
  $\mathcal{O}_{12,\;(1,1,2)}^{106}$ & {$S_{1}S_{3}S_{9}S_{26}$} \\
   \hline
  $\mathcal{O}_{12,\;(1,1,2)}^{107}$ & {$S_{2}S_{3}S_{9}S_{16}$, $S_{1}S_{2}S_{9}S_{16}$, $S_{2}S_{3}S_{8}S_{16}$, $S_{2}S_{3}S_{7}S_{16}$, $S_{1}S_{2}S_{8}S_{16}$, $S_{1}S_{2}S_{7}S_{16}$, $S_{2}S_{3}S_{9}S_{26}$, $S_{1}S_{2}S_{9}S_{26}$} \\
   \hline
 \end{longtable}
 \end{center}
 \begin{center}
 \begin{longtable}{| c  | m{13cm} |} 
   \hline
  $\mathcal{O}^j{(\mathcal{T}_ {10} ^{ 9 })}$ &  New Particles \\
   \hline
   \hline
   \endhead
 \multicolumn{2}{|c|}{\hspace{9cm} Continued on next page}
   \endfoot
  $\mathcal{O}_{12,\;(1,1,2)}^{108}$ & {$S_{2}S_{3}S_{8}S_{9}$} \\
   \hline
 \caption{Dimension 12 UV completions for topology $\mathcal{T}_ {10} ^{ 9 }$ (see Fig.~\ref{fig:d=10_topologies}).} 
 \label{tab:T12Top9B1L1D2}\\
   \endlastfoot
  $\mathcal{O}_{12,\;(1,1,2)}^{103}$ & {$S_{1}S_{2}S_{8}S_{9}$} \\
   \hline
  $\mathcal{O}_{12,\;(1,1,2)}^{104}$ & {$S_{1}S_{2}S_{8}S_{9}$, $S_{2}S_{3}S_{8}S_{9}$} \\
   \hline
  $\mathcal{O}_{12,\;(1,1,2)}^{105}$ & {$S_{1}S_{3}S_{9}S_{9}$} \\
   \hline
  $\mathcal{O}_{12,\;(1,1,2)}^{106}$ & {$S_{1}S_{3}S_{9}S_{9}$} \\
   \hline
  $\mathcal{O}_{12,\;(1,1,2)}^{107}$ & {$S_{2}S_{3}S_{8}S_{9}$, $S_{1}S_{2}S_{8}S_{9}$} \\
   \hline
 \end{longtable}
 \end{center}
 \begin{center}
 \begin{longtable}{| c  | m{13cm} |} 
   \hline
  $\mathcal{O}^j{(\mathcal{T}_ {10} ^{ 10 })}$ &  New Particles \\
   \hline
   \hline
   \endhead
 \multicolumn{2}{|c|}{\hspace{9cm} Continued on next page}
   \endfoot
  $\mathcal{O}_{12,\;(1,1,2)}^{108}$ & {$S_{2}S_{3}S_{8}S_{9}S_{14}$} \\
   \hline
 \caption{Dimension 12 UV completions for topology $\mathcal{T}_ {10} ^{ 10 }$ (see Fig.~\ref{fig:d=10_topologies}).} 
 \label{tab:T12Top10B1L1D2}\\
   \endlastfoot
  $\mathcal{O}_{12,\;(1,1,2)}^{103}$ & {$S_{1}S_{2}S_{8}S_{9}S_{14}$} \\
   \hline
  $\mathcal{O}_{12,\;(1,1,2)}^{104}$ & {$S_{1}S_{2}S_{8}S_{9}S_{14}$, $S_{2}S_{3}S_{8}S_{9}S_{14}$} \\
   \hline
  $\mathcal{O}_{12,\;(1,1,2)}^{105}$ & {$S_{1}S_{3}S_{9}S_{9}S_{15}$} \\
   \hline
  $\mathcal{O}_{12,\;(1,1,2)}^{106}$ & {$S_{1}S_{3}S_{9}S_{9}S_{15}$} \\
   \hline
  $\mathcal{O}_{12,\;(1,1,2)}^{107}$ & {$S_{2}S_{3}S_{8}S_{9}S_{14}$, $S_{1}S_{2}S_{8}S_{9}S_{14}$} \\
   \hline
 \end{longtable}
 \end{center}
 \begin{center}
 \begin{longtable}{| c  | m{13cm} |} 
   \hline
  $\mathcal{O}^j{(\mathcal{T}_ {10} ^{ 11 })}$ &  New Particles \\
   \hline
   \hline
   \endhead
 \multicolumn{2}{|c|}{\hspace{9cm} Continued on next page}
   \endfoot
  $\mathcal{O}_{12,\;(1,1,2)}^{108}$ & {$S_{2}S_{3}S_{5}S_{5}S_{8}$, $S_{2}S_{3}S_{5}S_{5}S_{7}$, $S_{2}S_{3}S_{5}S_{9}S_{11}$, $S_{2}S_{3}S_{5}S_{6}S_{9}$} \\
   \hline
 \caption{Dimension 12 UV completions for topology $\mathcal{T}_ {10} ^{ 11 }$ (see Fig.~\ref{fig:d=10_topologies}).} 
 \label{tab:T12Top11B1L1D2}\\
   \endlastfoot
  $\mathcal{O}_{12,\;(1,1,2)}^{103}$ & {$S_{1}S_{2}S_{9}S_{10}S_{11}$, $S_{1}S_{2}S_{5}S_{9}S_{11}$, $S_{1}S_{2}S_{8}S_{11}S_{11}$, $S_{1}S_{2}S_{7}S_{11}S_{11}$} \\
   \hline
  $\mathcal{O}_{12,\;(1,1,2)}^{104}$ & {$S_{1}S_{2}S_{8}S_{11}S_{11}$, $S_{1}S_{2}S_{7}S_{11}S_{11}$, $S_{2}S_{3}S_{5}S_{5}S_{8}$, $S_{2}S_{3}S_{5}S_{5}S_{7}$, $S_{1}S_{2}S_{9}S_{10}S_{11}$, $S_{2}S_{3}S_{5}S_{6}S_{9}$, $S_{2}S_{3}S_{5}S_{9}S_{11}$, $S_{1}S_{2}S_{5}S_{9}S_{11}$} \\
   \hline
  $\mathcal{O}_{12,\;(1,1,2)}^{105}$ & {$S_{1}S_{3}S_{5}S_{9}S_{11}$} \\
   \hline
  $\mathcal{O}_{12,\;(1,1,2)}^{106}$ & {$S_{1}S_{3}S_{5}S_{9}S_{11}$} \\
   \hline
  $\mathcal{O}_{12,\;(1,1,2)}^{107}$ & {$S_{2}S_{3}S_{5}S_{9}S_{11}$, $S_{1}S_{2}S_{5}S_{9}S_{11}$, $S_{1}S_{2}S_{9}S_{10}S_{11}$, $S_{2}S_{3}S_{5}S_{6}S_{9}$, $S_{2}S_{3}S_{5}S_{5}S_{8}$, $S_{2}S_{3}S_{5}S_{5}S_{7}$, $S_{1}S_{2}S_{8}S_{11}S_{11}$, $S_{1}S_{2}S_{7}S_{11}S_{11}$} \\
   \hline
 \end{longtable}
 \end{center}
 \begin{center}
 \begin{longtable}{| c  | m{13cm} |} 
   \hline
  $\mathcal{O}^j{(\mathcal{T}_ {10} ^{ 12 })}$ &  New Particles \\
   \hline
   \hline
   \endhead
 \multicolumn{2}{|c|}{\hspace{9cm} Continued on next page}
   \endfoot
  $\mathcal{O}_{12,\;(1,1,2)}^{108}$ & {$S_{2}S_{9}F_{6}F_{7}F_{8}$, $S_{3}S_{9}F_{6}F_{7}F_{8}$, $S_{3}S_{9}F_{2}F_{3}F_{3}$, $S_{2}S_{8}F_{6}F_{8}F_{8}$, $S_{2}S_{7}F_{6}F_{8}F_{8}$, $S_{3}S_{8}F_{6}F_{8}F_{8}$, $S_{3}S_{7}F_{6}F_{8}F_{8}$, $S_{3}S_{8}F_{2}F_{3}F_{3}$, $S_{3}S_{7}F_{2}F_{3}F_{3}$, $S_{2}S_{9}F_{6}F_{8}F_{13}$, $S_{3}S_{9}F_{6}F_{8}F_{13}$, $S_{3}S_{9}F_{2}F_{3}F_{10}$} \\
   \hline
 \caption{Dimension 12 UV completions for topology $\mathcal{T}_ {10} ^{ 12 }$ (see Fig.~\ref{fig:d=10_topologies}).} 
 \label{tab:T12Top12B1L1D2}\\
   \endlastfoot
  $\mathcal{O}_{12,\;(1,1,2)}^{103}$ & {$S_{2}S_{9}F_{5}F_{7}F_{8}$, $S_{1}S_{9}F_{5}F_{7}F_{8}$, $S_{1}S_{9}F_{10}F_{16}F_{49}$, $S_{2}S_{8}F_{5}F_{8}F_{8}$, $S_{2}S_{7}F_{5}F_{8}F_{8}$, $S_{1}S_{8}F_{5}F_{8}F_{8}$, $S_{1}S_{7}F_{5}F_{8}F_{8}$, $S_{1}S_{8}F_{10}F_{10}F_{16}$, $S_{1}S_{7}F_{10}F_{10}F_{16}$, $S_{2}S_{9}F_{5}F_{8}F_{13}$, $S_{1}S_{9}F_{5}F_{8}F_{13}$, $S_{1}S_{9}F_{3}F_{10}F_{16}$} \\
   \hline
  $\mathcal{O}_{12,\;(1,1,2)}^{104}$ & {$S_{1}S_{9}F_{8}F_{13}F_{18}$, $S_{3}S_{9}F_{8}F_{13}F_{18}$, $S_{1}S_{9}F_{6}F_{7}F_{8}$, $S_{2}S_{9}F_{6}F_{7}F_{8}$, $S_{3}S_{9}F_{1}F_{3}F_{10}$, $S_{2}S_{9}F_{1}F_{3}F_{10}$, $S_{1}S_{8}F_{13}F_{13}F_{18}$, $S_{1}S_{7}F_{13}F_{13}F_{18}$, $S_{3}S_{8}F_{13}F_{13}F_{18}$, $S_{3}S_{7}F_{13}F_{13}F_{18}$, $S_{1}S_{8}F_{6}F_{8}F_{8}$, $S_{1}S_{7}F_{6}F_{8}F_{8}$, $S_{2}S_{8}F_{6}F_{8}F_{8}$, $S_{2}S_{7}F_{6}F_{8}F_{8}$, $S_{3}S_{8}F_{1}F_{3}F_{3}$, $S_{3}S_{7}F_{1}F_{3}F_{3}$, $S_{2}S_{8}F_{1}F_{3}F_{3}$, $S_{2}S_{7}F_{1}F_{3}F_{3}$, $S_{1}S_{9}F_{13}F_{18}F_{50}$, $S_{3}S_{9}F_{13}F_{18}F_{50}$, $S_{1}S_{9}F_{6}F_{8}F_{13}$, $S_{2}S_{9}F_{6}F_{8}F_{13}$, $S_{3}S_{9}F_{1}F_{3}F_{3}$, $S_{2}S_{9}F_{1}F_{3}F_{3}$} \\
   \hline
  $\mathcal{O}_{12,\;(1,1,2)}^{105}$ & {$S_{1}S_{9}F_{8}F_{13}F_{18}$, $S_{3}S_{9}F_{8}F_{13}F_{18}$, $S_{3}S_{9}F_{2}F_{3}F_{3}$} \\
   \hline
  $\mathcal{O}_{12,\;(1,1,2)}^{106}$ & {$S_{3}S_{9}F_{7}F_{8}F_{17}$, $S_{1}S_{9}F_{7}F_{8}F_{17}$, $S_{1}S_{9}F_{3}F_{10}F_{16}$} \\
\hline
 \end{longtable}
 \end{center}
 \begin{center}
 \begin{longtable}{| c  | m{13cm} |} 
   \hline
  $\mathcal{O}^j{(\mathcal{T}_ {10} ^{ 13 })}$ &  New Particles \\
   \hline
   \hline
   \endhead
 \multicolumn{2}{|c|}{\hspace{9cm} Continued on next page}
   \endfoot
  $\mathcal{O}_{12,\;(1,1,2)}^{108}$ & {$S_{2}S_{3}S_{4}S_{5}S_{11}$, $S_{2}S_{2}S_{3}S_{5}S_{11}$, $S_{2}S_{3}S_{5}S_{6}S_{19}$, $S_{2}S_{3}S_{3}S_{5}S_{6}$, $S_{2}S_{3}S_{4}S_{5}S_{5}$, $S_{2}S_{2}S_{3}S_{5}S_{5}$, $S_{2}S_{3}S_{5}S_{5}S_{19}$, $S_{2}S_{3}S_{3}S_{5}S_{5}$} \\
   \hline
 \caption{Dimension 12 UV completions for topology $\mathcal{T}_ {10} ^{ 13 }$ (see Fig.~\ref{fig:d=10_topologies}).} 
 \label{tab:T12Top13B1L1D2}\\
   \endlastfoot
  $\mathcal{O}_{12,\;(1,1,2)}^{103}$ & {$S_{1}S_{2}S_{11}S_{11}S_{31}$, $S_{1}S_{1}S_{2}S_{11}S_{11}$, $S_{1}S_{2}S_{4}S_{11}S_{11}$, $S_{1}S_{2}S_{2}S_{11}S_{11}$, $S_{1}S_{2}S_{10}S_{11}S_{31}$, $S_{1}S_{1}S_{2}S_{10}S_{11}$, $S_{1}S_{2}S_{4}S_{5}S_{11}$, $S_{1}S_{2}S_{2}S_{5}S_{11}$} \\
   \hline
  $\mathcal{O}_{12,\;(1,1,2)}^{104}$ & {$S_{1}S_{2}S_{10}S_{11}S_{31}$, $S_{1}S_{1}S_{2}S_{10}S_{11}$, $S_{2}S_{3}S_{5}S_{6}S_{19}$, $S_{2}S_{3}S_{3}S_{5}S_{6}$, $S_{2}S_{3}S_{4}S_{5}S_{11}$, $S_{2}S_{2}S_{3}S_{5}S_{11}$, $S_{1}S_{2}S_{4}S_{5}S_{11}$, $S_{1}S_{2}S_{2}S_{5}S_{11}$, $S_{1}S_{2}S_{11}S_{11}S_{31}$, $S_{1}S_{1}S_{2}S_{11}S_{11}$, $S_{2}S_{3}S_{5}S_{5}S_{19}$, $S_{2}S_{3}S_{3}S_{5}S_{5}$, $S_{2}S_{3}S_{4}S_{5}S_{5}$, $S_{2}S_{2}S_{3}S_{5}S_{5}$, $S_{1}S_{2}S_{4}S_{11}S_{11}$, $S_{1}S_{2}S_{2}S_{11}S_{11}$} \\
   \hline
  $\mathcal{O}_{12,\;(1,1,2)}^{105}$ & {$S_{1}S_{3}S_{4}S_{5}S_{11}$, $S_{1}S_{2}S_{3}S_{5}S_{11}$} \\
   \hline
  $\mathcal{O}_{12,\;(1,1,2)}^{106}$ & {$S_{1}S_{3}S_{4}S_{5}S_{11}$, $S_{1}S_{2}S_{3}S_{5}S_{11}$} \\
   \hline
  $\mathcal{O}_{12,\;(1,1,2)}^{107}$ & {$S_{2}S_{3}S_{4}S_{5}S_{5}$, $S_{2}S_{2}S_{3}S_{5}S_{5}$, $S_{1}S_{2}S_{4}S_{11}S_{11}$, $S_{1}S_{2}S_{2}S_{11}S_{11}$, $S_{1}S_{2}S_{11}S_{11}S_{31}$, $S_{1}S_{1}S_{2}S_{11}S_{11}$, $S_{2}S_{3}S_{5}S_{5}S_{19}$, $S_{2}S_{3}S_{3}S_{5}S_{5}$, $S_{2}S_{3}S_{4}S_{5}S_{11}$, $S_{2}S_{2}S_{3}S_{5}S_{11}$, $S_{1}S_{2}S_{4}S_{5}S_{11}$, $S_{1}S_{2}S_{2}S_{5}S_{11}$, $S_{1}S_{2}S_{10}S_{11}S_{31}$, $S_{1}S_{1}S_{2}S_{10}S_{11}$, $S_{2}S_{3}S_{5}S_{6}S_{19}$, $S_{2}S_{3}S_{3}S_{5}S_{6}$} \\
\hline
 \end{longtable}
 \end{center}
 \begin{center}
 \begin{longtable}{| c  | m{13cm} |} 
   \hline
  $\mathcal{O}^j{(\mathcal{T}_ {10} ^{ 14 })}$ &  New Particles \\
   \hline
   \hline
   \endhead
 \multicolumn{2}{|c|}{\hspace{9cm} Continued on next page}
   \endfoot
  $\mathcal{O}_{12,\;(1,1,2)}^{108}$ & {$S_{2}S_{9}F_{7}F_{8}F_{13}$, $S_{3}S_{9}F_{3}F_{8}F_{13}$, $S_{3}S_{9}F_{3}F_{7}F_{10}$, $S_{2}S_{9}F_{7}F_{8}F_{8}$, $S_{3}S_{9}F_{3}F_{7}F_{8}$, $S_{3}S_{9}F_{3}F_{3}F_{8}$, $S_{2}S_{8}F_{8}F_{8}F_{8}$, $S_{2}S_{7}F_{8}F_{8}F_{8}$, $S_{3}S_{8}F_{3}F_{8}F_{8}$, $S_{3}S_{7}F_{3}F_{8}F_{8}$, $S_{3}S_{8}F_{3}F_{3}F_{8}$, $S_{3}S_{7}F_{3}F_{3}F_{8}$} \\
   \hline
 \caption{Dimension 12 UV completions for topology $\mathcal{T}_ {10} ^{ 14 }$ (see Fig.~\ref{fig:d=10_topologies}).} 
 \label{tab:T12Top14B1L1D2}\\
   \endlastfoot
  $\mathcal{O}_{12,\;(1,1,2)}^{103}$ & {$S_{2}S_{8}F_{8}F_{8}F_{8}$, $S_{2}S_{7}F_{8}F_{8}F_{8}$, $S_{1}S_{8}F_{8}F_{8}F_{10}$, $S_{1}S_{7}F_{8}F_{8}F_{10}$, $S_{1}S_{8}F_{8}F_{10}F_{10}$, $S_{1}S_{7}F_{8}F_{10}F_{10}$, $S_{2}S_{9}F_{8}F_{8}F_{13}$, $S_{1}S_{9}F_{8}F_{10}F_{13}$, $S_{1}S_{9}F_{3}F_{8}F_{10}$, $S_{2}S_{9}F_{7}F_{8}F_{13}$, $S_{1}S_{9}F_{3}F_{7}F_{8}$, $S_{1}S_{9}F_{10}F_{13}F_{49}$} \\
   \hline
  $\mathcal{O}_{12,\;(1,1,2)}^{104}$ & {$S_{1}S_{9}F_{7}F_{13}F_{50}$, $S_{3}S_{9}F_{10}F_{13}F_{50}$, $S_{1}S_{9}F_{8}F_{8}F_{13}$, $S_{2}S_{9}F_{7}F_{8}F_{13}$, $S_{2}S_{9}F_{8}F_{10}F_{13}$, $S_{3}S_{9}F_{3}F_{3}F_{8}$, $S_{2}S_{9}F_{3}F_{3}F_{7}$, $S_{3}S_{9}F_{3}F_{8}F_{13}$, $S_{1}S_{9}F_{7}F_{8}F_{13}$, $S_{2}S_{9}F_{7}F_{8}F_{8}$, $S_{2}S_{9}F_{3}F_{7}F_{8}$, $S_{3}S_{9}F_{3}F_{10}F_{13}$, $S_{2}S_{9}F_{3}F_{8}F_{10}$, $S_{1}S_{8}F_{8}F_{13}F_{13}$, $S_{1}S_{7}F_{8}F_{13}F_{13}$, $S_{3}S_{8}F_{3}F_{13}F_{13}$, $S_{3}S_{7}F_{3}F_{13}F_{13}$, $S_{1}S_{8}F_{8}F_{8}F_{13}$, $S_{1}S_{7}F_{8}F_{8}F_{13}$, $S_{2}S_{8}F_{8}F_{8}F_{8}$, $S_{2}S_{7}F_{8}F_{8}F_{8}$, $S_{2}S_{8}F_{3}F_{8}F_{8}$, $S_{2}S_{7}F_{3}F_{8}F_{8}$, $S_{3}S_{8}F_{3}F_{3}F_{13}$, $S_{3}S_{7}F_{3}F_{3}F_{13}$, $S_{2}S_{8}F_{3}F_{3}F_{8}$, $S_{2}S_{7}F_{3}F_{3}F_{8}$} \\
   \hline
  $\mathcal{O}_{12,\;(1,1,2)}^{105}$ & {$S_{1}S_{9}F_{8}F_{8}F_{13}$, $S_{3}S_{9}F_{3}F_{8}F_{13}$, $S_{3}S_{9}F_{3}F_{3}F_{8}$} \\
   \hline
  $\mathcal{O}_{12,\;(1,1,2)}^{106}$ & {$S_{3}S_{9}F_{7}F_{8}F_{8}$, $S_{1}S_{9}F_{3}F_{7}F_{8}$, $S_{1}S_{9}F_{3}F_{8}F_{10}$} \\
\hline
 \end{longtable}
 \end{center}
 \begin{center}
 \begin{longtable}{| c  | m{13cm} |} 
   \hline
  $\mathcal{O}^j{(\mathcal{T}_ {10} ^{ 15 })}$ &  New Particles \\
   \hline
   \hline
   \endhead
 \multicolumn{2}{|c|}{\hspace{9cm} Continued on next page}
   \endfoot
  $\mathcal{O}_{12,\;(1,1,2)}^{108}$ & {$S_{2}S_{5}F_{7}F_{8}F_{9}$, $S_{2}S_{5}F_{5}F_{7}F_{8}$, $S_{3}S_{5}F_{3}F_{4}F_{8}$, $S_{3}S_{5}F_{1}F_{3}F_{8}$, $S_{3}S_{5}F_{3}F_{7}F_{9}$, $S_{3}S_{5}F_{3}F_{5}F_{7}$, $S_{2}S_{5}F_{7}F_{8}F_{14}$, $S_{2}S_{5}F_{6}F_{7}F_{8}$, $S_{3}S_{5}F_{3}F_{7}F_{11}$, $S_{3}S_{5}F_{2}F_{3}F_{7}$, $S_{3}S_{5}F_{3}F_{8}F_{14}$, $S_{3}S_{5}F_{3}F_{6}F_{8}$, $S_{2}S_{11}F_{8}F_{8}F_{14}$, $S_{2}S_{11}F_{6}F_{8}F_{8}$, $S_{3}S_{6}F_{3}F_{8}F_{11}$, $S_{3}S_{6}F_{2}F_{3}F_{8}$, $S_{3}S_{6}F_{3}F_{8}F_{14}$, $S_{3}S_{6}F_{3}F_{6}F_{8}$, $S_{2}S_{5}F_{8}F_{8}F_{9}$, $S_{2}S_{5}F_{5}F_{8}F_{8}$, $S_{3}S_{5}F_{3}F_{8}F_{9}$, $S_{3}S_{5}F_{3}F_{5}F_{8}$} \\
   \hline
 \caption{Dimension 12 UV completions for topology $\mathcal{T}_ {10} ^{ 15 }$ (see Fig.~\ref{fig:d=10_topologies}).} 
 \label{tab:T12Top15B1L1D2}\\
   \endlastfoot
  $\mathcal{O}_{12,\;(1,1,2)}^{103}$ & {$S_{2}S_{11}F_{8}F_{8}F_{14}$, $S_{2}S_{11}F_{6}F_{8}F_{8}$, $S_{1}S_{11}F_{8}F_{10}F_{11}$, $S_{1}S_{11}F_{2}F_{8}F_{10}$, $S_{1}S_{11}F_{8}F_{10}F_{14}$, $S_{1}S_{11}F_{6}F_{8}F_{10}$, $S_{2}S_{5}F_{8}F_{8}F_{9}$, $S_{2}S_{5}F_{5}F_{8}F_{8}$, $S_{1}S_{10}F_{8}F_{10}F_{31}$, $S_{1}S_{10}F_{8}F_{10}F_{16}$, $S_{1}S_{10}F_{8}F_{9}F_{10}$, $S_{1}S_{10}F_{5}F_{8}F_{10}$, $S_{2}S_{11}F_{8}F_{9}F_{13}$, $S_{2}S_{11}F_{5}F_{8}F_{13}$, $S_{1}S_{11}F_{10}F_{13}F_{31}$, $S_{1}S_{11}F_{10}F_{13}F_{16}$, $S_{1}S_{11}F_{3}F_{8}F_{9}$, $S_{1}S_{11}F_{3}F_{5}F_{8}$, $S_{2}S_{11}F_{8}F_{13}F_{14}$, $S_{2}S_{11}F_{6}F_{8}F_{13}$, $S_{1}S_{11}F_{3}F_{8}F_{11}$, $S_{1}S_{11}F_{2}F_{3}F_{8}$, $S_{1}S_{11}F_{10}F_{13}F_{14}$, $S_{1}S_{11}F_{6}F_{10}F_{13}$} \\
\hline
 \end{longtable}
 \end{center}
 \begin{center}
 \begin{longtable}{| c  | m{13cm} |} 
   \hline
  $\mathcal{O}^j{(\mathcal{T}_ {10} ^{ 16 })}$ &  New Particles \\
   \hline
   \hline
   \endhead
 \multicolumn{2}{|c|}{\hspace{9cm} Continued on next page}
   \endfoot
  $\mathcal{O}_{12,\;(1,1,2)}^{108}$ & {$S_{2}F_{6}F_{7}F_{8}F_{9}$, $S_{2}F_{5}F_{6}F_{7}F_{8}$, $S_{3}F_{2}F_{3}F_{3}F_{4}$, $S_{3}F_{1}F_{2}F_{3}F_{3}$, $S_{3}F_{6}F_{7}F_{8}F_{9}$, $S_{3}F_{5}F_{6}F_{7}F_{8}$, $S_{2}F_{6}F_{8}F_{8}F_{9}$, $S_{2}F_{5}F_{6}F_{8}F_{8}$, $S_{3}F_{6}F_{8}F_{8}F_{9}$, $S_{3}F_{5}F_{6}F_{8}F_{8}$, $S_{2}F_{6}F_{8}F_{13}F_{14}$, $S_{2}F_{6}F_{6}F_{8}F_{13}$, $S_{3}F_{2}F_{3}F_{10}F_{11}$, $S_{3}F_{2}F_{2}F_{3}F_{10}$, $S_{3}F_{6}F_{8}F_{13}F_{14}$, $S_{3}F_{6}F_{6}F_{8}F_{13}$, $S_{2}F_{6}F_{8}F_{8}F_{14}$, $S_{2}F_{6}F_{6}F_{8}F_{8}$, $S_{3}F_{2}F_{3}F_{3}F_{11}$, $S_{3}F_{2}F_{2}F_{3}F_{3}$, $S_{3}F_{6}F_{8}F_{8}F_{14}$, $S_{3}F_{6}F_{6}F_{8}F_{8}$} \\
   \hline
 \caption{Dimension 12 UV completions for topology $\mathcal{T}_ {10} ^{ 16 }$ (see Fig.~\ref{fig:d=10_topologies}).} 
 \label{tab:T12Top16B1L1D2}\\
   \endlastfoot
  $\mathcal{O}_{12,\;(1,1,2)}^{103}$ & {$S_{2}F_{5}F_{8}F_{8}F_{9}$, $S_{2}F_{5}F_{5}F_{8}F_{8}$, $S_{1}F_{10}F_{10}F_{16}F_{31}$, $S_{1}F_{10}F_{10}F_{16}F_{16}$, $S_{1}F_{5}F_{8}F_{8}F_{9}$, $S_{1}F_{5}F_{5}F_{8}F_{8}$, $S_{2}F_{5}F_{7}F_{8}F_{9}$, $S_{2}F_{5}F_{5}F_{7}F_{8}$, $S_{1}F_{10}F_{16}F_{31}F_{49}$, $S_{1}F_{10}F_{16}F_{16}F_{49}$, $S_{1}F_{5}F_{7}F_{8}F_{9}$, $S_{1}F_{5}F_{5}F_{7}F_{8}$, $S_{2}F_{5}F_{8}F_{8}F_{14}$, $S_{2}F_{5}F_{6}F_{8}F_{8}$, $S_{1}F_{10}F_{10}F_{11}F_{16}$, $S_{1}F_{2}F_{10}F_{10}F_{16}$, $S_{1}F_{5}F_{8}F_{8}F_{14}$, $S_{1}F_{5}F_{6}F_{8}F_{8}$, $S_{2}F_{5}F_{8}F_{13}F_{14}$, $S_{2}F_{5}F_{6}F_{8}F_{13}$, $S_{1}F_{3}F_{10}F_{11}F_{16}$, $S_{1}F_{2}F_{3}F_{10}F_{16}$, $S_{1}F_{5}F_{8}F_{13}F_{14}$, $S_{1}F_{5}F_{6}F_{8}F_{13}$} \\
\hline
 \end{longtable}
 \end{center}
 \begin{center}
 \begin{longtable}{| c  | m{13cm} |} 
   \hline
  $\mathcal{O}^j{(\mathcal{T}_ {10} ^{ 17 })}$ &  New Particles \\
   \hline
   \hline
   \endhead
 \multicolumn{2}{|c|}{\hspace{9cm} Continued on next page}
   \endfoot
  $\mathcal{O}_{12,\;(1,1,2)}^{108}$ & {$S_{2}S_{3}S_{4}S_{9}S_{11}$, $S_{2}S_{3}S_{6}S_{9}S_{19}$, $S_{2}S_{3}S_{5}S_{8}S_{19}$, $S_{2}S_{3}S_{3}S_{5}S_{7}$, $S_{2}S_{3}S_{4}S_{5}S_{8}$, $S_{2}S_{2}S_{3}S_{5}S_{7}$, $S_{2}S_{3}S_{4}S_{5}S_{9}$, $S_{2}S_{3}S_{5}S_{9}S_{19}$} \\
   \hline
 \caption{Dimension 12 UV completions for topology $\mathcal{T}_ {10} ^{ 17 }$ (see Fig.~\ref{fig:d=10_topologies}).} 
 \label{tab:T12Top17B1L1D2}\\
   \endlastfoot
  $\mathcal{O}_{12,\;(1,1,2)}^{103}$ & {$S_{1}S_{2}S_{4}S_{9}S_{11}$, $S_{1}S_{2}S_{9}S_{11}S_{31}$, $S_{1}S_{2}S_{8}S_{11}S_{31}$, $S_{1}S_{1}S_{2}S_{7}S_{11}$, $S_{1}S_{2}S_{4}S_{8}S_{11}$, $S_{1}S_{2}S_{2}S_{7}S_{11}$, $S_{1}S_{2}S_{4}S_{5}S_{9}$, $S_{1}S_{2}S_{9}S_{10}S_{31}$} \\
   \hline
  $\mathcal{O}_{12,\;(1,1,2)}^{104}$ & {$S_{1}S_{2}S_{9}S_{10}S_{31}$, $S_{2}S_{3}S_{6}S_{9}S_{19}$, $S_{2}S_{3}S_{4}S_{9}S_{11}$, $S_{1}S_{2}S_{4}S_{5}S_{9}$, $S_{1}S_{2}S_{4}S_{8}S_{11}$, $S_{1}S_{2}S_{2}S_{7}S_{11}$, $S_{2}S_{3}S_{4}S_{5}S_{8}$, $S_{2}S_{2}S_{3}S_{5}S_{7}$, $S_{2}S_{3}S_{5}S_{8}S_{19}$, $S_{2}S_{3}S_{3}S_{5}S_{7}$, $S_{1}S_{2}S_{8}S_{11}S_{31}$, $S_{1}S_{1}S_{2}S_{7}S_{11}$, $S_{1}S_{2}S_{9}S_{11}S_{31}$, $S_{2}S_{3}S_{5}S_{9}S_{19}$, $S_{2}S_{3}S_{4}S_{5}S_{9}$, $S_{1}S_{2}S_{4}S_{9}S_{11}$} \\
   \hline
  $\mathcal{O}_{12,\;(1,1,2)}^{105}$ & {$S_{1}S_{3}S_{4}S_{9}S_{11}$, $S_{1}S_{3}S_{4}S_{5}S_{9}$} \\
   \hline
  $\mathcal{O}_{12,\;(1,1,2)}^{106}$ & {$S_{1}S_{3}S_{4}S_{5}S_{9}$, $S_{1}S_{3}S_{4}S_{9}S_{11}$} \\
   \hline
  $\mathcal{O}_{12,\;(1,1,2)}^{107}$ & {$S_{2}S_{3}S_{5}S_{9}S_{19}$, $S_{1}S_{2}S_{9}S_{11}S_{31}$, $S_{1}S_{2}S_{4}S_{9}S_{11}$, $S_{2}S_{3}S_{4}S_{5}S_{9}$, $S_{2}S_{3}S_{4}S_{5}S_{8}$, $S_{2}S_{2}S_{3}S_{5}S_{7}$, $S_{1}S_{2}S_{4}S_{8}S_{11}$, $S_{1}S_{2}S_{2}S_{7}S_{11}$, $S_{1}S_{2}S_{8}S_{11}S_{31}$, $S_{1}S_{1}S_{2}S_{7}S_{11}$, $S_{2}S_{3}S_{5}S_{8}S_{19}$, $S_{2}S_{3}S_{3}S_{5}S_{7}$, $S_{2}S_{3}S_{6}S_{9}S_{19}$, $S_{1}S_{2}S_{9}S_{10}S_{31}$, $S_{1}S_{2}S_{4}S_{5}S_{9}$, $S_{2}S_{3}S_{4}S_{9}S_{11}$} \\
\hline
 \end{longtable}
 \end{center}
 \begin{center}
 \begin{longtable}{| c  | m{13cm} |} 
   \hline
  $\mathcal{O}^j{(\mathcal{T}_ {10} ^{ 18 })}$ &  New Particles \\
   \hline
   \hline
   \endhead
 \multicolumn{2}{|c|}{\hspace{9cm} Continued on next page}
   \endfoot
  $\mathcal{O}_{12,\;(1,1,2)}^{108}$ & {$S_{3}S_{4}S_{9}S_{11}F_{3}$, $S_{3}S_{4}S_{9}S_{11}F_{7}$, $S_{2}S_{6}S_{9}S_{19}F_{7}$, $S_{3}S_{5}S_{8}S_{19}F_{3}$, $S_{3}S_{3}S_{5}S_{7}F_{3}$, $S_{3}S_{5}S_{8}S_{19}F_{8}$, $S_{3}S_{3}S_{5}S_{7}F_{8}$, $S_{2}S_{4}S_{5}S_{8}F_{8}$, $S_{2}S_{2}S_{5}S_{7}F_{8}$, $S_{3}S_{4}S_{5}S_{9}F_{3}$, $S_{3}S_{4}S_{5}S_{9}F_{8}$, $S_{2}S_{5}S_{9}S_{19}F_{8}$} \\
   \hline
 \caption{Dimension 12 UV completions for topology $\mathcal{T}_ {10} ^{ 18 }$ (see Fig.~\ref{fig:d=10_topologies}).} 
 \label{tab:T12Top18B1L1D2}\\
   \endlastfoot
  $\mathcal{O}_{12,\;(1,1,2)}^{103}$ & {$S_{1}S_{4}S_{9}S_{11}F_{10}$, $S_{1}S_{4}S_{9}S_{11}F_{8}$, $S_{2}S_{9}S_{11}S_{31}F_{8}$, $S_{1}S_{8}S_{11}S_{31}F_{10}$, $S_{1}S_{1}S_{7}S_{11}F_{10}$, $S_{1}S_{8}S_{11}S_{31}F_{8}$, $S_{1}S_{1}S_{7}S_{11}F_{8}$, $S_{2}S_{4}S_{8}S_{11}F_{8}$, $S_{2}S_{2}S_{7}S_{11}F_{8}$, $S_{1}S_{4}S_{5}S_{9}F_{3}$, $S_{1}S_{4}S_{5}S_{9}F_{13}$, $S_{2}S_{9}S_{10}S_{31}F_{13}$} \\
   \hline
  $\mathcal{O}_{12,\;(1,1,2)}^{104}$ & {$S_{2}S_{9}S_{10}S_{31}F_{10}$, $S_{2}S_{9}S_{10}S_{31}F_{7}$, $S_{2}S_{6}S_{9}S_{19}F_{7}$, $S_{3}S_{4}S_{9}S_{11}F_{10}$, $S_{3}S_{4}S_{9}S_{11}F_{8}$, $S_{1}S_{4}S_{5}S_{9}F_{7}$, $S_{1}S_{4}S_{5}S_{9}F_{8}$, $S_{2}S_{4}S_{8}S_{11}F_{3}$, $S_{2}S_{2}S_{7}S_{11}F_{3}$, $S_{2}S_{4}S_{8}S_{11}F_{8}$, $S_{2}S_{2}S_{7}S_{11}F_{8}$, $S_{2}S_{4}S_{5}S_{8}F_{8}$, $S_{2}S_{2}S_{5}S_{7}F_{8}$, $S_{3}S_{5}S_{8}S_{19}F_{3}$, $S_{3}S_{3}S_{5}S_{7}F_{3}$, $S_{3}S_{5}S_{8}S_{19}F_{13}$, $S_{3}S_{3}S_{5}S_{7}F_{13}$, $S_{1}S_{8}S_{11}S_{31}F_{8}$, $S_{1}S_{1}S_{7}S_{11}F_{8}$, $S_{1}S_{8}S_{11}S_{31}F_{13}$, $S_{1}S_{1}S_{7}S_{11}F_{13}$, $S_{2}S_{9}S_{11}S_{31}F_{3}$, $S_{2}S_{9}S_{11}S_{31}F_{8}$, $S_{2}S_{5}S_{9}S_{19}F_{8}$, $S_{3}S_{4}S_{5}S_{9}F_{3}$, $S_{3}S_{4}S_{5}S_{9}F_{13}$, $S_{1}S_{4}S_{9}S_{11}F_{8}$, $S_{1}S_{4}S_{9}S_{11}F_{13}$} \\
   \hline
  $\mathcal{O}_{12,\;(1,1,2)}^{105}$ & {$S_{3}S_{4}S_{9}S_{11}F_{3}$, $S_{3}S_{4}S_{9}S_{11}F_{8}$, $S_{1}S_{4}S_{5}S_{9}F_{8}$} \\
   \hline
  $\mathcal{O}_{12,\;(1,1,2)}^{106}$ & {$S_{1}S_{4}S_{5}S_{9}F_{3}$, $S_{1}S_{4}S_{5}S_{9}F_{8}$, $S_{3}S_{4}S_{9}S_{11}F_{8}$} \\
\hline
 \end{longtable}
 \end{center}
 \begin{center}
 \begin{longtable}{| c  | m{13cm} |} 
   \hline
  $\mathcal{O}^j{(\mathcal{T}_ {10} ^{ 19 })}$ &  New Particles \\
   \hline
   \hline
   \endhead
 \multicolumn{2}{|c|}{\hspace{9cm} Continued on next page}
   \endfoot
  $\mathcal{O}_{12,\;(1,1,2)}^{108}$ & {$S_{3}S_{4}S_{5}S_{11}F_{3}$, $S_{2}S_{3}S_{5}S_{11}F_{3}$, $S_{2}S_{5}S_{6}S_{19}F_{7}$, $S_{2}S_{3}S_{5}S_{6}F_{7}$, $S_{3}S_{4}S_{5}S_{11}F_{7}$, $S_{2}S_{3}S_{5}S_{11}F_{7}$, $S_{3}S_{5}S_{6}S_{19}F_{3}$, $S_{3}S_{3}S_{5}S_{6}F_{3}$, $S_{2}S_{4}S_{5}S_{11}F_{8}$, $S_{2}S_{2}S_{5}S_{11}F_{8}$, $S_{3}S_{5}S_{6}S_{19}F_{8}$, $S_{3}S_{3}S_{5}S_{6}F_{8}$, $S_{3}S_{4}S_{5}S_{5}F_{3}$, $S_{2}S_{3}S_{5}S_{5}F_{3}$, $S_{2}S_{5}S_{5}S_{19}F_{8}$, $S_{2}S_{3}S_{5}S_{5}F_{8}$, $S_{3}S_{4}S_{5}S_{5}F_{8}$, $S_{3}S_{5}S_{5}S_{19}F_{3}$, $S_{3}S_{3}S_{5}S_{5}F_{3}$, $S_{2}S_{4}S_{5}S_{5}F_{8}$, $S_{2}S_{2}S_{5}S_{5}F_{8}$, $S_{3}S_{5}S_{5}S_{19}F_{8}$, $S_{3}S_{3}S_{5}S_{5}F_{8}$} \\
   \hline
 \caption{Dimension 12 UV completions for topology $\mathcal{T}_ {10} ^{ 19 }$ (see Fig.~\ref{fig:d=10_topologies}).} 
 \label{tab:T12Top19B1L1D2}\\
   \endlastfoot
  $\mathcal{O}_{12,\;(1,1,2)}^{103}$ & {$S_{1}S_{11}S_{11}S_{31}F_{10}$, $S_{1}S_{1}S_{11}S_{11}F_{10}$, $S_{2}S_{4}S_{11}S_{11}F_{8}$, $S_{2}S_{2}S_{11}S_{11}F_{8}$, $S_{1}S_{11}S_{11}S_{31}F_{8}$, $S_{1}S_{1}S_{11}S_{11}F_{8}$, $S_{1}S_{4}S_{11}S_{11}F_{10}$, $S_{1}S_{2}S_{11}S_{11}F_{10}$, $S_{2}S_{11}S_{11}S_{31}F_{8}$, $S_{1}S_{2}S_{11}S_{11}F_{8}$, $S_{1}S_{4}S_{11}S_{11}F_{8}$, $S_{1}S_{10}S_{11}S_{31}F_{10}$, $S_{1}S_{1}S_{10}S_{11}F_{10}$, $S_{2}S_{4}S_{5}S_{11}F_{8}$, $S_{2}S_{2}S_{5}S_{11}F_{8}$, $S_{1}S_{10}S_{11}S_{31}F_{8}$, $S_{1}S_{1}S_{10}S_{11}F_{8}$, $S_{1}S_{4}S_{5}S_{11}F_{3}$, $S_{1}S_{2}S_{5}S_{11}F_{3}$, $S_{2}S_{10}S_{11}S_{31}F_{13}$, $S_{1}S_{2}S_{10}S_{11}F_{13}$, $S_{1}S_{4}S_{5}S_{11}F_{13}$, $S_{1}S_{2}S_{5}S_{11}F_{13}$} \\
\hline
 \end{longtable}
 \end{center}
 \begin{center}
 \begin{longtable}{| c  | m{13cm} |} 
   \hline
  $\mathcal{O}^j{(\mathcal{T}_ {10} ^{ 20 })}$ &  New Particles \\
   \hline
   \hline
   \endhead
 \multicolumn{2}{|c|}{\hspace{9cm} Continued on next page}
   \endfoot
  $\mathcal{O}_{12,\;(1,1,2)}^{108}$ & {$S_{2}S_{4}S_{11}F_{8}F_{8}$, $S_{2}S_{2}S_{11}F_{8}F_{8}$, $S_{3}S_{6}S_{19}F_{3}F_{8}$, $S_{3}S_{3}S_{6}F_{3}F_{8}$, $S_{2}S_{5}S_{19}F_{7}F_{8}$, $S_{2}S_{3}S_{5}F_{7}F_{8}$, $S_{3}S_{4}S_{5}F_{3}F_{8}$, $S_{2}S_{3}S_{5}F_{3}F_{8}$, $S_{3}S_{4}S_{5}F_{3}F_{7}$, $S_{2}S_{3}S_{5}F_{3}F_{7}$, $S_{2}S_{4}S_{5}F_{8}F_{8}$, $S_{2}S_{2}S_{5}F_{8}F_{8}$, $S_{3}S_{5}S_{19}F_{3}F_{8}$, $S_{3}S_{3}S_{5}F_{3}F_{8}$} \\
   \hline
 \caption{Dimension 12 UV completions for topology $\mathcal{T}_ {10} ^{ 20 }$ (see Fig.~\ref{fig:d=10_topologies}).} 
 \label{tab:T12Top20B1L1D2}\\
   \endlastfoot
  $\mathcal{O}_{12,\;(1,1,2)}^{103}$ & {$S_{2}S_{4}S_{11}F_{8}F_{8}$, $S_{2}S_{2}S_{11}F_{8}F_{8}$, $S_{1}S_{11}S_{31}F_{8}F_{10}$, $S_{1}S_{1}S_{11}F_{8}F_{10}$, $S_{2}S_{11}S_{31}F_{8}F_{13}$, $S_{1}S_{2}S_{11}F_{8}F_{13}$, $S_{1}S_{4}S_{11}F_{10}F_{13}$, $S_{1}S_{2}S_{11}F_{10}F_{13}$, $S_{1}S_{4}S_{11}F_{3}F_{8}$, $S_{1}S_{2}S_{11}F_{3}F_{8}$, $S_{2}S_{4}S_{5}F_{8}F_{8}$, $S_{2}S_{2}S_{5}F_{8}F_{8}$, $S_{1}S_{10}S_{31}F_{8}F_{10}$, $S_{1}S_{1}S_{10}F_{8}F_{10}$} \\
   \hline
  $\mathcal{O}_{12,\;(1,1,2)}^{104}$ & {$S_{1}S_{10}S_{31}F_{8}F_{13}$, $S_{1}S_{1}S_{10}F_{8}F_{13}$, $S_{3}S_{6}S_{19}F_{3}F_{13}$, $S_{3}S_{3}S_{6}F_{3}F_{13}$, $S_{2}S_{4}S_{11}F_{8}F_{8}$, $S_{2}S_{2}S_{11}F_{8}F_{8}$, $S_{2}S_{4}S_{5}F_{3}F_{8}$, $S_{2}S_{2}S_{5}F_{3}F_{8}$, $S_{1}S_{4}S_{11}F_{7}F_{13}$, $S_{1}S_{2}S_{11}F_{7}F_{13}$, $S_{3}S_{4}S_{5}F_{10}F_{13}$, $S_{2}S_{3}S_{5}F_{10}F_{13}$, $S_{1}S_{4}S_{11}F_{8}F_{8}$, $S_{1}S_{2}S_{11}F_{8}F_{8}$, $S_{2}S_{5}S_{19}F_{7}F_{8}$, $S_{2}S_{3}S_{5}F_{7}F_{8}$, $S_{2}S_{11}S_{31}F_{8}F_{10}$, $S_{1}S_{2}S_{11}F_{8}F_{10}$, $S_{3}S_{4}S_{5}F_{3}F_{8}$, $S_{2}S_{3}S_{5}F_{3}F_{8}$, $S_{2}S_{11}S_{31}F_{3}F_{7}$, $S_{1}S_{2}S_{11}F_{3}F_{7}$, $S_{1}S_{11}S_{31}F_{8}F_{13}$, $S_{1}S_{1}S_{11}F_{8}F_{13}$, $S_{3}S_{5}S_{19}F_{3}F_{13}$, $S_{3}S_{3}S_{5}F_{3}F_{13}$, $S_{2}S_{4}S_{5}F_{8}F_{8}$, $S_{2}S_{2}S_{5}F_{8}F_{8}$, $S_{2}S_{4}S_{11}F_{3}F_{8}$, $S_{2}S_{2}S_{11}F_{3}F_{8}$} \\
   \hline
  $\mathcal{O}_{12,\;(1,1,2)}^{105}$ & {$S_{1}S_{4}S_{11}F_{8}F_{8}$, $S_{1}S_{2}S_{11}F_{8}F_{8}$, $S_{3}S_{4}S_{5}F_{3}F_{8}$, $S_{2}S_{3}S_{5}F_{3}F_{8}$} \\
   \hline
  $\mathcal{O}_{12,\;(1,1,2)}^{106}$ & {$S_{3}S_{4}S_{5}F_{8}F_{8}$, $S_{2}S_{3}S_{5}F_{8}F_{8}$, $S_{1}S_{4}S_{11}F_{3}F_{8}$, $S_{1}S_{2}S_{11}F_{3}F_{8}$} \\
\hline
 \end{longtable}
 \end{center}
 \begin{center}
 \begin{longtable}{| c  | m{13cm} |} 
   \hline
  $\mathcal{O}^j{(\mathcal{T}_ {10} ^{ 21 })}$ &  New Particles \\
   \hline
   \hline
   \endhead
 \multicolumn{2}{|c|}{\hspace{9cm} Continued on next page}
   \endfoot
  $\mathcal{O}_{12,\;(1,1,2)}^{108}$ & {$S_{4}S_{11}F_{3}F_{8}F_{8}$, $S_{2}S_{11}F_{3}F_{8}F_{8}$, $S_{4}S_{11}F_{7}F_{8}F_{8}$, $S_{2}S_{11}F_{7}F_{8}F_{8}$, $S_{6}S_{19}F_{3}F_{7}F_{8}$, $S_{3}S_{6}F_{3}F_{7}F_{8}$, $S_{5}S_{19}F_{3}F_{7}F_{8}$, $S_{3}S_{5}F_{3}F_{7}F_{8}$, $S_{5}S_{19}F_{7}F_{8}F_{8}$, $S_{3}S_{5}F_{7}F_{8}F_{8}$, $S_{4}S_{5}F_{3}F_{8}F_{8}$, $S_{2}S_{5}F_{3}F_{8}F_{8}$, $S_{4}S_{5}F_{3}F_{7}F_{8}$, $S_{2}S_{5}F_{3}F_{7}F_{8}$, $S_{4}S_{5}F_{8}F_{8}F_{8}$, $S_{2}S_{5}F_{8}F_{8}F_{8}$, $S_{5}S_{19}F_{3}F_{8}F_{8}$, $S_{3}S_{5}F_{3}F_{8}F_{8}$} \\
   \hline
 \caption{Dimension 12 UV completions for topology $\mathcal{T}_ {10} ^{ 21 }$ (see Fig.~\ref{fig:d=10_topologies}).} 
 \label{tab:T12Top21B1L1D2}\\
   \endlastfoot
  $\mathcal{O}_{12,\;(1,1,2)}^{103}$ & {$S_{4}S_{11}F_{8}F_{8}F_{10}$, $S_{2}S_{11}F_{8}F_{8}F_{10}$, $S_{4}S_{11}F_{8}F_{8}F_{8}$, $S_{2}S_{11}F_{8}F_{8}F_{8}$, $S_{11}S_{31}F_{8}F_{8}F_{10}$, $S_{1}S_{11}F_{8}F_{8}F_{10}$, $S_{11}S_{31}F_{8}F_{10}F_{13}$, $S_{1}S_{11}F_{8}F_{10}F_{13}$, $S_{11}S_{31}F_{8}F_{8}F_{13}$, $S_{1}S_{11}F_{8}F_{8}F_{13}$, $S_{4}S_{11}F_{8}F_{10}F_{13}$, $S_{2}S_{11}F_{8}F_{10}F_{13}$, $S_{4}S_{11}F_{3}F_{8}F_{8}$, $S_{2}S_{11}F_{3}F_{8}F_{8}$, $S_{4}S_{5}F_{3}F_{8}F_{8}$, $S_{2}S_{5}F_{3}F_{8}F_{8}$, $S_{4}S_{5}F_{8}F_{8}F_{13}$, $S_{2}S_{5}F_{8}F_{8}F_{13}$, $S_{10}S_{31}F_{8}F_{10}F_{13}$, $S_{1}S_{10}F_{8}F_{10}F_{13}$} \\
\hline
 \end{longtable}
 \end{center}
 \begin{center}
 \begin{longtable}{| c  | m{13cm} |} 
   \hline
  $\mathcal{O}^j{(\mathcal{T}_ {10} ^{ 22 })}$ &  New Particles \\
   \hline
   \hline
   \endhead
 \multicolumn{2}{|c|}{\hspace{9cm} Continued on next page}
   \endfoot
  $\mathcal{O}_{12,\;(1,1,2)}^{108}$ & {$S_{2}S_{3}S_{4}S_{9}$, $S_{2}S_{3}S_{9}S_{19}$, $S_{2}S_{3}S_{8}S_{19}$, $S_{2}S_{3}S_{4}S_{8}$} \\
   \hline
 \caption{Dimension 12 UV completions for topology $\mathcal{T}_ {10} ^{ 22 }$ (see Fig.~\ref{fig:d=10_topologies}).} 
 \label{tab:T12Top22B1L1D2}\\
   \endlastfoot
  $\mathcal{O}_{12,\;(1,1,2)}^{103}$ & {$S_{1}S_{2}S_{8}S_{31}$, $S_{1}S_{2}S_{4}S_{8}$, $S_{1}S_{2}S_{4}S_{9}$, $S_{1}S_{2}S_{9}S_{31}$} \\
   \hline
  $\mathcal{O}_{12,\;(1,1,2)}^{104}$ & {$S_{1}S_{2}S_{9}S_{31}$, $S_{2}S_{3}S_{9}S_{19}$, $S_{2}S_{3}S_{4}S_{9}$, $S_{1}S_{2}S_{4}S_{9}$, $S_{1}S_{2}S_{4}S_{8}$, $S_{2}S_{3}S_{4}S_{8}$, $S_{2}S_{3}S_{8}S_{19}$, $S_{1}S_{2}S_{8}S_{31}$} \\
   \hline
  $\mathcal{O}_{12,\;(1,1,2)}^{105}$ & {$S_{1}S_{3}S_{4}S_{9}$} \\
   \hline
  $\mathcal{O}_{12,\;(1,1,2)}^{106}$ & {$S_{1}S_{3}S_{4}S_{9}$} \\
   \hline
  $\mathcal{O}_{12,\;(1,1,2)}^{107}$ & {$S_{2}S_{3}S_{4}S_{8}$, $S_{1}S_{2}S_{4}S_{8}$, $S_{1}S_{2}S_{8}S_{31}$, $S_{2}S_{3}S_{8}S_{19}$, $S_{2}S_{3}S_{9}S_{19}$, $S_{1}S_{2}S_{9}S_{31}$, $S_{1}S_{2}S_{4}S_{9}$, $S_{2}S_{3}S_{4}S_{9}$} \\
   \hline
 \end{longtable}
 \end{center}
 \begin{center}
 \begin{longtable}{| c  | m{13cm} |} 
   \hline
  $\mathcal{O}^j{(\mathcal{T}_ {10} ^{ 23 })}$ &  New Particles \\
   \hline
   \hline
   \endhead
 \multicolumn{2}{|c|}{\hspace{9cm} Continued on next page}
   \endfoot
  $\mathcal{O}_{12,\;(1,1,2)}^{108}$ & {$S_{2}S_{3}S_{4}S_{5}$, $S_{2}S_{2}S_{3}S_{5}$, $S_{2}S_{3}S_{5}S_{19}$, $S_{2}S_{3}S_{3}S_{5}$, $S_{2}S_{3}S_{6}S_{19}$, $S_{2}S_{3}S_{4}S_{11}$} \\
   \hline
 \caption{Dimension 12 UV completions for topology $\mathcal{T}_ {10} ^{ 23 }$ (see Fig.~\ref{fig:d=10_topologies}).} 
 \label{tab:T12Top23B1L1D2}\\
   \endlastfoot
  $\mathcal{O}_{12,\;(1,1,2)}^{103}$ & {$S_{1}S_{2}S_{10}S_{31}$, $S_{1}S_{2}S_{4}S_{5}$, $S_{1}S_{2}S_{11}S_{31}$, $S_{1}S_{2}S_{4}S_{11}$, $S_{1}S_{2}S_{2}S_{11}$, $S_{1}S_{1}S_{2}S_{11}$} \\
   \hline
  $\mathcal{O}_{12,\;(1,1,2)}^{104}$ & {$S_{1}S_{2}S_{11}S_{31}$, $S_{1}S_{1}S_{2}S_{11}$, $S_{2}S_{3}S_{5}S_{19}$, $S_{2}S_{3}S_{3}S_{5}$, $S_{2}S_{3}S_{4}S_{5}$, $S_{2}S_{2}S_{3}S_{5}$, $S_{1}S_{2}S_{4}S_{11}$, $S_{1}S_{2}S_{2}S_{11}$, $S_{1}S_{2}S_{4}S_{5}$, $S_{2}S_{3}S_{4}S_{11}$, $S_{2}S_{3}S_{6}S_{19}$, $S_{1}S_{2}S_{10}S_{31}$} \\
   \hline
  $\mathcal{O}_{12,\;(1,1,2)}^{105}$ & {$S_{1}S_{3}S_{4}S_{5}$, $S_{1}S_{3}S_{4}S_{11}$} \\
   \hline
  $\mathcal{O}_{12,\;(1,1,2)}^{106}$ & {$S_{1}S_{3}S_{4}S_{11}$, $S_{1}S_{3}S_{4}S_{5}$} \\
   \hline
  $\mathcal{O}_{12,\;(1,1,2)}^{107}$ & {$S_{2}S_{3}S_{4}S_{11}$, $S_{1}S_{2}S_{4}S_{5}$, $S_{1}S_{2}S_{10}S_{31}$, $S_{2}S_{3}S_{6}S_{19}$, $S_{2}S_{3}S_{4}S_{5}$, $S_{1}S_{2}S_{4}S_{11}$, $S_{1}S_{2}S_{11}S_{31}$, $S_{2}S_{3}S_{5}S_{19}$, $S_{2}S_{3}S_{3}S_{5}$, $S_{1}S_{1}S_{2}S_{11}$, $S_{1}S_{2}S_{2}S_{11}$, $S_{2}S_{2}S_{3}S_{5}$} \\
   \hline
 \end{longtable}
 \end{center}
 \begin{center}
 \begin{longtable}{| c  | m{13cm} |} 
   \hline
  $\mathcal{O}^j{(\mathcal{T}_ {10} ^{ 24 })}$ &  New Particles \\
   \hline
   \hline
   \endhead
 \multicolumn{2}{|c|}{\hspace{9cm} Continued on next page}
   \endfoot
  $\mathcal{O}_{12,\;(1,1,2)}^{108}$ & {$S_{2}S_{3}S_{4}S_{8}S_{9}$, $S_{2}S_{3}S_{8}S_{9}S_{19}$} \\
   \hline
 \caption{Dimension 12 UV completions for topology $\mathcal{T}_ {10} ^{ 24 }$ (see Fig.~\ref{fig:d=10_topologies}).} 
 \label{tab:T12Top24B1L1D2}\\
   \endlastfoot
  $\mathcal{O}_{12,\;(1,1,2)}^{103}$ & {$S_{1}S_{2}S_{8}S_{9}S_{31}$, $S_{1}S_{2}S_{4}S_{8}S_{9}$} \\
   \hline
  $\mathcal{O}_{12,\;(1,1,2)}^{104}$ & {$S_{1}S_{2}S_{8}S_{9}S_{31}$, $S_{2}S_{3}S_{8}S_{9}S_{19}$, $S_{2}S_{3}S_{4}S_{8}S_{9}$, $S_{1}S_{2}S_{4}S_{8}S_{9}$} \\
   \hline
  $\mathcal{O}_{12,\;(1,1,2)}^{105}$ & {$S_{1}S_{3}S_{4}S_{9}S_{9}$} \\
   \hline
  $\mathcal{O}_{12,\;(1,1,2)}^{106}$ & {$S_{1}S_{3}S_{4}S_{9}S_{9}$} \\
   \hline
  $\mathcal{O}_{12,\;(1,1,2)}^{107}$ & {$S_{2}S_{3}S_{4}S_{8}S_{9}$, $S_{1}S_{2}S_{4}S_{8}S_{9}$, $S_{1}S_{2}S_{8}S_{9}S_{31}$, $S_{2}S_{3}S_{8}S_{9}S_{19}$} \\
   \hline
 \end{longtable}
 \end{center}
 \begin{center}
 \begin{longtable}{| c  | m{13cm} |} 
   \hline
  $\mathcal{O}^j{(\mathcal{T}_ {10} ^{ 25 })}$ &  New Particles \\
   \hline
   \hline
   \endhead
 \multicolumn{2}{|c|}{\hspace{9cm} Continued on next page}
   \endfoot
  $\mathcal{O}_{12,\;(1,1,2)}^{108}$ & {$S_{2}S_{3}S_{5}S_{8}S_{16}$, $S_{2}S_{3}S_{5}S_{7}S_{16}$, $S_{2}S_{3}S_{9}S_{11}S_{26}$, $S_{2}S_{3}S_{6}S_{9}S_{26}$, $S_{2}S_{3}S_{5}S_{9}S_{16}$} \\
   \hline
 \caption{Dimension 12 UV completions for topology $\mathcal{T}_ {10} ^{ 25 }$ (see Fig.~\ref{fig:d=10_topologies}).} 
 \label{tab:T12Top25B1L1D2}\\
   \endlastfoot
  $\mathcal{O}_{12,\;(1,1,2)}^{103}$ & {$S_{1}S_{2}S_{9}S_{11}S_{16}$, $S_{1}S_{2}S_{5}S_{9}S_{26}$, $S_{1}S_{2}S_{9}S_{10}S_{26}$, $S_{1}S_{2}S_{8}S_{11}S_{16}$, $S_{1}S_{2}S_{7}S_{11}S_{16}$} \\
   \hline
  $\mathcal{O}_{12,\;(1,1,2)}^{104}$ & {$S_{1}S_{2}S_{8}S_{11}S_{16}$, $S_{1}S_{2}S_{7}S_{11}S_{16}$, $S_{2}S_{3}S_{5}S_{8}S_{16}$, $S_{2}S_{3}S_{5}S_{7}S_{16}$, $S_{1}S_{2}S_{9}S_{10}S_{26}$, $S_{2}S_{3}S_{6}S_{9}S_{26}$, $S_{2}S_{3}S_{9}S_{11}S_{26}$, $S_{1}S_{2}S_{5}S_{9}S_{26}$, $S_{1}S_{2}S_{9}S_{11}S_{16}$, $S_{2}S_{3}S_{5}S_{9}S_{16}$} \\
   \hline
  $\mathcal{O}_{12,\;(1,1,2)}^{105}$ & {$S_{1}S_{3}S_{9}S_{11}S_{26}$, $S_{1}S_{3}S_{5}S_{9}S_{26}$} \\
   \hline
  $\mathcal{O}_{12,\;(1,1,2)}^{106}$ & {$S_{1}S_{3}S_{5}S_{9}S_{26}$, $S_{1}S_{3}S_{9}S_{11}S_{26}$} \\
   \hline
  $\mathcal{O}_{12,\;(1,1,2)}^{107}$ & {$S_{2}S_{3}S_{5}S_{9}S_{16}$, $S_{1}S_{2}S_{9}S_{11}S_{16}$, $S_{2}S_{3}S_{6}S_{9}S_{26}$, $S_{1}S_{2}S_{9}S_{10}S_{26}$, $S_{1}S_{2}S_{5}S_{9}S_{26}$, $S_{2}S_{3}S_{9}S_{11}S_{26}$, $S_{2}S_{3}S_{5}S_{8}S_{16}$, $S_{2}S_{3}S_{5}S_{7}S_{16}$, $S_{1}S_{2}S_{8}S_{11}S_{16}$, $S_{1}S_{2}S_{7}S_{11}S_{16}$} \\
   \hline
 \end{longtable}
 \end{center}
 \begin{center}
 \begin{longtable}{| c  | m{13cm} |} 
   \hline
  $\mathcal{O}^j{(\mathcal{T}_ {10} ^{ 26 })}$ &  New Particles \\
   \hline
   \hline
   \endhead
 \multicolumn{2}{|c|}{\hspace{9cm} Continued on next page}
   \endfoot
  $\mathcal{O}_{12,\;(1,1,2)}^{108}$ & {$S_{2}S_{8}F_{6}F_{8}F_{9}$, $S_{2}S_{7}F_{5}F_{6}F_{8}$, $S_{3}S_{8}F_{6}F_{8}F_{9}$, $S_{3}S_{7}F_{5}F_{6}F_{8}$, $S_{3}S_{8}F_{2}F_{3}F_{4}$, $S_{3}S_{7}F_{1}F_{2}F_{3}$, $S_{2}S_{9}F_{6}F_{13}F_{14}$, $S_{3}S_{9}F_{6}F_{13}F_{14}$, $S_{3}S_{9}F_{2}F_{10}F_{11}$, $S_{2}S_{9}F_{6}F_{8}F_{14}$, $S_{3}S_{9}F_{6}F_{8}F_{14}$, $S_{3}S_{9}F_{2}F_{3}F_{11}$} \\
   \hline
 \caption{Dimension 12 UV completions for topology $\mathcal{T}_ {10} ^{ 26 }$ (see Fig.~\ref{fig:d=10_topologies}).} 
 \label{tab:T12Top26B1L1D2}\\
   \endlastfoot
  $\mathcal{O}_{12,\;(1,1,2)}^{103}$ & {$S_{2}S_{9}F_{5}F_{8}F_{9}$, $S_{1}S_{9}F_{5}F_{8}F_{9}$, $S_{1}S_{9}F_{10}F_{16}F_{31}$, $S_{2}S_{9}F_{5}F_{7}F_{9}$, $S_{1}S_{9}F_{5}F_{7}F_{9}$, $S_{1}S_{9}F_{16}F_{31}F_{49}$, $S_{2}S_{8}F_{5}F_{8}F_{14}$, $S_{2}S_{7}F_{5}F_{6}F_{8}$, $S_{1}S_{8}F_{5}F_{8}F_{14}$, $S_{1}S_{7}F_{5}F_{6}F_{8}$, $S_{1}S_{8}F_{10}F_{11}F_{16}$, $S_{1}S_{7}F_{2}F_{10}F_{16}$} \\
   \hline
  $\mathcal{O}_{12,\;(1,1,2)}^{104}$ & {$S_{1}S_{8}F_{13}F_{14}F_{18}$, $S_{1}S_{7}F_{6}F_{13}F_{18}$, $S_{3}S_{8}F_{13}F_{14}F_{18}$, $S_{3}S_{7}F_{6}F_{13}F_{18}$, $S_{1}S_{8}F_{6}F_{8}F_{9}$, $S_{1}S_{7}F_{5}F_{6}F_{8}$, $S_{2}S_{8}F_{6}F_{8}F_{9}$, $S_{2}S_{7}F_{5}F_{6}F_{8}$, $S_{3}S_{8}F_{1}F_{3}F_{11}$, $S_{3}S_{7}F_{1}F_{2}F_{3}$, $S_{2}S_{8}F_{1}F_{3}F_{11}$, $S_{2}S_{7}F_{1}F_{2}F_{3}$, $S_{1}S_{9}F_{18}F_{37}F_{50}$, $S_{3}S_{9}F_{18}F_{37}F_{50}$, $S_{1}S_{9}F_{6}F_{13}F_{14}$, $S_{2}S_{9}F_{6}F_{13}F_{14}$, $S_{3}S_{9}F_{1}F_{3}F_{4}$, $S_{2}S_{9}F_{1}F_{3}F_{4}$, $S_{1}S_{9}F_{13}F_{18}F_{37}$, $S_{3}S_{9}F_{13}F_{18}F_{37}$, $S_{1}S_{9}F_{6}F_{8}F_{14}$, $S_{2}S_{9}F_{6}F_{8}F_{14}$} \\
   \hline
  $\mathcal{O}_{12,\;(1,1,2)}^{105}$ & {$S_{1}S_{9}F_{13}F_{14}F_{18}$, $S_{3}S_{9}F_{13}F_{14}F_{18}$, $S_{3}S_{9}F_{2}F_{3}F_{4}$} \\
   \hline
  $\mathcal{O}_{12,\;(1,1,2)}^{106}$ & {$S_{3}S_{9}F_{7}F_{9}F_{17}$, $S_{1}S_{9}F_{7}F_{9}F_{17}$, $S_{1}S_{9}F_{10}F_{11}F_{16}$} \\
\hline
 \end{longtable}
 \end{center}
 \begin{center}
 \begin{longtable}{| c  | m{13cm} |} 
   \hline
  $\mathcal{O}^j{(\mathcal{T}_ {10} ^{ 27 })}$ &  New Particles \\
   \hline
   \hline
   \endhead
 \multicolumn{2}{|c|}{\hspace{9cm} Continued on next page}
   \endfoot
  $\mathcal{O}_{12,\;(1,1,2)}^{108}$ & {$S_{3}S_{4}S_{9}F_{3}F_{8}$, $S_{3}S_{4}S_{9}F_{3}F_{7}$, $S_{2}S_{9}S_{19}F_{7}F_{8}$, $S_{3}S_{8}S_{19}F_{3}F_{8}$, $S_{3}S_{3}S_{7}F_{3}F_{8}$, $S_{2}S_{4}S_{8}F_{8}F_{8}$, $S_{2}S_{2}S_{7}F_{8}F_{8}$} \\
   \hline
 \caption{Dimension 12 UV completions for topology $\mathcal{T}_ {10} ^{ 27 }$ (see Fig.~\ref{fig:d=10_topologies}).} 
 \label{tab:T12Top27B1L1D2}\\
   \endlastfoot
  $\mathcal{O}_{12,\;(1,1,2)}^{103}$ & {$S_{1}S_{8}S_{31}F_{8}F_{10}$, $S_{1}S_{1}S_{7}F_{8}F_{10}$, $S_{2}S_{4}S_{8}F_{8}F_{8}$, $S_{2}S_{2}S_{7}F_{8}F_{8}$, $S_{1}S_{4}S_{9}F_{10}F_{13}$, $S_{1}S_{4}S_{9}F_{3}F_{8}$, $S_{2}S_{9}S_{31}F_{8}F_{13}$} \\
   \hline
  $\mathcal{O}_{12,\;(1,1,2)}^{104}$ & {$S_{2}S_{9}S_{31}F_{8}F_{10}$, $S_{2}S_{9}S_{31}F_{3}F_{7}$, $S_{2}S_{9}S_{19}F_{7}F_{8}$, $S_{3}S_{4}S_{9}F_{10}F_{13}$, $S_{3}S_{4}S_{9}F_{3}F_{8}$, $S_{1}S_{4}S_{9}F_{7}F_{13}$, $S_{1}S_{4}S_{9}F_{8}F_{8}$, $S_{2}S_{4}S_{8}F_{3}F_{8}$, $S_{2}S_{2}S_{7}F_{3}F_{8}$, $S_{2}S_{4}S_{8}F_{8}F_{8}$, $S_{2}S_{2}S_{7}F_{8}F_{8}$, $S_{3}S_{8}S_{19}F_{3}F_{13}$, $S_{3}S_{3}S_{7}F_{3}F_{13}$, $S_{1}S_{8}S_{31}F_{8}F_{13}$, $S_{1}S_{1}S_{7}F_{8}F_{13}$} \\
   \hline
  $\mathcal{O}_{12,\;(1,1,2)}^{105}$ & {$S_{3}S_{4}S_{9}F_{3}F_{8}$, $S_{1}S_{4}S_{9}F_{8}F_{8}$} \\
   \hline
  $\mathcal{O}_{12,\;(1,1,2)}^{106}$ & {$S_{1}S_{4}S_{9}F_{3}F_{8}$, $S_{3}S_{4}S_{9}F_{8}F_{8}$} \\
   \hline
  $\mathcal{O}_{12,\;(1,1,2)}^{107}$ & {$S_{2}S_{4}S_{8}F_{3}F_{8}$, $S_{2}S_{2}S_{7}F_{3}F_{8}$, $S_{2}S_{4}S_{8}F_{8}F_{8}$, $S_{2}S_{2}S_{7}F_{8}F_{8}$, $S_{1}S_{8}S_{31}F_{3}F_{7}$, $S_{1}S_{1}S_{7}F_{3}F_{7}$, $S_{3}S_{8}S_{19}F_{7}F_{8}$, $S_{3}S_{3}S_{7}F_{7}F_{8}$, $S_{2}S_{9}S_{19}F_{3}F_{13}$, $S_{2}S_{9}S_{19}F_{3}F_{8}$, $S_{2}S_{9}S_{31}F_{8}F_{13}$, $S_{1}S_{4}S_{9}F_{3}F_{8}$, $S_{1}S_{4}S_{9}F_{3}F_{7}$, $S_{3}S_{4}S_{9}F_{8}F_{8}$, $S_{3}S_{4}S_{9}F_{7}F_{13}$} \\
\hline
 \end{longtable}
 \end{center}
 \begin{center}
 \begin{longtable}{| c  | m{13cm} |} 
   \hline
  $\mathcal{O}^j{(\mathcal{T}_ {10} ^{ 28 })}$ &  New Particles \\
   \hline
   \hline
   \endhead
 \multicolumn{2}{|c|}{\hspace{9cm} Continued on next page}
   \endfoot
  $\mathcal{O}_{12,\;(1,1,2)}^{108}$ & {$S_{3}F_{3}F_{4}F_{8}F_{14}$, $S_{3}F_{1}F_{3}F_{6}F_{8}$, $S_{3}F_{3}F_{7}F_{9}F_{11}$, $S_{3}F_{2}F_{3}F_{5}F_{7}$, $S_{2}F_{7}F_{8}F_{9}F_{14}$, $S_{2}F_{5}F_{6}F_{7}F_{8}$, $S_{3}F_{3}F_{8}F_{9}F_{11}$, $S_{3}F_{2}F_{3}F_{5}F_{8}$, $S_{2}F_{8}F_{8}F_{9}F_{14}$, $S_{2}F_{5}F_{6}F_{8}F_{8}$} \\
   \hline
 \caption{Dimension 12 UV completions for topology $\mathcal{T}_ {10} ^{ 28 }$ (see Fig.~\ref{fig:d=10_topologies}).} 
 \label{tab:T12Top28B1L1D2}\\
   \endlastfoot
  $\mathcal{O}_{12,\;(1,1,2)}^{103}$ & {$S_{1}F_{8}F_{10}F_{14}F_{31}$, $S_{1}F_{6}F_{8}F_{10}F_{16}$, $S_{1}F_{8}F_{9}F_{10}F_{11}$, $S_{1}F_{2}F_{5}F_{8}F_{10}$, $S_{2}F_{8}F_{8}F_{9}F_{14}$, $S_{2}F_{5}F_{6}F_{8}F_{8}$, $S_{1}F_{10}F_{13}F_{14}F_{31}$, $S_{1}F_{6}F_{10}F_{13}F_{16}$, $S_{1}F_{3}F_{8}F_{9}F_{11}$, $S_{1}F_{2}F_{3}F_{5}F_{8}$, $S_{2}F_{8}F_{9}F_{13}F_{14}$, $S_{2}F_{5}F_{6}F_{8}F_{13}$} \\
   \hline
  $\mathcal{O}_{12,\;(1,1,2)}^{104}$ & {$S_{2}F_{8}F_{10}F_{11}F_{14}$, $S_{2}F_{2}F_{6}F_{8}F_{10}$, $S_{2}F_{3}F_{4}F_{7}F_{9}$, $S_{2}F_{1}F_{3}F_{5}F_{7}$, $S_{2}F_{7}F_{8}F_{9}F_{14}$, $S_{2}F_{5}F_{6}F_{7}F_{8}$, $S_{3}F_{10}F_{11}F_{13}F_{37}$, $S_{3}F_{2}F_{10}F_{13}F_{18}$, $S_{3}F_{3}F_{4}F_{8}F_{14}$, $S_{3}F_{1}F_{3}F_{6}F_{8}$, $S_{1}F_{7}F_{9}F_{13}F_{37}$, $S_{1}F_{5}F_{7}F_{13}F_{18}$, $S_{1}F_{8}F_{8}F_{14}F_{14}$, $S_{1}F_{6}F_{6}F_{8}F_{8}$, $S_{2}F_{3}F_{8}F_{11}F_{14}$, $S_{2}F_{2}F_{3}F_{6}F_{8}$, $S_{2}F_{3}F_{4}F_{8}F_{9}$, $S_{2}F_{1}F_{3}F_{5}F_{8}$, $S_{2}F_{8}F_{8}F_{9}F_{14}$, $S_{2}F_{5}F_{6}F_{8}F_{8}$, $S_{3}F_{3}F_{11}F_{13}F_{37}$, $S_{3}F_{2}F_{3}F_{13}F_{18}$, $S_{3}F_{3}F_{4}F_{13}F_{14}$, $S_{3}F_{1}F_{3}F_{6}F_{13}$, $S_{1}F_{8}F_{9}F_{13}F_{37}$, $S_{1}F_{5}F_{8}F_{13}F_{18}$, $S_{1}F_{8}F_{13}F_{14}F_{14}$, $S_{1}F_{6}F_{6}F_{8}F_{13}$} \\
   \hline
  $\mathcal{O}_{12,\;(1,1,2)}^{105}$ & {$S_{3}F_{3}F_{4}F_{8}F_{14}$, $S_{3}F_{1}F_{3}F_{6}F_{8}$, $S_{1}F_{8}F_{8}F_{14}F_{14}$, $S_{1}F_{6}F_{6}F_{8}F_{8}$} \\
   \hline
  $\mathcal{O}_{12,\;(1,1,2)}^{106}$ & {$S_{1}F_{3}F_{8}F_{9}F_{11}$, $S_{1}F_{2}F_{3}F_{5}F_{8}$, $S_{3}F_{8}F_{8}F_{9}F_{9}$, $S_{3}F_{5}F_{5}F_{8}F_{8}$} \\
\hline
 \end{longtable}
 \end{center}
 \begin{center}
 \begin{longtable}{| c  | m{13cm} |} 
   \hline
  $\mathcal{O}^j{(\mathcal{T}_ {10} ^{ 29 })}$ &  New Particles \\
   \hline
   \hline
   \endhead
 \multicolumn{2}{|c|}{\hspace{9cm} Continued on next page}
   \endfoot
  $\mathcal{O}_{12,\;(1,1,2)}^{108}$ & {$S_{4}S_{5}S_{11}F_{3}F_{8}$, $S_{2}S_{5}S_{11}F_{3}F_{8}$, $S_{5}S_{6}S_{19}F_{3}F_{7}$, $S_{3}S_{5}S_{6}F_{3}F_{7}$, $S_{4}S_{5}S_{11}F_{7}F_{8}$, $S_{2}S_{5}S_{11}F_{7}F_{8}$, $S_{5}S_{6}S_{19}F_{7}F_{8}$, $S_{3}S_{5}S_{6}F_{7}F_{8}$, $S_{4}S_{5}S_{5}F_{3}F_{8}$, $S_{2}S_{5}S_{5}F_{3}F_{8}$, $S_{5}S_{5}S_{19}F_{3}F_{8}$, $S_{3}S_{5}S_{5}F_{3}F_{8}$, $S_{4}S_{5}S_{5}F_{8}F_{8}$, $S_{2}S_{5}S_{5}F_{8}F_{8}$, $S_{5}S_{5}S_{19}F_{8}F_{8}$, $S_{3}S_{5}S_{5}F_{8}F_{8}$} \\
   \hline
 \caption{Dimension 12 UV completions for topology $\mathcal{T}_ {10} ^{ 29 }$ (see Fig.~\ref{fig:d=10_topologies}).} 
 \label{tab:T12Top29B1L1D2}\\
   \endlastfoot
  $\mathcal{O}_{12,\;(1,1,2)}^{103}$ & {$S_{11}S_{11}S_{31}F_{8}F_{10}$, $S_{1}S_{11}S_{11}F_{8}F_{10}$, $S_{4}S_{11}S_{11}F_{8}F_{10}$, $S_{2}S_{11}S_{11}F_{8}F_{10}$, $S_{11}S_{11}S_{31}F_{8}F_{8}$, $S_{1}S_{11}S_{11}F_{8}F_{8}$, $S_{4}S_{11}S_{11}F_{8}F_{8}$, $S_{2}S_{11}S_{11}F_{8}F_{8}$, $S_{10}S_{11}S_{31}F_{10}F_{13}$, $S_{1}S_{10}S_{11}F_{10}F_{13}$, $S_{4}S_{5}S_{11}F_{3}F_{8}$, $S_{2}S_{5}S_{11}F_{3}F_{8}$, $S_{10}S_{11}S_{31}F_{8}F_{13}$, $S_{1}S_{10}S_{11}F_{8}F_{13}$, $S_{4}S_{5}S_{11}F_{8}F_{13}$, $S_{2}S_{5}S_{11}F_{8}F_{13}$} \\
\hline
 \end{longtable}
 \end{center}
 \begin{center}
 \begin{longtable}{| c  | m{13cm} |} 
   \hline
  $\mathcal{O}^j{(\mathcal{T}_ {10} ^{ 30 })}$ &  New Particles \\
   \hline
   \hline
   \endhead
 \multicolumn{2}{|c|}{\hspace{9cm} Continued on next page}
   \endfoot
  $\mathcal{O}_{12,\;(1,1,2)}^{108}$ & {$S_{3}F_{3}F_{3}F_{4}F_{8}$, $S_{3}F_{1}F_{3}F_{3}F_{8}$, $S_{2}F_{7}F_{8}F_{8}F_{9}$, $S_{2}F_{5}F_{7}F_{8}F_{8}$, $S_{3}F_{3}F_{7}F_{8}F_{9}$, $S_{3}F_{3}F_{5}F_{7}F_{8}$, $S_{2}F_{8}F_{8}F_{8}F_{9}$, $S_{2}F_{5}F_{8}F_{8}F_{8}$, $S_{3}F_{3}F_{8}F_{8}F_{9}$, $S_{3}F_{3}F_{5}F_{8}F_{8}$, $S_{3}F_{3}F_{7}F_{10}F_{11}$, $S_{3}F_{2}F_{3}F_{7}F_{10}$, $S_{2}F_{7}F_{8}F_{13}F_{14}$, $S_{2}F_{6}F_{7}F_{8}F_{13}$, $S_{3}F_{3}F_{8}F_{13}F_{14}$, $S_{3}F_{3}F_{6}F_{8}F_{13}$, $S_{3}F_{3}F_{3}F_{8}F_{11}$, $S_{3}F_{2}F_{3}F_{3}F_{8}$, $S_{2}F_{8}F_{8}F_{8}F_{14}$, $S_{2}F_{6}F_{8}F_{8}F_{8}$, $S_{3}F_{3}F_{8}F_{8}F_{14}$, $S_{3}F_{3}F_{6}F_{8}F_{8}$} \\
   \hline
 \caption{Dimension 12 UV completions for topology $\mathcal{T}_ {10} ^{ 30 }$ (see Fig.~\ref{fig:d=10_topologies}).} 
 \label{tab:T12Top30B1L1D2}\\
   \endlastfoot
  $\mathcal{O}_{12,\;(1,1,2)}^{103}$ & {$S_{1}F_{8}F_{10}F_{10}F_{31}$, $S_{1}F_{8}F_{10}F_{10}F_{16}$, $S_{2}F_{8}F_{8}F_{8}F_{9}$, $S_{2}F_{5}F_{8}F_{8}F_{8}$, $S_{1}F_{8}F_{8}F_{9}F_{10}$, $S_{1}F_{5}F_{8}F_{8}F_{10}$, $S_{1}F_{10}F_{13}F_{31}F_{49}$, $S_{1}F_{10}F_{13}F_{16}F_{49}$, $S_{2}F_{7}F_{8}F_{9}F_{13}$, $S_{2}F_{5}F_{7}F_{8}F_{13}$, $S_{1}F_{3}F_{7}F_{8}F_{9}$, $S_{1}F_{3}F_{5}F_{7}F_{8}$, $S_{1}F_{8}F_{10}F_{10}F_{11}$, $S_{1}F_{2}F_{8}F_{10}F_{10}$, $S_{2}F_{8}F_{8}F_{8}F_{14}$, $S_{2}F_{6}F_{8}F_{8}F_{8}$, $S_{1}F_{8}F_{8}F_{10}F_{14}$, $S_{1}F_{6}F_{8}F_{8}F_{10}$, $S_{1}F_{3}F_{8}F_{10}F_{11}$, $S_{1}F_{2}F_{3}F_{8}F_{10}$, $S_{2}F_{8}F_{8}F_{13}F_{14}$, $S_{2}F_{6}F_{8}F_{8}F_{13}$, $S_{1}F_{8}F_{10}F_{13}F_{14}$, $S_{1}F_{6}F_{8}F_{10}F_{13}$} \\
\hline
 \end{longtable}
 \end{center}
 \begin{center}
 \begin{longtable}{| c  | m{13cm} |} 
   \hline
  $\mathcal{O}^j{(\mathcal{T}_ {10} ^{ 31 })}$ &  New Particles \\
   \hline
   \hline
   \endhead
 \multicolumn{2}{|c|}{\hspace{9cm} Continued on next page}
   \endfoot
  $\mathcal{O}_{12,\;(1,1,2)}^{108}$ & {$S_{3}S_{8}F_{3}F_{4}F_{8}$, $S_{3}S_{7}F_{1}F_{3}F_{8}$, $S_{2}S_{8}F_{8}F_{8}F_{9}$, $S_{2}S_{7}F_{5}F_{8}F_{8}$, $S_{3}S_{8}F_{3}F_{8}F_{9}$, $S_{3}S_{7}F_{3}F_{5}F_{8}$, $S_{3}S_{9}F_{7}F_{10}F_{11}$, $S_{2}S_{9}F_{7}F_{13}F_{14}$, $S_{3}S_{9}F_{3}F_{13}F_{14}$, $S_{3}S_{9}F_{3}F_{8}F_{11}$, $S_{2}S_{9}F_{8}F_{8}F_{14}$, $S_{3}S_{9}F_{3}F_{8}F_{14}$} \\
   \hline
 \caption{Dimension 12 UV completions for topology $\mathcal{T}_ {10} ^{ 31 }$ (see Fig.~\ref{fig:d=10_topologies}).} 
 \label{tab:T12Top31B1L1D2}\\
   \endlastfoot
  $\mathcal{O}_{12,\;(1,1,2)}^{103}$ & {$S_{1}S_{9}F_{8}F_{10}F_{31}$, $S_{2}S_{9}F_{8}F_{8}F_{9}$, $S_{1}S_{9}F_{8}F_{9}F_{10}$, $S_{1}S_{9}F_{13}F_{31}F_{49}$, $S_{2}S_{9}F_{7}F_{9}F_{13}$, $S_{1}S_{9}F_{3}F_{7}F_{9}$, $S_{1}S_{8}F_{8}F_{10}F_{11}$, $S_{1}S_{7}F_{2}F_{8}F_{10}$, $S_{2}S_{8}F_{8}F_{8}F_{14}$, $S_{2}S_{7}F_{6}F_{8}F_{8}$, $S_{1}S_{8}F_{8}F_{10}F_{14}$, $S_{1}S_{7}F_{6}F_{8}F_{10}$} \\
   \hline
  $\mathcal{O}_{12,\;(1,1,2)}^{104}$ & {$S_{3}S_{8}F_{3}F_{11}F_{13}$, $S_{3}S_{7}F_{2}F_{3}F_{13}$, $S_{1}S_{8}F_{8}F_{9}F_{13}$, $S_{1}S_{7}F_{5}F_{8}F_{13}$, $S_{2}S_{8}F_{3}F_{8}F_{11}$, $S_{2}S_{7}F_{2}F_{3}F_{8}$, $S_{2}S_{8}F_{8}F_{8}F_{9}$, $S_{2}S_{7}F_{5}F_{8}F_{8}$, $S_{1}S_{8}F_{8}F_{13}F_{14}$, $S_{1}S_{7}F_{6}F_{8}F_{13}$, $S_{2}S_{8}F_{3}F_{8}F_{9}$, $S_{2}S_{7}F_{3}F_{5}F_{8}$, $S_{3}S_{8}F_{3}F_{13}F_{14}$, $S_{3}S_{7}F_{3}F_{6}F_{13}$, $S_{3}S_{9}F_{3}F_{4}F_{8}$, $S_{1}S_{9}F_{8}F_{13}F_{14}$, $S_{2}S_{9}F_{3}F_{4}F_{7}$, $S_{2}S_{9}F_{7}F_{13}F_{14}$, $S_{1}S_{9}F_{7}F_{37}F_{50}$, $S_{2}S_{9}F_{10}F_{13}F_{14}$, $S_{3}S_{9}F_{10}F_{37}F_{50}$, $S_{3}S_{9}F_{3}F_{4}F_{13}$, $S_{2}S_{9}F_{3}F_{4}F_{8}$, $S_{2}S_{9}F_{8}F_{8}F_{14}$, $S_{1}S_{9}F_{8}F_{13}F_{37}$, $S_{2}S_{9}F_{3}F_{8}F_{14}$, $S_{3}S_{9}F_{3}F_{13}F_{37}$} \\
   \hline
  $\mathcal{O}_{12,\;(1,1,2)}^{105}$ & {$S_{3}S_{9}F_{3}F_{4}F_{8}$, $S_{1}S_{9}F_{8}F_{13}F_{14}$, $S_{3}S_{9}F_{3}F_{13}F_{14}$} \\
   \hline
  $\mathcal{O}_{12,\;(1,1,2)}^{106}$ & {$S_{1}S_{9}F_{8}F_{10}F_{11}$, $S_{3}S_{9}F_{7}F_{8}F_{9}$, $S_{1}S_{9}F_{3}F_{7}F_{9}$} \\
\hline
 \end{longtable}
 \end{center}
 \begin{center}
 \begin{longtable}{| c  | m{13cm} |} 
   \hline
  $\mathcal{O}^j{(\mathcal{T}_ {10} ^{ 32 })}$ &  New Particles \\
   \hline
   \hline
   \endhead
 \multicolumn{2}{|c|}{\hspace{9cm} Continued on next page}
   \endfoot
  $\mathcal{O}_{12,\;(1,1,2)}^{108}$ & {$S_{2}S_{9}F_{6}F_{7}F_{9}$, $S_{3}S_{9}F_{2}F_{3}F_{4}$, $S_{3}S_{9}F_{6}F_{7}F_{9}$, $S_{2}S_{8}F_{6}F_{8}F_{14}$, $S_{2}S_{7}F_{6}F_{6}F_{8}$, $S_{3}S_{8}F_{2}F_{3}F_{11}$, $S_{3}S_{7}F_{2}F_{2}F_{3}$, $S_{3}S_{8}F_{6}F_{8}F_{14}$, $S_{3}S_{7}F_{6}F_{6}F_{8}$, $S_{2}S_{9}F_{6}F_{8}F_{9}$, $S_{3}S_{9}F_{6}F_{8}F_{9}$} \\
   \hline
 \caption{Dimension 12 UV completions for topology $\mathcal{T}_ {10} ^{ 32 }$ (see Fig.~\ref{fig:d=10_topologies}).} 
 \label{tab:T12Top32B1L1D2}\\
   \endlastfoot
  $\mathcal{O}_{12,\;(1,1,2)}^{103}$ & {$S_{2}S_{9}F_{5}F_{8}F_{14}$, $S_{1}S_{9}F_{10}F_{11}F_{16}$, $S_{1}S_{9}F_{5}F_{8}F_{14}$, $S_{2}S_{8}F_{5}F_{8}F_{9}$, $S_{2}S_{7}F_{5}F_{5}F_{8}$, $S_{1}S_{8}F_{10}F_{16}F_{31}$, $S_{1}S_{7}F_{10}F_{16}F_{16}$, $S_{1}S_{8}F_{5}F_{8}F_{9}$, $S_{1}S_{7}F_{5}F_{5}F_{8}$, $S_{2}S_{9}F_{5}F_{13}F_{14}$, $S_{1}S_{9}F_{3}F_{11}F_{16}$, $S_{1}S_{9}F_{5}F_{13}F_{14}$} \\
   \hline
  $\mathcal{O}_{12,\;(1,1,2)}^{104}$ & {$S_{1}S_{9}F_{6}F_{7}F_{9}$, $S_{3}S_{9}F_{1}F_{10}F_{11}$, $S_{1}S_{9}F_{8}F_{14}F_{18}$, $S_{2}S_{9}F_{6}F_{7}F_{9}$, $S_{2}S_{9}F_{1}F_{10}F_{11}$, $S_{3}S_{9}F_{8}F_{14}F_{18}$, $S_{1}S_{8}F_{6}F_{8}F_{14}$, $S_{1}S_{7}F_{6}F_{6}F_{8}$, $S_{3}S_{8}F_{1}F_{3}F_{4}$, $S_{3}S_{7}F_{1}F_{1}F_{3}$, $S_{1}S_{8}F_{13}F_{18}F_{37}$, $S_{1}S_{7}F_{13}F_{18}F_{18}$, $S_{2}S_{8}F_{6}F_{8}F_{14}$, $S_{2}S_{7}F_{6}F_{6}F_{8}$, $S_{2}S_{8}F_{1}F_{3}F_{4}$, $S_{2}S_{7}F_{1}F_{1}F_{3}$, $S_{3}S_{8}F_{13}F_{18}F_{37}$, $S_{3}S_{7}F_{13}F_{18}F_{18}$, $S_{1}S_{9}F_{6}F_{8}F_{9}$, $S_{3}S_{9}F_{1}F_{3}F_{11}$, $S_{1}S_{9}F_{13}F_{14}F_{18}$, $S_{2}S_{9}F_{6}F_{8}F_{9}$, $S_{2}S_{9}F_{1}F_{3}F_{11}$, $S_{3}S_{9}F_{13}F_{14}F_{18}$} \\
   \hline
  $\mathcal{O}_{12,\;(1,1,2)}^{105}$ & {$S_{1}S_{9}F_{8}F_{14}F_{18}$, $S_{3}S_{9}F_{2}F_{3}F_{4}$, $S_{3}S_{9}F_{8}F_{14}F_{18}$} \\
   \hline
  $\mathcal{O}_{12,\;(1,1,2)}^{106}$ & {$S_{3}S_{9}F_{8}F_{9}F_{17}$, $S_{1}S_{9}F_{3}F_{11}F_{16}$, $S_{1}S_{9}F_{8}F_{9}F_{17}$} \\
\hline
 \end{longtable}
 \end{center}
 \begin{center}
 \begin{longtable}{| c  | m{13cm} |} 
   \hline
  $\mathcal{O}^j{(\mathcal{T}_ {10} ^{ 33 })}$ &  New Particles \\
   \hline
   \hline
   \endhead
 \multicolumn{2}{|c|}{\hspace{9cm} Continued on next page}
   \endfoot
  $\mathcal{O}_{12,\;(1,1,2)}^{108}$ & {$S_{2}S_{5}F_{7}F_{8}F_{9}$, $S_{2}S_{5}F_{5}F_{7}F_{8}$, $S_{3}S_{5}F_{3}F_{3}F_{4}$, $S_{3}S_{5}F_{1}F_{3}F_{3}$, $S_{3}S_{5}F_{7}F_{8}F_{9}$, $S_{3}S_{5}F_{5}F_{7}F_{8}$, $S_{2}S_{11}F_{8}F_{13}F_{14}$, $S_{2}S_{11}F_{6}F_{8}F_{13}$, $S_{3}S_{6}F_{3}F_{10}F_{11}$, $S_{3}S_{6}F_{2}F_{3}F_{10}$, $S_{3}S_{6}F_{8}F_{13}F_{14}$, $S_{3}S_{6}F_{6}F_{8}F_{13}$, $S_{2}S_{5}F_{8}F_{8}F_{9}$, $S_{2}S_{5}F_{5}F_{8}F_{8}$, $S_{3}S_{5}F_{8}F_{8}F_{9}$, $S_{3}S_{5}F_{5}F_{8}F_{8}$, $S_{2}S_{5}F_{8}F_{8}F_{14}$, $S_{2}S_{5}F_{6}F_{8}F_{8}$, $S_{3}S_{5}F_{3}F_{3}F_{11}$, $S_{3}S_{5}F_{2}F_{3}F_{3}$, $S_{3}S_{5}F_{8}F_{8}F_{14}$, $S_{3}S_{5}F_{6}F_{8}F_{8}$} \\
   \hline
 \caption{Dimension 12 UV completions for topology $\mathcal{T}_ {10} ^{ 33 }$ (see Fig.~\ref{fig:d=10_topologies}).} 
 \label{tab:T12Top33B1L1D2}\\
   \endlastfoot
  $\mathcal{O}_{12,\;(1,1,2)}^{103}$ & {$S_{2}S_{11}F_{8}F_{8}F_{9}$, $S_{2}S_{11}F_{5}F_{8}F_{8}$, $S_{1}S_{11}F_{10}F_{10}F_{31}$, $S_{1}S_{11}F_{10}F_{10}F_{16}$, $S_{1}S_{11}F_{8}F_{8}F_{9}$, $S_{1}S_{11}F_{5}F_{8}F_{8}$, $S_{2}S_{11}F_{8}F_{8}F_{14}$, $S_{2}S_{11}F_{6}F_{8}F_{8}$, $S_{1}S_{11}F_{10}F_{10}F_{11}$, $S_{1}S_{11}F_{2}F_{10}F_{10}$, $S_{1}S_{11}F_{8}F_{8}F_{14}$, $S_{1}S_{11}F_{6}F_{8}F_{8}$, $S_{2}S_{5}F_{7}F_{8}F_{9}$, $S_{2}S_{5}F_{5}F_{7}F_{8}$, $S_{1}S_{10}F_{10}F_{31}F_{49}$, $S_{1}S_{10}F_{10}F_{16}F_{49}$, $S_{1}S_{10}F_{7}F_{8}F_{9}$, $S_{1}S_{10}F_{5}F_{7}F_{8}$, $S_{2}S_{11}F_{8}F_{13}F_{14}$, $S_{2}S_{11}F_{6}F_{8}F_{13}$, $S_{1}S_{11}F_{3}F_{10}F_{11}$, $S_{1}S_{11}F_{2}F_{3}F_{10}$, $S_{1}S_{11}F_{8}F_{13}F_{14}$, $S_{1}S_{11}F_{6}F_{8}F_{13}$} \\
\hline
 \end{longtable}
 \end{center}
\input{2Der_UniqueOnly/Tables_dim12/Table_dim12_Top34_B1_L1.tex}
\input{2Der_UniqueOnly/Tables_dim12/Table_dim12_Top35_B1_L1.tex}
\input{2Der_UniqueOnly/Tables_dim12/Table_dim12_Top36_B1_L1.tex}
\input{2Der_UniqueOnly/Tables_dim12/Table_dim12_Top37_B1_L1.tex}
\input{2Der_UniqueOnly/Tables_dim12/Table_dim12_Top38_B1_L1.tex}
\input{2Der_UniqueOnly/Tables_dim12/Table_dim12_Top39_B1_L1.tex}
\input{2Der_UniqueOnly/Tables_dim12/Table_dim12_Top40_B1_L1.tex}
\input{2Der_UniqueOnly/Tables_dim12/Table_dim12_Top41_B1_L1.tex}
\input{2Der_UniqueOnly/Tables_dim12/Table_dim12_Top42_B1_L1.tex}
\input{2Der_UniqueOnly/Tables_dim12/Table_dim12_Top43_B1_L1.tex}
\input{2Der_UniqueOnly/Tables_dim12/Table_dim12_Top44_B1_L1.tex}
\input{2Der_UniqueOnly/Tables_dim12/Table_dim12_Top45_B1_L1.tex}
\input{2Der_UniqueOnly/Tables_dim12/Table_dim12_Top46_B1_L1.tex}
\input{2Der_UniqueOnly/Tables_dim12/Table_dim12_Top47_B1_L1.tex}
\input{2Der_UniqueOnly/Tables_dim12/Table_dim12_Top48_B1_L1.tex}
\input{2Der_UniqueOnly/Tables_dim12/Table_dim12_Top49_B1_L1.tex}
\input{2Der_UniqueOnly/Tables_dim12/Table_dim12_Top50_B1_L1.tex}
\input{2Der_UniqueOnly/Tables_dim12/Table_dim12_Top51_B1_L1.tex}
\input{2Der_UniqueOnly/Tables_dim12/Table_dim12_Top52_B1_L1.tex}
\input{2Der_UniqueOnly/Tables_dim12/Table_dim12_Top53_B1_L1.tex}
\input{2Der_UniqueOnly/Tables_dim12/Table_dim12_Top54_B1_L1.tex}
\input{2Der_UniqueOnly/Tables_dim12/Table_dim12_Top55_B1_L1.tex}
\input{2Der_UniqueOnly/Tables_dim12/Table_dim12_Top56_B1_L1.tex}
\input{2Der_UniqueOnly/Tables_dim12/Table_dim12_Top57_B1_L1.tex}
\input{2Der_UniqueOnly/Tables_dim12/Table_dim12_Top58_B1_L1.tex}
\input{2Der_UniqueOnly/Tables_dim12/Table_dim12_Top59_B1_L1.tex}
\input{2Der_UniqueOnly/Tables_dim12/Table_dim12_Top60_B1_L1.tex}
\input{2Der_UniqueOnly/Tables_dim12/Table_dim12_Top61_B1_L1.tex}
\input{2Der_UniqueOnly/Tables_dim12/Table_dim12_Top62_B1_L1.tex}
\input{2Der_UniqueOnly/Tables_dim12/Table_dim12_Top63_B1_L1.tex}
\input{2Der_UniqueOnly/Tables_dim12/Table_dim12_Top64_B1_L1.tex}
\input{2Der_UniqueOnly/Tables_dim12/Table_dim12_Top65_B1_L1.tex}
\input{2Der_UniqueOnly/Tables_dim12/Table_dim12_Top66_B1_L1.tex}
\input{2Der_UniqueOnly/Tables_dim12/Table_dim12_Top67_B1_L1.tex}
\input{2Der_UniqueOnly/Tables_dim12/Table_dim12_Top68_B1_L1.tex}
\input{2Der_UniqueOnly/Tables_dim12/Table_dim12_Top69_B1_L1.tex}
\input{2Der_UniqueOnly/Tables_dim12/Table_dim12_Top70_B1_L1.tex}
\input{2Der_UniqueOnly/Tables_dim12/Table_dim12_Top71_B1_L1.tex}
\input{2Der_UniqueOnly/Tables_dim12/Table_dim12_Top72_B1_L1.tex}
\input{2Der_UniqueOnly/Tables_dim12/Table_dim12_Top73_B1_L1.tex}
\input{2Der_UniqueOnly/Tables_dim12/Table_dim12_Top74_B1_L1.tex}
\input{2Der_UniqueOnly/Tables_dim12/Table_dim12_Top75_B1_L1.tex}
\input{2Der_UniqueOnly/Tables_dim12/Table_dim12_Top76_B1_L1.tex}
\input{2Der_UniqueOnly/Tables_dim12/Table_dim12_Top77_B1_L1.tex}
\input{2Der_UniqueOnly/Tables_dim12/Table_dim12_Top78_B1_L1.tex}
\input{2Der_UniqueOnly/Tables_dim12/Table_dim12_Top79_B1_L1.tex}
\input{2Der_UniqueOnly/Tables_dim12/Table_dim12_Top80_B1_L1.tex}
\input{2Der_UniqueOnly/Tables_dim12/Table_dim12_Top81_B1_L1.tex}
\input{2Der_UniqueOnly/Tables_dim12/Table_dim12_Top82_B1_L1.tex}
\input{2Der_UniqueOnly/Tables_dim12/Table_dim12_Top83_B1_L1.tex}
\input{2Der_UniqueOnly/Tables_dim12/Table_dim12_Top84_B1_L1.tex}
\input{2Der_UniqueOnly/Tables_dim12/Table_dim12_Top85_B1_L1.tex}
\input{2Der_UniqueOnly/Tables_dim12/Table_dim12_Top86_B1_L1.tex}
\input{2Der_UniqueOnly/Tables_dim12/Table_dim12_Top87_B1_L1.tex}
\input{2Der_UniqueOnly/Tables_dim12/Table_dim12_Top88_B1_L1.tex}
\input{2Der_UniqueOnly/Tables_dim12/Table_dim12_Top89_B1_L1.tex}

\begin{center}
\begin{longtable}{| c | m{13cm} |}
   \hline
   $\mathcal{O}^j(\mathcal{T}_{9}^{12})$ & New Particles \\
   \hline
   \hline
   \endhead
   \multicolumn{2}{|c|}{\hspace{9cm} Continued on next page} \\
   \endfoot
  $\mathcal{O}_{12,(1,1,3)}^{65}$ & {$S_{3}S_{5}F_{3}F_{5}$, $S_{3}S_{5}F_{5}F_{8}$, $S_{3}S_{5}F_{3}F_{7}$, $S_{2}S_{5}F_{3}F_{8}$, $S_{3}S_{5}F_{7}F_{8}$, $S_{2}S_{5}F_{5}F_{8}$} \\
   \hline
   \caption{Dimension 12 UV completions for topology $\mathcal{T}_{9}^{12}$  (see Fig.~\ref{fig:d=9_topologies}).}
   \label{tab:T12Top12B1L1D3}\\
   \endlastfoot
  $\mathcal{O}_{12,(1,1,3)}^{1}$ & {} \\
   \hline
  $\mathcal{O}_{12,(1,1,3)}^{2}$ & {} \\
   \hline
  $\mathcal{O}_{12,(1,1,3)}^{3}$ & {} \\
   \hline
  $\mathcal{O}_{12,(1,1,3)}^{4}$ & {} \\
   \hline
  $\mathcal{O}_{12,(1,1,3)}^{5}$ & {} \\
   \hline
  $\mathcal{O}_{12,(1,1,3)}^{6}$ & {} \\
   \hline
  $\mathcal{O}_{12,(1,1,3)}^{7}$ & {} \\
   \hline
  $\mathcal{O}_{12,(1,1,3)}^{8}$ & {} \\
   \hline
  $\mathcal{O}_{12,(1,1,3)}^{9}$ & {} \\
   \hline
  $\mathcal{O}_{12,(1,1,3)}^{10}$ & {} \\
   \hline
  $\mathcal{O}_{12,(1,1,3)}^{11}$ & {} \\
   \hline
  $\mathcal{O}_{12,(1,1,3)}^{12}$ & {} \\
   \hline
  $\mathcal{O}_{12,(1,1,3)}^{13}$ & {} \\
   \hline
  $\mathcal{O}_{12,(1,1,3)}^{14}$ & {} \\
   \hline
  $\mathcal{O}_{12,(1,1,3)}^{15}$ & {} \\
   \hline
  $\mathcal{O}_{12,(1,1,3)}^{16}$ & {} \\
   \hline
  $\mathcal{O}_{12,(1,1,3)}^{17}$ & {} \\
   \hline
  $\mathcal{O}_{12,(1,1,3)}^{18}$ & {} \\
   \hline
  $\mathcal{O}_{12,(1,1,3)}^{19}$ & {} \\
   \hline
  $\mathcal{O}_{12,(1,1,3)}^{20}$ & {} \\
   \hline
  $\mathcal{O}_{12,(1,1,3)}^{21}$ & {} \\
   \hline
  $\mathcal{O}_{12,(1,1,3)}^{22}$ & {} \\
   \hline
  $\mathcal{O}_{12,(1,1,3)}^{23}$ & {} \\
   \hline
  $\mathcal{O}_{12,(1,1,3)}^{24}$ & {} \\
   \hline
  $\mathcal{O}_{12,(1,1,3)}^{25}$ & {} \\
   \hline
  $\mathcal{O}_{12,(1,1,3)}^{26}$ & {} \\
   \hline
  $\mathcal{O}_{12,(1,1,3)}^{27}$ & {} \\
   \hline
  $\mathcal{O}_{12,(1,1,3)}^{28}$ & {} \\
   \hline
  $\mathcal{O}_{12,(1,1,3)}^{29}$ & {} \\
   \hline
  $\mathcal{O}_{12,(1,1,3)}^{30}$ & {} \\
   \hline
  $\mathcal{O}_{12,(1,1,3)}^{31}$ & {} \\
   \hline
  $\mathcal{O}_{12,(1,1,3)}^{32}$ & {} \\
   \hline
  $\mathcal{O}_{12,(1,1,3)}^{33}$ & {} \\
   \hline
  $\mathcal{O}_{12,(1,1,3)}^{34}$ & {} \\
   \hline
  $\mathcal{O}_{12,(1,1,3)}^{35}$ & {} \\
   \hline
  $\mathcal{O}_{12,(1,1,3)}^{36}$ & {} \\
   \hline
  $\mathcal{O}_{12,(1,1,3)}^{37}$ & {} \\
   \hline
  $\mathcal{O}_{12,(1,1,3)}^{38}$ & {} \\
   \hline
  $\mathcal{O}_{12,(1,1,3)}^{39}$ & {} \\
   \hline
  $\mathcal{O}_{12,(1,1,3)}^{40}$ & {} \\
   \hline
  $\mathcal{O}_{12,(1,1,3)}^{41}$ & {} \\
   \hline
  $\mathcal{O}_{12,(1,1,3)}^{42}$ & {} \\
   \hline
  $\mathcal{O}_{12,(1,1,3)}^{43}$ & {} \\
   \hline
  $\mathcal{O}_{12,(1,1,3)}^{44}$ & {} \\
   \hline
  $\mathcal{O}_{12,(1,1,3)}^{45}$ & {} \\
   \hline
  $\mathcal{O}_{12,(1,1,3)}^{46}$ & {} \\
   \hline
  $\mathcal{O}_{12,(1,1,3)}^{47}$ & {} \\
   \hline
  $\mathcal{O}_{12,(1,1,3)}^{48}$ & {} \\
   \hline
  $\mathcal{O}_{12,(1,1,3)}^{49}$ & {} \\
   \hline
  $\mathcal{O}_{12,(1,1,3)}^{50}$ & {} \\
   \hline
  $\mathcal{O}_{12,(1,1,3)}^{51}$ & {} \\
   \hline
  $\mathcal{O}_{12,(1,1,3)}^{52}$ & {$S_{2}S_{11}F_{2}F_{8}$, $S_{2}S_{11}F_{8}F_{8}$} \\
   \hline
  $\mathcal{O}_{12,(1,1,3)}^{53}$ & {} \\
   \hline
  $\mathcal{O}_{12,(1,1,3)}^{54}$ & {$S_{3}S_{5}F_{1}F_{7}$, $S_{2}S_{11}F_{1}F_{8}$, $S_{3}S_{5}F_{7}F_{13}$, $S_{3}S_{5}F_{1}F_{8}$, $S_{3}S_{5}F_{8}F_{13}$} \\
   \hline
  $\mathcal{O}_{12,(1,1,3)}^{55}$ & {$S_{1}S_{11}F_{6}F_{10}$, $S_{2}S_{11}F_{3}F_{8}$, $S_{1}S_{11}F_{10}F_{13}$, $S_{2}S_{11}F_{6}F_{8}$, $S_{1}S_{11}F_{6}F_{8}$, $S_{1}S_{11}F_{8}F_{13}$} \\
   \hline
  $\mathcal{O}_{12,(1,1,3)}^{56}$ & {} \\
   \hline
  $\mathcal{O}_{12,(1,1,3)}^{57}$ & {} \\
   \hline
  $\mathcal{O}_{12,(1,1,3)}^{58}$ & {$S_{2}S_{5}F_{2}F_{8}$, $S_{1}S_{11}F_{2}F_{8}$, $S_{1}S_{11}F_{2}F_{13}$, $S_{1}S_{11}F_{7}F_{8}$, $S_{1}S_{11}F_{7}F_{13}$} \\
   \hline
  $\mathcal{O}_{12,(1,1,3)}^{59}$ & {$S_{2}S_{11}F_{3}F_{5}$, $S_{2}S_{11}F_{5}F_{8}$, $S_{3}S_{5}F_{3}F_{5}$, $S_{3}S_{5}F_{5}F_{13}$, $S_{3}S_{5}F_{3}F_{7}$, $S_{1}S_{11}F_{8}F_{10}$, $S_{2}S_{11}F_{3}F_{8}$, $S_{1}S_{11}F_{10}F_{13}$, $S_{3}S_{5}F_{7}F_{13}$, $S_{1}S_{11}F_{5}F_{8}$, $S_{1}S_{11}F_{5}F_{13}$} \\
   \hline
  $\mathcal{O}_{12,(1,1,3)}^{60}$ & {$S_{2}S_{5}F_{1}F_{8}$, $S_{2}S_{5}F_{8}F_{8}$} \\
   \hline
  $\mathcal{O}_{12,(1,1,3)}^{61}$ & {$S_{2}S_{11}F_{3}F_{6}$, $S_{2}S_{5}F_{3}F_{8}$, $S_{2}S_{11}F_{3}F_{8}$, $S_{2}S_{5}F_{6}F_{8}$, $S_{2}S_{11}F_{6}F_{8}$, $S_{2}S_{11}F_{8}F_{8}$} \\
   \hline
  $\mathcal{O}_{12,(1,1,3)}^{62}$ & {$S_{2}S_{5}F_{3}F_{5}$, $S_{2}S_{11}F_{3}F_{8}$, $S_{2}S_{5}F_{3}F_{8}$, $S_{2}S_{11}F_{5}F_{8}$, $S_{2}S_{5}F_{5}F_{8}$, $S_{2}S_{5}F_{8}F_{8}$} \\
   \hline
  $\mathcal{O}_{12,(1,1,3)}^{63}$ & {} \\
   \hline
  $\mathcal{O}_{12,(1,1,3)}^{64}$ & {$S_{2}S_{5}F_{3}F_{6}$, $S_{3}S_{5}F_{3}F_{7}$, $S_{3}S_{5}F_{6}F_{7}$, $S_{1}S_{11}F_{3}F_{6}$, $S_{3}S_{5}F_{3}F_{8}$, $S_{1}S_{11}F_{3}F_{13}$, $S_{3}S_{5}F_{6}F_{8}$, $S_{1}S_{11}F_{6}F_{7}$, $S_{2}S_{5}F_{6}F_{8}$, $S_{1}S_{11}F_{7}F_{13}$} \\
   \hline
\end{longtable}
\end{center}

\begin{center}
\begin{longtable}{| c | m{13cm} |}
   \hline
   $\mathcal{O}^j(\mathcal{T}_{9}^{18})$ & New Particles \\
   \hline
   \hline
   \endhead
   \multicolumn{2}{|c|}{\hspace{9cm} Continued on next page} \\
   \endfoot
  $\mathcal{O}_{12,(1,1,3)}^{65}$ & {} \\
   \hline
   \caption{Dimension 12 UV completions for topology $\mathcal{T}_{9}^{18}$ (see Fig.~\ref{fig:d=9_topologies}).}
   \label{tab:T12Top18B1L1D3}\\
   \endlastfoot
  $\mathcal{O}_{12,(1,1,3)}^{1}$ & {} \\
   \hline
  $\mathcal{O}_{12,(1,1,3)}^{2}$ & {} \\
   \hline
  $\mathcal{O}_{12,(1,1,3)}^{3}$ & {} \\
   \hline
  $\mathcal{O}_{12,(1,1,3)}^{4}$ & {} \\
   \hline
  $\mathcal{O}_{12,(1,1,3)}^{5}$ & {} \\
   \hline
  $\mathcal{O}_{12,(1,1,3)}^{6}$ & {} \\
   \hline
  $\mathcal{O}_{12,(1,1,3)}^{7}$ & {} \\
   \hline
  $\mathcal{O}_{12,(1,1,3)}^{8}$ & {} \\
   \hline
  $\mathcal{O}_{12,(1,1,3)}^{9}$ & {} \\
   \hline
  $\mathcal{O}_{12,(1,1,3)}^{10}$ & {} \\
   \hline
  $\mathcal{O}_{12,(1,1,3)}^{11}$ & {} \\
   \hline
  $\mathcal{O}_{12,(1,1,3)}^{12}$ & {} \\
   \hline
  $\mathcal{O}_{12,(1,1,3)}^{13}$ & {} \\
   \hline
  $\mathcal{O}_{12,(1,1,3)}^{14}$ & {} \\
   \hline
  $\mathcal{O}_{12,(1,1,3)}^{15}$ & {} \\
   \hline
  $\mathcal{O}_{12,(1,1,3)}^{16}$ & {} \\
   \hline
  $\mathcal{O}_{12,(1,1,3)}^{17}$ & {} \\
   \hline
  $\mathcal{O}_{12,(1,1,3)}^{18}$ & {} \\
   \hline
  $\mathcal{O}_{12,(1,1,3)}^{19}$ & {} \\
   \hline
  $\mathcal{O}_{12,(1,1,3)}^{20}$ & {} \\
   \hline
  $\mathcal{O}_{12,(1,1,3)}^{21}$ & {} \\
   \hline
  $\mathcal{O}_{12,(1,1,3)}^{22}$ & {} \\
   \hline
  $\mathcal{O}_{12,(1,1,3)}^{23}$ & {} \\
   \hline
  $\mathcal{O}_{12,(1,1,3)}^{24}$ & {} \\
   \hline
  $\mathcal{O}_{12,(1,1,3)}^{25}$ & {} \\
   \hline
  $\mathcal{O}_{12,(1,1,3)}^{26}$ & {} \\
   \hline
  $\mathcal{O}_{12,(1,1,3)}^{27}$ & {} \\
   \hline
  $\mathcal{O}_{12,(1,1,3)}^{28}$ & {} \\
   \hline
  $\mathcal{O}_{12,(1,1,3)}^{29}$ & {} \\
   \hline
  $\mathcal{O}_{12,(1,1,3)}^{30}$ & {} \\
   \hline
  $\mathcal{O}_{12,(1,1,3)}^{31}$ & {} \\
   \hline
  $\mathcal{O}_{12,(1,1,3)}^{32}$ & {} \\
   \hline
  $\mathcal{O}_{12,(1,1,3)}^{33}$ & {} \\
   \hline
  $\mathcal{O}_{12,(1,1,3)}^{34}$ & {} \\
   \hline
  $\mathcal{O}_{12,(1,1,3)}^{35}$ & {} \\
   \hline
  $\mathcal{O}_{12,(1,1,3)}^{36}$ & {} \\
   \hline
  $\mathcal{O}_{12,(1,1,3)}^{37}$ & {} \\
   \hline
  $\mathcal{O}_{12,(1,1,3)}^{38}$ & {} \\
   \hline
  $\mathcal{O}_{12,(1,1,3)}^{39}$ & {} \\
   \hline
  $\mathcal{O}_{12,(1,1,3)}^{40}$ & {} \\
   \hline
  $\mathcal{O}_{12,(1,1,3)}^{41}$ & {} \\
   \hline
  $\mathcal{O}_{12,(1,1,3)}^{42}$ & {} \\
   \hline
  $\mathcal{O}_{12,(1,1,3)}^{43}$ & {} \\
   \hline
  $\mathcal{O}_{12,(1,1,3)}^{44}$ & {} \\
   \hline
  $\mathcal{O}_{12,(1,1,3)}^{45}$ & {} \\
   \hline
  $\mathcal{O}_{12,(1,1,3)}^{46}$ & {} \\
   \hline
  $\mathcal{O}_{12,(1,1,3)}^{47}$ & {} \\
   \hline
  $\mathcal{O}_{12,(1,1,3)}^{48}$ & {} \\
   \hline
  $\mathcal{O}_{12,(1,1,3)}^{49}$ & {} \\
   \hline
  $\mathcal{O}_{12,(1,1,3)}^{50}$ & {} \\
   \hline
  $\mathcal{O}_{12,(1,1,3)}^{51}$ & {} \\
   \hline
  $\mathcal{O}_{12,(1,1,3)}^{52}$ & {$S_{1}S_{9}S_{26}F_{7}$, $S_{1}S_{9}S_{26}F_{16}$} \\
   \hline
  $\mathcal{O}_{12,(1,1,3)}^{53}$ & {} \\
   \hline
  $\mathcal{O}_{12,(1,1,3)}^{54}$ & {} \\
   \hline
  $\mathcal{O}_{12,(1,1,3)}^{55}$ & {} \\
   \hline
  $\mathcal{O}_{12,(1,1,3)}^{56}$ & {} \\
   \hline
  $\mathcal{O}_{12,(1,1,3)}^{57}$ & {} \\
   \hline
  $\mathcal{O}_{12,(1,1,3)}^{58}$ & {} \\
   \hline
  $\mathcal{O}_{12,(1,1,3)}^{59}$ & {} \\
   \hline
  $\mathcal{O}_{12,(1,1,3)}^{60}$ & {$S_{3}S_{9}S_{26}F_{13}$, $S_{3}S_{9}S_{26}F_{2}$} \\
   \hline
  $\mathcal{O}_{12,(1,1,3)}^{61}$ & {$S_{1}S_{9}S_{26}F_{13}$, $S_{1}S_{9}S_{26}F_{18}$, $S_{3}S_{9}S_{26}F_{18}$, $S_{3}S_{9}S_{26}F_{3}$} \\
   \hline
  $\mathcal{O}_{12,(1,1,3)}^{62}$ & {$S_{3}S_{9}S_{26}F_{7}$, $S_{3}S_{9}S_{26}F_{17}$, $S_{1}S_{9}S_{26}F_{17}$, $S_{1}S_{9}S_{26}F_{10}$} \\
   \hline
  $\mathcal{O}_{12,(1,1,3)}^{63}$ & {} \\
   \hline
  $\mathcal{O}_{12,(1,1,3)}^{64}$ & {} \\
   \hline
\end{longtable}
\end{center}

\begin{center}
\begin{longtable}{| c | m{13cm} |}
   \hline
   $\mathcal{O}^j(\mathcal{T}_{9}^{19})$ & New Particles \\
   \hline
   \hline
   \endhead
   \multicolumn{2}{|c|}{\hspace{9cm} Continued on next page} \\
   \endfoot
  $\mathcal{O}_{12,(1,1,3)}^{65}$ & {$S_{3}F_{3}F_{5}F_{8}$, $S_{2}F_{3}F_{8}F_{8}$, $S_{3}F_{3}F_{7}F_{8}$, $S_{2}F_{5}F_{8}F_{8}$} \\
   \hline
   \caption{Dimension 12 UV completions for topology $\mathcal{T}_{9}^{19}$ (see Fig.~\ref{fig:d=9_topologies}).}
   \label{tab:T12Top19B1L1D3}\\
   \endlastfoot
  $\mathcal{O}_{12,(1,1,3)}^{1}$ & {} \\
   \hline
  $\mathcal{O}_{12,(1,1,3)}^{2}$ & {} \\
   \hline
  $\mathcal{O}_{12,(1,1,3)}^{3}$ & {} \\
   \hline
  $\mathcal{O}_{12,(1,1,3)}^{4}$ & {} \\
   \hline
  $\mathcal{O}_{12,(1,1,3)}^{5}$ & {} \\
   \hline
  $\mathcal{O}_{12,(1,1,3)}^{6}$ & {} \\
   \hline
  $\mathcal{O}_{12,(1,1,3)}^{7}$ & {} \\
   \hline
  $\mathcal{O}_{12,(1,1,3)}^{8}$ & {} \\
   \hline
  $\mathcal{O}_{12,(1,1,3)}^{9}$ & {} \\
   \hline
  $\mathcal{O}_{12,(1,1,3)}^{10}$ & {} \\
   \hline
  $\mathcal{O}_{12,(1,1,3)}^{11}$ & {} \\
   \hline
  $\mathcal{O}_{12,(1,1,3)}^{12}$ & {} \\
   \hline
  $\mathcal{O}_{12,(1,1,3)}^{13}$ & {} \\
   \hline
  $\mathcal{O}_{12,(1,1,3)}^{14}$ & {} \\
   \hline
  $\mathcal{O}_{12,(1,1,3)}^{15}$ & {} \\
   \hline
  $\mathcal{O}_{12,(1,1,3)}^{16}$ & {} \\
   \hline
  $\mathcal{O}_{12,(1,1,3)}^{17}$ & {} \\
   \hline
  $\mathcal{O}_{12,(1,1,3)}^{18}$ & {} \\
   \hline
  $\mathcal{O}_{12,(1,1,3)}^{19}$ & {} \\
   \hline
  $\mathcal{O}_{12,(1,1,3)}^{20}$ & {} \\
   \hline
  $\mathcal{O}_{12,(1,1,3)}^{21}$ & {} \\
   \hline
  $\mathcal{O}_{12,(1,1,3)}^{22}$ & {} \\
   \hline
  $\mathcal{O}_{12,(1,1,3)}^{23}$ & {} \\
   \hline
  $\mathcal{O}_{12,(1,1,3)}^{24}$ & {} \\
   \hline
  $\mathcal{O}_{12,(1,1,3)}^{25}$ & {} \\
   \hline
  $\mathcal{O}_{12,(1,1,3)}^{26}$ & {} \\
   \hline
  $\mathcal{O}_{12,(1,1,3)}^{27}$ & {} \\
   \hline
  $\mathcal{O}_{12,(1,1,3)}^{28}$ & {} \\
   \hline
  $\mathcal{O}_{12,(1,1,3)}^{29}$ & {} \\
   \hline
  $\mathcal{O}_{12,(1,1,3)}^{30}$ & {} \\
   \hline
  $\mathcal{O}_{12,(1,1,3)}^{31}$ & {} \\
   \hline
  $\mathcal{O}_{12,(1,1,3)}^{32}$ & {} \\
   \hline
  $\mathcal{O}_{12,(1,1,3)}^{33}$ & {} \\
   \hline
  $\mathcal{O}_{12,(1,1,3)}^{34}$ & {} \\
   \hline
  $\mathcal{O}_{12,(1,1,3)}^{35}$ & {} \\
   \hline
  $\mathcal{O}_{12,(1,1,3)}^{36}$ & {} \\
   \hline
  $\mathcal{O}_{12,(1,1,3)}^{37}$ & {} \\
   \hline
  $\mathcal{O}_{12,(1,1,3)}^{38}$ & {} \\
   \hline
  $\mathcal{O}_{12,(1,1,3)}^{39}$ & {} \\
   \hline
  $\mathcal{O}_{12,(1,1,3)}^{40}$ & {} \\
   \hline
  $\mathcal{O}_{12,(1,1,3)}^{41}$ & {} \\
   \hline
  $\mathcal{O}_{12,(1,1,3)}^{42}$ & {} \\
   \hline
  $\mathcal{O}_{12,(1,1,3)}^{43}$ & {} \\
   \hline
  $\mathcal{O}_{12,(1,1,3)}^{44}$ & {} \\
   \hline
  $\mathcal{O}_{12,(1,1,3)}^{45}$ & {} \\
   \hline
  $\mathcal{O}_{12,(1,1,3)}^{46}$ & {} \\
   \hline
  $\mathcal{O}_{12,(1,1,3)}^{47}$ & {} \\
   \hline
  $\mathcal{O}_{12,(1,1,3)}^{48}$ & {} \\
   \hline
  $\mathcal{O}_{12,(1,1,3)}^{49}$ & {} \\
   \hline
  $\mathcal{O}_{12,(1,1,3)}^{50}$ & {} \\
   \hline
  $\mathcal{O}_{12,(1,1,3)}^{51}$ & {} \\
   \hline
  $\mathcal{O}_{12,(1,1,3)}^{52}$ & {} \\
   \hline
  $\mathcal{O}_{12,(1,1,3)}^{53}$ & {} \\
   \hline
  $\mathcal{O}_{12,(1,1,3)}^{54}$ & {} \\
   \hline
  $\mathcal{O}_{12,(1,1,3)}^{55}$ & {} \\
   \hline
  $\mathcal{O}_{12,(1,1,3)}^{56}$ & {} \\
   \hline
  $\mathcal{O}_{12,(1,1,3)}^{57}$ & {} \\
   \hline
  $\mathcal{O}_{12,(1,1,3)}^{58}$ & {} \\
   \hline
  $\mathcal{O}_{12,(1,1,3)}^{59}$ & {} \\
   \hline
  $\mathcal{O}_{12,(1,1,3)}^{60}$ & {} \\
   \hline
  $\mathcal{O}_{12,(1,1,3)}^{61}$ & {} \\
   \hline
  $\mathcal{O}_{12,(1,1,3)}^{62}$ & {} \\
   \hline
  $\mathcal{O}_{12,(1,1,3)}^{63}$ & {} \\
   \hline
  $\mathcal{O}_{12,(1,1,3)}^{64}$ & {} \\
   \hline
\end{longtable}
\end{center}

  \subsection{\texorpdfstring{$\Delta B=1, \Delta L=-3$}{Delta B=1,Delta L=-3}}

We have one-derivative operators,
   \begin{multicols}{3}
 \begin{spacing}{1.5}
 \noindent
 \begin{math}
\mathcal{O}_{12, \, (1,-3,1)}^{1}\equiv D\bar{H}\bar{H}ddd\bar{L}\bar{L}\bar{e}\,,\\
\mathcal{O}_{12, \, (1,-3,1)}^{2}\equiv D\bar{H}\bar{H}ddQ\bar{L}\bar{L}\bar{L}\,,\\
\mathcal{O}_{12, \, (1,-3,1)}^{3}\equiv D\bar{H}\bar{H}ddd\bar{e}\bar{e}\bar{\nu}\,,\\
\mathcal{O}_{12, \, (1,-3,1)}^{4}\equiv D\bar{H}\bar{H}ddu\bar{L}\bar{L}\bar{\nu}\,,\\
\mathcal{O}_{12, \, (1,-3,1)}^{5}\equiv DH\bar{H}ddd\bar{L}\bar{L}\bar{\nu}\,,\\
\mathcal{O}_{12, \, (1,-3,1)}^{6}\equiv D\bar{H}\bar{H}ddQ\bar{L}\bar{e}\bar{\nu}\,,\\
\mathcal{O}_{12, \, (1,-3,1)}^{7}\equiv D\bar{H}\bar{H}dQQ\bar{L}\bar{L}\bar{\nu}\,,\\
\mathcal{O}_{12, \, (1,-3,1)}^{8}\equiv D\bar{H}\bar{H}ddu\bar{e}\bar{\nu}\bar{\nu}\,,\\
\mathcal{O}_{12, \, (1,-3,1)}^{9}\equiv DH\bar{H}ddd\bar{e}\bar{\nu}\bar{\nu}\,,\\
\mathcal{O}_{12, \, (1,-3,1)}^{10}\equiv DH\bar{H}ddQ\bar{L}\bar{\nu}\bar{\nu}\,,\\
\mathcal{O}_{12, \, (1,-3,1)}^{11}\equiv D\bar{H}\bar{H}duQ\bar{L}\bar{\nu}\bar{\nu}\,,\\
\mathcal{O}_{12, \, (1,-3,1)}^{12}\equiv D\bar{H}\bar{H}dQQ\bar{e}\bar{\nu}\bar{\nu}\,,\\
\mathcal{O}_{12, \, (1,-3,1)}^{13}\equiv D\bar{H}\bar{H}QQQ\bar{L}\bar{\nu}\bar{\nu}\,,\\
\mathcal{O}_{12, \, (1,-3,1)}^{14}\equiv D\bar{H}\bar{H}duu\bar{\nu}\bar{\nu}\bar{\nu}\,,\\
\mathcal{O}_{12, \, (1,-3,1)}^{15}\equiv DH\bar{H}dQQ\bar{\nu}\bar{\nu}\bar{\nu}\,,\\
\mathcal{O}_{12, \, (1,-3,1)}^{16}\equiv DH\bar{H}ddu\bar{\nu}\bar{\nu}\bar{\nu}\,,\\
\mathcal{O}_{12, \, (1,-3,1)}^{17}\equiv DHHddd\bar{\nu}\bar{\nu}\bar{\nu}\,,\\
\mathcal{O}_{12, \, (1,-3,1)}^{18}\equiv D\bar{H}\bar{H}uQQ\bar{\nu}\bar{\nu}\bar{\nu}\,,
\label{eq:1Ddequal12basisB1L-3}
 \end{math}
 \end{spacing}
 \end{multicols}
\noindent
and two-derivative operators,
 \begin{multicols}{3}
 \begin{spacing}{1.5}
 \noindent
 \begin{math}
\mathcal{O}_{12, \,(1 , -3 ,2)}^{1} \equiv DD\bar{H}ddd\bar{L}\bar{L}\bar{L} \,, \\
 \mathcal{O}_{12, \,(1 , -3 ,2)}^{2} \equiv DD\bar{H}ddd\bar{e}\bar{L}\bar{\nu} \,, \\
 \mathcal{O}_{12, \,(1 , -3 ,2)}^{3} \equiv DD\bar{H}ddQ\bar{L}\bar{L}\bar{\nu} \,, \\
 \mathcal{O}_{12, \,(1 , -3 ,2)}^{4} \equiv DD\bar{H}ddu\bar{L}\bar{\nu}\bar{\nu} \,, \\
 \mathcal{O}_{12, \,(1 , -3 ,2)}^{5} \equiv DDHddd\bar{L}\bar{\nu}\bar{\nu} \,, \\
 \mathcal{O}_{12, \,(1 , -3 ,2)}^{6} \equiv DD\bar{H}ddQ\bar{e}\bar{\nu}\bar{\nu} \,, \\
 \mathcal{O}_{12, \,(1 , -3 ,2)}^{7} \equiv DD\bar{H}dQQ\bar{L}\bar{\nu}\bar{\nu} \,, \\
 \mathcal{O}_{12, \,(1 , -3 ,2)}^{8} \equiv DDHddQ\bar{\nu}\bar{\nu}\bar{\nu} \,, \\
 \mathcal{O}_{12, \,(1 , -3 ,2)}^{9} \equiv DD\bar{H}duQ\bar{\nu}\bar{\nu}\bar{\nu} \,, \\
 \mathcal{O}_{12, \,(1 , -3 ,2)}^{10} \equiv DD\bar{H}QQQ\bar{\nu}\bar{\nu}\bar{\nu} \, .
 \label{eq:2Ddequal12basisB1L-3}
 \end{math}
 \end{spacing}
 \end{multicols}
\noindent
All three-derivative operators at mass dimension $d=12$ can be obtained from the dimension-10 one-derivative basis in Eq.~\eqref{eq:1Ddequal10basisB1L-3}, according to $\mathcal{O}_{12,(1,-3,3)}^{j} \equiv D^2\mathcal{O}_{10,(1,-3,1)}^{j},$ $j\in\{1\isep 5\}.$


\pagebreak

 \bibliographystyle{JHEP}
 \bibliography{biblio.bib}

\end{document}